\documentstyle[amssymb,12pt,epsfig,a4]{book}

\renewcommand{\baselinestretch}{1.3}

\newfont{\abg}{cmsl12}    
\def\bfB{\mbox{{\abg {\bf B}}}}
\def\gtrue{\mbox{{\abg {\bf g}}}} 
\def\bfg{\gtrue}
\def\bfh{\mbox{{\abg {\bf h}}}}
\newcommand{\pl}{\label}

\newcommand{\captionstyle}{\small}

\begin{document}
\pagestyle{myheadings}

\def\Di{\displaystyle}
\def\nn{\nonumber \\}
\def\sr{\stackrel}
\def\be{\begin{equation}}
\def\ee{\end{equation}}
\def\ba{\begin{eqnarray}}
\def\ea{\end{eqnarray}}
\def\re{(\ref }
\def\rz#1 {(\ref{#1}) }   \def\ry#1 {(\ref{#1})}
\def\el#1 {\pl{#1}\end{equation}}

\let\a=\alpha \let\b=\beta \let\g=\gamma \let\d=\delta
\let\e=\varepsilon \let\ep=\epsilon \let\z=\zeta \let\h=\eta \let\th=\theta
\let\dh=\vartheta \let\k=\kappa \let\l=\lambda \let\m=\mu
\let\n=\nu \let\x=\xi \let\p=\pi \let\r=\rho \let\s=\sigma
\let\t=\tau \let\o=\omega \let\c=\chi \let\ps=\psi
\let\ph=\varphi 
\let\Ps=\Psi
\let\O=\Omega \let\S=\Sigma \let\P=\Pi
\let\Th=\Theta \let\L=\Lambda \let \G=\Gamma \let\D=\Delta
\def\q{\quad} \def\qq{\qquad}
\def\w{\wedge}
\def\2{{1\over2}} \def\4{{1\over4}}
\def\5{\bar} \def\6{\partial}

\def\({\left(} \def\){\right)} \def\<{\langle} \def\>{\rangle}
\def\lb{\left\{} \def\rb{\right\}}
\def\[{\lbrack} \def\]{\rbrack}
\let\lra=\leftrightarrow \let\LRA=\Leftrightarrow
\def\ul{\underline}
\def\wt{\widetilde}
\let\Ra=\Rightarrow \let\ra=\rightarrow
\let\la=\leftarrow \let\La=\Leftarrow

\let\mytilde=\widetilde 

\let\ap=\approx \let\eq=\equiv \let\hc=\dagger
\let\ti=\tilde \let\bl=\biggl \let\br=\biggr
\def\CL{{\cal L}} \def\CX{{\cal X}} \def\CA{{\cal A}} \def\CE{{\cal E}}
\def\CF{{\cal F}} \def\CD{{\cal D}} \def\rd{\rm d}
\def\rD{\rm D} \def\CH{\cal H} \def\CT{{\cal T}} \def\CM{{\cal M}}
\def\CI{{\cal I}} \newcommand{\dR}{\mbox{{\sl I \hspace{-0.8em} R}}}
\newcommand{\dN}{\mbox{{\sl I \hspace{-0.8em} N}}}
\def\CP{{\cal P}} \def\CS{{\cal S}} \def\C{{\cal C}}
\def\CR{{\cal R}}
\def\CV{{\cal V}}

\def\bfPs{{\bf{\Ps}}}
\def\Dsl{\not{\!\! D}}
\def\dsl{\not{\! \partial}}
\def\tDsl{\not{\!\! \widetilde{D}}}
\def\bfe{{\bf{e}}}
\def\dlr{\stackrel{\lra}{\6}}
\def\dl{\overleftarrow{\6}}
\def\dr{\overrightarrow{\6}}
 \setcounter{footnote}{0}

\def\0{{\underline{0}}}
\def\1{{\underline{1}}}
\newcommand{\done}{\mbox{{\sl {\bf {\sl 1}} \hspace{-1em} I}}}
\def\CC{{\cal C}}
\def\7{\mbox{$\2$}}
\def\U{{\bf{U}}}\def\V{{\bf{V}}}\def\Z{{\bf{Z}}}
\def\I{\re}\def\II{\re}
\def\sgn{\mbox{sgn}}

\def\smovs{\hbox{$\{\hbox{sector-moves}\}$}}
\def\osmovs{\hbox{$\{\hbox{sector-moves}\}^\uparrow$}}
\def\oG{\hbox{${\cal G}_+^\uparrow$}}
\def\bpmovs{\hbox{$\{\hbox{bifurcation point moves}\}$}}


\def\R{R}

\def\Rboost{\hbox{${\R}^{\rm (boost)}$}}
\def\Zflip{\hbox{${{\Z}_2}^{\rm (flip)}$}}
\def\ZPT{\hbox{${{\Z}_2}^{\rm (PT)}$}}

\def\attn{{\Large $\bigtriangledown
    \mbox{\scriptsize\hspace{-.276cm}!}$\hspace{.1cm}}}
\newcommand{\sckai}{s_i\hspace*{-1.4ex}\raisebox{1.3ex}{\scriptsize
    $k$} }
\newcommand{\sczai}{s_i\hspace*{-1.3ex}\raisebox{1.3ex}{\scriptsize
    $0$} }
\newcommand{\sclai}{s_i\hspace*{-1.3ex}\raisebox{1.3ex}{\scriptsize
    $l$}\,\,\,\, }
\newcommand{\sckak}{s_k\hspace*{-1.7ex}\raisebox{1.3ex}{\scriptsize
    $k$}\,\, }
\newcommand{\scoaf}{s_4\hspace*{-1.3ex}\raisebox{1.3ex}{\scriptsize
    $1$} }
\newcommand{\scoat}{s_2\hspace*{-1.3ex}\raisebox{1.3ex}{\scriptsize
    $1$} }
\newcommand{\scoaz}{s_0\hspace*{-1.3ex}\raisebox{1.3ex}{\scriptsize
    $1$} }
\newcommand{\sckakmo}{s_{k-1}\hspace*{-3.9ex}\raisebox{1.3ex}{\scriptsize
    $k$}\hspace{2.8ex}} \newcommand{\sclakinv}{s^{\smash
    -1}_i\hspace*{-2.7ex}\raisebox{1.3ex}{\scriptsize
    $l$}\hspace{2.2ex}} \newcommand{\sciakinv}{s^{\smash
    -1}_k\hspace*{-2.9ex}\raisebox{1.3ex}{\scriptsize
    $i$}\hspace{2.2ex} }

\newcount\nummer \nummer=0

\newcommand{\ins}{\rfloor}
\newcommand{\dZ}{\mbox{$\Bbb Z$}}   


\def\nullextr{{\bf II},20--22}
\def\eomRTa{{\bf I},30}
\def\OIIhchart{{\bf II},3}
\def\CasXsquX3{{\bf I},33,43}
\def\Kruskalref{{\bf II},33}
\def\saddleextr{{\bf II},29}
\def\saddleextrlength{{\bf II},30}
\def\flipref{{\bf II},17}
\def\JTref{{\bf II},11}
\def\KVref{{\bf II},13}
\def\JTetcref{{\bf II},11--13}
\def\deSitterref{{\bf II},10}
\def\Modelref{{\bf II},4}
\def\CcritKV{{\bf II},37}


\thispagestyle{empty}
\begin{center}
\vspace*{2cm}
{
\huge \bf
Gravity\\
in Two Spacetime Dimensions\\
}
\vspace{4cm}
{ \large
Der Fakult\"at f\"ur Mathematik, Informatik und Naturwissenschaften \\[4mm]
der Rheinisch--Westf\"alischen Technischen Hochschule Aachen \\[4mm]
vorgelegte Habilitationsschrift zur Erlangung der venia legendi\\[4mm]
}
\vspace{4cm}
{
{von }\\[5mm]
{\large Dipl.--Ing. Dr. Thomas Strobl }\\[5mm]
{aus }\\[5mm]
{\large Wien}\\[5mm]
}
\vfill
\end{center}
\newpage \thispagestyle{empty} \cleardoublepage

\tableofcontents
\chapter{Introduction} There is no doubt that, at the fundamental
level, theoretical physics is in the need of reconciling General
Relativity and quantum theory. This becomes obvious already from
Einstein's equations \be G_{\m\n} = \frac{8 \pi G}{c^2} T_{\m\n} \, .
\pl{Einsteineqs} \ee Here $G_{\m\n}$ is the Einstein tensor,
determined solely in terms of the metric $g$, which governs the
`gravitational force', and $T_{\m\n}$ is the energy momentum tensor of
matter fields. We know that matter fields are governed by a quantum
field theory, the standard model of strong and electroweak
interactions. Equations \re{Einsteineqs}), however, contain purely {\em
  classical}, commuting quantities. On the other hand, Einstein's
equations show that, in the presence of matter, $T_{\m\n} \neq 0$, the
metric $g$ {\em can not\/} be strictly Minkowskian (since then
$G_{\m\n}\neq 0$); it can be so at best to some approximation. A
Minkowski background metric is used, however, in standard quantum
field theory of elementary particle physics.

Certainly, no current experiments in our laboratories come only close
to explicitly display the above contradiction experimentally. Still,
from the point of view of a unified and mathematically consistent
theory of all interactions, a new theory is needed.

There is a variety of (at least apparently) different contemporary
approaches to improve on this situation: It ranges from noncommutative
geometry \cite{noncomm}, over twistors \cite{Penrosetwistor}, over the
attempt to quantize Einstein gravity nonperturbatively in a canonical
fashion \cite{Ashtekarbuchundloopreview}, to the advent of string
theory \cite{Polchinski}, to name just the most prominent directions.
The at the moment probably most promising of these attempts is the
string approach. But also string theory is surely not formulated in a
fully satisfactory and self-consistent way yet.

Black holes appear to provide an important link between gravity and
quantum theory. First, near a black hole (and in particular near the
singularity) the gravitational field cannot be ignored or approximated
sufficiently well by a flat, Minkowskian metric. Second, the discovery
of Hawking radiation \cite{Hawkingrad} demonstrated that quantum
effects cannot be ignored for black holes; rather they lead to a
(slow, but still nonvanishing) evaporation of the classically
indissoluble objects. Moreover, theoretical considerations allow to
establish a striking analogy between the laws of thermodynamics and
the laws governing black holes in {\em classical\/} general relativity
\cite{BHlaws}. Further Gedankenexperiments \cite{Bekenstein} show that
the second law of thermodynamics should be altered to include an
additive contribution to the entropy, assigned to the black hole.
Already on dimensional grounds, the Planck length $l_P = \sqrt{\hbar
  G/c^3}$ enters the generalized second law of thermodynamics
postulated in this way. It is felt by many contemporary theoreticians
that black holes will be the gate to some fundamentally new picture of
our world, bringing together seemingly separate subjects such as
gravity, quantum theory, and thermodynamics.

As a side-remark let me mention that by now certainly black holes are
no more purely theoretical objects. There is already overwhelming
experimental evidence for the existence of black holes in the center
of many galaxies, including ours (cf., e.g.,
\cite{Muenchnerodersonstwernoch}). Also black holes have been found as
partners in binary systems \cite{Doppelsternsysteme}.

It is one of the famous results of General Relativity, called the
no--hair theorem \cite{nohair}, that within Einstein--Maxwell theory 
all asymptotically flat, stationary black holes are determined
uniquely by specification of their mass $m$, their charge $q$, and
their angular momentum $J$. This scenario extends to some other matter
couplings such as to the case of additional scalar fields (with convex
potential). Still, the `no--hair conjecture' does not hold in full
generality: counter examples have been found recently (cf references
in \cite{nohair}) such as, e.g., in the Einstein--Yang--Mills theory
with gauge group $SU(2)$ (note that these solutions are unstable,
however).  Thus, (in those cases where the no--hair theorem or
conjecture holds) the space of (stationary) black hole solutions
(modulo gauge transformations) is {\em finite}\/ dimensional only.

The formulation of gravity in two spacetime dimensions allows to
isolate precisely these most interesting objects of General
Relativity: Upon definition of appropriate two--dimensional
gravitational actions, cf.\ chapters 2 and 3 below, the space of
their classical solutions does not only {\em contain}\/
(two--dimensional) black holes, it consists {\em only\/} of such
spacetimes, except possibly for naked singularity solutions and a
vacuum ground state. Many of the considerations performed for
their four--dimensional analogs, moreover, like the prediction of
Hawking radiation as well as the analogy to thermodynamics, may be
transferred also into the two--dimensional setting. Finally, also
here the original theory is invariant with respect to
diffeomorphisms (the group of coordinate transformations), such
that a transition to the quantum theory is plagued by similar
conceptual difficulties as in four spacetime dimensions (like the
`problem of time' \cite{Isham} to name just one of them).

During about the last decade, two--dimensional gravity models thus
turned out as an interesting and promising theoretical laboratory to
test ideas on how to approach quantization of gravity. Hundreds of
works appeared within this field. Still, on many of the main,
fundamental questions, even in this simplified context, no final
conclusion or consensus has been reached by now.

The present work focuses on the {\em foundation\/} of such
investigations of the two--dimensional world of gravity. It is felt
that a good and sound understanding of the classical aspects, which,
as we will show, are also far from trivial, provides a necessary basis
for tackling issues of quantum gravity and quantum black holes.
Emphasis has been put on exactly solvable models in this context.

Only at the very end of this report we dare a glance into the 2d
quantum world. In particular, we will apply {\em one\/} particular
quantization scheme to pure gravity Yang--Mills systems, comparing the
final result with the preceding classical considerations. We did not
attempt to also include a decent account of all developments on the
quantization of 2d gravity models. The thermodynamics and  Hawking
evaporation of such models, which fill large portions of the 2d gravity
activity, are not discussed in this report (except for some remarks). 


The organization of the present survey is as follows: In chapter 2 we
provide some of the basic geometrical ingredients and then construct
and motivate a wide class of two--dimensional models of (pure)
gravity.  In the subsequent chapter this class of models is even
enlarged further, while additional motivation is found, e.g., in
regarding invariant sectors of four--dimensional gravity as
effectively two--dimensional theories.  The addition of various types
of matter fields is discussed in this chapter, too.  In chapter 4, a
$\s$--model is presented, which allows for a unifying and elegant
treatment of the whole class of gravity models discussed before
(including Yang--Mills fields, but preferably in the absence of
further matter). This formulation proves a powerful tool for the
analysis on the classical as well as on the quantum level in
subsequent chapters.

Chapters 5 and 6 contain a careful discussion of various aspects of
the classical solutions. In particular, for the gravity Yang--Mills
models a complete classification of all possible global solutions is
presented. The resulting spacetimes are in part highly nontrivial,
including multi black holes, twists of the causal structure, and
spacetimes of almost arbitrarily complicated (two--dimensional)
topology.

The final chapter turns to the quantization of these models. Here the
main part focuses on a canonical, Hamiltonian approach of
quantization.  The physical quantum states are found and compared to
the space of classical solutions.  Open issues, like the construction
of an inner product between physical states, are addressed. Topics as
approaches to Hawking radiation 
of 2d black holes are touched on only briefly. 

\vspace{2mm}

The present work would not have been possible without, in particular,
two people, who collaborated with me over several years within the
subject: Peter Schaller and Thomas Kl\"osch. The joint discovery of
Poisson $\s$--models with P.\ Schaller \cite{PSM1}, on the one
hand, and the exhaustive investigation of global classical features of
spacetimes performed in \cite{IbisIII+}, on the other hand, need a
word of mention here. I am grateful to these two people for permitting
me to adapt some of this material for the present survey.


%
%
\chapter{2d geometry and gravitational actions \pl{Chap2dgeom}}
In this chapter we introduce and motivate 2d actions of
$f(R)$--type gravity theories and extensions with nontrivial
torsion. The actions discussed here will be purely geometrical,
i.e.\ they are functionals of geometrical quantities only.
Eventual dilaton--like fields enter as auxiliary fields only and,
in contrast to the models discussed in the subsequent chapter,
they are not necessary for the formulation of the action. The
relation of the actions in terms of purely geometrical quantities
and their reformulation in terms of auxiliary fields is made
clear.

The present chapter provides the  notation and conventions used
within this work. It also contains some relations peculiar to two
dimensions (some further 2d peculiarities may be found in section
\ref{Seccarter} below).

In the Introduction we already provided some motivation for
studying two--dimensional gravity theories. Some additional, in
part complementary reasons may be found in sections \ref{Secwhy}
and \ref{Sectopological}, where we also address the question of
why the consideration of a whole {\em class\/} of gravity models
is of interest, too.

\section{From four to lower dimensions \pl{Seclower}}
One of the appealing features of Einstein's theory of gravitation is
that the `mediator of the gravitational force' (thinking in terms of
the old, Newtonian language) is an intrinsically geometrical object
characterizing spacetime itself, namely its own metric $g$.
Ultimately, this is a consequence of the principle of equivalence, the
conceptual basis of Einstein's theory of general relativity. The
Einstein--Hilbert action, governing the dynamics of the metric $g$, is
the simplest conceivable geometrical object yielding a coordinate
independent local action:\footnote{Within the context of determining
  field equations of a given action, we will freely drop boundary
  contributions within this report. Thus we also did not include the
  standard boundary term in the action below.}  \be \frac{1}{16 \pi G}
\int d^4 x \sqrt{-\mbox{det}\, g} \, R \, ,
\label{Einstein} \ee
where $R$ is the (intrinsic) curvature scalar of the (metrical and
torsion--free) connection, determined uniquely by $g$.

Generically, when studying toy models of a physical theory by
reducing the dimension $D$ of spacetime, there are two possible
routes: The first one is a Kaluza--Klein procedure, where, in its
simplest form, one assumes the basic fields of the theory, which
here is the metric, to be `independent' of $n$ coordinate
directions; then the action in $D$ dimensions {\em induces}\/ an
action in $D-n$ dimensions.  The other standard (and simpler)
route is to replace the $D$--dimensional action by its
lower--dimensional analog.

While the latter approach truly yields a {\em toy}\/ model, which
will have similarities but also pronounced differences due to the
generic simplification encountered when reducing the dimension of
spacetime, the former approach captures a sector of the {\em
  physical}\/ $D$-- respectively four--dimensional theory. Models and
actions resulting from a Kaluza--Klein approach are thus of
particular interest. Such models will be discussed, among others,
in the following chapter (cf in particular the beginning of the
next section as well as section \ref{Secmore}). Here we first turn
to the second approach.

While almost any 4d action has a meaningful 2d analog, the action
\re{Einstein}) does not (at least when one requires the functional
to yield some nontrivial field equations). Variation of the
$D$--dimensional analog of \re{Einstein}), $\int d^D x
\sqrt{-\mbox{det} \, g} \, R$, with respect to the metric, yields
the Einstein tensor \be G_{\m\n} \equiv R_{\m\n} - \2 g_{\m\n} R
\, . \label{G} \ee In the absence of additional matter fields the
Einstein equations are thus tantamount to requiring this tensor to
vanish.

In two spacetime dimensions ($D=2$), however, the Einstein tensor
\re{G}) vanishes {\em identically}. This may be seen, e.g., as
follows: The general symmetries $R_{\a\b\g\d}=R_{[\a\b][\g\d]}$ of
the curvature tensor, the brackets indicating antisymmetrization,
allow one to conclude \be R_{\a\b\g\d} = - \frac{R}{2} \e_{\a\b}
\e_{\g\d} \, , \pl{R} \ee in two dimensions. Here $\e_{\a\b}$ are
the components of the $\e$--{\em tensor}, related to the
(numerical) $\e$--symbol $\e(\a\b)$ by means of $\e_{\a\b} =
\sqrt{-\mbox{det} \, g} \, \e(\a\b)$, where we will stick to the
convention $\e(01) = + 1$ throughout this work.  Using $\e_{\a\b}
\e_{\g\d} = g_{\a\d}g_{\b\g}- g_{\a\g}g_{\b\d}$, we then indeed
obtain $R_{\a\b}\equiv R^\g{}_{\a \g \b} = \2 R g_{\a\b} \LRA
G_{\a\b} = 0$ as a 2d identity, as we wanted to establish.

There is a simple alternative to see that the Einstein--Hilbert term does
not yield any field equations in two spacetime dimensions:
$\sqrt{-g} \, R$ is a total divergence in $D=2$.  (Here, as well
as in what follows, we use the abbreviation $\sqrt{-\mbox{det} \,
g}=:\sqrt{-g}$.) This fact will become obvious in Einstein--Cartan
variables, to be introduced further below.

Let us remark, however, that \be \int_M d^2 x \sqrt{-g} R \pl{td}
\ee is sometimes used as a 2d action within, e.g., the Euclidean path
integral, yielding a topological invariant of the manifold $M$, 
namely its Euler characteristic. 

The lowest dimension where an action of the form \re{Einstein})
remains to make sense (also classically) is $D=3$. But even in
three spacetime dimensions the Einstein--Hilbert term does not
yield {\em propagating}\/ gravitational modes.  While in four
spacetime dimensions the vacuum Einstein equations $R_{\m\n}=0$
allow for nontrivial gravitational excitations (like gravitational
waves), in $D=3$ they imply that the spacetime is {\em flat}:
$R_{\m\n\r\s}=0$.

Before returning to the question of a geometrical action in two
dimensions, we briefly dwell on the consequences of equation \re{R}) for the
geometry of any 2d spacetime. As is obvious from this relation, in
$D=2$ the spacetime has to be flat already when the Ricci {\em
  scalar}\/ $R$ vanishes.  Moreover, all (algebraic) curvature invariants are
determined completely by means of $R$.

This does {\em not}\/ mean, however, that the geometry (i.e.\ the
geometrical content of the spacetime modulo coordinate
transformations) is completely fixed by means of the function $R$
already. We provide an explicit example for this: With an
appropriate choice of coordinates $x^0,x^1$ any 2d metric may be
brought into the local form (cf section \ref{Secattain}) \be ds^2
= 2 dx^0dx^1 + k(x^0,x^1) (dx^1)^2 \, . \pl{LC} \ee In terms of
this metric the curvature scalar is $R= {\partial_0}^2 k$. Thus,
for the particular choice $k=h_0(x^0)+ h_1(x^1)$, the function
$h_1$ drops out of the curvature (this applies also to the full
tensor \re{R}), since det$g = -1$ for a metric of the form
\re{LC})). However, while a metric with $h_1 \equiv 0$ evidently
has the Killing vector field (generator of an isometry) $\6_1$, it
may be shown by explicit calculation (but cf also below) that a
metric with $h_1'\neq 0$ has {\em no}\/ Killing field, provided
only that $h_0'''\neq 0$.\footnote{If, on the other hand, $h_0'''=
0$, the function $h_1(x^1)$ can be {\em always}\/ eliminated by a
change of coordinates. After all, $g$ then is the metric of a
spacetime of constant curvature and spaces of a fixed constant
curvature are unique up to diffeomorphisms.} (Here and in what
follows a prime denotes differentiation with respect to the
argument of the respective function.) As the existence of a
Killing field is a geometrical feature of a spacetime, independent
of the choice of coordinates representing a metric, the above
family of spacetimes contain at least two geometrically distinct
ones, despite the fact that all of these spacetimes have the same
curvature scalar $R$.

\section{Purely geometrical 2d actions \pl{Secpurely}}
The simplest, purely geometrical candidate 2d gravitational action
is of the form \be \int d^2 x \sqrt{-g} (\a R^2 -2 \L) \, ,
\pl{R2} \ee where $\a$ is a coupling constant and $-2\L$ a
cosmological constant. Note that while in four spacetime
dimensions there are several independent quadratic curvature
invariants, here all of them are proportional to just $R^2$. As in
any dimension, an action of the form \re{R2}) yields field
equations which are {\em fourth}\/ order differential equations
for the metric $g$ ($R$ containing two derivatives). Introducing
an auxiliary field $\phi$, the number of derivatives may be
reduced to two: One easily establishes the equivalence \be \2 \int
d^2x \sqrt{-g} \, R^2 \sim \int d^2x \sqrt{-g}\, [\phi R - \2
\phi^2] \, , \pl{R2dil} \ee which holds on the classical as well
as on the quantum level (Gaussian integral), since the field
equations obtained from variation with respect to $\phi$ are
purely algebraic in $\phi$.

More generally we may consider the purely geometrical action \be L
= \int_M d^2x \sqrt{-g} f(R) \, , \pl{fR} \ee where $f$ is some
arbitrary (sufficiently well-behaved) function of the curvature
scalar. Such actions were proposed, and in part studied, by
Schmidt \cite{Schmidt91} (cf also references
therein; for a later work cf \cite{SolofR}).\footnote{After 
completion of most parts of the present
work a more recent account of Schmidt on $f(R)$--theories and
their classical solutions was published \cite{Schmidtneu}, which
in part parallels some of the discussion to follow.} 

Note that even in two dimensions this is {\em by no means}\/ the
most general conceivable diffeomorphism invariant local action.
Only when supplementing $R$ with $R_{(;a;b \cdots ;d)}$, where the
round brackets indicate symmetrization, and allowing all possible
contractions between any of these terms, the most general
diffeomorphism invariant local action functional of a 2d metric
$g$ can be constructed (cf \cite{Iyer}). Such 
gravitational actions will not be considered in the present work
(and the author is also not familiar with literature where such
possibilities were considered in any detail).

To illustrate the potential relevance of such terms from the
geometrical point of view, however, we briefly return to the
example of two nonisometrical spacetimes \re{LC}) defined by
$k=h_0(x^0)$ with $h_0'''\neq 0$ on the one hand and $k=h_0(x^0)+
h_1(x^1)$, $h_1'\neq 0$,  on the other hand. As discussed above,
here the invariant $R$ (and thus any purely algebraic invariant
constructed from the Riemann tensor!)  is not `fine' enough to
distinguish between the two spacetimes. However, using e.g.\
$(\nabla R)^2 \equiv R_{;a}R^{;a} = - k [(\6_0)^3 k]^2$ in
addition to $R$ it {\em is\/} possible to separate the two
spacetimes (coordinate--independently): A Killing vector field has
to annihilate all (algebraic and nonalgebraic) curvature
invariants simultaneously; $\partial_1$, however, annihilates
$(\nabla R)^2$ only for the first of the two metrics considered.

It is also possible to reformulate the action \re{fR}) by means of
an auxiliary field $\phi$ (at least locally, cf the remarks
below). Consider an action of the form \be L= \int_M d^2x
\sqrt{-g} [\phi R - V(\phi)] \, , \pl{Proto} \ee where $V$ is
again some sufficiently well-behaved function. The variation of
\re{Proto}) with respect to $\phi$ is algebraic in $\phi$ and thus
also here $\phi$ may be eliminated by implementing the solution
back into \re{Proto}) (provided only $V'' \neq 0$). This will
yield an action of the form \re{fR}).

We remark on this occasion that in subsequent chapters we will
refer to the auxiliary field $\phi$ as the {\em dilaton}\/ field;
note that in the present context it has been introduced as a
purely {\em
  auxiliary}\/ field within the framework of an entirely geometrical
action (that is, a diffeomorphism invariant, local action functional of
the metric only).

Assuming that $f$ and $V$ are (at least) $C^2$--functions and that
their second derivatives are nonzero, any $f$ gives rise to a $V$
and vice versa. If $V$ is linear (as is the case in the so--called
Jackiw--Teitelboim model \cite{JTmodel} discussed in the
subsequent chapter, cf equation \re{JT})), clearly $\phi$ cannot
be eliminated by means of its field equations and thus this choice
of $V$ cannot give rise to an action of the type \re{fR}). This is
reflected also in the relation $f''(V(\phi))=1/V''(\phi)$, which
may be derived provided $V'' \neq 0$ and which is seen to become
singular precisely when $V''$ becomes zero (again primes denote
differentiation with respect to the argument of the respective
function). It is interesting to observe vice versa that the choice
$f=R$, which was found above to yield no field equations, --- or,
more generally, the choice $f : = a + b R$, $a,b \in \dR$, which
is singular for $a\neq 0$ ---, also corresponds to a singular
limit of $V$.

Note that in {\em general}\/ the above bijection between $f$ and
$V$ works only within some neighborhood of a given value of
$\phi$. The bijection is global, iff one restricts to the class of
convex functions ($R^2$--gravity providing the prototype of
these). However, we might be interested, in a first step, in {\em
local}\/ solutions to the field equations of a model \re{fR}) with
some nonconvex function $f$. If the zeros of $f''$ are isolated,
the above method may still be applicable patchwise: For all
classical solutions of the model \re{fR}) where $R$ does not {\em
constantly}\/ take the value of a critical point of $f'$,
smoothness of the solution guarantees that any point on spacetime
where the value of $R$ is noncritical, has a neighborhood such
that within this neighborhood the above mentioned bijection works.
Thus there then exists a model \re{Proto}) with {\em
  identical}\/ local solutions.\footnote{The space of possible values
  of $R$ respectively $\phi$ may be considered as a target space.
  Smoothness of solutions then always allows to transfer
  considerations which are local on the spacetime to considerations
  local on the target space and vice versa.  This point of view may
  become more transparent when discussing $\s$--models explicitly in
  the sequel of this work.} In this way much more general models
\re{fR}) allow a discussion by means of the alternative
description \re{Proto}). 
The global information possibly lost within such a
transition has then to be restored in a second step. E.g.\ there
could be additional solutions to the field equations \re{fR}) with
a constant critical value of $R$; the possibility for such
additional de Sitter solutions has to be checked.  Otherwise only
the local solutions with values of $\phi$ below and above critical
points (or lines) have to be patched together. With these
cautionary remarks taken into account (which, in principle, may be
adapted also to the quantum regime), it is then possible to trade
in a Lagrangian \re{fR}) against a Lagrangian \re{Proto}) --- and
vice versa!\footnote{Excluded are now only those
  (in part singular) cases where $f''$ and/or $V''$ vanish {\em
    identically}\/ within some range.}

\section{Einstein--Cartan variables}
For later purposes we also need a Einstein--Cartan or Palatini 
formulation of the above gravity theories. We first set the
notation and conventions: Let us denote spacetime indices by Greek
letters from the middle of the alphabet, taking values zero and
one, $\m,\n,\ldots \, \in \{ 0,1\}$, and Lorentz indices (indices
referring to an orthonormal frame) from the beginning of the Latin
alphabet, with either $a,b,\ldots \in
\{\underline{0},\underline{1}\}$ or, when using a null basis,
$a,b,\ldots \in \{+,-\}$. (Indices from the beginning of the Greek
alphabet, such as those used in equation \re{R}), denote indices
in a general basis --- except in the context of 2d supergravity
theories, where they will label spinor indices.)

Our metric $g$ has signature $(+,-)$. In terms of the {\em
  vielbein}\/ or {\em zweibein}, we then have \be g=e^\0 e^\0 - e^\1
e^\1 = 2 e^+ e^- \, , \pl{metric} \ee where here the product between
the oneforms is understood as {\em symmetrized}\/ tensor product. In
\re{metric}) we made use of the convention $e^\pm := (e^\0 \pm e^\1
)/\sqrt{2}$ for introducing a light cone basis; in terms of this
basis, raising and lowering indices is accomplished simply by
replacing a lower $+$ ($-$) by an upper $-$ ($+$) and vice versa ($e^+
= e_-$ etc.).

The sign ambiguity in the metric induced volume form, i.e.\ in the
$\e$--tensor, will be fixed by choosing \be \e = e^\0 e^\1 =e^-e^+
\pl{epsilon} \, . \ee {\em Here}, and in any other product of
forms except for those yielding a metric, the product is understood as
the {\em antisymmetrized}\/ tensor product, i.e.\ as the wedge
product: $e^\0 e^\1 \equiv e^\0 \wedge e^\1$. By means of the standard
decomposition of forms, $\e = \2 \e_{ab} e^a e^b$, we conclude from
\re{epsilon}) that $\e_{\0\1}=\e^{+-}=+1$.

Beside the zweibein $e^a \equiv e^a_\m dx^\m$, the Einstein--Cartan
or spin connection $\o^a{}_b\equiv \o^a{}_{b\m}dx^\m$ is of
importance (here $x^\m$ are coordinates in some arbitrary chart of
the two--dimensional spacetime $M$.) Metricity of the connection
translates into antisymmetry with respect to its two Lorentz
indices: $\o_{ab} + \o_{ba} =0$. This, in turn, implies that the
spin connection is a $SO(D-1,1)$ Lie algebra valued
oneform. In $D=2$ the Lorentz group is just one--dimensional and
there is just one independent connection one--form $\o$: $\o_{ab}
= \e_{ab} \, \o$. Under a local Lorentz transformation, i.e.\ a
change of the orthonormal basis, $e^\pm \to \exp(\pm \a) \,
e^\pm$, where $\a$ is an arbitrary (smooth) function on $M$, $\o
\to \o + d\a$, which is the standard gauge transformation of an
abelian gauge field. The curvature of spacetime is characterized
by the Lie algebra valued twoform $\O^a{}_b \equiv \2
R^a{}_{b\m\n} dx^\m dx^\n$, which is nothing but the curvature of
the gauge field $\o^a{}_b$. Since in $D=2$ this gauge field is
abelian, the nonlinear terms in the connection do not contribute
and one simply finds $\2 R^a{}_{b\m\n} dx^\m dx^\n = \e^a{}_b \,
d\o$.  Combining this equation with equation \re{R}), we infer \be
R\e = -2 d\o \, . \pl{2dR} \ee

Using the Hodge duality operation $\ast$, defined (in the present
context of Lo\-ren\-tzian signature) by means of the relations $\ast 1
= \e$, $\ast \e = -1$, and $\ast  e^a = \e^a{}_b e^b \LRA \ast
e^\pm = \pm e^\pm$, equation \re{2dR}) is seen to be equivalent to
$R=2\ast d\o$.

The variables $e^a$ and $\o$ form a description of gravity
equivalent to the metrical one only when, in addition to the local
Lorentz symmetry, they are also subject to a constraint. Beside
curvature there is the geometrical notion of torsion, describing
the (internal) `twist' of the 2d surface. In Einstein--Cartan
formulation, it is associated to the Lorentz vector valued torsion
two--form \be De^a \equiv de^a + \e^a{}_b \o e^b \, .\pl{torsion}
\ee The standard connection used in Einstein's theory of gravity
(as well as in all the actions discussed so far) is torsion--free:
$De^a \equiv 0$. It is a well--known fact that the latter equation
(together with metricity) determines the spin connection
$\o^a{}_b$ (and thus also the curvature) uniquely in terms of the
{\em vielbein}. In two dimensions this relation may be brought
into a particularly simple form: \be \o = (\ast de^a)\, e_a \, .
\pl{torsionzero} \ee (Since both the curvature as well as the
metric $g$ are invariant with respect to local Lorentz
transformations, the curvature is determined implicitly by means
of $g$ alone also when using the Einstein--Cartan formalism.)

We are now also in the position to show that in two dimensions
$\sqrt{-g} R$ is a total divergence. Indeed this is {\em
evident}\/ now in view of $\e = \sqrt{-g} \, d^2x$ (where $d^2 x
\equiv dx^0 dx^1$) and equation \re{2dR}). (Note, however, that
here $\o$ is understood as an implicit functional of the metric on
behalf of equation \re{torsionzero}); it does not seem to be
possible to express $\sqrt{-g}R$ explicitly as a covariant total
differential of $g$ itself.)

It is a noteworthy feature of the 4d Einstein-Hilbert
action that, when expressed in terms of the {\em
vielbein}\/ and spin connection as independent
variables, the torsion zero condition follows
automatically upon variation of the action with respect to
$e^a$ (provided the theory is not coupled to spinors,
however!). This does {\em not}\/ hold in the case of the
2d actions discussed in the present chapter. We thus may
use the Einstein--Cartan formulation {\em either}\/ viewing
$\o$ as an abbreviation for the right--hand side of
equation \re{torsionzero}) {\em or}\/  implementing the
torsion zero condition by means of extra Lagrange
multiplier fields, making sure that introduction of the
latter does not alter the theory, certainly. We will
follow the second route mainly.

In this approach the $R^2$ action \re{R2}) becomes (for $\a :=
-(1/4)$): \be \int (\ast d\o) d\o -2 \L \e + X_a De^a \, ,
\pl{R2form} \ee with $\e$ given in equation \re{epsilon}), $De^a$
in equation \re{torsion}), and $X^a$ being the Lagrange
multipliers mentioned above. Variation with respect to $\o$ yields
an equation that may be solved uniquely for $X^a$ (this equation
is algebraic in $X^a$, but, however, no more in $\o$). So $X^a$
and $\o$ may be eliminated simultaneously within the action
\re{R2form}), with $\o$ in the first term being given by equation
\re{torsionzero}) --- and thus indeed no additional degrees of
freedom are introduced in this formulation as compared to
\re{R2}). Note, however, that since the equations for $X^a$ are
not purely algebraic in $\o$, the equivalence is guaranteed only
on the classical level; on the quantum level it needs to be
checked separately (e.g.\ by means of a careful path integral
argument) --- or, one uses one of the two formulations \re{R2})
and \re{R2form}) to {\em define}\/ the quantum theory for both of
them.

Similarly the general action \re{fR}) may be rewritten in terms of
Einstein--Cartan variables. In its (at least locally) equivalent form
\re{Proto}), one finds for $L$ (now interpreted as functional of $e^a,
\o,\phi$, and $X^a$): \be L = -2 \int \phi d\o + \2 V(\phi) \e + X_a
De^a \, . \pl{Protoform} \ee Note that irrespective of the choice of
$V$ (resp.\ $f$ in \re{fR})), this action is of first order form,
i.e.\ (in the variables now used) its field equations are all at most first
order differential equations.

\section{Extension to theories with nontrivial torsion \pl{Sectorsion}}
When working with Einstein--Cartan variables it may appear unnatural
to impose the zero torsion condition by hand. In particular, using
vielbein and spin connection as independent variables in the
Einstein--Cartan formulation of pure $D=4$ Einstein gravity, torsion
zero {\em follows\/} as one of the (vacuum) Einstein field equations.
Thus, also in the 2d setting, we may be interested in a purely
geometrical, diffeomorphism and Lorentz invariant local action
constructed (merely) in terms of the {\em independent}\/ one--forms
$e^a$ and $\o$. The most general action of this type which yields
second order differential equations of motion is of the form \be
L^{KV} = \int [-{1 \over 4} d\o \w \ast d\o - {1\over 2\a} De^a \w
\ast De_a + {\L \over \a^2} \e] \, , \pl{KV} \ee and was proposed by
Katanaev and Volovich \cite{KV1}.

More generally, we may consider the higher order theories
 \be L^{grav}
= \int_M d^2x \sqrt{-g} F(\t^2,R) \, , \pl{Faction} \ee where
$\t^a \equiv \ast De^a$, $\t^2 \equiv \t^a\t_a$, and $F$ is some
(reasonable) function. This is the most general geometrical action
constructed in terms of {\em algebraic}\/ curvature and torsion
invariants. We remind the reader that in \re{Faction}) $\o$ is
considered as variable {\em independent}\/ of the zweibein. In
contrast, in equation \re{fR}), when formulated in Einstein--Cartan
variables, $\o$ is {\em either}\/ to be understood as an
abbreviation for the right--hand side of \re{torsionzero}) {\em
or}\/ torsion zero has to be implemented by {\em additional}\/
Lagrange multiplier fields $X^a$. Thus, a model defined by
\re{Faction}) with a function $F$ independent of $\t^2$ does in
general {\em not\/} coincide with a model of the form \re{fR}),
but constitutes another, {\em different}\/ model (where the
torsion is not forced to vanish).

Suppose first that $F$ depends nontrivially on $\t^2$. Then also
here it is possible to reexpress the action in first order form by
introducing auxiliary fields.  Indeed, at least on a local and
classical level, an action \re{Faction}) (provided it is
sufficiently generic, cf the discussion below) can be cast into
the form: \be L^{grav} = \int_M \,\, X_a De^a + \phi d\o +
W(X^2,\phi) \, \varepsilon \,\, , \pl{grav} \ee where $X^2 \equiv
X_aX^a \equiv 2X^+X^-$ and $W$ is some two--argument function of
the indicated variables. Nontrivial dependence of $F$ on $\t^2$
will be seen to require also $W$ to depend nontrivially on $X^2$.
On the other hand, we now realize that a function $W$ depending
merely on $\phi$ allows to cover the torsion--free case
\re{Protoform}) resp.\ \re{fR}), too. Thus when working with
\re{grav}) we simultaneously may cover the torsion--free case
(with $f'' \not \equiv 0$, but anything else is of no interest
anyway, cf our previous discussion) as well as \re{Faction}) with
nontrivial explicit torsion dependence --- and further conditions,
which we will specify now.

To decide under what precise conditions models of the type
\re{Faction}) may be described equivalently by means of \re{grav})
and vice versa, the following relation between the functions $F$
and $W$ is helpful: They are nothing but Legendre transforms of
one another! Indeed, using equation \re{2dR}) and $De^a = - \t^a
\varepsilon$, and just naively equating the two actions
(multiplying one of them by the irrelevant factor $-1$), we find
$F(\t^2,R)= X_a\t^a + \2 \phi R - W(X^2,\phi)$. The process of
eliminating $X^a$ and $\2 \phi$ from the right--hand side by means
of the equations of motion, moreover, coincides precisely with the
standard Legendre procedure. Thus (global/local) equivalence holds
provided the {\em Hessian}\/ of $F$ and $W$ (interpreted as
function of {\em three}\/ arguments $X^a$ and $\phi/2$!)  is
(globally/locally) nondegenerate.

The KV model \re{KV}) provides an example where the equivalence to
\re{grav}) clearly holds on a classical and quantum level
globally. It results upon the quadratic choice \be W^{KV}= -
\frac{\a}{2}  X^2 - \phi^2 +\frac{\L}{\a^2} \, .  \el WKV

However, also more general cases, where the Hessians vanish only
on lower dimensional submanifolds, may be treated interchangably,
if the treatment takes care of the global relation, too. Note that
this holds also on the quantum level as the equations for $X^a$
and $\phi$ are purely algebraic. As the treatment in the
subsequent chapters of this work will be careful to not miss any
global effects, such `weakly degenerate' cases are covered, too.

The above restriction on the Hessians of $W$ and $F$ as a condition
for the bijection between respective models confirms also the
difference noted between a model of the type \re{fR}) on the one hand
and \re{Faction}) with an $F$ that is torsion independent on the other
hand. The latter yields a strictly degenerate Hessian of $F$.  A model
\re{fR}), in contrast, is described by a strictly degenerate $W$
(namely one that is independent of $X^2$).

Adapting the argumentation, the present perspective also
illuminates the relation of \re{fR}) to \re{Protoform}) (i.e.\ to
$W$ with trivial $X^2$ dependence): In this case $X^a$ ceases to
be connected to $\t^a$ via a Legendre transformation ($\t^a \equiv
0$!), but it acts as Lagrange multiplier to enforce
\re{torsionzero}).  After the (nonalgebraic) elimination of $\o$
and $X^a$,
we are back to the problem of relating \re{fR}) to \re{Proto}). The
one--argument functions $f$ and $V$ are now recognized as Legendre
transforms of one another.

On the other extreme side not covered in the bijection of $F$ and
$W$ is a function $F$ that is independent of $\t^2$. Such models
result from \re{grav}) with an $X^2$--independent $W$ when {\em
dropping}\/ the term $X_a De^a$ altogether: \be L = \int_M \,\,
\phi \, d\o + W(\phi) \, \varepsilon \,\, , \pl{gravtriv} \ee In
view of the simplicity of \re{gravtriv}), it is plausible that
such theories are quite different from the theories \re{grav}). As
we will show in section \ref{Sectriv} below, the field equations
of \re{gravtriv}) lead to constant curvature spaces with {\em
unrestricted}\/ metric $g$. Such theories appear somewhat `ill'
therefore.

Models of the general type \re{Faction}) (restricted to the case of
nontrivial torsion dependence of $F$) were considered recently in
\cite{Hehl} and, in its equivalent form \re{grav}), in
\cite{PartI} (but cf also \cite{Ikedaold2,Ikeda,PSMold}).  Particular
cases thereof (in particular the KV--model, equation \re{KV})) have been
considered also before \cite{KVall,Kummertorsion,SolodukhinKV,
HehlKV,Ikedaold1,Kummerlectures}.

\section{Why so many different Lagrangians \pl{Secwhy}}
We thus arrived at a wealth of acceptable two--dimensional purely
geometrical actions, which, within the 2d setting, may be taken to
play the role of the Einstein--Hilbert term \re{Einstein}) of 4d
gravity. The question may arise why we do not restrict our
attention to just one of the possible choices for a 2d gravity
action. The reason for considering this large amount of different
actions is at least three--fold:

First, as discussed above, in two spacetime dimensions the naive
analog  \re{td}) of \re{Einstein}) is meaningless (but cf also the
remarks following equation \re{td})). We thus used the
`equivalence principle' (in the sense that we restricted 
attention to a diffeomorphism invariant, geometrical, and local
functional) as the main guiding principle for the construction of
the action governing the dynamics of the gravitational variables.
This opened a wide territory for possible Lagrangians. Restricting
to the use of merely algebraic invariants, leads to the class of
theories \re{fR}) or, if allowing for nonzero torsion, to
\re{Faction}). (In $D=2$ the class of `geometric--algebraic'
(local) actions is thus parameterized by just one function of one
respectively two arguments.)

In four spacetime dimensions certainly there are even much more
possibilities for the construction of covariant actions. However,
there we also have the experiment and this singles out the
simplest among the possible actions (cf, however,  possible
quantum corrections mentioned below).

Second, in $D=4$ gravity relatively little is known about higher order
actions. The discussion of the theory resulting from the
Einstein--Hilbert term alone is difficult enough already: compared
to the size of the total space of all possible classical solutions
(even if restricted to physically reasonable ones), only very few
exact solutions are known and under control. Moreover, the
quantization of the Einstein--Hilbert term alone, although subject
of intensive endeavours (cf, e.g., the loop approach
\cite{Ashtekarbuchundloopreview}), is still beyond our abilities.
In $D=2$, on the other hand, it {\em is}\/ technically possible to
obtain several exact results (and we will present a lot of them in
the sequel) about {\em all}\/ of the actions discussed in this
(and the beginning of the following) chapter.

Third, there are several concepts in gravity which are not very
sensitive to the particular choice of an action. One such an example
is the existence of black hole (or otherwise singular) solutions; only
very particular Lagrangians will exclude such solutions. On the other
hand, there are some results proven specifically for the 4d Einstein
theory \re{Einstein}). Quantum contributions are expected to lead to
corrections to this Lagrangian and it is not clear if those results
still apply. Thus there is some interest in the generality of some
statements, applying to a whole class of Lagrangians.

Here we have in mind e.g.\ the area theorem \cite{Hawkingareathm}
proven for 4d Einstein gravity. It states that, under some
(reasonable) conditions, the total area of the horizon of a black
hole  cannot decrease. This statement covers complicated dynamical
processes such as collisions of black holes, which are otherwise
hardly tractable even by approximations on the computer. The area
theorem is just one of the cornerstones of the notable analogy
\cite{BHlaws} between general relativity and thermodynamics
mentioned in the Introduction, the horizon area in the former
theory playing the role of entropy in the latter one. It is not at
all clear, if, and in what way, such an analogy may be maintained
when the Einstein--Hilbert term \re{Einstein}) receives small,
higher order corrections as predicted e.g.\ by string theory (cf
also section \ref{Secstring} below). It turns out that, phrased
appropriately general, most of the analogy between gravity and
thermodynamics may be found even in the context of the
two--dimensional theories considered above and in the subsequent
chapter (cf, e.g., \cite{Frolov}). Certainly, it is no more the
area of a black hole horizon which can play the role of entropy;
after all, in two spacetime dimensions the horizon consists of
just a point. However, there is a precisely formulated proposal of
Wald \cite{Wald,Iyer} for what the entropy of a black hole should be
proportional to, which is devised for an arbitrary diffeomorphism
invariant action in any spacetime dimension. In the specific case
of the Einstein--Hilbert term this quantity reduces to the area of
the horizon, but already higher order corrections destroy this
geometrical interpretation. The formalism of Wald also yields a
nonvanishing entropy for 2d black holes (coinciding with the
quantities found (or postulated) in \cite{Frolov}). It is then
e.g.\ an interesting question to see, in which subclass of 2d
models this quantity also obeys a second law (generalizing the
above area theorem). 

Last but not least, although different members of the class of
models \re{fR}) and \re{Faction}) have many features in common,
there are also some qualitatively new properties arising in some
of them which are absent in one or the other particular case. Such
properties turn up in particular when dealing with global aspects
of the theories under discussion. To sharpen ones techniques in
dealing with gravity theories in general, we do not want to limit
the class of 2d models too much. After all, the problems arising
e.g.\ in the quantization of 4d gravity may be expected to be
still much more complicated than the most involved 2d cases.

In the subsequent chapter, we will thus even enlarge the number of
models to consider. The models of the present chapter will then be
referred to as `geometrical' ones, as all of them may be
formulated entirely in terms of geometrical variables only.

\section{Why consider low--dimensional, topological models \pl{Sectopological}}
Although not obvious at first sight, {\em none}\/ of the
gravitational actions discussed so far, as well as also {\em
none\/} of the (matterless) generalized dilaton theories discussed
in the subsequent chapter below, turn out to allow for propagating
modes of the gravitational variables. The space of solutions to
the respective field equations of a fixed model {\em
  modulo}\/ its local symmetries, called the moduli space, is {\em
  finite}--dimensional only.  Such actions are ususally called {\em
  topological}.\footnote{The notion of a {\em topological field
    theory}\/ is not completely rigid (similar to the notion of {\em
    solitons}, e.g.), and the finite dimensionality of the moduli
  space is just one of its criteria. A working definition for
  such theories is provided, e.g., in \cite{Blau}. In view of their
  Poisson $\s$--formulation discussed below, it seems justified,
  however, to truly call the 2d gravitational actions topological.}

This is, of course, {\em not}\/ to be confused with an action that
is a topological invariant itself, like equation \re{td}). A
topological invariant used as an action, yields {\em no}\/ field
equations at all, and, due to the resulting unlimited local
symmetry in the absence of further fields, the respective moduli
space is {\em zero\/}--dimensional.

One of the reasons for calling a given field theory {\em
  topological}\/ is the fact that several features of its moduli space
are sensitive to the topology of the underlying spacetime only or,
better, the underlying base manifold. An example for this is the
{\em dimension\/} of the moduli space. Still, this does {\em
not\/} mean, that, when viewed as gravity theory as done within
this work, its observables yield topological invariants of the
base manifold only. The base manifold is not equipped with a
rigid, fixed metric, certainly; rather, a given solution induces a
metric on it. A typical observable would then be something like
the total size of the (say compact) universe (for some
specifically fixed moment in time and measured by means of the
induced metric on that hypersurface). Different values of this
observable yield gauge inequivalent solutions, i.e.\ different
points in the moduli space.

We mentioned already in section \ref{Seclower} that Einstein
gravity in 2+1 dimensions (with or without cosmological term) is
topological, too. In fact, most of the 1+1 theories show a more
interesting spacetime structure than 2+1 Einstein gravity. In the
latter theory spacetime is either flat locally (vanishing
cosmological constant $\L$) or it has constant curvature ($\L \neq
0$). Due to the freedom in constructing nontrivial global
solutions it is still possible to find solutions in this 2+1
theory which show aspects of black holes \cite{BTZ} (defined by
the existence of regions causally disconnected to `null
infinity'). However, the (global) constancy of curvature in these
solutions is quite counterintuitive to what one naively would
regard a black hole (with curvature singularities etc.).

A simple choice of $f$ in the 2d theory \re{fR}) such as $f :=
|R|^{2/3}$, on the other hand, may be seen to yield 2d metrics which
are {\em identical}\/ to the `$(r,t)$'--part of the well--known
Schwarzschild spacetime. Other choices of $f$ reproduce the
Reissner--Nordstr\"om
 spacetime or even more involved spacetimes (cf section
\ref{Secmaxext} below for a detailed exposition). {}From this
perspective, taken together with the advantages resulting from the
existence of a whole class of tractable 2d actions (cf discussion
in the previous subsection), we find a reduction to two spacetime
dimensions more promising than one to three
dimensions.\footnote{There are, however,
  certainly also aspects of 2+1 gravity which are not found in a 2d
  theory. One of these is, e.g., the possibility for a black hole to
  carry a nonzero angular momentum.  Moreover, also in $D=3$ more
  involved gravity actions could be considered; to the best of my
  knowledge, this has not been pursued much, however. ---  For a recent
  book surveying the status of the art in (standard) 2+1 gravity the
  reader is referred to \cite{Carlipbook}.}

Despite the fact that the moduli space, i.e.\ the space of
physically inequivalent solutions, is finite dimensional, the pure
low--dimensional gravity models are {\em not}\/ just mechanical:
First, to start with one is given a field theory with the
(infinite dimensional) diffeomorphism group as (part of) the local
symmetries. The action of this group is markedly different from
that of a gauge group in a principle fiber bundle like in a
Yang--Mills theory (this is true for any dimension of the base
manifold), and it is this complication which is one of the main
reasons for the difficulty in quantizing gravity. Second, at least
for a generic choice of the `potentials' $W$ or $V$ (in \re{grav})
and \re{Proto}), resp.), the moduli space has quite a nontrivial
topology, including cusps and other relics of the nontrivial
structure of the gauge orbits. Although the gauge orbit structure
is certainly incomparably simpler than its counterpart in four
dimensions, it is still capable of posing technical problems {\em
not}\/ present in purely mechanical models.\footnote{Here we have
in mind, e.g., the appearance of winding
  numbers within some {\em region}\/ of the support of the physical
  wave functions in the quantum treatment of chapter \ref{Sectowards}.
  These winding
  numbers are a relic of the underlying field theory, despite
  topological, and pose a qualitatively new technical difficulty
  within defining an inner product. This point will be discussed 
  in some detail in section \ref{Secdirac} below.} Such topological
field theories may thus serve as toy models to further develop our
technical skills in what we call quantization of a diffeomorphism
invariant theory.

Furthermore, besides the clearly nonnegligible difficulties in
quantizing selfinteracting field theories, much of the
complication occurring upon the quantization of a gauge theory
(and in particular one like gravity) stems from the nontrivial
structure of the gauge orbits. Topological field theories do not
contain selfinteracting, propagating physical modes. They do,
however, show an in part quite nontrivial structure of the gauge
orbits, still leaving (global) physical degrees of freedom, to be
handled with care. They thus allow us to sharpen our understanding
and techniques in dealing with an essential part of the problem of
quantizing physical gauge theories.

If matter fields are added, moreover, in general one obtains
propagating modes. (An exception to this are Yang--Mills fields,
which also have no propagating modes in two dimensions, still
rendering the model `almost topological´ \cite{Blau}.) In the 2d
gravity--matter theory with propagating modes, it is suggestive
although not compelling to ascribe the propagating modes to the
matter fields. In the presence of an addditional scalar field $f$,
e.g., the function $f$ may also be used to fix part of the
diffeomorphism invariance (by using it as one of the local
coordinates); in this case, part of the metric $g$ carries
propagating modes of the total theory.

It is appropriate to remark on this occasion that the sector of 4d
Einstein gravity characterized by spacetime solutions with two
commuting Killing vector fields {\em is}\/ (effectively) a
two--dimensional gravity--matter theory {\em and}\/ it has
propagating modes (it contains e.g.\ plane parallel gravitational
waves). The action resulting from the appropriate (Kaluza Klein)
reduction is found to consist of a 2d gravitational part of
generalized dilaton form (cf section \ref{Secgendil} below;
qualitatively this part of the action is similar to the action
\re{Proto}) discussed above) and a `matter' part, consisting of a 
nonminimally coupled scalar field (cf, e.g.,
\cite{Schmidt98} as well as section \ref{Secmore} below). In this
case, both the metric {\em and}\/ the matter fields in the lower
dimension correspond to components of one and the same
four--dimensional metric.

A possibility of generating propagating modes in purely
geometrical 2d actions may be to consider Lagrangians constructed
from curvature invariants that are not restricted to be purely
algebraic (cf remarks in section \ref{Secpurely}). Such theories
were considered, e.g., in \cite{Schmidtrefsweg}, but not much
seems to be known about them. Furthermore, in addition to allowing
the connection to carry nontrivial torsion (as in section
\ref{Sectorsion}), one might also give up its metricity.

On the other hand, there does not seem to be urgent need for
considering such extensions. Indeed, besides the arguments given
already above in favor of discussing 2d gravity theories even if
their gravitational part does not contain propagating modes, there
is also one more, maybe most important argument: The phenomenon of
Hawking radiation (or, to be more careful, many aspects of it and,
in particular, its very existence) does {\em not}\/ seem to be
sensitive to the reduction in the dimension. Coupling a scalar
field to a (topological) 2d gravity action permits much the
same discussion (cf, e.g., \cite{Stromingerrev}) as the one
performed by Hawking in his famous paper \cite{Hawkingrad}.
Questions such as whether the black hole loses all its mass in the
process of Hawking radiation or whether it rather reaches some
final endstate of nonzero mass can be posed in two spacetime
dimensions as well --- and the chance of finding answers are much
higher than in the analogous four--dimensional problem. The 2d
investigation may show also that this question is not really
well--posed or that it has to be rephrased appropriately to really
make sense. 2d models may thus help to sharpen our conceptual
ideas of how to approach complicated problems in the interface
between black holes and quantum mechanics.

\section{Role of the dilaton within the present chapter}
Within this chapter the fields $\phi$ and $X^a$ have been
introduced merely as auxiliary fields. All the actions considered
thus far allowed a formulation in terms of purely geometrical
quantities. Moreover, except for \re{Protoform}), the fields
$\phi$ and $X^a$ always have a nice geometrical significance: They
are specific functions of curvature and torsion --- the form of
the function is determined by the choice of the Lagrangian and its
Legendre transformation. E.g., in the simplest case of a quadratic
Lagrangian, equation \re{KV}) above, they are just proportional to
$R$ and $\ast De^a$, respectively.

In the torsion--free case \re{Protoform}), the fields $X^a$ do not
have a similar geometrical significance (at least as far as I am
aware of)
--- while the dilaton $\phi$ still is a specific function of the
curvature $R$. In these models the introduction of the auxiliary
fields $X^a$ became necessary only when switching from the
metrical variable $g$ to the Einstein--Cartan variables, treating
$\o$ as a variable {\em independent}\/ of  $e^a$. (Using $\o$ as
an abbreviation for \re{torsionzero}), on the other hand, the
Lagrange multipliers $X^a$ are absent as well.)

In the context of the present chapter, matter Lagrangians will
necessarily be {\em independent}\/ of $\phi$ and $X^a$. This
changes in the subsequent section, where the field $\phi$ will get
another, more prominent interpretation (it is also then when it
receives its name `dilaton´). {}From the perspective of that chapter
it will appear natural to regard $g$ {\em and}\/ $\phi$ as the
basic gravitational variables of the theory. Correspondingly,
matter actions can then have $\phi$--dependent coupling
`constants´. As this just generalizes (and thus includes) all
matter couplings conceivable from the perspective of the present
chapter, the discussion of matter actions will be taken up in that
context.

We finally remark that the geometrical interpretation of $\phi$
found within this chapter may also help to acquaint the reader
with the existence of this field in most works on 2d gravity.
Spherically symmetric reduction of 4d Einstein gravity, considered
below, provides a prominent alternative perspective on the role of
$\phi$. Both point of views need extrapolation when discussing the
most general 2d dilaton theory below.

\chapter{Generalized dilaton theories and matter actions \pl{Chapgen}}
The first section of the present chapter deals with further
alternatives to the models found in the previous chapter. The most
general dilaton theory will be found as a generalization of
spherically reduced 4d gravity and the effective theory for
strings in a two--dimensional target space. Appropriate field
redefinitions (cf section \ref{Sectwo}) relate them to models
found in the previous chapter. In the subsequent two sections
standard matter fields are introduced and the field equations of
the resulting coupled matter--gravity theories are provided. In
section \ref{Secmore} effectively two--dimensional theories
resulting from other than spherical reduction of Einstein gravity
are mentioned. Generically these reductions of pure gravity
contain additional matter fields in the lower dimension, thus
fitting into the framework of theories discussed by then. Section
\ref{Secsugra}, finally, deals with supersymmetric extensions of
general dilaton theories.

\section{\sloppy Generalized dilaton action replacing the Einstein--Hilbert
  term \pl{Secgen}}
\subsection{Spherically reduced gravity \pl{Secspher}}
A natural approach to find a gravity action in two dimensions is
to dimensionally reduce the four-dimensional Einstein--Hilbert
action \re{Einstein}). In the present context the simplest
possibility is a spherical reduction. Up to choice of coordinates
any spherically symmetric four--dimensional line element
$(ds^2)_{(4)}$ takes the form \be (ds^2)_{(4)}=\bfg_{\mu
\nu}(x^\m) dx^\m dx^\n -\Phi^2(x^\m) \, (d \vartheta^2 +
sin^2\vartheta d \varphi^2),\quad \m, \n \in \{0,1\}\,\,.
\pl{ansatz} \ee Two out of four coordinates in the
four--dimensional spacetime have been adapted to the symmetry,
namely $\varphi$ and $\vartheta$, the other two coordinates $x^\m$
have been left unrestricted. The symmetry of the four--dimensional
line element requires that $\bfg$ and $\Phi$ depend only  on the
two coordinates $x^\m$. Although components of one and the same
four--dimensional metric, they thus may be interpreted as a metric
$\bfg$ and a scalar field $\Phi$ on a two--dimensional spacetime
with coordinates $x^\m$. The appearance of additional fields
beside the metric is a generic feature of a dimensional reduction.
Similarly Kaluza and Klein arrived at the coupled Einstein--Maxwell
theory in four dimensions by starting {}from the pure
Einstein--Hilbert term in a five dimensional spacetime \cite{KK}.

Implementing the ansatz \re{ansatz}) into the Einstein--Hilbert
action \re{Einstein}) and integrating over the angular variables
$\varphi$ and $\vartheta$, one obtains ($G:=1$) \cite{Hajetal}:
\be L^{spher}= \int_M d^2x \sqrt{-\bfg} \, \, \left[\frac{1}{4}
\Phi^2 R(\bfg) - \2  \bfg^{\m\n} \6_\m \Phi\6_\n \Phi -
\frac{1}{2}\right] . \el spher Here $R(\bfg)$ denotes the Ricci
scalar of the (torsion--free, metrical) connection of the
two-dimensional metric $\bfg$. Note the similarity of this
two--dimensional action with the action \re{Proto}) found in the
previous section. (The reason for using slightly different
symbols, namely $\bfg$ and $\Phi$ instead of $g$ and $\phi$, will
become apparent in section \ref{Sectwo} below.) The presence of
the kinetic term for the `dilaton' $\Phi$ may suggest that now the
2d theory might contain also propagating modes. However, the
residual diffeomorphism invariance in the two coordinates $x^\m$
is still large enough to eliminate these modes.  In fact, we know {}from
the four--dimensional theory that, up to diffeomorphisms, all
(simply connected) solutions of \re{spher}) are parameterized by
just one parameter, the Schwarzschild mass $m$.

To be precise, implementing an ansatz such as \re{ansatz}) into an
action may not commute with determining the respective field
equations. The consistency check works here, however, and the field
equations of the effective 2d action \re{spher}) coincide with the
four--dimensional Einstein equations restricted to spherical symmetry
by means of \re{ansatz}). We remark on this occasion that according to
its definition, $\Phi$ is restricted to be positive in \re{spher}).
Such a restriction on the dilaton may be avoided easily by using a
variable $\tilde \Phi$ instead with, e.g., $\Phi = \exp (\tilde \Phi)$
(the corresponding 2d action results {}from insertion of this equation
into \re{spher})).

Dynamically propagating modes are obtained in the spherical sector
of Einstein gravity by adding the action of a scalar field $f$ to
\re{Einstein}): \be \2 \int d^4 x \sqrt{- g^{(4)}} (\nabla^{(4)}
f)^2 \, .  \ee Implementation of \re{ansatz}) and requiring $f$ to
be spherically invariant (i.e.\ independent of $\vartheta$ and
$\varphi$), one obtains the effective 2d matter action \be \int_M
d^2 x \sqrt{-\bfg} (2 \pi \Phi^2) (\nabla f)^2 \, , \pl{spherf}
\ee with $(\nabla f)^2 \equiv \bfg^{\m\n} \6_mf \6_\n f$. Note
that, viewed as a two--dimensional field theory, the coupling of
the scalar field $f$ to the gravity action \re{spher}) is {\em
nonminimal}. The prefactor proportional to $\Phi^2$ is a
consequence of the  reduction of the volume element in the 4d
action. An additional cosmological constant term in the
Einstein--Hilbert part \re{Einstein}) similarly leads to the
addition of $\int d^2x \sqrt{-\bfg} \L \Phi^2$ to \re{spher}),
which is no more just a cosmological constant term {}from the point
of view of two--dimensional geometry.

We remark that the coupled model \re{spher}) and \re{spherf}) is
highly nontrivial and of substantial interest. Recent computer
simulations \cite{Chop} indicated e.g.\ that it shows some
kind of critical behavior: Only when the energy concentration of
the scalar field $f$ within some initially almost flat spacetime
surpasses some critical value, a black hole is formed during the
subsequent collapse. Below the threshold the spacetime remains
regular. Slightly above the threshold, the mass of the emerging
black hole follows a power law with a critical exponent of about
$0.37$. (Further literature on this phenomenon, called `Choptuik
effect', may be found in \cite{Chopviele}.)

On the semiclassical level, on the other hand, the model should show
Hawking radiation. Indeed, when requiring the scalar field in Hawking's
paper \cite{Hawkingrad} to be spherically symmetric, one obtains just
this two--dimensional model and Hawking radiation is still predicted
by his work.

These are just two of several motivations for the interest in the
effective two--dimensional theory \re{spher}) and \re{spherf}) on the
classical as well as on the quantum level. For the concrete {\em
  analysis}\/ of the model, however, it is no more important to
remember that the phenomena described by this model are phenomena
of four--dimensional physics (only for the final physical
interpretation it is of relevance certainly); the theory may be
regarded  as an inherently two--dimensional one on this level.

\subsection{String inspired dilaton gravity \pl{Secstring}}
We briefly change the scenery and turn to string theory. String theory
is probably the most promising current direction in our attempt to
unify all fundamental interactions, including gravity, into a
consistent quantum theory. We do not and can not recapitulate even the
basics of string theory here, referring for this to \cite{GSW} or the
more up--to--date book \cite{Pol}. However, since the name {\em
  dilaton gravity}\/ as well as much of the activity in
two--dimensional gravity theories over the recent decade have their
origin in a discovery \cite{dil,Wittdil} of a certain black hole
solution within the context of string theory, we will at least briefly
outline the relation. Moreover, some of the terminology used below
will be encountered again in the subsequent chapter on Poisson
$\s$--models and it may be useful to acquaint the reader already in
the present context with some of the notions used there. Finally,
regarding string theory as an at least potentially physical theory,
its relation to low--dimensional gravity theories provides further
motivation for studying the latter.

In the simplest scenario, a string is a one--dimensional, open or
closed object moving in spacetime. Its (lowest) excitations are
supposed to describe the (known) elementary particles. The geodesic
movement of a point particle follows {}from extremizing an action
functional defined by the metric induced length of the world line of
the particle. Similarly, the movement of a string is governed by an
action minimizing its (two--dimensional) world surface (usually called
the {\em worldsheet}). Denote by $z^m$, $m = 1,2$, local coordinates
on the worldsheet and by $X^\m$, $\m = 1,\ldots,D$, coordinates on the
$D$--dimensional spacetime.  The latter may well be curved, carrying a
metric $G_{\m\n}(X)$. The dynamical fields are the $X^\m(z)$.  The
world surface $X^\m(z^m)$ of a string provides a map from the
worldsheet into spacetime.  The spacetime is thus the (possibly
curved) {\em target space}\/ of the theory. Introducing an {\em
  auxiliary}\/ metric $\g$ on the two--dimensional worldsheet, the
action for minimal worldsheet surfaces may be put into the form
\cite{Polyakov}: \be L^{string} = \frac{1}{4\pi \a'} \int d^2 z
\sqrt{|\g|} \g^{mn} \6_m X^\m \6_n X^\n \, G_{\m\n}(X) \, ,
\pl{string} \ee where $\a'$, needed to make the action dimensionless,
is proportional to the square of the Planck length $\sqrt{\hbar G/c^3}$.
This is the action of a two--dimensional (in general) nonlinear
$\s$--model, where, however, the metric $\g$ is not fixed.

The inclusion of the world sheet metric $\g$ does not add any new
content to the theory (at least locally and classically).  Indeed the
action \re{string}) is invariant with respect to 2d diffeomorphisms as
well as Weyl rescalings $\g \to \O^2 \g$, $\O$ being an arbitrary
(nonvanishing and smooth) function on the worldsheet; the three local
symmetries allow to locally gauge away all components of $\g$ (putting
it e.g.\ into 2d Minkowski form).\footnote{This is no more possible on
  the global level, the space of gauge-inequivalent metrics forming
  Teichm\"uller space then.}

The quantization of \re{string}) is not possible in full generality.
It may be achieved only for exceptional choices of the {\em target}\/
space, such as, e.g., for flat Minkowski space. But also in the latter
case, attempts to quantize the theory are plagued by anomalies of the
classical Weyl--invariance of \re{string}); only in the critical
dimension $D$ they are absent, which, in the bosonic case, is $D=26$.

In the spectrum of the excitations of the closed string a massless spin two
particle is found, which is interpreted as the graviton (in
spacetime). It constitutes the perturbation around the Minkowski
target space one started with and the string may thus be understood
to generate its own background gravitational field $G_{\m\n}(X)$.

Finally, also the quantization in noncritical spacetime dimensions
$D$ has been achieved. It was  more or less required only that the
string is represented by a 2d conformal field theory with vanishing
central charge. For a
2d $\s$--model of the general type \re{string}) this leads to the requirement
that its $\beta$--function(s) (interpreting $G$ as an infinity of
coupling constants) has/have to vanish. Actually, consistency of the
previously given physical picture motivates  to incorporate also the other
massless excitations of the string as its background in the
action. In the case of the closed string these fields are given by a spin zero
particle, the dilaton $\Phi(X)$, and  a two--form gauge field
$B_{\m\n}(X)$. The  latter
may be contracted with $dX^\m \wedge dX^\n$ to provide a possible 2d
action to be added to the $\s$--model action \re{string}). For
dimensional reasons the action is proportional $(1/\a')$ again.

The dilaton, on the other hand, is argued to give rise to a
contribution \be (1/2\pi) \int d^2x \sqrt{|\g|} R(\g) \Phi(X(x)) \, .
\label{Rdil} \ee Here $R(\g)$ is the Ricci scalar of the worldsheet
metric $\g$.  Note that this contribution is no more Weyl invariant on
the classical level. However, the action \re{Rdil}) is already
dimensionless without the use of $\a'$, and its contribution to Weyl
transformations of $\g$ has to cancel only the one--loop contribution
of \re{string}) (and the gauge field part to the $\s$--model). Quantum
Weyl invariance, on the other hand, is guaranteed by vanishing
$\beta$--functions. For a $\s$--model of the above kind, and with $B$
set to zero for simplicity, vanishing $\b$--functions give the
following restrictions on the background fields in leading order of
$\a'$ \cite{GSW}: \ba R_{\m\n} - \nabla_\m\nabla_\n \Phi + O(\a') &=&
0 \pl{beta1}\, , \\ \frac{D-26}{3 \a'} + R + 4 (\nabla \Phi)^2 - 4
\Box \Phi + O(\a') &=& 0 \pl{beta2} \, , \ea where $\Box \equiv
G^{\m\n} \nabla_\m \nabla_\n$.  These equations are obtained in a
$D$--dimensional target spacetime. In the critical dimension ($D=26$)
and with $\Phi := 0$, the above equations are identified
with nothing but the vacuum Einstein equations for $G$!  (For $D \neq
26$ the equations require a nontrivial dilaton vacuum for consistency,
cf, e.g., \cite{Pol}.)  Taking into account also higher order
contributions, string theory thus predicts specific corrections of
order $\hbar$ to the Einstein equations, as was already mentioned
briefly in section \ref{Secwhy} above.

An action giving rise to the field equations \re{beta1}), \re{beta2}),
may, furthermore, be identified with the low energy effective action
of string theory. Note, however, that this action is an action for
fields on the {\em target space\/} (which, after all, is our spacetime), its
basic fields appear in the quantization of the string action only as
massless excitations.

In string theory, too, it was found worthwhile to explore the
spacetime dimension $D=2$ \cite{dil,Wittdil}. There then is no gauge
field $B_{\m\n}$ (its curvature vanishes identically in $D=2$ and its
contribution to \re{beta1}) and \re{beta2}) thus drops out
automatically), and, dropping again the tachyon present in bosonic
string theory, the $D=2$ field equations take the above form with
$(D-26)/3 = -8$. The effective action yielding the equations \re{beta1})
and \re{beta2}) to leading order in $\a'$ then takes the form \be
L^{dil}(G,\Phi) = \int_M d^2X \sqrt{-G} \, \exp (-2 \Phi(X)) \, \,
\left[R(G)-4 (\nabla \Phi)^2 +\L \right] \pl{dil0} \, , \ee with $\L =
(D-26)/6\a'$. It was argued by Witten \cite{Wittdil} that a $D=2$
theory with a target space geometry governed by precisely these
equations has a realization as an $SO(2,1)/U(1)$--WZW exact conformal
field theory.  Enthusiam arose when the existence of black hole
solutions very similar to the ($(r,t)$--part of the) Schwarzschild
metric was discovered \cite{dil,Wittdil}.

\vspace{2mm}

The model \re{dil0}) was analysed from the string perspective
thereafter, e.g., in \cite{Verlindeetal}. However, it soon
received attention also as a two--dimensional gravity model
itself, irrespective of its origin, for reasons which will
be made clear now.

We first rewrite the action \re{dil0})
identically in a notation more adapted to the previous
subsection: Replacing the symbol $G$ of the $D=2$ spacetime metric
by $\bfg$, as before, and, similarly, the coordinates $X^\m$,
$\m=0,1$, of (the target spacetime $M$) by $x^\m$, equation
\re{dil0}) reads \be L^{dil}(\bfg,\Phi) = \int_M d^2x
\sqrt{-\det \bfg} \, \exp (-2 \Phi) \, \, \left[R-4
  (\nabla \Phi)^2 +\L \right] \pl{dil1} \, .  \ee By means of the
field redefinition $\bar \Phi := 2 \sqrt{2} \exp(- \Phi), \, {\bar
  \Phi} >0$, this may be put into a form more reminiscent of the
spherically symmetric action \re{spher}): \be L^{dil}(g,\bar \Phi)
= \int_M d^2x \sqrt{-\bfg} \, \, \left[\frac{1}{8} {\bar \Phi}^2
  R(\bfg)-\2 \bfg^{\m\n} \6_\m \bar \Phi\6_\n \bar \Phi + \frac{1}{8}
  \bar \Phi^2 \L \right] \, . \el dil2

Regarding \re{dil1}) resp.\ \re{dil2}) as a two--dimensional model in
its own right (similar to \re{spher}) above, although with a very
different motivation), one may now try to couple scalar fields $f_i$ to this
action.  As the theory in question is inherently two--dimensional,
minimal coupling is most natural \cite{CGHS}: \ba
L^{CGHS}(\bfg,\Phi,f_i) & =& L^{dil}(\bfg,\Phi) + L^{mat}(f_i,\bfg) \,
\, ,
\pl{CGHS} \\
\\
L^{mat}(f_i,\bfg) &=&\frac{1}{2} \int_M d^2x \sqrt{- \bfg} \, \,
\sum_{i=1}^N \bfg^{\m\n}
\6_\m f_i \6_\n f_i \, \, .  \pl{mat} \ea

The above model, known as the CGHS model, certainly contains
propagating modes. Despite its similarities with the analytically
hardly accessible
coupled model of the preceding subsection, Callan, Giddings, Harvey,
and Strominger  managed to show in their by now seminal work \cite{CGHS}
that, at least on the classical level, the present model may be solved
{\em exactly}. The complete classical solution may be written in
explicit form (this result is recapitulated in section
\ref{Secmatsol}).  Furthermore, the CGHS--model (and
modifications of it resulting from adding counter terms) proved to be
an interesting laboratory for semiclassical considerations, devised to
possibly give answers to questions such as what the fate of a black
hole in the process of Hawking radiation could be
\cite{CGHS,Stromingerrev}.

In view of the exact classical solvability, solvability on the quantum
level was addressed soon as well \cite{Mikovic,KucharCGHS,
  JackiwCGHS}. The considerations in this direction do not seem to have
come to a final conclusion yet, however (cf corresponding remarks at
the end of chapter \ref{Sectowards}).

The above mentioned modifications of the CGHS model discussed in the
literature in the context of Hawking radiation (cf, e.g.,
\cite{Thorl,Stromingerrev}) consisted mainly in the replacement of some
of the dilaton dependent prefactors in \re{dil1}) or \re{dil2}). This
was motivated in part by the freedom in adding counter terms to the
action in a semiclassical treatment, and the modifications were
interpreted as of order $\hbar$.  In the following subsection a
general model will be provided, which, among others, comprises also
all these possibilities.


\subsection{Most general dilaton theory \pl{Secgendil}}
All the torsion--free gravitational 2d actions discussed so far, in
this as well as in the previous chapter, are particular cases of
the general action \be L^{gdil} (\bfg, \Phi)=\frac{1}{2} \int_M
d^2x \sqrt{-\bfg} \,\, \left[\U(\Phi) R(\bfg) - \V(\Phi)+ \Z(\Phi)
\bfg^{\m\n} \6_\m
  \Phi\6_\n \Phi \right] \, . \el gdil The $f(R)$--models of the
previous chapter are seen to result from an identically vanishing
$\Z(\Phi)$ and a linear function $\U$. Spherically reduced gravity
as well as the string inspired model are obtained by choices of
the three `potentials' $\U$, $\V$, and $\Z$ with nonvanishing
$\Z$, cf equations \re{spher}) and \re{dil1}) or \re{dil2});
adding counter terms to the latter model, moreover, the potentials
receive corrections in form of a power series in $\hbar$.

The action \re{gdil}), called generalized dilaton action, was
suggested in \cite{Banks} first. It is the most general
diffeomorphism invariant action yielding second order differential
equations for a 2d metric $\bfg$ and a scalar (dilaton) field
$\Phi$. $\U$, $\V$, and $\Z$ are three functions parameterizing
different actions.  For a particular model they are thus regarded
to be fixed (except possibly for changes under
re\-nor\-ma\-li\-za\-tion). Similar to the discussion of actions
\re{fR}), the potentials are assumed to be sufficiently regular.
More specific restrictions are required when necessary (cf, e.g.,
section \ref{Sectwo} below), but, in a `first approximation', they
may be regarded as arbitrary smooth functions (within some domain
of definition).

The model \re{gdil}) appears to be much more general than, e.g.,
\re{Proto}) or \re{fR}). However, as we will now show, by
appropriate transformation of field variables, the action can be
simplified always, e.g., to the form \re{Proto}), so that the true
`master action', allowing to describe all  relevant gravitational
actions, with or without torsion, is \re{grav}).

Relevant matter couplings are provided after this step in section
\ref{Secmat} below.

\subsection{Two possible routes to a simplified, equivalent model \pl{Sectwo}}
\subsubsection{Conformal rescalings of the metric \pl{Secconf}}
In the following we will restrict ourselves to the case that $\U$ has
an inverse function $\U^{-1}$ everywhere on its domain of definition.
This excludes extrema of the function $\U$. If $\U$ has extrema, the
transformations below may be used locally only. Then similar remarks
as those at the end of section \ref{Secpurely} apply here as well, i.e.\
one either has to restrict the ensuing analysis to a local level, or,
the transformations are used on an intermediary level only (and the
global information is restored when transforming back to the original
variables).

For simplicity, we will, furthermore, assume that $\U$, $\U^{-1}$,
$\V$, and $\Z$ are all $C^\infty$. (These conditions can be
relaxed easily.)

A conformal transformation of the Ricci scalar produces an additive
term with two derivatives on the conformal exponent (cf, e.g.,
\cite{Waldbuch})\footnote{But note the different sign convention for
  $R$ in \cite{Waldbuch}; the conventions for the curvature and the
  Ricci tensor chosen in this report, and fixed in section
  \ref{Seclower} above, coincide with those used, e.g., in
  \cite{MTW}.}.  Therefore it is near at hand to get rid of the
kinetic term for $\Phi$ in $I_{gdil}$, equation \re{gdil}), by a
$\Phi$-{\em dependent}\/ conformal transformation.  This was first
proposed by H.\ Verlinde \cite{Verlinde} for string inspired dilaton
gravity and then generalized to $I_{gdil}$ in \cite{BanksKunst}.  We
thus change variables from $\gtrue$ to an (in this context)
`auxiliary' metric $g$ defined by: \be g := \O^2(\Phi) \,\, \gtrue
\, , \quad \O(u) \equiv \exp \[-\int^u \frac{\Z(z)}{2\U'(z)} dz +
const. \] \, \pl{gneu} , \ee where the integration constant has been
displayed explicitly to keep track of the corresponding ambiguity.
Note that the integral exists on behalf of our assumptions above.

Now, after a partial integration, the action $I_{gdil}$ as a
functional of $g$ and $\Phi$ again has the form of equation
\re{gdil}), but with a potential $\Z$ that is identically zero,
while $\U$ remains unchanged and $\V$ is replaced by
$\V(\Phi)/\O^2(\Phi)$. Due to the resulting absence of the
$\Z$--term, we now may use \be \phi := \U(\Phi) \pl{fi} \ee
instead of $\Phi$ so as to also trivialize the potential $\U$
which becomes the identity map then. (Again this is possible due
to our assumptions on the potential $\U$.)

Thus, in the new field variables $g$ and $\phi$, the action
\re{gdil}) takes the form \be L^{gdil} (g, \phi)=
\frac{1}{2}\int_M d^2x \sqrt{-g} \,\, \left[\phi R(g) -
V(\phi)\right] \, , \el Proto2 where the function $V$ is defined
by $V(\U(z)) = \V(z)/\Omega^2(z)$ (different constants chosen in
the definition of $\O$ in \re{gneu}) are thus seen to rescale the
potential $V$).  The action \re{Proto2}) is --- up to an overall
factor of 1/2 --- identical to \re{Proto}). This explains the
change in notation for the 2d metric on $M$ between the previous
and the present chapter.  Note that in this context, the variables
$g$ and $\phi$ should not be regarded as the metric and dilaton on
the 2d spacetime.  So to say, `by chance' they may also have this
interpretation, {\em when}\/ starting with {\em
  another}\/ 2d model. Rather, they are coordinates in the space of
fields simplifying the Lagrangian, similar to the use of, e.g., normal
coordinates for a Lagrangian of coupled oscillators. The information
on the two out of originally three functions in \re{gdil}) is stored
in the transformation formulas between the old and the new variables.

There is no possibility to get rid also of the remaining potential
$V$ in \re{Proto2}) by any further field dependent conformal
transformation of the metric. However, in the context of the next
chapter, another (local) transformation of variables will allow to
also eliminate $V$. In that context the action and the resulting
field equations will have been trivialized. (Basically, completed
by the final step provided below in the context of $\s$--models,
these changes of variables yield a B\"acklund--like 
transformation  for the original model \re{gdil}) with its
originally nontrivial field equations equations \re{eomg1}) and
\re{eomPhi}) below.)

For the special case of string inspired dilaton theory \re{dil1})
(or \re{dil2})), on the other hand, the final step mentioned above
is not necessary. The transition to $g$ and $\phi$ is already
sufficient to ensure that $g$ becomes flat in this specific case!
Indeed, with an appropriate choice of $const$ in the definition of
$\O$, \re{dil1}) yields $\O^2(\Phi)=\U(\Phi)$, so that due to
$\V(\Phi) =-\L \U(\Phi)$, we find  $V(\phi)=-\L=const.$ Variation
of \re{Proto2}) with respect to $\phi$ now shows that the
auxiliary metric $g$ has to be flat.

By means of this transformation, the solvability of the CGHS model
becomes transparent as well. In fact, the action \re{mat}) is
conformally (or Weyl) invariant, it therefore takes the same form
with respect to $g$ as with respect to $\bfg$. Thus, up to a
choice of spacetime coordinates, also here $g$ takes standard
Minkowski form. The field equations resulting from the variation
with respect to the $f_i$ reduce to the ones of $N$ massless
scalar fields in Minkowski spacetime! One  is left only to realize
that due to the diffeomorphism invariance only one of the three
field equations $\delta L^{CGHS}/\delta g_{\m\n}(x) =0$ is
independent \cite{Sundermeyer}
 and that this one may be solved always for
$\phi$, which entered the action as Lagrange multiplier field.

Still, the representation of $L^{CGHS}$ in terms of $g$ and $f_i$
does not imply that the scalar fields $f_i$ and the original
metric $\bfg$ decouple completely. The situation may again be compared
with the
introduction of normal coordinates for coupled harmonic
oscillators. To trace the coupling explicitly, one notices that
the transition from $g$ to $\bfg$ involves $\phi$, which in turn
is coupled directly (via $g$) to  the scalar fields $f_i$.

Before we close this section, let us return to the case of spherically
symmetric 4d gravity \re{spher}). An appropriate choice of the integration
constant in \re{gneu}) yields $\O^2 = \Phi/2=\sqrt{\phi}$ and
$V(\phi)=1/2\sqrt{\phi}$. Thus, the space of solutions to \rz spher
will be reproduced from the simpler action \rz Proto2 with this
potential $V$, if, at the end, according to \ry gneu , $\bfg :=
g/\sqrt{\phi}$.

Some cautionary remark is in place at this point: In both
examples, i.e.\ in the string inspired as well as in the
spherically symmetric model, the conditions placed on the
potentials at the beginning of this section are fulfilled globally
(recall $\Phi > 0$ in the latter example). In both cases $\phi >0$
and $\phi \to 0$ will correspond to the place of the curvature
singularity. However, in terms of the `auxiliary' variables $\phi$
and $g$, only in the spherically symmetric case, where $V \to
\infty$ for $\phi \to 0$, this singularity is still visible. In
the string inspired model, on the other hand, $g$ was seen to be
(globally) flat, since the potential $V$ is constant in this case,
and $\phi$ will be perfectly well--behaved also for negative
values. A maximal extension of a local solution in terms of the
variables $g$ and $\phi$ thus will {\em
  differ}\/ from the maximal extension in terms of $\bfg$ and $\Phi$.
However, equivalence is obtained, when the restriction $\phi > 0$
is recalled in the former extension; it is {\em this\/}
restriction which transforms the global structure of 2d Minkowski
space into the one of a black hole.

Finally, for later purposes it will be worthwhile to also display
the transformation \re{gneu}) in terms of Einstein--Cartan
variables. We denote those of the auxiliary metric by $e^a$ and
$\o$ as before and those of $\bfg$ by ${\bf e}^a$ and $\varpi$ (so
that $\gtrue = 2{\bf e}^+{\bf e}^-$). Distributing the conformal
factor in equation \re{gneu}) equally on $\bfe^+$ and $\bfe^-$ and
using equation \re{torsionzero}), one easily obtains: \ba e^\pm
&:= &\O \, {\bf e}^\pm \pl{econf} \\ \o(e) &=& \varpi({\bf e}) -
\ast d(\ln \O) \, . \pl{oconf} \ea Here we displayed the
torsion--free connections; the connection on the left--hand side is
torsion--free with respect to $e^a$, while $\varpi$ on the
right--hand side is torsion--free with respect to ${\bf e}^a$. Note
that in the second term of equation \re{oconf}) it is irrelevant
if $\bfg$ or $g$ is used for the Hodge dual operation (this
applies to one--forms only).

Equation \re{oconf}) may be used to establish also $R(\bfg) = \O^2
[ R(g) - \Box \ln \O^2]$ with $\Box \equiv g^{\m\n} \nabla_\m
\nabla_\n$, which was used on an intermediary level to find the
conformal factor \re{gneu}) canceling the kinetic term in
\re{gdil}).

We finally remark for later use and for historical reasons that
the so--called Jackiw-Teitelboim model of 2d (anti)deSitter
gravity \cite{JTmodel} \be L^{JT}=-\frac12\int_M d^2x\sqrt{-\det
g} \; \phi \; (R-\L) \el JT is also contained as a particular
member of the family \re{Proto2}) (or \re{Proto}), but {\em not\/}
\re{fR})).

\subsubsection{Equivalence to theories with torsion}
In the first part of this subsection it was shown that by means of
a field dependent conformal transformation, accompanied by a
(target space like) reparameterization of the dilaton, the general
model \re{gdil}) may be mapped to its simplified version
\re{Proto2}). As emphasized above, however, the difference between
the original metric $\bfg$ and its auxiliary counterpart $g$ has
to be kept in mind strictly. Here we present another approach,
where actually no field variables need to be changed (except,
again, for a reparameterization of the dilaton) and the action is
still of the form \re{grav}). This point of view was advocated
first by \cite{Kummertorsion} (but cf also a footnote in
\cite{Kawai}) and, later, but apparently independently, by
\cite{Hehl}. 

Consider a theory \re{grav}) with a potential $W$ linear in
$X^2\equiv X^aX_a$ and integrate out the fields $X^a$. The
resulting action has the form \be L^{tor}=\int \o d\phi +
[G_1(\phi) + G_2(\phi) \t^2] \e \, \, , \pl{Ltor} \ee where, as
before, $\t^a = \ast De^a$ is the Hodge dual of the torsion
two--form, $\t^2 \equiv \t^a\t_a$, and $G_1$ and $G_2$ are
(basically) arbitrary functions of the dilaton.  (In such steps we
freely drop exact two--forms, i.e.\ `total divergences', as they
do not contribute to the field equations.)

In equation \re{Ltor}) the action $L^{tor}$ is regarded as a
functional of $e^a$ and $\o$. However, it is well-known, that
instead of the spin connection one may also use torsion
as independent variables beside the zweibein. Generalizing
equation \re{torsionzero}) to the case of nonvanishing torsion,
one obtains \be \o = (\ast de^a - \t^a) e_a :\equiv \o(e) - \t^a
e_a \, , \pl{omegatorsion} \ee where $\o(e)$ is used to denote the
(unique) torsion free part of the connection $\o$ (i.e.\ its
Riemannian part, while the second term \re{omegatorsion}) may be
referred to as its post--Riemannian part).

Next we observe that \re{Ltor}) is quadratic and purely algebraic
in $\t^a$. Integrating out torsion, we easily arrive at \be
L^{tor}[e^a] = \int \phi \, d \o(e) + G_1(\phi) \, \e +
\frac{1}{4G_2(\phi)} \, d\phi \ast d\phi \, , \ee which, as
indicated, is now a functional of the zweibein or metric only.
Introducing a new dilaton field $\Phi$ satisfying $\U(\Phi) =
-\phi$ for a chosen function $\U$, the first term takes the form
$\frac{1}{2}\int \U(\Phi) R(g) \, \sqrt{-g} \, d^2x$, where $R(g)$
is the curvature determined by $g=2e^+e^-$. With the
identification \ba &\bfg=g \quad \mbox{or} \quad {\bf e}^a = e^a\,
, \; \varpi({\bf
  e})=\o(e)& \nn &\Phi =\U^{-1} (-\phi) & \pl{equivtor}
\ea the resulting action is now seen to be precisely of the form
\re{gdil}), with $\V$, $\Z$ being determined uniquely in terms of
$G_1$, $G_2$, and $\U$.

We leave it to the reader to determine more precisely the
conditions for an equivalence and (local/global) bijection of the
two models in question, since the argumentation is  analogous to
that in previous sections. We remark here only that obviously
$G_2$  has to be nonzero to accommodate for a model with a kinetic
term $\Z \neq 0$ in the generalized dilaton theory \re{gdil}),
that is to say, the model, equivalent in the sense \re{equivtor}),
necessarily has a nontrivial torsion--dependence ($F_2 \neq 0$ in
equation \re{tor2}) below).

We thus have obtained an equivalence of \re{gdil}) to the purely
geometric model \be L^{tor} = \int \sqrt{-g} [F_1(R) + F_2(R)
\t^2] \pl{tor2} \, , \ee quadratic in torsion. Modulo possibly
global differences (and with some exceptional cases not covered by
the bijection), the two models have {\em identical}\/ solutions
for the metric $g=\bfg$. This is quite in contrast to the
preceding method \re{gneu}), where the difference between the
metric $g$ in \re{Proto2}) and the metric $\bfg$ in \re{gdil}) has
been repeatedly emphasized. On the other hand, although the
metric, and thus (the metrical part of) the spacetime, of the two
theories \re{tor2}) and \re{gdil}) are identical, the spacetime of
the former theory is equipped also with generically nonvanishing
torsion, while the spacetime \re{gdil}) carries the scalar dilaton
field $\Phi$, not present in the action \re{tor2}). (This is a
consequence of the step of integrating out the respective fields,
rendering an equivalent action for the {\em remaining\/} fields.)

In any case, using the transformation \re{gneu}) or the one above,
the study of the generalized dilaton theory may always be
transferred to a study of \re{grav}). The latter was found to also
describe all the geometrical models, with our without torsion,
presented in the previous chapter. Our further analysis of the
gravitational part of the theory in subsequent sections will thus
focus on the general action \re{grav}). At least in the generic
case, and with care concerning global issues, any result obtained
for that action, can then be transferred to the original
gravitational theory one might have started with instead.

\section{Matter actions \pl{Secmat}}
We now intend to present several matter actions which may
be added to the Lagrangian \re{gdil}) of generalized dilaton gravity.
The resulting field equations will then be provided in the subsequent
subsection. As far as possible, they will be formulated in terms of
the energy momentum tensor of the matter fields so as to allow simple
further generalizations.

The matter Lagrangians will be written in terms of the `original
variables' present in \re{gdil}). For a geometric, torsion--free
action \re{fR}) resp.\ \re{Proto}), on the other hand, the
transformations \re{gneu}) and \re{fi}) are trivial, and the bold
face variables $\bfg$, $\Phi$ may be identified with $g$, $\phi$
of chapter \ref{Chap2dgeom}. In that case, the matter actions will
be required to be independent of the dilaton, so that the Legendre
transformation from $\phi$ to $R$ is still possible. Similar
statements hold for theories with nontrivial torsion \re{Faction})
resp.\ \re{grav}).

Generalizing the standard actions for real massless scalar fields
$f$ and fermionic fields ${\bf{\Ps}}$ in curved spacetime
\cite{Birrel} by allowing dilaton dependent couplings
$\beta(\Phi)$ and $\gamma(\Phi)$, one has: \ba L^{scal}&=&\2 \int
d^2x \, \beta(\Phi) \, \sqrt{- \gtrue} \; \gtrue^{\m\n} \6_{\m}f
\6_{\n}f \, , \pl{scal} \\ L^{ferm}[{\bf e},{\bf{\Ps}}]&=&\2
\int_{M} d^2x \, \gamma(\Phi) \, \det ({\bf e}_\r^b) \, \left( i
{\overline{{\bf{\Ps}}}} \s^{a}{\bf
  e}_{a}^{\m } D_{\m } {\bf{\Ps}} + \mbox{herm.conj.}\right)  \, .
\pl{ferm1} \ea As usual, formulation of a fermion action {\em
  requires}\/ the use of Einstein--Cartan variables in a curved
spacetime, which for the metric $\bfg$ are denoted by ${\bf e}^a$ and
$\varpi$ again.

In the above, ${\bf e}_a^\m $, is the inverse of ${\bf e}_\m^a$,
the component-matrix of the zweibein ${\bf e}^\pm={\bf e}^\pm_\m
dx^\m$.  $D_\m$ denotes the covariant derivative: $D_\m = \6_\m +
\2 \varpi_\m \s^3$. In the absence of torsion, $\varpi_\m$ is
determined through ${\bf e}^a$ by means of equation
\re{torsionzero}). (Otherwise, i.e.\ in the context of theories
with torsion, the latter equation is replaced by equation
\re{omegatorsion}), written in boldface indices, or, as
generically is preferable in this context, the connection is
regarded as an independent variable then.) We remark on this
occasion, however, that in two spacetime dimensions the spin
connection drops out of the action \re{ferm1}); we will return to
this feature below. In the presence of an additional $U(1)$ gauge
field $\cal A$, $D_\m$ is {\em understood}\/ to contain also the
standard ${\cal A}_\m$ part (cf below). The basic elements of the
Clifford algebra have been represented by $\s^\pm \equiv \2(\s^1
\pm i \s^2)$, $\s^1,\s^2$ being Pauli-matrices, and
$\overline{{\bf{\Ps}}} \equiv {\bf{\Ps}}^\dagger \s^1$, where
${\bf{\Ps}}$ is a two-component complex column vector.  (On the
classical level the entries of this vector may in general be taken
anti--commuting or commuting, as one prefers; we will treat them
as commuting variables, although several final formulas will be
insensitive to that choice.)

The global $U(1)$-symmetry of $L^{ferm}$ may be turned into a local
one by the standard procedure: $D_\m \rightarrow D_\m + i\, {\cal
  A}_\m$, where $\cal A$ is an (abelian) connection one-form. The dynamics
of $\cal A$ is generated by \be L^{U(1)} = \frac{1}{4} \int_{M}
d^2x \sqrt{-\det \gtrue} \; \alpha(\Phi) {\cal F}_{\m \n} {\cal
F}^{\m \n} \; , \pl{U1} \ee where ${\cal F} = d {\cal A}$ and
$\alpha(\Phi)$ provides an again dilaton dependent coupling.

In section \re{Secfield}) below we will derive the field equations
of the action \be L^{tot} = L^{gdil} + L^{scal} + L^{ferm} +
L^{U(1)} \, .  \pl{I}\ee This will be done in the original
variables corresponding to general potentials in $L^{gdil}$. When
trying to find solutions to these field equations (which will be
possible for a subclass only, cf e.g.\ the remarks at the end of
section \ref{Secspher}, we will apply the simplifying trick of the
preceding subsection, mapping $L^{gdil}$ to its simplified brother
\re{Proto2}) by a change of variables \re{fi}) and \re{gneu}) (or
\re{econf}), (\ref{oconf})). We thus need to extend this
transformation to the matter part as well.

As noted already before, the action for scalar fields is invariant
with respect to conformal or Weyl transformations \re{gneu}) of the
metric. As a consequence we cannot get rid of a nonminimal coupling
($\b' \neq 0$), if present in the original model.

In contrast to scalar fields $f$, which remain unchanged under a
conformal transformation \re{gneu}), $f \to f$, spinors are
usually transformed according to $\bfPs \to \O^{1/2} \bfPs$. Here
it is also possible (and advisable) to get rid of $\gamma(\Phi)$
by means of the further redefinition: \be \psi :=
\sqrt{\frac{|\gamma(\Phi)|}{\O(\Phi)}} \, \,\, \bfPs \, \pl{psi}.
\ee (Inspection of \re{ferm1}) shows that any unwanted derivative
term in the first part of the action, resulting from the action of
$D_\m$ on a {\em real\/} multiplicative factor in front of the
fermion, cancels against its respective hermitian conjugated term;
this is also the reason why $\varpi$ drops out from the action
\re{ferm1}) but a gauge field $\CA$ does not.)

In the new variables $g=2e^+e^-$, $\phi$, and $\psi$, the action
$L^{tot}$ takes the form\footnote{Here and in what follows
  we assume $\gamma >0$; for $\g<0$, $B_\m^a$ is to be replaced by
  $-B_\m^a$ in equation \re{J}).} \ba L^{tot}[g,\phi,f,\psi,A]&= & \2 \int
d^2x \, \sqrt{- g} \; \big[ \phi R(g) - V(\phi) + \wt \b(\phi) \,
\6^\m f \6_\m f + \nn & & \mbox{}+ 2 e_a^{\m} B_\m^{a} + \2
\wt{\a}(\phi) \CF^{\m\n}\CF_{\m\n} \big] \, , \pl{J} \ea where we
used the following abbreviations/redefinitions: \ba & B_\m^{a}
\equiv \overline{\psi} \s^{a} (i \! \dlr_\m - \CA_\m) \psi \, ,
\quad \! V(\U(\Phi)) =  \frac{\V(\Phi)}{\O^2(\Phi)} \, , &\\ &
\wt{\a}(\U(\Phi)) = \O^2(\Phi)\, \a(\Phi) \, , \quad \wt
\b(\U(\Phi)) = \b(\Phi) \, \; .& \pl{wtbeta} \ea Here $\dlr_\m
\equiv (\dr_{\! \m} - \dl_{\! \m})/2$, with $\dl_{\!  \m}$ acting
to its left and $\dr_{\! \m} = \partial_\m$, as usual, to its
right. Indices are raised by means of the `auxiliary metric' $g$
and we made use of $ \sqrt{- g} = \det (e_\m^a)$. Note that
\re{J}) is in the same form as \re{I}), just with the appropriate
replacement of variables and a specification of `potentials'.
Therefore the field equations in the new variables follow from the
field equations of the general model \re{I}), provided in the
following subsection, by specialization.

In the above we used the field dependent conformal transformations
\re{gneu}), extended to fermions by means of \re{psi}), to
simplify the form of the action. To avoid confusion, let us add a
remark on the behavior of $L^{tot}$ with respect to general
conformal transformations: The actions for scalar and fermion
fields $f$ and $\bfPs$, respectively, are conformally invariant.
Hereby $f$ remains unmodified ($f$ has conformal weight zero) and
$\bfPs$ transforms with the inverse fourth root of the conformal
factor ($\bfPs$ has conformal weight minus one half). Still, the
total action is not invariant under conformal (or Weyl)
transformations due to the presence of the {\em
  gravitational}\/ and the $U(1)$ part of the action.\footnote{The
  action for gauge fields is conformally invariant only in four
  spacetime dimensions (and this on the classical level only); the
  trace of its energy momentum tensor, equation \re{TU1}) below, vanishes
  merely for $d=4$.}

The action $L^{tot}$ has several gauge symmetries, on the other hand:
it is invariant with respect to diffeomorphisms, local Lorentz
transformations in the spin and frame bundle, and $U(1)$ gauge
transformations.

We finally provide also the generalization of the action \re{U1})
to general nonabelian gauge fields.

Allowing for a dilaton-dependent coupling constant $\a(\Phi)$
again, the action for such a system has the form \be
 L^{YM}= \int {\a(\Phi) \over 4 } tr (\CF \wedge\ast \CF) \, ,
 \el YMaction
where $\CF = d \CA + \CA \wedge\CA$ is the curvature two-form of
the Lie algebra valued connection one-form $\CA$, the trace is
taken in some faithful representation of the chosen Lie algebra,
and $\ast$ is the Hodge dual operation with respect to the {\em
dynamical}\/ metric.

When talking about gravity Yang-Mills (YM) systems in this report,
we always will have in mind some Lagrangian consisting of  the sum
of \re{YMaction}) and a gravitational action, i.e.\ \re{gdil}) in
the context of dilaton gravity and \re{grav}) in the context of a
geometrical theory (in the latter context $\a$ should be
independent of the dilaton, although, mathematically, we can also
deal with cases where $\a$  depends on $\phi$ and $X^2$). Note
that in the context of \re{gdil}) the Hodge dual operation is
understood with respect to ${\bf g}$, while in the geometrical
context it is understood with respect to $g$. For two-forms this
makes a difference, which, however, may be absorbed by redefining
the function $\a$. So, at least after a redefinition of the
fields, the gravity YM action always has the form \be L^{gravYM} =
L^{grav} + L^{YM} \pl{GYM} \, , \ee where $L^{grav}$ is given by
\re{grav}) above and, in general, (the appropriately redefined)
$\a$ is some function of $\phi$.

The Lagrangian \re{I}) may be generalized easily to the nonabelian
case. In \re{ferm1}) $\bfPs$ would denote a multiplet of fermion
fields then, with $D_\m$ the corresponding covariant derivative.

Similarly, certainly also the scalar fields could be coupled to
the gauge fields. This may be done already in the abelian case
upon transition to complex valued scalar fields $f$.


\section{General field equations \pl{Secfield}}
In this section we provide the classical field equations of the
gravity--matter system \re{I}). In particular, the gravitational
theory is taken to be torsion free; otherwise additional
contributions arise in the Dirac equation below. The geometrical
theories \re{fR}) are covered in their equivalent formulation
\re{Proto}) (replacing the variables by boldfaced ones); the
matter action then should  be independent of $\Phi$, though. The
field equations of the general model $L^{tot}$ expressed in terms
of the auxiliary variables $g$ and $\phi$, cf equation \re{J}),
are easily obtained by specialization. The generalization to more
general matter fields will be obvious, too.

The variation of $L^{scal}$ and $L^{ferm}$ with respect to its
matter content yields the following generalizations of the
massless Klein--Gordon and Dirac equation: \ba \Box f + \(\ln |\b
|\)' \6^\m \Phi \6_\m f &=& 0 \pl{KG}\\ \tDsl \bfPs &=& 0
\pl{Dirac} \ea where $\Box = \nabla^\m \nabla_\m$, with
$\nabla_\m$ the Riemannian covariant derivative, and where again
prime denotes differentiation with respect to the argument of the
respective function. Furthermore, $$\tDsl = \bfe_a^\m \s^a
\wt{D}_\m \, , \quad \wt{D}_\mu = \6_\m - \2 \s^3 {\varpi}_\m + i
\wt{\CA}_\m \, , \quad \wt{\CA}_\m \equiv {\CA}_\m -i \6_\m \ln
\sqrt{|\gamma(\Phi)|} \, .  $$ So both equations \re{KG}) and
\re{Dirac}) are modified by a term resulting from the dilaton
dependent coupling. For the fermions, however, the modification
has a particularly simple form: One merely has to add an imaginary
part to the $U(1)$ connection $\CA$. Moreover, since this
imaginary part is an exact form, the modification may equally well
be absorbed into the fermion field with a redefined absolute
value: $\wt{\bfPs} := \sqrt{|\gamma(\Phi)|} \bfPs$ satisfies the
{\em standard}\/ Dirac equation in curved spacetime, $\Dsl
\wt{\bfPs} = 0$.

This is related to the considerations in the preceding subsection,
where it was found that a nonminimal coupling for fermions can be
transformed away, while a similar procedure is not possible for
nonminimally coupled scalars. Let us remark, however, that
although $\varpi$ was found to drop out of the action \re{ferm1}),
the field equations certainly {\em do\/} contain the spin
connection within the covariant derivative (they would not be
covariant otherwise). Upon variation of $\bfPs$, it enters when
using the torsion zero condition to replace a derivative of the
zweibein resulting from partial integration.

Variation of the action $L^{tot}$, equation \re{I}), with respect
to the gauge field $\CA$ yields: \be \nabla_\n \[ \a(\Phi)
\CF^{\m\n} \] = \g(\Phi) {\overline{{\bf{\Ps}}}} \bfe_a^\m \s^{a}
{\bf{\Ps}} \pl{eomA} \, . \ee Note that if an action contains a
gauge field $\CA$, the solutions of the coupled system do {\em
not}\/ contain those of the system without gauge field as a
subsector: if $\CA \equiv 0$, the above equation implies $\bfPs
\equiv 0$! It is an additional equation without counterpart in the
system without a gauge field. This is to be kept in mind when
comparing solutions with and without gauge fields.

We now come to the variation of $L^{tot}$ with respect to the metric
$\bfg_{\m\n}$ (or the zweibein $\bfe_\m^a$).  One finds 
\be \(
{\bfg}_{\m\n} \Box - \nabla_\m \nabla_\n\) \, \U(\Phi) + \2 \bfg_{\m\n}
\, \(\V(\Phi)- \Z(\Phi) (\nabla \Phi)^2  \) + \Z(\Phi) \6_\m \Phi
\6_\n \Phi = {\bf T}_{\m\n} \, . \pl{eomg1} \ee Here ${\bf T}_{\m\n}$
is the energy momentum tensor of the matter fields, ${\bf T}_{\m\n}=
{\bf T}^{scal}_{\m\n}+ {\bf T}^{ferm}_{\m\n} + {\bf T}^{U(1)}_{\m\n}$,
where \ba {\bf T}^{scal}_{\m\n} &=& \b (\Phi) \, \[\6_\m f \6_\n f -
\2 \bfg_{\m\n} \6_\m f \6^\m f \] \pl{Tscal} \, , \\ {\bf
  T}^{U(1)}_{\m\n} &=& \a(\Phi) \,
\[\CF_{\m \rho} \CF_{\n}{}^\rho - \frac{1}{4} \bfg_{\m\n} \CF_{\lambda
  \rho} \CF^{\lambda\rho} \] \pl{TU1} \, . \ea The energy momentum
tensor for the fermion fields is defined by ${\bf T}^{ferm}_{\m\n}
= \bfe_{a (\m} \, \(\d L^{ferm}/\d \bfe_{a}^{\n)}\) / \det (
\bfe_\rho^b)$, where the brackets around the indices $\m$ and $\n$
indicate symmetrization. For an action that depends on the
vielbein only via the combination $\gtrue_{\m\n}=\bfe_\m^a
\bfe_{a\n}$, which is {\em not\/} the case for fermions but, e.g.,
for scalar fields, this reduces to the standard expression; for
instance ${\bf
  T}^{scal}_{\m\n}$, defined in the above manner, just reduces to
${\bf T}^{scal}_{\m\n} = 2 \(\delta L^{scal}/\delta
\gtrue^{\m\n}\)/\sqrt{-\gtrue}$. (In {\em such\/} a case the
symmetrization in the indizes $\m$ and $\n$ would not be
necessary.)

The fermionic part of the action, equation \re{ferm1}), is of the
form $\int d^2x \, \det ( \bfe_\rho^b)$
$\bfe_a^\m \, \bfB_\m^a$,
resulting in ${\bf T}_{\m\n} = \bfe_{a(\m} \bfB_{\n)}^a -
\bfg_{\m\n} \, \bfe_{a}^{\rho} \bfB_{\rho}^a$ since $\bfB_\m^a$
does not depend on the geometrical variables; this is a particular
feature of two spacetime dimensions, since only there the
spin connection $\varpi$ drops out from the ferm\-ionic
action altogether. As such ${\bf T}_{\m\n}$ is not tracefree.
However, it is straightforward to show that, as a consequence of
the field equations \re{Dirac}) with arbitrary $\gamma(\Phi)$,
$\bfe_{a}^{\rho} \bfB_{\rho}^a = 0$, so that ${\bf
  T}^{ferm}$ becomes tracefree `on--shell'. (We remark that the
symmetrization in the indices $\m$ and $\n$ is essential; the
unsymmetrized version remains nonsymmetrical also on--shell.)
The fermionic part of the energy momentum tensor may now be
written as \be {\bf T}^{ferm}_{\m\n} = \bfe_{a(\m} \bfB_{\n)}^a \,
, \quad \bfB_\n^a \equiv \gamma(\Phi) \,
\[{\overline{{\bf{\Ps}}}} \s^{a}
(i \dlr_\n - \CA_\n) {\bf{\Ps}} \] \, ,\pl{Tferm} \ee where
$\dlr_\m \equiv (\dr_{\! \m} - \dl_{\! \m})/2$, as before, with
$\dl_{\!  \m}$ acting to its left (but not on $\gamma(\Phi)$ or
$\bfe_{a\m}$, certainly).

The field equations \re{eomg1}) may be simplified by taking the trace
and eliminating $\Box \U(\Phi)$: \be -\nabla_\m \nabla_\n \U(\Phi) - \2
\bfg_{\m\n} \, \V(\Phi) + \Z(\Phi) \( \6_\m \Phi \6_\n \Phi -\2
\bfg_{\m\n} \, \6_\m \Phi \6^\m \Phi \) ={\bf T}_{\m\n} - \bfg_{\m\n}
\, {\bf T} \, . \pl{eomg2} \ee Here ${\bf T}={\bf T}^{U(1)}$ denotes
the trace of the $U(1)$ part of the energy momentum tensor, the other
two contributions to the trace being zero as a consequence of the
conformal invariance of the respective parts of the action.

Equations \re{eomg1}) or \re{eomg2}) are the analog of the
Einstein equations in four spacetime dimensions  (cf
\re{Einsteineqs}) in the Introduction). By comparison, the
left--hand side is found to replace the four-dimensional Einstein
or Ricci tensor, respectively. For a minimal coupling, the matter
field equations take the same form as in higher dimensions,
furthermore.

Finally, variation with respect to the dilaton $\Phi$ leads to the
equation 
\be - \U' \, R({\bf g}) + \Z' \, \6_\m \Phi \6^\m \Phi + 2\Z \, \Box \Phi
+\V' - \2 \a' \, \CF_{\m\n}\CF^{\m\n} - \b' \, \6^\m f \6_\m f -
2\gamma' \, {\bf e}_{a}^{\m } \frac{\bfB_\m^a}{\g}=0 \pl{eomPhi}
\, ,\ee where $\bfB_\m^a$ has been defined in equation \re{Tferm}).

The field equations \re{eomg1}) (or \re{eomg2})) hold also for
more general matter couplings. Only the second half of equation
\re{eomPhi}) has to be changed by the corresponding
$\Phi$--derivative of the respective matter part under
consideration.

\section{More on dimensionally reduced theories \pl{Secmore}}
There are several possibilities to generalize the dimensional
reduction of Einstein gravity from four to two dimensions as presented
in section \ref{Secspher}.

First, we may consider, e.g., the more general case of warped
products. In the spherically symmetric case the four--dimensional
spacetime splits into a two-dimensional `$(r,t)$'-part (our $M$ in
section \ref{Secspher}) and at each point of this two-surface a
sphere is attached, the volume of which, however, depends on the
position within  $M$. Thus in this case the four-manifold splits
into a product manifold $M \times S^2$ (where, in the maximal
extension, $M \sim \dR^2$) only topologically, the metric being
`warped'. More generally, one may regard the following ansatz for
the four-metric: \be \left( \bfg^{(4)}_{ij}\right)(x,y)=\left(
  \begin{array}{cc} {\bar \Phi}^2(y) \, g_{\m\n}(x) & 0 \\ 0 &
   - \Phi^2(x) \, h_{\r\s}(y)    \end{array} \right) \, , \el warped
where $x$ and $y$ are coordinates on two two-manifolds. Now, by
means of the vacuum Einstein equations it may be established
\cite{Kummerwarped} that either ${\bar \Phi}$ or $\Phi$ has to be
constant. Without loss of generality we thus may restrict
attention to ${\bar \Phi}=1$. Other components of the Einstein
equations then allow to show that the metric $h$ necessarily has
constant (2d) curvature. In this way several solutions in addition
to the Schwarzschild spacetime may be described. Although the
direct implementation of the ansatz \re{warped}) into the
Einstein--Hilbert action does {\em not}\/ lead to an action
yielding the correct field equations on the two-manifolds (the
contradictory statement in the preprint-version of
\cite{Kummerwarped} was removed in a revised version), restriction
to a specific 2d constant curvature space will yield a 2d dilaton
action of the form \re{gdil}). In this way it should be possible
to obtain the list of global warped solutions in
\cite{Kummerwarped} also by means of specializing the results in
chapter \ref{Chapglobal}  below to the respective case. For
further recent work on warped solutions cf
\cite{Kummerwarpedzitat13andSchmidtMignemipapier}.

Alternatively, the spherically reduced theory may be regarded as a
particular case of requiring the four-metric in Einstein gravity
to be invariant with respect to a symmetry group. Requiring, e.g.,
that $\bfg^{(4)}$ has two commuting hypersurface-orthogonal Killing
vector fields, the four-metric may be put into the form (cf \cite{Schmidt98}
and references therein) 
\begin{equation}
 \left( \bfg^{(4)}_{ij} \right) = \left(
  \begin{array}{cc} g_{\m\n}(x) & 0 \\ 0 &
  -  \exp{\Phi(x)} \, h_{\r\s}(x)    \end{array} \right) \, ,
\nonumber
\end{equation}
\be h_{\r\s}(x)=\left(\begin{array}{cc}{\exp{(f(x))}} & 0 \\ 0 &\exp{(-f(x))}
 \end{array} \right) \, ,
\el 2KV where $x$ are the remaining two coordinates on the
four-manifold. Implementation of this ansatz into the Einstein--Hilbert
action, yields a generalized dilaton theory \re{gdil}) coupled
nonminimally to the above scalar field $f$ as in \re{scal})! Up to an
overall-factor, the coefficient functions in the coupled dilaton
theory come out as \cite{Schmidt98}: \be \U = \exp \Phi \, , \; \V =
0\, , \;\Z=-\2 \exp \Phi \, , \; \b= 2 \exp \Phi\, . \pl{Koeffkt} \ee
This sector of the theory thus has one propagating mode described by
$f$. Although now formulated as an inherently two-dimensional theory,
it is equivalent to a sector of Einstein vacuum gravity containing,
e.g., gravitational plane waves.

There are several other possibilities for two commuting Killing vector
fields (cf, e.g., \cite{Waldbuch}).  In each of these cases one
obtains a two-dimensional field theory with propagating modes. One of
the simplest of these sectors is described by cylindrical
gravitational waves, discussed already many years ago by Kucha\v{r}
\cite{Kucharcyl}; but cf also \cite{cylandere} for more recent work on
this subject. In a series of works by Nicolai, Korotkin and, in part,
Samtleben \cite{Nicolaialle}, in a similar manner, a dilaton theory
\re{gdil}) with simply $\U = \Phi$, $\V=\Z=0$ {\em coupled}\/ to a
nonlinear, dilaton-dependent $\s$-model is obtained. It turns out
that the resulting system, despite containing propagating modes, is an
integrable model and the machinery of integrable systems on flat
backgrounds can be adapted to the curved case, too. Unfortunately we
will not have the possibility to go further into this material within
the present work. It is, however, an interesting arena of 2d gravity
models, describing a sector of physical Einstein gravity in terms of
integrable systems. The relation to generalized dilaton gravity with
matter fields may also further strengthen our interest in such
theories, moreover. (Cf also \cite{restliche2KVs} for further related
work.)

Up to now in this subsection we discussed only a dimensional reduction
of {\em four-dimensional vacuum}\/ Einstein gravity. It is also
possible, e.g., to play similar games in reducing $D$--dimensional
Einstein theory (with $D \ge 4$) to two dimensions. Again one obtains
dilaton theories coupled to matter, at least in some cases (cf, e.g.,
\cite{Schmidt9709071,Schmidt98}). Otherwise, we may be interested
also in, e.g., a spherical reduction of 4d Einstein gravity coupled to
matter fields. We provided already an example of this at the end of
section \ref{Secspher}.

Of particular interest is also the spherical reduction of
Yang-Mills (YM) fields. Here some care is needed. For a general
gauge group it is, e.g., too restrictive to require that the
four-dimensional gauge field has to be strictly invariant with
respect to rotations. By an appropriate choice of gauge, the
latter requirement would lead to an effectively two-dimensional YM
theory \re{YMaction}), with, again, a dilaton dependent coupling
$\a$. This covers a subsector of the spherically symmetric 4d YM
theory only.

To cover the {\em whole}\/ space of spherically symmetric Einstein--YM
solutions, we are allowed to require only that the gauge field is
invariant with respect to rotations {\em up to a gauge
  transformation}. In the nonabelian case this leaves additional,
two-dimensional scalar (Higgs) fields (cf, e.g., \cite{Straumann}).
In the abelian case, however, invariance of the connection one-form
may be required strictly without loss of generality and a 2d
gravity-U(1) gauge theory is induced.\footnote{I am grateful to M.\ 
  Bojowald for discussion on this issue.}

There are finally several other possible routes of attaching a
physical significance to one or the other two-dimensional model by
means of dimensional reductions of physical or tentatively physical
higher-dimensional theories. For some possibilities arising in the
context of reducing effective string theories cf, e.g.,
\cite{HarveyStromTasi,Cadoni9904Cavag9709}, and references
therein.

Finally, let us mention also \cite{Frolovdil}, where it is shown that a
dilaton black hole metric is induced on the worldsheet of a string
if it moves in the spacetime of a  four-dimensional black hole. The 2d
Hawking radiation receives here an interesting interpretation in terms
of the higher dimensional theory.

\section{Extension to 2d supergravity \pl{Secsugra}}
The supersymmetric extension of a general dilaton theory
\re{gdil}) was first considered  in \cite{Parker,Susyauchzuerst} (cf also
\cite{Susyandere,Rivelles,Ertl} for supersymmetrization of particular
models and \cite{Susyrelated} for recent, in part related work).
It is obtained in a  straightforward manner by using the
superfield formalism of \cite{Howe}. In this framework the action
takes the same form \re{gdil}), where, however, each term is
replaced by an appropriately constrained supersymmetric extension
and, simultaneously, the volume form $d^2x$ is replaced by its
({\em  worldsheet}) superspace analog $d^2x \, d^2\theta$. A term such as
$\U(\Phi)$, e.g., is replaced by $\U(\bar \Phi)$, where $\bar
\Phi$ is the superfield $\bar \Phi = \Phi + i \bar \theta \xi +i
\bar \theta \theta f$, with $\Phi$ being the bosonic dilaton
field, $\xi$ a Majorana spinorial superpartner, and $f$ an
auxiliary bosonic scalar.

While by means of the superfield formalism of \cite{Howe} the
supersymmetric extension of any 2d {\em torsion--free}\/ gravity
theory is defined uniquely, for many practical calculations it is
necessary to reexpress the supersymmetric extension of the action
in terms of its component fields ($\Phi$, $\xi$, $f$ etc.). The
resulting action and all the more its field equations become
lengthy and their analysis involved.

We will not provide any further details on this approach in the
present treatise. However, for later use we will provide the
supersymmetric extension of the generalized dilaton theory in its
simpler formulation \re{Proto2}) or \re{Protoform}). It was provided
in component form in a recent paper by Izquierdo
\cite{Izq}:\footnote{In fact, after completion of the part of the
  present work on supergravity (cf also \cite{susy}), I became aware
  that the action \re{Izq}) was found already in
  \cite{Ikedasusy,Ikeda} (which was not noticed in \cite{Izq} or also
  in \cite{Ertl}, where the latter of these two works was cited in a
  different context only).  \cite{Ikedasusy,Ikeda} contain also parts
  of the considerations of \cite{susy}, but, such as in the bosonic case,
  the hidden structure of a (graded) Poisson manifold is not noted
  (and consequently these references contain no statement about, e.g.,
  the classical solutions).}  \ba L^{susygdil}&=&X_a \left( De^a + 2i
  ({\bar\psi}\gamma^a\psi) \right) + \phi d\omega - 4 uu' \e + 4i u
({\bar\psi}\gamma_3\psi)+ \nn && + \, i u' ({\bar\chi} e^a \gamma_a
\psi) + i {\bar\chi} \left( d\psi + \frac{1}{2}\omega \gamma^3 \psi
\right) + \frac{iu''}{8} \e \, ({\bar\chi}\chi) \pl{Izq} \ea Beside
the bosonic variables encountered already before, here $\psi$ and
$\chi$ are a oneform--valued and zeroform--valued Majorana spinors,
respectively, both fields being of odd Grassmann parity. $u$, on the
other hand, is a function of the dilaton $\phi$. As the comparison of
the bosonic part of \re{Izq}) shows, it relates to $V$ in
\re{Protoform}) through $V = -4 (u^2)'$, the prime denoting
differentiation with respect to the argument $\phi$. This
parameterization of the potential $V$ is found more convenient for the
supersymmetric extension (cf terms linear in $u$).

We do not demonstrate equivalence of \re{Izq}) with the one
obtained in the superfield formalism of \re{Proto2}), but rely on
the analysis of \cite{Izq} for this. On the other hand, it may be
established by direct calculation that \re{Izq}) provides a
supersymmetrization of \re{Protoform}). Indeed, the following
local (infinitesimal) supersymmetry transformations may be
verified to alter \re{Izq}) by a total divergence only: \ba \d e^a
&=& 4i {\bar\psi} \g^a \ep \nn \d \psi &=& d \ep + \mbox{$\2$} \o
\g^3 \ep + u' e^a \g_a \ep \nn \d \o &=& -8iu' {\bar \ep} \g^3
\psi - i u''{\bar \ep} \g^a \chi e_a \pl{Izqtrafo} \\ \d \phi &=&
-\mbox{$\2$}i{\bar\chi} \g^3 \ep \nn \d X^a &=& -iu'{\bar\chi}\g^a
\ep \nn \d \chi &=& -8u\g^3 \ep - 4X_a \g^a \ep \, , \nonumber \ea
where $\ep$ is a spinor field on M.

At latest at this point we have to admit that in the context of
supergravity (but {\em only}\/ in this context) we changed
conventions so as to conform to \cite{Izq}: the Lorentz metric has
signature $(-,+)$ and  $\e^{{\underline{0}}{\underline{1}}} = +1$
now (in contrast to what follows from \re{metric}) and
\re{epsilon}), respectively). The gamma matrices used here,
furthermore, are related to the usual Pauli matrices $\s_i$
according to: $\g^{\underline{0}} = -i \s_2$, $\g^{\underline{1}}
= \s_1$, and $\g^3 \equiv \g^{\underline{0}}\g^{\underline{1}}=
\s_3$. Changes in the conventions used to establish that
\re{Izqtrafo}) are supersymmetry transformations are quite
delicate, and we did not want to take the risk of introducing
wrong signs by adopting to the conventions used otherwise in this
report.

The supersymmetric extension of \re{gdil}) appears quite
nontrivial. One may even suspect that it will introduce
propagating modes. However, as we will establish in chapter
\ref{Chaplocal} by using the approach of the next chapter, this is
not the case. In fact, as the results of chapter \ref{Chaplocal}
will show that on-shell (i.e.\ by using the field equations) and
up to local symmetries {\em all}\/ the fermionic variables may be
put to zero! Thus, we will establish that the supersymmetric
extension of generalized dilaton gravity is trivial on-shell. This
was by no means expected, e.g., in \cite{Parker}.

On the other hand, the supersymmetric extension may still be of
some practical value. In \cite{Parker} it was used, e.g., to
establish that some notion of total mass (of, say, eventual black
holes of the spacetime) is bounded from below under certain
conditions. This result applies to the presence of quite a general
matter content of the theory, which, in particular, is {\em not}\/
required to be supersymmetric.

It is also possible to supersymmetrize the matter part of the
general dilaton theory. For example, for \re{scal}) with $\b:
\equiv 1$, i.e.\ for \be L^{scal}=\2 \int d^2x  \, \sqrt{-g} \;
g^{\m\n} \6_{\m}f \6_{\n}f \,  \pl{scal2} \ee in the present
context,  such an extension takes the form \cite{Izq}: \be
L^{susyscal}= L^{scal} - \mbox{$\frac{i}{4}$} \int d^2x  \,
\sqrt{-g} \, \left[\bar \l \g^\m D_\m \l + 4 \6_\m f {\bar
\psi}_\n \g^\m \g^\n \l + i{\bar \psi}_\n \g^\m \g^\n  \psi_\m
{\bar \l} \l \right] \, ,\pl{susyscal}\ee where $\g^\m \equiv
e_a{}^\m \g^a$, $e_a{}^\m$ being the inverse matrix to $e_\m{}^a$.
Here $\l$ is the superpartner to $f$. The supersymmetry
transformations \re{Izqtrafo}) are then accompanied by \be \d f =
i \bar \ep \l \quad , \quad \; \d \l = -2 \6_\m f \g^\m \ep + 2i
\bar \l \psi_\m \g^\m \ep \, . \ee For the purely dilatonic part
of the action, the supersymmetric extension does not introduce new
degrees of freedom, since the newly introduced superpartners
either are determined by means of the field equations or are gauge
equivalent to zero (as we will show in the subsequent chapters).
In \re{susyscal}) a spinor field $\l$ is added to the coupled
dilaton--scalar field action. If there are no further hidden
symmetries (such as, e.g.,  the superconformal symmetries arising
in superstring theory), new propagating modes are introduced by a
supersymmetric extension of \re{gdil}) coupled to
\re{scal2}). This issue should be analyzed further.

We finally remark that the superfield approach of \cite{Howe} is
restricted to torsion--free gravity theories. Previous attempts to
adapt the formalism to supersymmetrize, e.g., the KV model
\re{KV}) failed thus far (but cf also \cite{Ertl} for some
first steps). Below we will provide a different approach, inspired
by combining the formulation of Izquierdo with the one of the
following chapter, which will allow for a systematic extension to
also include the supersymmetrization of arbitrary theories of the
form \re{Faction}). However, also here, in the absence of
additional matter fields, on-shell equivalence to the bosonic
theory will be proven, at least on the classical level.

\chapter{2d gravity--Yang--Mills systems in terms of Poisson
  $\s$--models \pl{SecPSM}}
Poisson $\s$--models (PSMs) turn out to greatly simplify the analysis
of the topological (nonpropagating) part of the gravity theories
discussed so far (this includes {\em all\/} the purely gravitational
theories as well as additional Yang--Mills fields). This applies both
to the classical level and all the more to the quantum level.

{}{}From the point of view of the classical local solutions in generalized
dilaton gravity, the formulation of the two--dimensional gravity
action \re{gdil}) in terms of a PSM allows to get rid of the last remaining
potential $V$ left in \re{Proto2}) after the field redefinitions
introduced in section \ref{Secconf}. The solution of the
field equations of the purely gravitational part of the action is then
trivial in this formulation. Still, as will be shown in chapter
\ref{Chapglobal} the original model can yield a large 
variety of nontrivial two--dimensional spacetimes, including various
types of black holes. 

The definition of PSMs uses the notion of Poisson manifolds, 
a generalization of symplectic manifolds. Definition and
essential features of Poisson manifolds will be recapitulated in
the subsequent section (cf \cite{Quantumgroups,Mad} for further details).
The ensuing section contains the general definition of the
$\s$--model, which may be associated to any Poisson manifold.
Thereafter the relation to the previously introduced
gravity Yang--Mills actions is displayed. Supersymmetric
extensions of these actions may be covered, when using graded
Poisson manifolds as target space.

The final subsection briefly addresses extensions occurring for
propagating bosonic and fermionic modes. Here the framework allows to
trivialize the field equations only in the chiral case.

Let us finally remark that notion of PSMs is dispensable (although
also here in part illuminative) when restricting attention to the
string inspired dilaton theory \re{dil1}). As remarked already in the
previous chapter, the transformations \re{gneu}) are sufficient in
this case to essentially trivialize the resulting field equations.

Getting acquainted with PSMs pays for itself when one is
interested in a model characterized by a more generic potential in
\re{Proto2}) or \re{grav}), or when one wants to analyze the
whole class of pure (super)gravity YM models presented thus far. On the
other hand, Poisson $\s$--models may find interest also in their own
right within mathematical physics. Only recently, e.g., they were
found useful \cite{Cat} in the context of deformation quantization
\cite{Defoquant} of finite dimensional Poisson manifolds.

\section{Poisson manifolds \pl{Secpoisson}}
Denote by $N$ a manifold equipped with a Poisson bracket relation
$\{.,.\}$ between functions on $N$. By definition the Poisson bracket
is bilinear and antisymmetric. It is, moreover, subject to the
Leibnitz rule $\{f,gh\}=\{f,g\}h
+g\{f,h\}$ and the Jacobi identity $\{f,\{g,h\}\}+cycl.=0$.

In local coordinates $X^i$ on $N$ a Poisson bracket may be expressed
in terms of a two-tensor $\CP^{ij}(X):=\{X^i,X^j\}$.  With the
Leibnitz rule this relation determines the Poisson structure uniquely:
\be \{f,g\}=\CP^{ij}f_{,i}g_{,j} \, \, ,  \el Tens where $f_{,i} \equiv
\6 f/ \6 X^i$ and summation convention is understood. {}From \rz Tens it
is obvious that $\CP^{ij}$ transforms covariantly under coordinate
transformations. In terms of $\CP^{ij}$ the Jacobi identity reads \be
\CP^{ij}{,_k}\CP^{kl}+cycl.(i,j,l)=0 \, .  \el Jaco

There is another consequence of the Leibnitz rule: The Poisson bracket
of the constant function with anything else vanishes. In the case of a
general Poisson manifold there may be also other functions which
share this property. They are called Casimir functions.

An example for a $2n$-dimensional Poisson manifold is provided by the
phase space of an $n$-particle system: Denote by $Q^\alpha$ the positions of
the particle and by $P_\alpha$ their momenta. The Poisson structure is then
given by $\{Q^\a,P_\b \}=\delta^\a_\b$ and $\{Q^\a, Q^\b \} = 0 =
\{P_\a,  P_\b \}$.

With \rz Tens it is easy to see that   the constant function is the only
Casimir function in this example, which obviously holds whenever
$\CP^{ij}$ is
nondegenerate. Actually, Poisson manifolds with nondegenerate $\CP$
are symplectic
manifolds,  the symplectic two-form being equal to the inverse
of the Poisson tensor $\CP$.  For them the above example is in some
sense generic: Locally {\em any} symplectic manifold allows
so-called Darboux coordinates such that the Poisson tensor takes
the form of the $2n\times 2n$ matrix:
\be
  \left(
  \begin{array}{rr}
  0&\done\\-\done&0
  \end{array}
  \right) \quad ,
\el Darb
where $\done$ denotes the $n \times n$ unit matrix.

Another example for a Poisson structure is given by the Poisson
bracket
\be
  \{X^i,X^j\}=\sum_{k=1}^3 \varepsilon^{ijk} X^k
\el Brac
 on $\dR^3$, where $\varepsilon^{ijk}$ is the completely antisymmetric
three-tensor.
Obviously the Poisson structure \rz Brac is degenerate:
Darboux coordinates cannot exist on a manifold of odd dimension.
Indeed $R^2:= \sum_{i=1}^3 X^iX^i$ is a Casimir function.

In the example \rz Brac the Poisson structure can be restricted
consistently to any two-sphere given by the choice of a constant value
for the radius function $R$. The restricted Poisson structure is
nondegenerate. It is easy to convince oneself that Darboux coordinates
on a two-sphere of radius $R_0$ are provided by one of the coordinates
$X^i$ and the azimuthal angle around the respective axis; e.g.\ with
$Z := X^3$ and $\Phi := \arctan (X^2/X^1)$ it follows from \rz Brac
that $\{ \Phi , Z \}=1$.  The physical phase spaces corresponding to
these two-spheres are the classical analogs of spin systems.

The picture outlined here generalizes in a straightforward way: A
Poisson manifold with degenerate Poisson structure foliates (more
accurately: stratifies, cf below) into a family of lower
dimensional manifolds (symplectic leaves), each of which is
characterized by assigning constant values to the Casimir functions
and is equipped with a nondegenerate Poisson structure. To be more
precise: Some of the symplectic leaves may have lower dimension than
the generic ones, such as the origin $X^i=0$ in the example \ry Brac ,
and in some of these cases (not so for \ry Brac ) level surfaces of
Casimir functions correspond to different symplectic leaves.  Moreover,
a symplectic leaf does not need to be a (regular) submanifold of $N$, i.e.\ it
might wind around densely in some higher dimensional subregion of $N$
(as a higher-dimensional generalization of what one knows from paths
on a torus in classical mechanics, cf, e.g., \cite{Thi}). In that
case constant values of Casimir functions characterize parts of such
leaves only.  So what we are dealing with is a stratification of $N$
rather than a foliation. Nevertheless we will stick to the more common
nomenclature introduced above.

In the neighborhood of a generic point in $N$ we then may always find
coordinates $\wt X^i := (C^I,Q^\a,P_\b)$ such that $C^I$ are Casimirs
and the $Q$s and $P$s are pairs of canonically conjugate variables
\cite{Weinstein}. 
The Poisson tensor then takes the standard form \re{Darb}) extended by
$k=n-r$ rows and columns with zeros; here $k$ denotes the number of
independent Casimir functions, $r$ is the rank of the Poisson matrix
$\CP^{ij}$, and $n=\mbox{dim} N$. Such coordinates $\wt X^i$ will be
called Casimir-Darboux (CD) coordinates further on. In the example
above, $(R,\Phi,Z)$ provides a CD coordinate system.

As an example for a nonlinear Poisson structure let us take the
quadratic modification \be \{X,Y\}= - Z^2 + \frac{1}{4} \; , \quad \{
Y, Z \} = X \; , \quad \{Z, X \} = Y \el PR2 of \rz Brac on $\dR^3$
with coordinates $(X,Y,Z)$, which will become relevant in the ana\-ly\-sis
of (Euclidean) $R^2$-gravity (with cosmological constant).  It is
straightforward to check that \be C_{R^2}:= X^2 + Y^2 -
\mbox{$\frac{2}{3}$} Z^3 + \7 Z  \el CR2 is a Casimir function of
the bracket \ry PR2 .  $\Phi := \arctan (Y/X)$ and $Z$, on the other
hand, are still canonically conjugates (i.e.\ $\{\Phi, Z\}=1$). Thus in
this case $(C_{R^2}, \Phi,Z)$ provides a local CD coordinate system.

It is a nice excercise to visualize the 'foliation' of $N=\dR^3$ into
symplectic leaves. This is done most easily by rotating the square
root of the positive parts of the function \be h(Z) :=
\mbox{$\frac{2}{3}$} Z^3 - \7 Z+ C_{R^2} \el hachtung 
around the $Z$-axis
(since $h(Z) = X^2+Y^2$).  The resulting picture is the following: For
values of $C_{R^2}$ smaller than $-1/6$ the symplectic leaves are
diffeomorphic to the $(X,Y)$-plane. At $C_{R^2} =-1/6$ there is an
additional pointlike leaf at $(X,Y,Z)=(0,0,-\7)$.  Indeed, this is one
of the two points where the right--hand side of \rz PR2 (and thus also
the Poisson tensor $\CP$) vanishes. Within the range $(-1/6,1/6)$ of
the Casimir, the pointlike leaf turns into a ellipsoid--like surface
($S^2$ topologically), again accompanied by a (tolopological) `plane'
(situated at larger values of $Z$).  For $C_{R^2}=1/6$ the $S^2$ and
the plane touch at $(X,Y,Z)=(0,0,\7)$. This value of $C_{R^2}$
corresponds to three symplectic leaves: An $S^2$ with one pole removed
($\sim$ plane), this pole $(X,Y,Z)=(0,0,\7)$, and a `plane without
origin' (topologically a cylinder).  For larger values of the Casimir,
finally, one has only planar leaves again.

For later purposes we need some generalizations of the above Poisson
brackets. First, the $su(2)$ bracket \re{Brac}) generalizes to
arbitrary Lie algebras. Whenever $\CP(X)$ is linear in $X \in \dR^n$,
$\CP^{ij}=f^{ij}{}_k X^k$, the Jacobi identity \re{Jaco}) is easily found to
reduce to the standard Jacobi identity for structure constants
$f^{ij}{}_k$. $N$ may then be identified with a Lie algebra (or,
better, the dual of a Lie algebra; but within our applications below
these two will be identified by means of the Killing metric).

Next, the nonlinear bracket \re{PR2}) on $N=\dR^3$ may be generalized
by replacing the right--hand side of its first equation by some {\em
  arbitrary}\/ function $W(X^2+Y^2,Z)$. (This is also the
most general modification which is compatible with the remaining
two brackets.)

The previous Poisson bracket was invariant with respect to
rotations around the $Z$-axis. Switching to Lorentz boosts
instead, and denoting the three coordinates in $N$ by $X^-$,
$X^+$, and $\phi$ (or $X^3$), the analogous Poisson tensor takes
the form \be \left(\CP^{ij}\right)(X) = \left(
  \begin{array}{ccc} 0 & -W & -X^- \\ W & 0 & X^+\\X^- & -X^+ & 0
  \end{array} \right) \, , \el P where
$W$ is some arbitrary ($C^2$--)function of $X^2\equiv 2X^-X^+$ and
$\phi$.  This bracket will play a role in the context of the gravity 
model \re{grav}).

For a general function $W$, a Casimir function of the above brackets
cannot be written in explicit form. This changes, e.g., in the case
when $W$ depends on $\phi \equiv X^3$ only, $W=V(\phi)/2$. Then \be C=
2X^+X^- - \int^\phi V(z)dz \pl{Ctorless} \ee is easily seen to provide
a Casimir function to the above (`Lorentzian') bracket \re{P}). For its
Euclidean analog we only need to replace $2X^+X^-$ by $X^2+Y^2$
and rename $\phi$ by $Z$.

Independently of the explicit form of $C$, a CD coordinate system 
for the general bracket \re{P}) and its Euclidean analog is
provided, e.g., by \be (C,\pm \ln|X^\pm|,\phi) \, , \pl{CDLor} \ee
(for either choice of the signs) and by $(C,\arctan (Y/X),Z)$ in
the Lorentzian and Euclidean case, respectively. In the Lorentzian
case this is proven at once upon inspection of $\{X^\pm , \phi \}=
\pm X^\pm$, following from \re{P}) by definition. The angle $\arctan
(Y/X)$ in the Euclidean case, on the other hand, may be viewed as a
Wick rotation of $(\ln |X^+|-\ln |X^-|)/2$. A further
analysis of the Poisson bracket \re{P}) will be provided in section
\ref{Hui}.

Note that at points characterized by \be X^+ = X^- = W = 0 \pl{crit}
\ee (and analogously in the Euclidean case) the tensor $\CP$ has rank
zero. We will call these points critical and denote the corresponding
values of the Casimir and the dilaton by $C_{crit}$ and $\phi_{crit}$,
respectively (there may be several critical values, certainly,
according to the number of zeros of the function $W(0,\phi)$).  All
noncritical points are `generic' in the sense that in some
neighborhood of them it is always possible to construct a CD
coordinate system (cf also the explicit formulas above).  In general
there will be both zero-- and two--dimensional leaves associated to a
critical value $C_{crit}$ of a Casimir function $C$.  The latter of
these certainly also allow local CD coordinates.

The above examples provided Poisson brackets on linear spaces
only. A (nonlinear) group manifold, on the other hand, provides an
example for a degenerate Poisson bracket on a nonlinear space
(cf, e.g., \cite{Quantumgroups} for further details or
\cite{Schladming} for an explicit exposition of such a bracket
for the group $SU(2)$).

\section{Defining the $\s$--model \pl{Secdefining}}
We now come to the definition of the $\s$--model.
Let $M$ be the two-dimensional worldsheet of a field theory
(with local coordinates $x^\m$).  As dynamical fields we take
(among others, cf below) coordinates $X^i$ on some (arbitrary)
Poisson manifold $N$.  $N$ is thus the target space of the theory.
Instead of being equipped with a metric, as the target space in
string theory  (cf section \ref{Secstring}), it now is equipped
only with a Poisson bracket resp.\ a Poisson bivector
$\CP^{ij}(X)$. The fields $X^i(x)$ are not yet sufficient to
define a $\s$--model. The reason for this is that the tensor $\CP$, in 
contrast to a metric $G_{ij}$ or an antisymmetric tensor field
$B_{ij}(X)$ (cf section \ref{Secstring}),  has only
contravariant (`upper') indices, which cannot be contracted
with $dX^i$ (in the absence of an additional metric $G$ on the tartget
space, which $N$ was not required to be equipped with).

We thus need additional {\em fields}\/ to define a 
$\s$--model. As such we choose the following objects: Let $(A_i)$ (or
$A=A_i dX^i$) denote a one--form on the world sheet taking values in
$T^* N$ (more precisely, a one--form on the world sheet which is
simultaneously the pullback of a section of $T^* N$ by the map
$X(x)$). So, $A_i=A_{i\m}dx^\m$ is a one--form on $M$ and $i$ is a
{\em covariant}\/ index in $N$. This object may be used for
saturating free indices of $\CP$.

With this input, there naturally exists the following $\s$--model
\cite{PSM1}:\footnote{Independently from \cite{PSM1} and
  \cite{PSMold}, the action
  \re{PSM}) was considered also in \cite{Ikeda}. The relation
  \re{Jaco}) was derived by the requirement for maximal local
  symmetries of the model (cf the discussion below). However, the
  implications of \re{Jaco}), leading to the profitable use of the
  notion of a target space and adapted curvilinear coordinates
  thereon, was realized in \cite{PSM1} only (and exploited there,
  too). Other related work is 
  \cite{Ikedaold2,Schladming,Anton,PhD,PSMDubna}. 
  Recently, the $\s$-model was rediscovered 
  in \cite{Cat}, where it was used as topological field theory to derive
  formulas \cite{Kontsevich} of relevance for the quantization of
  general (finite dimensional) Poisson manifolds.}
\be L^{PSM}=\int_M A_i dX^i + \2 \CP^{ij}(X) A_i A_j \,
, \pl{PSM} \ee where certainly the wedge product between forms on the
world sheet is to be understood again.  Note that the definition of
this action did not require any metric on the (two--dimensional) base
manifold. This is one of the characteristic features of a topological
field theory \cite{Blau}.  Moreover, it is {\em covariant}\/ with
respect to diffeomorphisms on the target space.  This is one of the
essential features of a $\s$--model.


To prevent potential confusion, we also write out the action \re{PSM})
in full: \be L^{PSM} = \int_M dx^\m \wedge dx^\n \, [A_{\m i}(x) \6_\n
X^i(x) + \2 \CP^{ij}(X(x)) A_{\m i}(x) A_{\n j}(x)] \, .  \el
actioncoor Here one need not care about any geometrical significance
of the quantities involved; (at least locally) the fields are simply
$A_{\m i}(x)$ and $X^i(x)$.  $(\CP^{ij})$, on the other hand, is some
fixed, prescribed matrix depending explicitly on $X$ (in general at
least).

In a more abstract language, on the other hand, the action may be
understood also as follows: It is a functional of a map $\CX$ from $M$ to
$N$ and of a one--one--form $A$ on the product $B = M \times N$
(equipped with its natural grading). In local coordinates $A= A_{\m
  i}dx^\m \wedge dX^i$. The map $\CX$ may be interpreted also as a map
from $M$ to the tensor product $B$ (extended trivially on the
first factor). The pullback of $A$ with respect to this map yields
a two--form on $M$; integrating over this two--form gives rise to
the first term in the action \re{PSM}). On the other hand, $\CP$
may be interpreted also as a bivector on $B$; contracting it twice
with $A$, yields a two--zero form on $B$. Its pullback with
respect to $\CX$ provides the second contribution to the action.

Note that this formulation of the $\s$--model requires $B$ to be a
trivial product (only then there exists the natural grading). In the
case of nontrivial fibrations of $B$, the framework of standard fiber
bundles seems inapplicable in the present context. (Using the language
of \cite{vanNieu}, this is the case because in general \re{PSM}) is
the action of a {\em nonlinear}\/ gauge theory; cf equations
\re{syma}) and \re{symb}) below.)  An attempt to still formulate the
$\s$--model for nontrivial bundles $B$ has been provided in
\cite{PSM1}. Due to the remaining ambiguity left in the definition of
$A$ in that framework, however, the formulation presented there calls
for further refinement (or a completely different approach adapted to
(appropriately) generalized) nontrivial bundles).

It is a nice exercise to check that, as a consequence of \re{Jaco}),
the infinitesimal transformations \ba \d_\ep X^i(x) &=& \ep_j(x)
\, \CP^{ji}(X(x)) \pl{syma} \\ \d_\ep A_i &=& d\ep_i(x) + \CP^{lm}{}_{,i}
\, A_l \, \ep_m \, \pl{symb} \ea change the action \re{PSM}) by a total
divergence, $\int_M d(\ep_i dX^i)$, only.  These are thus local
symmetries of the action. They are obviously nonlinear generalizations
of standard gauge transformations: For a linear Poisson tensor the
second line reduces to an infinitesimal gauge transformation of a Lie
algebra valued $A=(A_i)$, $\d_\ep A = D_A \ep$, while the first line
corresponds to infinitesimal (co)adjoint transformations of a
Lie--algebra valued field $X=(X^i)$.

Variation of \re{PSM}) leads to the field equations \ba dX^i +
\CP^{ij} A_j&=&0 \pl{eoma}\\ dA_i + \2 \CP^{lm}{}_{,i} A_l \wedge A_m
 &=&0 \, . \pl{eomb} \ea The second of these
equations is recognized as a generalized zero curvature condition,
reducing to exactly $F=0$ in the case of a linear $\CP$; similarly, 
in the linear case, the first equation reduces to $D_A X=0$.
For a linear $\CP$ such
equations result from a two--dimensional, so--called $BF$--theory
\cite{Blau}; indeed, after a partial integration in the first
term, the action \re{PSM}) is seen to reduce to $\int_M X^iF_i$
in this case, $(F_i)$ denoting the curvature of the gauge
field.\footnote{This works in the linear case only: Defining a
generalized curvature two--form by the left--hand side of equation
\re{eomb}), the action \re{PSM}) takes the above form, iff $\CP$
is linear.}


The symmetries \re{syma}), \re{symb}) are {\em all}\/ the independent
symmetries of the model \re{PSM}). As a consequence, the obvious
diffeomorphism invariance of the action is already incorporated in
these equations: For any given vector field $\xi =\xi^\m(x) \6 /
\6 x^\m$ generating diffeomorphisms on the worldsheet manifold
$M$, the (field dependent) choice $\ep_i := \xi^\m A_{i\m}$ in
(\ref{syma}, \ref{symb}) results in \ba \d_{i_\xi A} X^i
&\equiv& \CL_\xi X^i - i_\xi (dX^i +  \CP^{ij} A_j) \nn \d_{i_\xi A}
A_i   &\equiv& \CL_\xi A_i - i_\xi (dA_i + \2 \CP^{lm}{}_{,i} A_l
\wedge A_m) \pl{diff} \, ,\ea where $\CL_\xi$ denotes the Lie
derivative along $\xi$. Obviously the additional terms on the
right--hand side of \re{diff}) vanish for any solution of the field
equations \re{eoma}, \ref{eomb}).

Let us remark that in the general case the field equations \re{eoma},
\ref{eomb}) are highly nonlinear and, without making use of features
induced by the Jacobi identity, finding the general solution to the
field equations would be illusive! In the previous subsection the
Jacobi identity \re{Jaco}) for a bivector $\CP^{ij}$ was seen to be
the characteristic feature for it to define a Poisson bracket.
According to a theorem by Weinstein \cite{Weinstein} (cf the previous
subsection) we then know that locally on the Poisson manifold there
exist adapted CD coordinates. Denoting those by $\wt X^i \equiv (\wt
X^I,\wt X^\a, \wt X^{\underline{\a}})$, the action functional
\re{PSM}) is locally seen to take the trivial form \be L^{PSM} =
\int_M \wt A_i d \wt X^i + \wt A_\a \wt A_{\underline{\a}} \qquad
\mbox{(locally!)} \, , \pl{trivact} \ee where a sum over pairs of
indices $(\a, {\underline{\a}})$ is understood and $\wt A_i$ denotes
the components of $A = A_i dX^i = \wt A_i d \wt X^i$ in this (target
space) coordinate system.

The field equations from this action are determined easily. Due to
the target space covariance of the theory, they alternatively may
be derived from specialization of the general field equations
\re{eoma}, \ref{eomb}) for the case of a Poisson tensor in
Casimir--Darboux form. In any case these field  equations are
also solved very easily. Moreover, further application of the also greatly
simplified local symmetries \re{syma}, \ref{symb}) show that
locally all the free functions in the action may be gauged away.
{\em On the local level}\/ only constant values of the Casimir
functions $\wt X^I$ remain to  parameterize the general solutions
up to gauge transformations.

One should stress, however, that in general this route of solving the
field equations works locally only. Only for thoses cases where CD
coordinates exist globally on the target space, we may restrict our
attention to \re{trivact}) instead of \re{PSM}).  Otherwise further
work is necessary, as will be seen more explicitly in chapter
\ref{Chapglobal} below (for the subclass of PSMs describing gravity
models).

The action \re{PSM}) may be extended further without spoiling its
symmetries, if, e.g., a Casimir function on $N$ and a volume form on
$M$ are chosen. Such an extension is necessary when a 2d Yang--Mills
theory on a fixed spacetime background is described. A similar
extension allows to cover also the $G/G$ WZW model, where $N$ is seen
to be a group manifold endowed with its Sklyanin Poisson bracket. We
refer to \cite{Schladming,Anton} for the details. Here we now turn to
the case of the previously discussed gravity models.

\section{Generalized dilaton gravity as Poisson $\s$--model}
As demonstrated in the previous two chapters, the action \re{grav})
may be viewed as the most general 2d gravity action (under current
consideration). The generalized dilaton theory \re{gdil}), e.g.,
takes this form after appropriate field redefinitions, cf section
\ref{Sectwo}. It thus suffices to demonstrate the equivalence of
\re{grav}) with some class of PSMs.

We first collect the one-- and zero--forms of
\re{grav}) into the following multiplets: \be (X^i) := (X^-,X^+,
\phi) \equiv (X^a,\phi) \; , \quad (A_i) := (e^+,e^-,\o) \equiv
(e_a,\o) \pl{multi1} \; . \ee We then perform a partial
integration in the terms containing derivatives on the one--forms.
Now, comparison with \re{PSM}) shows that the resulting action
indeed takes the form of a PSM, and the respective Poisson tensor
is found to coincide with \re{P})!

Note that when this action is taken in CD coordinates, equation
\re{trivact}), none of the three original potentials in
\re{gdil}) is present anymore. All the non\-lineari\-ties of
the (corresponding) field equations have been put into a clever
choice of field variables. In this way we now succeeded to
complete the transformation  of the original system
\re{gdil}) or, likewise, of \re{grav}) to a trivial one.

It is certainly also possible to express \re{gdil}) directly in
terms of a PSM. For this purpose we only need to combine equations
\re{multi1}) with equations \re{econf},\ref{oconf}) to obtain: \be
(X^i) := (X^a, \U(\Phi)) \; , \quad (A_i)= (\exp
\[-\int^\Phi \frac{\Z(z)}{2\U'(z)} dz \] {\bf e}_a, \varpi
+\frac{\Z(\Phi)}{2\U'(\Phi)} \ast d \Phi) \, \pl{multi2} .\ee Now
there is no talk of conformal transformations anymore. equation
\re{multi2}) provides a definition of new field variables such
that (upon elimination of $X^a$ and $A_3$) the action \re{PSM})
yields \re{gdil}) identically. The Poisson structure to be used in
this case is \re{P}) with $W=V(X^3)/2$, 
where the function $V$ is
defined in the text following Equation \re{Proto2}). (As remarked on
similar occasions above, for smooth potentials $\U,\V,\Z$ the new
variables \re{multi2}) and the potential $V$ exist in regions
where $\U'\neq 0$.)

One might even go one step further and apply the transformation to CD
coordinates directly to the right--hand sides of \re{multi2}).  Then
the resulting equation provides the trivializing transformation in one
step. (The applicability is then, however, restricted to smaller local
patches in general.)\footnote{There are cases such as spherically
  symmetric gravity, where CD coordinates exist globally (cf, e.g.,
  section \ref{Secabelschwarz}). Then such a transformation may be
  useful also on the level of the action.}

Some remark on coordinate transformations in the target space is
in place. The theory \re{PSM}) is formulated independently of the
choice of particular coordinates on the target space. I.e., two
Poisson tensors being related by a diffeomorphism define the {\em
same}\/ PSM. On the other hand, an {\em identification}\/ such as
\re{multi1}) or \re{multi2}) singles out one particular coordinate
system on the target space. It is, e.g., the third component of
$A$ and not its first one in the chosen coordinate system $X^i$, which
corresponds to the spin connection $\o$ of 
\re{Protoform}).

This implies, on the other hand, that a diffeomorphism on the target
space can mediate between two {\em different}\/ gravity models. For
instance, the `Lorentz--covariant' coordinate change $X^a \to
F(X^2,\phi) \, X^a$, $\phi \to \phi$ for a two--argument function $F$,
is found to change (only) the `potential'´´ $W$ in \re{P}). A
simple calculation yields: $W \to W F^2 + 2 W X^2 \, {(\ln F)}^. +
X^2 F'$, where the dot/prime denotes differentiation of the function
under consideration with respect to its first/second argument (and, in
the resulting functions, the first argument is taken as the old $X^2$
to be expressed in terms of the new $X^2$ and $\phi$). Let us specify
this example to the case where $F$ depends on $\phi$ only for
simplicity. Suppose furthermore that the original function $W$ is
independent of its first argument as well, characterizing a
torsion--free geometrical model \re{fR}). Then the above diffeomorphism
on the target space is seen to produce a potential $W$ linear in its
first argument (provided only $F' \neq 0$). Identifying the components
$A_i$ in the {\em new}\/ target space coordinates with
Einstein--Cartan variables according to \re{multi1}), one obtains by
the above diffeomorphism a theory of the type \re{Faction}) quadratic
in torsion from the torsion--free theory \re{fR}).\footnote{I am
  grateful to W.\ Kummer for discussions leading to the above
  observation.}

We finally remark that replacing the Poisson bracket \re{P}) by
its Euclidean analog provided in the paragraph before that
formula, leads to the corresponding Euclidean gravity
theory. This will be of interest in particular on the quantum
level.

\section{Extension to include 2d supergravity \pl{Secextsugra}}
To incorporate also the supersymmetric extension of a generalized
dilaton (or any other gravity theory \re{Faction})), the notion of
PSMs has to be extended to the case of graded Poisson manifolds.  We
now will first provide the appropriate generalization, and then
demonstrate that the action \re{Izq}) indeed fits into the framework
of (graded) PSMs.  Finally we will briefly dwell on the structure
found on the target space of the theory and comment on straightforward
procedures to construct supersymmetric extensions of the more general
gravity theory \re{Faction}).\footnote{As remarked already in section
  \ref{Secsugra} above, parts of the considerations of the present
  section may be found also in \cite{Ikedasusy,Ikeda}.}

So in this section we allow $N$ to carry a $Z_2$--grading, i.e.\
some of the fields $X^i$ and $A_{\m i}$ may be Grassmann valued,
$\s_i$ denoting their respective parity (so $X^i X^j =
(-1)^{\s_i\s_j} X^jX^i$ etc.). $N$ is equipped with a {\em
graded}\/ Poisson bracket \be \{ X^i , X^j \} \equiv \CP^{ij} \, .
\pl{bracketsus} \ee Note that now the Poisson bracket is
anti--symmetric only for the case that at least one of its entries
is an even (commuting) quantity; in general one has
$\CP^{ij}=(-1)^{\s_i\s_j+1} \CP^{ji}$ (while $\CP^{ij}$ itself has
grading $\s_i + \s_j$). In terms of the two--tensor $\CP^{ij}$ the
standard, graded Jacobi identity (cf, e.g., \cite{Henneaux}) may
be brought into the form \be (-1)^{\s_i\s_k} \( \CP^{ij}
\frac{\dl}{\partial X^s} \) \CP^{sk} + cycl(ijk) = 0 \,
, \pl{Jacobisus} \ee where a sum over the index $s$ is
understood (but not over $i$ or $k$ in the first of the three
cyclic terms). In terms of left derivatives $\dr = \partial$ (in
contrast to the above right derivatives $\dl$) this equation may be
written equivalently as $(-1)^{\s_i\s_k}\CP^{is} \partial_s
\CP^{jk} + cycl(ijk) = 0$.

An appropriate graded generalization of the action \re{PSM}) is
\be L = \int A_i \wedge dX^i - \2 A_i \wedge A_j \, \CP^{ji} \, ,
\pl{PSMsus}\ee where the order of the terms and indices has
been chosen so as to avoid unnecessary signs in the considerations to
follow, while simultaneously $L$ coincides with \re{PSM}) in the
purely bosonic case.

Note that when using, e.g., the action \re{PSM}) as
{\em definition\/} of the graded PSM, one would need to require the tensor
$\CP^{ij}$ to satisfy some {\em other}\/ relation than the
standard graded Jacobi identity (so as to allow for the
existence of the local symmetries etc.), resulting from the latter
by some unusual changes in signs. Also, explicit signs in the
field equations and symmetries cannot be avoided with such a
choice of the action, as opposed to \re{PSMsus}).

The field equations and the local symmetries of the PSM,
generalizing the previous bosonic formulas, are found to be: \be
dX^i - A_j \CP^{ji} = 0 \,\, , \; \, \, dA_i - \2 A_k A_l
(\CP^{lk} \dl_i) =0 \,\, ,\pl{eomsus} \ee \be \delta
X^i = \ep_j \CP^{ji} \; , \quad \delta A_i = d \ep_i - A_j \ep_k
(\CP^{kj} \dl_i) \pl{symmsus} \, \ee where $\ep_i$
is of the same Grassmann parity as $X^i$. It is again a simple
exercise to show that due to \re{Jacobisus}) the action changes only
by a total divergence: $\delta L = \int d (\ep_i dX^i)$.

Thus there is a local symmetry for any pair of fields $(X^i,
A_i)$. Since the action is of first order in these fields, this
implies that there are at most a finite number of physical (gauge
invariant) degrees of freedom again. (More precisely, being a one--form,
$A_i$ has two components for each value of $i$; however, the `time
component' $A_{0i}$ of the one--form $A_i$ enters a Hamiltonian
formulation as Lagrange multiplier for the constraints only and
therefore it must not be included in the above naive counting (cf
section \ref{Secham} for further details).)

The following observations will be found helpful later on: Starting
from an action \re{PSMsus}) with some {\em arbitrary} matrix
$\CP^{ij}$, there exist well-defined requirements on the action which
uniquely lead to the Jacobi identity \re{Jacobisus}).  E.g., the
constraints found in a Hamiltonian treatment of a PSM--type action (cf
section \ref{Secham}) are first class, {\em iff}\/ $\CP^{ij}$
satisfies the Jacobi identity. In this way the Jacobi identity is
found to be the necessary and sufficient requirement for a Poisson
$\s$-model to have a maximal number of local gauge symmetries. If it
is not satisfied, on the other hand, an action of the PSM form will
have propagating modes.

It may be seen also that \re{Jacobisus}) is the necessary and
sufficient condition for the field equations of a Lagrangian
\re{PSMsus}) to form a so-called free differential algebra (FDA)
\cite{FDA}:\footnote{The following considerations are inspired by those of
  Izquierdo \cite{Izq}, but cf also a footnote in \cite{AntonKlim},
  where a similar (although not quite correct) observation has been
  made, too.} Applying an exterior derivative to the first set of the
field equations \re{eomsus}), we obtain $dA_j \, \CP^{ji} - A_j \,
(\CP^{ji} \dl_k) \, dX^k=0$. Eliminating $dA_j$ and $dX^k$ by means of
\re{eomsus}), we end up with an expression bilinear in the $A's$. By
definition of an FDA, the resulting equations have to be fulfilled
{\em identically}, without any restriction to the one--forms $A_i$. It
is a simple exercise, recommended to the reader, to show that this
requirement is fulfilled if and only if $\CP^{ij}$ satisfies equations
\re{Jacobisus}). Application of an exterior derivative to the second
set of equations \re{eomsus}), on the other hand, does not lead to any
further restrictions; they are fulfilled identically on behalf of
\re{Jacobisus}) already.

Now we are in the position to show that the action \re{Izq}) fits
into our framework. We first collect the one-- and zero--forms of
\re{Izq})  into two multiplets: \be (X^i) := (X^a, \chi^\alpha,
\phi) \; , \quad (A_i) := (e_a, i \bar \psi_\a, \o)
\pl{multisus} \; . \ee After a simple partial integration the
action \re{Izq}) is seen to take the form \re{PSMsus}) and the
coefficient matrix $\CP^{ij}$ may be read off by straightforward
comparison:  \ba &\CP^{ab} = -\e^{ab} (4 u u' + \frac{1}{8i} u''
\chi_\a \chi^\a ) \; , \quad \CP^{a \a } = u' (\gamma^a \chi)^\a
\; , & \nn &\CP^{\a\b} = - 8i u (\gamma^3)^{\a\b} - 4i X_a
(\gamma^a)^{\a\b} \, , & \pl{PBsus}
\\ & \{ X^a, \phi \} = \e^a{}_b X^b \; , \quad \{ \chi^\a, \phi \}
=-\2 (\gamma^3 \chi)^\a & \nonumber \ea where $(\gamma^a \chi)^\a
\equiv (\gamma^a)^\a{}_\b \chi^\b$, spinor indices have been
raised and lowered by means of $\e^{\a \b}$ ($\chi_\a = \e_{\a \b}
\chi^\b$ with $\e^{01}:=1$), and, in the last line, the
identification \re{bracketsus}) was used.

We now need to show only that the matrix \re{PBsus}) satisfies the
Jacobi identity \re{Jacobisus}). The simplest way of doing so is
direct verification. However, in \cite{Izq} it was proven already that
the field equations of the action \re{Izq}) form an FDA.  Thus, {\em
  using}\/ the result of \cite{Izq}, the validity of the graded Jacobi
identities is proven upon our general observation above, namely that
the FDA property of the field equations of an action of the {\em
  form\/} (\ref{PSMsus}) is tantamount to the validity of
\re{Jacobisus}).\footnote{In order to prove as in \cite{Izq} that the
  field equations of the action \re{Izq}) form an FDA, the simplest
  method is to to check the validity of \re{Jacobisus}) for the
  bracket \re{PBsus}), and to subsequently provide the general proof
  for the equivalence of the latter relations with the FDA-property of
  \re{eomsus}).}

Up to now it may seem that we have not gained much in
reformulating the theory \re{Izq}) in terms of a graded PSM.
However, the first simplification to be observed is the notational
convenience gained. Comparison of the actions, field equations and
local symmetries may convince the reader. (For example, equation  
\re{Izqtrafo}) is (only) {\em part\/} of the symmetries \re{symmsus});
cf below.)
Second, the observation
that the coefficient matrix $\CP^{ij}$ satisfies the Jacobi
identity opens the door to methods otherwise hardly accessible. It
will be by such  methods that we will enable us to show on-shell
triviality of the supersymmetric extension of \re{gdil}) (in the
sense explained above) in chapter \ref{Chaplocal} below.

We come back to the symmetries.  In the context of \re{Izq}) resp.\ 
\re{PBsus}) there are {\em five}\/ independent local symmetries
contained in single line \re{symmsus}): $\ep_\phi$ generates local
Lorentz symmetries (cf last line in equation \re{PBsus})).  The
Grassmann spinor $\ep_\a$, on the other hand, generates precisely the
local supersymmetry transformations \re{Izqtrafo}). The remaining two
symmetries correspond to the obvious diffeomorphism invariance of the
action (similar to equations \re{diff}) above; but cf also the remarks
in section \ref{Secsugraloes} concerning degenerate metrics).

We return to the structure found in the {\em target space\/} $N$ of the
theory. The target space is spanned by a Lorentz vector $X^a$, a
Majorana spinor $\chi^\a$, and the dilaton $\phi$. $X^a$ and
$\chi^\a$ may be combined into a $(1+1)$--dimensional superspace.
$\phi$, on the other hand, generates Lorentz boosts in this
superspace by means of the Poisson brackets \re{PBsus}), last
line. Indeed, $\e^a{}_b$ is the Lie algebra element of the
(one--dimensional) Lorentz group in the fundamental representation
and $\gamma^3$ is easily identified with the generator of Lorentz
boosts in a two--dimensional spinor space: $\gamma^3 \equiv
\g^{\underline{0}} \g^{\underline{1}} = \2 [\g^{\underline{0}},
\g^{\underline{1}}]$ (irrespective of the choice of presentation
for the generators $\g^{\underline{0}}$ and $\g^{\underline{1}}$
in the Clifford algebra).

This structure in the target space will remain also within
generalizations to more general 2d supergravity theories including
the supersymmetrization of theories with nontrivial torsion
\re{grav}), for which a supersymmetrization has not been provided
in the literature yet. Using the same fields \re{multisus}) as
before, such supersymmetric theories may be found by searching for
more general solutions to the Jacobi identity \re{Jacobisus}) than
the one given in \re{PBsus}). However, the then yet unknown
Poisson tensor should be restricted to agree with the last line in
\re{PBsus}). This is due to the relation of the Poisson bracket on
the target space to the local symmetries \re{symmsus}), and thus
implicitly required by local Lorentz invariance, present in any
gravity theory in Einstein--Cartan formulation.

It is worth noting that merely upon restriction of the Poisson tensor
by the last line of \re{PBsus}), the Jacobi identities \re{Jacobisus}
with one of the indices corresponding to $\phi$ requires the Poisson
tensor components $\CP^{ab}$, $\CP^{a \a}$, and $\CP^{\a\b}$ to
transform covariantly under Lorentz transformations! E.g.\ for $\CP^{a
  \a}$ the Jacobi identities require $\{ \CP^{a \a}, \phi \} =
\e^a{}_b\CP^{b\a} -\2 (\g^3)^\a{}_\b \CP^{a\b}$.  Thus to obtain the
most general supergravity theory that fits into the present framework,
we can proceed as follows: It must be possible to build the unknown
tensor components of $\CP$ by means of the Lorentz covariant
quantities $X^a$, $\chi^\a$, $\e^{ab}$, $(\gamma^a)^{\a\b}$, and
$(\g^3)^{\a\b}$ ($\e^{\a\b}$ is incorporated automatically by raising
and lowering spinor indices) with coefficients that are Lorentz
invariant functions, i.e.\ functions of $X^2\equiv X^aX_a$,
$\chi_\a\chi^\a$, and $\phi$. E.g., the antisymmetric tensor
$\CP^{ab}$ {\em must}\/ be of the form $\CP^{ab}= \e^{ab} (F_1 +
\chi_\a \chi^\a F_2)$, where $F_{1,2}$ are functions of the two
arguments $X^2$ and $\phi$. The remaining Jacobi identities then
reduce to a (comparatively simple) set of differential equations for
these coefficient functions.

Proceeding in this way, e.g., by replacing all (five) coefficients
in its first two lines of the bracket \re{PBsus}) by yet undetermined
coefficient functions of $X^2$ and $\phi$, one can show that the
remaining Jacobi identities \re{Jacobisus}) force the coefficients to
agree with those provided already in \re{PBsus}) (except for a
simultaneous global prefactor). More general theories can thus be
obtained only by using further covariant entities to build $\CP^{a
  \a}$ and (possibly also) $\CP^{\a \b}$. Indeed, Lagrangians
quadratic in torsion require an extra additive term
$F(X^aX_a,\phi) X^a (\gamma^3)^\a{}_\b\chi^\b$ in $\CP^{a\a}$
\cite{Kummerpriv,prep}, also perfectly compatible with Lorentz
covariance.

Alternatively to searching for general solutions of the Jacobi
identities with the above restrictions, there is a more efficient
procedure, which allows to cover at least some of the theories with
torsion (but presumably even all of \re{grav})): At the end of the
previous subsection we emphasized already that diffeomorphisms on
the target space connect (generically) different gravity theories
with one another. However, the validity of the Jacobi identities
is certainly independent of the choice of coordinates. Thus, much more
general solutions to the Jacobi identities than equation \re{PBsus})
may be found by applying a (`Lorentz covariant') diffeomorphism
to the bracket \re{PBsus}). An example of such a diffeomorphism was provided
at the end of the previous subsection.

In \cite{prep} more general supergravity theories will be
constructed by the methods outlined here, probably covering the
whole class of models \re{grav}) resp.\ \re{Faction}). By
construction, the resulting theories will be invariant with respect to
superdiffeomorphisms incorporated within \re{symmsus}), thus allowing
for an interpretation as supergravity theory.

The method of (Lorentz invariant) diffeomorphisms on the target
space can be extended also to the whole five--dimensional space
$(X^i)$, certainly. In this way {\em different}\/
supersymmetrizations of one and the same bosonic theory will be
obtained. (We expect, e.g., that the two different
supersymmetrizations of the string inspired dilaton theory
presented in \cite{Rivelles} are related to one another
 by such a transformation.)

\section{Including matter fields \pl{Secinclumat}}
In the preceding sections we showed that all
the pure (super)gravity models discussed so far may be
reformulated in terms of particular PSMs. We now want to discuss
modifications necessary when matter fields are added.

\subsection{Additional Yang--Mills fields}
In two spacetime dimensions scalar and fermionic matter have
propagating modes. However, Yang-Mills (YM) fields do not.
YM theories on a fixed metric background are known to be 
almost topological field theories \cite{Blau}, becoming truly
topological in the (appropriately defined) limit $\a \to \infty$
of the coupling constant in \re{YMaction}). There thus is a chance
that YM fields fit into the framework of PSMs without changing the
form of the PSM action, just by defining an extended Poisson bracket.
This indeed is the case as we will now show.

To begin with we bring  \rz YMaction into first order form: \be
L^{YM} \sim {L^{YM}}'= \int \left( E^i \CF_i + \frac{1}{\a(\phi)}
E^i E_i \, \varepsilon \right) \, , \el equiYM where the indices
$i$ are raised and lowered by means of the Killing metric and
$\varepsilon \equiv e^- \wedge e^+$. The equivalence of $L^{YM}$
with ${L^{YM}}'$ is seen by integrating out the `electrical'
fields $E$, in complete analogy to equation \re{R2dil}).

In the case of a YM action on a fixed metric background, the first
term is recognized to be of $BF$-form (cf our brief discussion
following equation \re{eomb})) and the second term, containing the
Casimir $E_iE^i$ is just one of these possible extensions of a PSM
we referred to at the end of section \ref{Secdefining}.

In fact, when dealing with the coupled gravity-YM system, an extension
of the PSM in the above way is not necessary; the volume form
$\varepsilon$ is dynamical in this context and generated by one of the
terms in $\CP^{ij}A_i A_j$ as we will now show in detail. To avoid
eventual notational confusion let us rename $X^\pm$ into
$\varphi^\pm$.  Then $L^{grav} + {L^{YM}}'$ reads \be {L^{gravYM}}=
\int \varphi^a De_a + \phi \, d \o + E^j \CF_j +
\left[W((\varphi)^2,\phi)+ \frac{1}{\a(\phi)} \, E^j E_j\right]
\varepsilon \, , \el GYMprime where $\CF_j \equiv d \CA_j + f^{kl}{}_j
\CA_k \CA_l$, indices $a$ run over $+$ and $-$, and 
indices $j$ run from 1 to $n$, $n$ being the dimension of the chosen
Lie group. After partial integrations (dropping the corresponding
surface terms) and the identifications \be X^i := (\varphi^a,\phi,E^j)
\; , \quad A_i := (e_a,\o,\CA_j)\,, \el coor \rz GYMprime turns out to
be of Poisson $\s$-form \rz PSM on an $n+3$-dimensional target space
with Poisson brackets: \ba & \{\varphi^+,\varphi^-\}= W + (1/\a) \,
E^j E_j \; , \quad \{\varphi^\pm,\phi\}=\pm \varphi^\pm \; , & \nn &
\{\varphi^\pm,E^j\}= 0=\{\phi,E^j\}\; , & \nn &
\{E^j,E^k\}=f^{jk}{}_lE^l \; . & \pl{PoiGYM} \ea

Note that, in the present context, \re{syma}) and \re{symb})
entail diffeomorphisms, Lorentz transformations, {\em and}\/
nonabelian gauge transformations all at once and at the same
footing. Although certainly in $D=2$ the gravity and YM forces are
nonpropagating, one thus finds an appealing unification of gravity--YM
interactions here.

The bracket \re{PoiGYM}) will be further analyzed when
constructing the local solutions to  the field equations of
\re{GYMprime}) in chapter \ref{Chaplocal} below.

\subsection{Fermionic matter}
We now turn to the inclusion of fermionic matter fields with
Lagrangian \re{ferm1}) (without a $U(1)$ field). Redefining
field variables as in sections \ref{Sectwo} and \ref{Secmat} (cf,
in particular, equation \re{J})), and applying the 2d identity $\det
(e_\r^b) \, e_{a}^{\m}= \e(ab)\e(\m\n) \, e_\n^b$, where $\e(\cdot
\, \cdot)$ denotes the 2d antisymmetric symbol without any metric
dependence, the fermion part of the action takes the simple form
\be \int A_i \wedge J^i \; , \pl{fermmod} \ee where we have, by way of
exception, not suppressed the wedge in this formula, and where the
components of the current one-form $J^i$ are given by $J^j_\m =
(B^-_\m,-B^+_\m,0) = (i\psi_R^\ast {\stackrel{\lra}{\6}}_\m
\psi_R, -i\psi_L^\ast {\stackrel{\lra}{\6}}_\m \psi_L,0)$. Here
$\psi_{R,L}$ denote the posi\-tive/negative chirality components
of $\psi$. Thus the total action $L^{gravferm}=L^{gdil} +
I^{ferm}$ becomes \be L^{gravferm}= \int_{M} A_i
(dX^{i}+J^{i})+\2 \CP^{ij}A_i A_j \pl{PSJ} \ee with $i,j \, \in \,
\{-,+,3\}$ and $J$ as given above. When staying on the purely
classical level, the fermionic variables may be taken commuting
and $J$ may be simplified further by the following
parameterization: \be \psi \equiv \(
\begin{array}{c} \psi_{R}\\ \psi_{L}
\end{array} \) :=
\( \begin{array}{c} r \, \exp(-i \rho)\\l \, \exp(i \lambda)
\end{array} \)  \quad \Ra \quad
J^i = (r^2d\rho, l^2 d\lambda, 0) \; . \pl{Jpar} \ee An extension
to a multiplet of fermions and including YM-fields
proceeds along similar lines. We refer to \cite{Heiko} for the details.

Although the modification \re{fermmod}) is apparently simple, in
particular in view of the also simple form of \re{Jpar}), it will
turn out, however, that the framework of PSMs leads only to a
partial solvability of the field equations in the general case
(according to present day knowledge at least). A simplification of
the Poisson tensor by choosing adapted (CD) coordinates on the
target space, leads to a less trivial form of the current $J$ in
general. Only in the chiral case ($\psi_L=0 \lra l=0$) a
trivialization of $\CP$ by means of choosing field variables
induced by the CD coordinates \re{CDLor}) leaves the current $J$
(basically) unmodified (cf chapter \ref{Chaplocal} for details).

\subsection{Scalar fields}
Finally, we turn to massless scalar fields with Lagrangian
\re{scal}). We first bring this action into first order form: \be
L^{scal}=\2 \int_{M} \b(\Phi) df \wedge \ast df \cong \int_{M}
B \wedge df + \frac{1}{2\b(\Phi)} B \wedge \ast B \pl{IB} \, \, ,
\ee where again we displayed the wedges and we introduced a
one--form $B$ that equals $-\beta(\Phi) * d f$ on shell.  `$\ast$'
denotes the Hodge dual operation with respect to the dynamical
metric $\gtrue$. However, due to the conformal invariance of the
action, and the fact that $f$ carries conformal weight zero, we
may equally well take $g$ instead, defined in equation \re{gneu}).
Next we split $B$ into its self--dual and its anti self--dual
part. Due to $\ast A_\mp = \pm A_\mp$, which is equivalent to
$\ast {\bf e}^\pm = \pm \ast {\bf e}^\pm$, this splitting is
achieved by means of the decomposition: \be B = R A_- + L A_+
\pl{B} \,\, , \ee where the notation $R$ and $L$ stands for
`right--moving' and `left--moving', respectively; in
particular, the above $R$ has nothing to do with the curvature
scalar of $g$ or $\gtrue$. (We hope this does not
lead to any confusion.)

Combining equations \re{B}) and \re{IB}) and using $\b(\Phi)=\wt
\b(\phi)$, equation
 \re{wtbeta}), we obtain \be \int_{M} A_-  R df +
A_+ L df - \frac{RL}{\wt \b(\phi)} A_- A_+ \pl{IB2} \ee to be
added to \re{PSM}). Comparing this with equations
\re{PSJ},\ref{Jpar}), we see that the first two terms in \re{IB2})
give rise to a current $J^i$ which is precisely of the form
\re{Jpar}). Here this current is even simpler: the scalar field
$f$ plays the role of the two phases $\rho$ and $\lambda$, which
are {\em equal}\/ (while $r^2$ corresponds to $R$ and $l^2$ to
$L$).  The third and remaining term in \re{IB2}), on the other
hand, implies a complication: it mixes $R$- and $L$-fields and in
this respect resembles a (dilaton dependent) mass term for the
fermions.

In principle one can absorb this last term into the Poisson structure;
it still satisfies the Jacobi identity \re{Jaco}) for $i=1,2,3$, with
the fields or coordinates $R$ and $L$ entering $\CP$ as {\em
  parameters\/} then.  However, when changing to CD-coordinates of
this Poisson tensor, fields such as $\wt A_i$ depend implicitly on $R$
and $L$ also and thus may not be varied independently of the latter.
This route does not seem advisable therefore.


%


\chapter{Classical solutions on a local level}
\pl{Chaplocal}
In this and the following chapter we will consider the classical
solutions of the gravity models presented above. In the present
chapter we will focus on the local solutions. In the subsequent
chapter the maximal extensions to global solutions will be
considered.
%
%
The more complicated the theory, the less knowledge on the global
solutions is accessible. In the case of pure gravity Yang--Mills
systems, e.g., the solutions are under control for {\em
arbitrary}\/ spacetime topologies and, indeed, for a model generic
enough, maximally extended solutions on (pointed) Riemann surfaces
of arbitrary genus can be constructed.  If additional matter
fields are present, on the other hand, one is happy to find the
general {\em local\/} solutions, for the cases where this is
possible at all.

We first deal with the local solutions for the topological, pure
gravity YM models. They are obtained by application of several methods,
which, in part, may be combined also. Thereafter, we focus on
classically solvable models in the presence of scalar and fermionic
matter. Here the general solution may be constructed for a subclass of
possible gravity actions only.

\section{Pure 2d Gravity--Yang--Mills models \pl{Secpurelocal}}
In gauge theories the choice of a gauge often
dramatically simplifies the classical field equations. As an example,
one may compare the solutions found for
the KV model equation (\ref{KV}) in \cite{Katloes} within the
(less favorable) conformal gauge with those found in \cite{Kumdom}
within an axial kind of gauge (similar to the light cone gauge below).
The field equations of torsion--free theories (and also of some theories
with torsion) have been solved, more or less elegantly, by a large
number of authors in a variety of ways, and also with different
generality of the model.  The derivation of the solutions below, in
particular within the next subsection, was made possible only by such
pioneering works. I tried, however, not to follow the historical
routes, but rather, the presentation is lead by the desire for brevity
and elegance of the derivations.\footnote{For papers dealing
with local solutions of the field equations cf
\cite{Banks,Kunstclass,PhD,PartI,Hehl,Schmidtneu} 
 and references therein.}

Alternatively to the use of gauge conditions, in the present
context we may use also adapted field variables. These are
suggested by the $\s$-model formulation presented in chapter
\ref{SecPSM}. (A somewhat similar, but less developed approach has
been used also for the KV model in \cite{SolodukhinKV,HehlKV}.)

The theories without torsion are the simplest ones. The gauge
fixing method shall be applied only to these theories, and,
moreover, already in their reformulated form, equation
(\ref{Proto2}): \be L(g, \phi)= \int_M d^2x \sqrt{-g} \,\,
\left[\phi R(g) - V(\phi)\right] \, . \el Proto3 Thus, strictly
speaking, also here adapted field variables are used. But this
step is not necessary; without it the formulas and intermediate
steps just become somewhat more extended.

The theories with torsion in their most general form considered within
this work are described by the Lagrangian (cf equation (\ref{grav})):
\be L^{grav} = \int_M \,\, X_a De^a + \phi d\o + W(X^2,\phi) \,
\varepsilon \,\, .\pl{grav2} \ee They are of increasing difficulty
depending on the $X^2$-dependence of $W$. In the simplest case
(linearity in $X^2$) they are theories quadratic in torsion, possibly
even with a constant coefficient in the linear term as within the
KV--model, cf equation (\ref{WKV}). For a general potential $W$ the
solution may not be written down in terms of elementary functions
only. However, it turns out that it is possible to determine the
general local solution explicitly in terms of the `Casimir function'
$C$ defined by the Poisson structure in the Poisson $\s$-model (PSM)
formulation. The function $C$ will be determined implicitly by an {\em
  ordinary, first order}\/ differential equation, which is found
systematically within the PSM approach. The solutions of this general
theory will be constructed in the framework of PSMs only, as this
still appears as the most efficient method.  Also additional YM fields and
an extension to supergravity will be considered only there.

\subsection{Gauge fixing method, torsion--free theories \pl{Secgauge}}
In this section we first make some general remarks on subsequently
used gauge conditions. In each case we will provide an argumentation
of why (and in which regions!) the respective gauge condition is
attainable. Solving the torsion--free theories in the context of the
metrical variables, we will start with showing that there exists a
Killing vector field. This permits stronger gauge conditions,
leading to a further simplified derivation. Within the context of
Einstein--Cartan variables the solution is obtained without prior
knowledge of the invariance property of the spacetimes; otherwise the
latter derivation would simplify further. We also include some remarks
on the transformation back to the original variables $\bfg$ and $\Phi$
of interest in those cases where one has started from an action \re{gdil}).

\subsubsection{Attainability of certain gauge conditions \pl{Secattain}}
As is well known, in two spacetime dimensions any metric is
conformally flat: Using null coordinates $u$ and $v$, $g$ necessarily
takes the (local) form (conformal gauge) \be g=\exp[\rho(u,v)] \, du
\, dv \, .
\pl{confgauge} \ee
Note, however, that the (local) conformal flatness of any 2d metric, following
from \re{confgauge}), does {\em not}\/ imply that the global causal
structure of a 2d spacetime is always the one of flat 2d Minkowski space
--- even in cases where the conformal gauge is attainable {\em
  globally}\/ on the regarded spacetime. This comes about as in
general the (maximal) domain of definition of the function $\rho$
generically is only a subset of $\dR^2$.  For Schwarzschild (SS)
this is a well-known feature of the Kruskal coordinates; for
Reissner-Nordstr\"om (RN), or more complicated 2d metrics, this same
mechanism applies in a less trivial way.

The above examples of (the two-dimensional part of) SS and RN are
cases where the conformal gauge is attainable globally (on the
regarded spacetime), i.e., the whole manifold may be covered within
one coordinate patch such that $g$ takes the form \re{confgauge}). In
general the conformal gauge may be achieved only locally (even on
simply connected spacetimes). As will be demonstrated in section
\ref{Secmaxext} (cf, e.g., Fig.~\ref{fig:fig3}) below, there are
(simply connected) spacetime manifolds where, within one chart of
coordinates, smoothness considerations may obstruct the global
attainability of \re{confgauge}).

Somewhat less popular than the conformal gauge \re{confgauge}), but for some
purposes better suited, is the chiral gauge
\be g=2dx^0dx^1 + k(x^0,x^1) \,
dx^1 dx^1 \, . \pl{lcgauge} \ee
The attainability, and the region of validity, of this gauge may be
seen as follows: First
we label one set of null lines on $M$ by the $x^1$--coordinate. This implies
$g_{00}=0$. Then we choose the second coordinate
$x^0$ to coincide with some affine parameter
along these null extremals. This may be seen to yield $g_{01}$
independent of $x^0$. As an affine parameter is determined only up to
linear transformations, which in our case may be
$x^1$-dependent,  one may absorb $g_{01}(x^1)$ into $x^0$
by an appropriate rescaling of the affine parameter.
(Alternatively, prescribe the  normalization (or synchronization) of the affine
parameters on any chosen line $x^0 = c=const$ by requiring that
there the `speeds' $\6_0|_{c,x^1}$ have unit inner product with $\6_1$.)
The result is \re{lcgauge}). Obviously this procedure works on any part
of the two-dimensional spacetime $M$ that may be foliated into
null-lines.

If $k$ in \re{lcgauge}) does not depend on $x^1$, this metric has a
Killing field $\6_1$.  But, remarkably, also the converse is true:
{\em Any\/} 2d metric with a Killing vector field permits coordinates
such that locally it takes the generalized Eddington-Finkelstein form
\be g=2dx^0dx^1+h(x^0)(dx^1)^2 \, .  \el 011h It is obtained by
labeling Killing trajectories and (locally) transversal null-lines by
$x^0$ and $x^1$, respectively, where $x^1$ is the Killing-flow
parameter and $x^0$ an affine parameter of the chosen null-lines
normalized as before.  In the construction above \re{011h}) results,
if the `normalization hypersurface' $x^0 = c$ is identified with a
Killing trajectory (locally transversal to the chosen null-lines) and
$x^1$ is taken as a Killing-flow parameter along this trajectory. For
a given metric $g$, the function $h$ in \re{011h}) is unique up to an
equivalence relation $h(x^0) \sim {1\over{a^2}}h(ax^0+b)$, $a,b=const$.

Only for Minkowski and deSitter space this is not quite true,
because they have more than one Killing field (in fact three).
Minkowski space is described by $h=ax^0+b$ and (anti-)deSitter
space of curvature $R$ by $h={R\over 2}(x^0)^2+ax^0+b$ (since
$R\equiv h''$). Different choices of $h(x^0)$ correspond to
different Killing fields $\6_1$. For instance, in the case of
Minkowski space, $h=const$ implies that $\6_1$ generates
translations (timelike, null, or spacelike, according to
$\mbox{sgn}\,h$), whereas $h$ linear in $x^0$ ($a\ne0$)
corresponds to boosts.  (For deSitter cf the paragraph around
\re{deSitter}) and Fig.~\ref{fig:fig9}.)

Note that even in the presence of a Killing field, in general, charts with
 \re{011h}) are applicable in smaller regions only than charts with
\re{lcgauge}). The Killing trajectories $x^0=const$ may become
parallel to the null-lines $x^1=const$ somewhere or they may run into
a point where the Killing vector field vanishes.

In the presence of a Killing field, instead of \re{011h}), we
may also choose a generalized SS form of $g$ locally: \be g= h(r) dt^2
-{1 \over h(r)} dr^2 \, , \el gSS where $h$ is the {\em same} function as
in \re{011h}). This gauge results from \re{011h}) upon the choice of
new coordinates \be r := x^0 \, , \;\; t := x^1 + f(x^0) \el rtKoord
with \be f(x^0) \equiv \int^{x^0} {du \over h(u)} +\hbox{const} \, . \el
fun Obviously, this transformation is applicable only
wherever $h \neq 0$, i.e.\ within regions where the Killing field is
nonnull [from \re{011h}) $h$ is recognized as the norm of the
Killing vector field].

We finally remark for the sake of completeness that in the presence of a
Killing vector field also the conformal gauge may be specified further. It
is then possible, again within regions of nonnull Killing field, to
require $\rho$ to be a function of the {\em single\/} argument $u+v$ (or
$u-v$) only. Since we will not need this result below, we refer for a
simple and elegant proof of this statement to \cite{Schmidt91}.


\subsubsection{Metrical variables \pl{Secmetric}}
The field equations following from the redefined dilaton action
\re{Proto3}) have already quite a simple form (cf equations
(\ref{eomg2}) and (\ref{eomPhi})) \be \nabla_\m \nabla_\n \phi + \7
V(\phi) \, g_{\m\n} =0 \,\; , \quad R = V'(\phi) \, .  \pl{metriceom}
\ee The first equation shows that $g$ has a Killing vector field $\ast
d\phi$ (provided $d\phi \neq 0$, certainly). This is a consequence of
the following small lemma: \be \nabla_\m \nabla_\n f = F \, g_{\m\n} \;\;
\Ra \CL_{\ast df} \, g = 0 \, \, \pl{LemmaF} \ee
for some function $F$, provided only $df \neq 0$. Proof:
Contract the assumption with $\varepsilon_{\r}{}^\n$, use the
covariant constancy of $\varepsilon$, $\nabla \varepsilon = 0$, and
symmetrize the relation obtained in its free indices; this yields
$\nabla_\m v_\n + \nabla_\n v_\m = 0$, the defining property of a
Killing vector, where $v^\m \equiv \varepsilon^{\m\n} \partial_\n f =-
(\ast df)^\m$. 

Thus the metric $g$ may be put into the form \re{011h}). Since,
moreover, in these coordinates the Killing vector $v=\ast d\phi$ has
the form $\6_1$ and, on the other hand, $v(\phi) = d\phi(v) = 0$,
$\phi$ can be a function of $x^0$ only. Then, however, the zero-zero
component of the first equation \re{metriceom}) reduces to merely
$\6_0\6_0 \phi = 0$. The residual gauge freedom of \re{011h}) may now
be employed to obtain \be \phi = x^0 \, . \pl{phix0} \ee

Implementation of \re{phix0}) and \re{011h}) into the second equation
\re{metriceom}), finally, leads to $h''(x^0) = V'(x^0)$, which is
integrated to \be h(x^0) = \int^{x^0} V(z) dz + C \, , \pl{h} \ee
where $C$ is a (meaningful) constant of integration.  In the case
of $R^2$-gravity \I{R2}), e.g., one has $V=2\L +\phi^2/4\a$ and
thus, with $\a := (-1/8)$, one finds \be h^{R^2}= -\mbox{${2\over
3}$} (x^0)^3 +2 \L x^0 +C \, . \pl{R2h}\ee

Equations \re{phix0}), \re{h}), and \re{011h}) provide the general local
solution of the field equations (in regions of applicability of the
gauge \re{011h})) {\em provided}\/ $d\phi \neq 0$.

Up to the choice of coordinates, the local solutions depend on one
integration constant $C$ only, which is a specific function of the
total mass (at least in cases where the latter may be defined in a
sensible way, cf also \cite{Lau} for details). We then obtain a
`generalized Birkhoff theorem' for the class of theories considered
(which includes spherically symmetric gravity as a particular case).

If, on the other hand, $\phi = \phi_0=const$, then equations
\re{metriceom}) reduce to \be V(\phi_0) \equiv 0, \; \: R = V'(\phi_0) =
const \, . \ee In this case the metric can be put into the form
\re{011h}) with, e.g., $h=V'(\phi_0) \: (x^0)^2$.  So there is {\em
  one}\/ more solution for each zero of the potential $V$. In the
previous example of $R^2$ gravity, for example, there are two such
solutions for $\sgn \, \L \neq \sgn \, \a$ ($R=\pm \sqrt{-2\L/\a}$), one
Minkowski solution for vanishing cosmological constant $\L$, and no
solution of this type for $\sgn \L = \sgn \a$.

\subsubsection{Transforming back to original dilaton variables}
The solution found above is already in its final form when one
considers $f(R)$-type theories, cf section \ref{Secpurely}.  For a
generalized dilaton theory with an action (\ref{gdil}), on the other
hand, the above solution provides the {\em auxiliary}\/ metric only,
while the `true metric', denoted by $\bfg$ in that chapter, has to be
obtained by a Weyl rescaling according to section \ref{Sectwo}.

Since the conformal factor is a function of $x^0$ only, too, the
metric $\bfg$ also has a Killing vector and may be brought into the
form
\be \bfg = 2 du dr + {\bfh}(r) du^2 \, . \pl{011href} \ee
We now want to provide the function ${\bfh}$ as well as $\Phi(r)$.

First, in terms of coordinates $x^0$, $\Phi$ is simply
$\Phi=\U^{-1}(x^0)$, where $\U^{-1}$ denotes the inverse function to
$\U$. With the function $\O$ as defined in equation (\ref{gneu})
in terms of the potentials $\U$ and $\Z$, we then have $\bfg =
g/\O^2(\U^{-1}(x^0))$. $\bfg$ thus may be put into light cone gauge by
the coordinate transformation \be r := r(x^0) \equiv \int^{x^0}
\frac{dx}{\O^2(\U^{-1}(x))} \; \, , \quad u := x^1 \, .  \ee From this
${\bfh}$ is read off as ${\bfh}(r)=h(x^0(r))/\O^2(\U^{-1}(x^0(r)))$.

Let us illustrate this by means of string inspired dilaton gravity
(\ref{dil2}).  There
$h^{dil}(x^0) =- \L x^0 + C$, $C$ denoting the integration constant,
and $x^0 = \phi = \U(\bar \Phi)={1\over 8}{\bar \Phi}^2=
\exp(-2\Phi) \in \dR_+$. As $\r=\ln  x^0$,
equation \re{gneu}) becomes $\bfg= g / x^0$.
The above coordinate transformation $r = \ln x^0$ yields
\be {\bfh}^{dil}(r) = -\L + C \exp(-r) \pl{hDil} \ee
with $r \in \dR$.

For spherically symmetric gravity, equation (\ref{spher}),
the analogous procedure yields 
${\bfh}^{SS}(r) = 1 + 2C /r$, leading to the identification $m=-C$ in
this case (upon comparison of \re{gSS}) with the above function $h$ and
the standard form of the Schwarzschild line element).

The transition from $g$ to $\bfg$, although conformal, may have
important implications on the global structure of the resulting
theory, namely if due to a divergent conformal factor the domain of
$\bfg$ is only part of the maximally extended domain of $g$.  E.g., in
the string inspired dilaton theory, section \ref{Secstring}, the
maximal extension of $g$ is Minkowski space with its diamond-like
Carter-Penrose-diagram. The Penrose diagram of $\bfg$, on the other
hand, found by studying the universal coverings of the local charts
obtained above (cf section \ref{Secmaxext} below), is of the
Schwarzschild-type.  Although $\phi = x^0$ was defined for positive
values only in this case, $g$ remains well-behaved for $x^0 \in \dR$
and may be extended to these values. The conformal factor in the
relation between $\bfg$ and $g$, $1/x^0$, on the other hand, blows up
at $x^0 =0$.  Correspondingly $R(\bfg)$ is seen to diverge at $x^0
\equiv \exp r = 0$ (use $R={\bfh}''(r)$ and equation (\ref{hDil})) and
$\bfg$ cannot be extended in the same way as $g$. As a result, the
form of the Carter--Penrose diagrams are different.

Such a behavior is generic also in the case that 
(\ref{fi}) maps $\Phi \in \dR$ to a part of $\dR$ only, say,
e.g., to $(a,\infty)$ with an increasing $\U$. The conformal
exponent $\ln \O$ will diverge at $x^0=a$, as $\U$ was required to
be a diffeomorphism, and for potentials $\V$ that do not diverge
too rapidly for $\phi \to - \infty$ the Carter-Penrose diagrams of
$g$ and $\bfg$ differ.

We finally remark that according to \I{gneu}) the separate
deSitter solutions remain  `deSitter' also in the context
of \I{gdil}). If $\U$ has extrema, furthermore, then in some cases
there will also arise solutions of constant $\Phi$ corresponding
to the critical values of $\U$.

\subsubsection{In Einstein--Cartan variables}
Describing the (general) metric \rz lcgauge by a {\em zweibein},
clearly one of the two one-forms $e^\pm$
has to have a vanishing coefficient function in front of
$dx^0$. A globally attainable choice of the Lorentz frame
then brings the {\em zweibein}\/ into  the form:
\be e^+ = dx^1 \; , \quad e^-=dx^0 + \mbox{$\2$} k\, dx^1  \, . \el gauge2

Let us now determine the function $k$ in the gauge \rz gauge2 by means
of the e.o.m.\ of \rz grav2 with $W=\7 V(\phi)$, which is the
Einstein--Cartan version of \re{Proto3}).  (As remarked already above,
we do not make use of \re{LemmaF}), or its Einstein--Cartan analog, in
this subsection, which would further simplify the subsequent
considerations, but, as demonstrated, which is not really necessary.)
Beside torsion zero and $-2d\o = V' \e$ (or, by use of (\ref{2dR}), 
$R=V'(\phi)$) the e.o.m.\ are
\ba d\phi + \varepsilon_{ab}X^a e^b &=& 0 \, \nn 
dX_a + \varepsilon_{ab} X^b \o + \mbox{$\2$} \varepsilon_{ab} \,
V(\phi) \, e^b &=&0 \pl{var} \, . \ea {}From $De^+=0$ we learn at once
that $\o_0=0$. Then the $x^0$--components of \rz var (with $a=-$)
become $\6_0 X^+ = 0$ and $\6_0 \phi = X^+$, respectively. This yields
\be X^+=F(x^1) \; , \quad \phi= F(x^1) x^0 + G(x^1) \, , \el Loes for
some functions $F$ and $G$. Suppose next that $X^+ \neq 0$. Then, by
the residual gauge transformations of \re{gauge2}) we can put $F$ to
one (by a Lorentz transformation and a subsequent diffeomorphism) and
$G$ to zero (by an $x^1$-dependent shift of the origin of $x^0$).
This yields $X^+=1$ and \re{phix0}).

Next, the equations \rz var may be seen to imply
$V(\phi)d\phi= d(X^2)$; thus $X^2 = 2 X^- = h(\phi)$ with the
function $h$ as defined in \ry h . The
one-components of the field equations \rz var with ($a=-$) now reduce
to,
\ba \mbox{$\2$} V(x^0) + \o_1 &=&0 \, , \pl{G+} \\
X^- - \mbox{$\2$}  k &=& 0 \, , \pl{G3}
\ea
which finally determines $\o_1$ and yields $k=h(x^0)$.

\subsection{Target space method, the general case\pl{PSMlocalsolution}}

\subsubsection{Torsion--free theories}
In section \ref{SecPSM} we have shown that pure gravity theories may
be described by a Poisson $\s$--model (PSM), equation \I{PSM}), with
Poisson bracket \I{P}). In terms of Casimir-Darboux (CD) coordinates
the action was seen to take the form \I{trivact}). In the present case
with a three-dimensional target space and one Casimir coordinate
$C=\wt X^1$, the field equations of  \I{trivact}) take the form (they
also follow from specializing \I{eoma}) and \I{eomb}), certainly):
\ba d \wt X^1 =0 \; , && d A_{\wt 1}=0  \pl{eom2a} \\
A_{\wt 2} = d \wt X^3 \; , && A_{\wt 3} = -d \wt X^2 \; , \pl{eom2b}
\ea while the remaining two field equations $d A_{\wt 2} =d A_{\wt
  3}=0$ are redundant obviously. In this form the solution of the
field equations becomes a triviality: Locally \re{eom2a}) is
equivalent to \be \wt X^1= \mbox{const ,} \quad A_{\wt 1}= df \el
Loesung where $f$ is some arbitrary function on $M$, while \re{eom2b})
determines $A_{\wt 2}$ and $A_{\wt 3}$ in terms of the otherwise
unrestricted functions $\wt X^2$, $\wt X^3$.

This is already the general local solution to the field equations (within
patches of applicability of the chosen CD coordinates). Still, the
solution should be transformed back to the gravitational variables.

We thus need the relation between $A_i$, $X^i$, used in the
identification \I{multi1}) and in the definition of the bracket \I{P}),
on the one hand, and, on the other hand, $A_{\wt i}$ and $\wt X^i$,
for which the solution is available now.  According to equations
\I{CDLor}) and \I{Ctorless}) a possible choice of CD target space
coordinates is \be \wt X^i :=
\left({\2}\Big[X^2\!-\!\!\int\limits^{\;\;\;\phi}\!\! V(z) dz\Big]
  \;,\; \ln|X^+| \;,\; \phi\right) \, , \el Xtildeneu valid in patches
on $N$ with $X^+ \neq 0$. {}From $A_i= {\6 \wt X^j \over \6 X^i}
A_{\wt j}$, we then easily infer \be e^+ \equiv A_- = X^+ A_{\wt 1} \,
, \quad e^- \equiv A_+ = {1 \over X^+ } A_{\wt 2} + X^- A_{\wt 1} ,.
\el relationneu By means of \re{eom2b}) and (\ref{Loesung}) the metric
$g=2e^+e^-$ thus becomes \be g=2 d\phi df + X^2 \, df df \, , \el
gX3neu where $X^2 = \int^{\phi} \, V(z) dz + const= h(\phi)$
according to \re{Xtildeneu}) and \re{Loesung}).

Using, finally, $\phi$ and $f$ as coordinates $x^0$ and $x^1$ on $M$,
this takes the form \re{011h}) with the function $h$ as defined
above (and coinciding with equation \re{h})).

\subsubsection{Theories with general torsion dependence}
\pl{Hui}
Let us now discuss the solution to the field equations for a general
potential $W$ in \I{grav}), using this opportunity to present also a
systematic construction of a CD--coordinate system.

By definition a Casimir coordinate $\wt X^1$ is characterized by the
equation $\{ \wt X^1, \cdot \}=0$ $\Leftrightarrow (\6 \wt X^1 / \6
X^i) \CP^{ij}=0$.  For $j=3$ the latter implies that $\wt X^1$ has to
be a Lorentz invariant function of $X^\pm$, i.e.  \be \wt X^1=\2
C[X^2,\phi] \el tildeX1 for some two-argument function $C=C(u,v)$.
For $j=\pm$ we then obtain \be 2 W(u,v) C,_u + C,_v =0 \, \el CDiff
where the comma denotes differentiation with respect to the
corresponding argument of $C$. As \rz CDiff is a first order
differential equation, it may be solved for any given potential $W$
locally, illustrating the general feature of a local foliation of
Poisson manifolds for the special case \ry P . An important
consequence of \rz CDiff is the relation $C,_u \neq 0$. This follows
as on the target space (!) we have $dC \neq 0$ (by definition of a
target space coordinate function): according to equation \ry CDiff ,
$C,_u =0$ at some point would imply $C,_v=0$, in contradiction to $dC
\neq 0$. Certainly, as $C,_u$ is a function on $N$, in contrast to
$dC$ it remains nonzero also upon restriction to submanifolds
$C=const$.

Trying $C,_u(u,v) \equiv 1$ as a simplest nontrivial choice, we see
that the defining equation  for $C$ may be integrated immediately in
the torsion--free case $W(u,v)=V(v)/2$ (and only then) to yield
\be C= X^2 -  \int^{\phi} V(z) dz \, ,  \el CV
in coincidence with the first entry of \re{Xtildeneu}) or with
equation \re{Ctorless}).

Using the method of characteristics, the general case
\rz CDiff may be reduced to an ordinary
first order differential equation: We may express the lines of constant
values of the function $C$ in the form
\be {du \over dv}=2W(u,v) \, .\el Diff
The constant of integration of this  equation is a
function of  $C$ in general; however, as clearly any function of a Casimir is
 a Casimir again, we may just identify the integration constant with $C$.
To illustrate this, we choose
\be 2W(u,v) := V(v) + T(v) u  \, \, . \el T
We then obtain from \rz Diff
\be u = \left[ \int^v V(z) \exp \left(-\int^z T(y) dy \right) dz +
const(C) \right]
\, \exp \left(\int^v T(x) dx \right) \, , \el u
where the lower boundaries in the integrations over $T$ coincide.
Upon the choice $const(C) :=C$, \rz u gives
\be C (u,v) = u \, \exp \left(-\int^v T(x) dx \right) -
\int^v V(z) \exp \left(-\int^z T(y) dy \right) dz   \, . \el C
The integrations on the right-hand side should be  understood as definite
integrals with somehow fixed, $C$--independent lower boundaries. Different
choices for these boundaries rescale $C$ linearly.

Let us specialize \rz C to some cases of particular interest.  In the
case $T \equiv 0$, describing torsion--free gravity it yields
\re{CV}).  For the Katanaev-Volovich (KV) model of 2D gravity with
torsion, equation \I{KV}), as a second example, \rz C and \rz WKV
yield upon appropriate choices for the constants of integration and a
rescaling by $\a^3$: \be C^{KV}= -2 \a^2 \exp (\a \phi) \left( W^{KV}
  +{2 \phi \over \a} - {2 \over \a^2} \right) \, . \el CKV Here $C,_u
(u,v) = \a^3 \exp (\a v)$.

The remaining task is to find Darboux coordinates. On patches with
either $X^+ \neq 0$ or $X^- \neq 0$ this is a triviality almost:
According to the defining relations \I{P}) of the Poisson brackets, we
have $\{X^\pm,\phi\}=\pm X^\pm$, so obviously $\pm \ln |X^\pm|$ and
$\phi$ are conjugates. Altogether therefore \be \wt X^i :=
(\mbox{$\2$}C,\pm \ln|X^\pm|,\phi) \, \el Xtilde forms a CD coordinate
system on regions of $N$ with $X^\pm \neq 0$, respectively. 

These considerations complete our discussion of the Poisson bracket
\re{P}) of section \ref{Secpoisson}. Note that, taken all together, 
 the present formalism reduces the {\em field theoretical\/} task of finding
 solutions to Lagrangian field equations to a problem on a three--
 (and thus finite--) dimensional space $N$. In the case of torsion--free
 theories, the latter problem is quite trivial, moreover, (cf our
 discussion above as well as the one in section  \ref{Secpoisson}). 
In the case of a theory of general torsion dependence, equation \re{grav}),
on the other hand, the (more involved) problem of solving the 
field equations has been reduced to the determination of $C$, which is
as much as can be done in this case. 

So, now we just have to repeat the steps following equation
\re{Xtildeneu}) in the more general setting of a potential $W(u,v)$.
However, as we do not want to restrict ourselves to potentials e.g.\ 
purely linear in $u$, as in equation \re{T}), we do not know $C$ in
explicit form. Nevertheless, also in the case of a completely general
Lagrangian \I{grav}), the metric takes the form \re{011h}) locally.
Moreover, $h$ may be determined in terms of the Casimir function
$C(u,v)$, and the Killing field $\6/\6_1$ will be shown to be a
symmetry direction of all of the solution. (This is not a triviality
as, e.g., the connection $\o$ is {\em not\/} determined (up to Lorentz
transformations) by the metric $g$ for nonvanishing torsion.)

 For the sake
of brevity we display the calculation for both of the sets
\re{Xtilde}) of CD--coordinates simultaneously. This means that in the
following we restrict our attention to local solutions on $M$ that map
into regions of $N$ with $X^+ \neq 0$ {\em or} $X^- \neq 0$.

Inserting into the analog of equations \re{relationneu}) the solutions
\re{eom2b}), (\ref{Loesung}) and reexpressing thereby $\wt X^i$ in
terms of the original fields $X^i$ again, \re{Xtilde}), we obtain
\ba e^\pm \equiv A_\mp &=& C,_u X^\pm df \nn
e^\mp \equiv A_\pm &=& \pm {1 \over X^\pm } d\phi + C,_u X^\mp df \nn
\o \equiv A_3 &=& \mp d(ln |X^\pm|) - C,_u W  df \, ,
\pl{relation} \ea
where in the last line we used \re{CDiff}).
For the metric $g=2e^+e^-$ this yields
\be g=\pm 2C,_u \, d\phi  df + \bar h(\phi,C) \,  df df \,,
\el gX3
with \be \bar h(\phi,C) :=  C,_u{}^2 \, \cdot \,  X^2 \; .\el hbar
Here $X^2$ is a function  of $\phi$ and $C$ by inverting
the field equation 
$C(X^2,\phi) = const$, whereas $C,_u$ in \rz hbar
and \rz gX3 is, more explicitly,  $C,_u \left(X^2(\phi,C),\phi\right)$.
Note that according to the context, $C$
either denotes a function of $X^i$ or the constant which it equals due to
the first equation \re{eom2a}).

Now again we want to fix a gauge.  {}From \rz gX3 we learn that
 $C,_u d\phi \wedge df \neq 0$, because otherwise  $\det g = 0$. This
 implies that we may choose $C,_u d\phi$ and $df$ as coordinate
 differentials on $M$. Let us therefore fix the diffeomorphism invariance
of the underlying gravity theory by setting
\ba \int^{\phi} C,_u[X^2(z,C),z] dz &:=& x^0 \, , \pl{x0} \\ f &:=&
\pm x^1 \, .\pl{x1} \ea In this gauge, $g$ is seen to take the form
\re{011h}) again with
\be h(x^0) = \bar
h(\phi(x^0),C) \, , \ee where $\phi(x^0)$ denotes the inverse of
\re{x0}).
The local Lorentz invariance may be fixed by means of
 \be X^\pm := \pm 1 \, ,\el Lorentz
finally. Besides \re{Lorentz}, \ref{x0})
the complete set of fields then takes the form
\be  e^\pm = C,_u dx^1 \, , \; e^\mp = {dx^0 \over C,_u} + \mbox{$\2$} X^2 \,
C,_u dx^1 \, , \;
\o =  \mp W \, C,_u dx^1 \, , \el inKarte
and $X^\mp= \mp\mbox{$\2$} X^2$. Here again $X^2$, $C,_u$, and $W$
depend on $\phi(x^0)$ and $C=const$ only, $C$ being the only
integration constant left in the local solutions.

In the torsion-free case, considered already above, $C,_u=1$,
$\phi(x^0)=x^0$, and, cf equations \re{hbar}, \ref{h}, \ref{CV}),
$X^2=h(x^0)=\int^{x^0}V(z)dz +C$, thus reproducing the results
obtained before. For the KV--model \I{KV}), as an example for a theory
with torsion, on the other hand, the above formulas yield $x^0 = \a^2
\exp (\a \phi)$, $x^0 \in \dR^+$, and \be h^{KV}(x^0) = {1 \over \a}
\left\{C x^0 - 2 (x^0)^2 \, \left[(\ln x^0-1)^2+1- \Lambda\right]
\right\} \, . \el hKV   Certainly $R=h''(x^0)$ does not
hold any more here; instead we find \be R = -{4 \over \a} \ln \left({x^0
    \over \a^2} \right) \, \, . \pl{RKV} \ee This is obtained
most easily by concluding $R = -4\phi$ from (the Hodge dual of) the
three--component of \re{eomb}).

In the above we captured the solutions within regions of $M$ where
either $X^+ \neq 0$ or $X^-\neq 0$. Clearly, in regions where $X^2
=2X^+X^-\neq 0$ the two charts \re{inKarte}) must be related to each
other by a gauge transformation, i.e., up to a Lorentz transformation,
by a diffeomorphism. However, one of these two charts extends smoothly
into regions with only $X^+\neq 0$ (but possibly with zeros of $X^-$),
and $(+\leftrightarrow -)$ for the other chart.\footnote{In the subsequent
  chapter, 
  regions with $X^2\neq 0$ will be called `sectors', while patches
  with merely $X^+\neq 0$ {\em or} $X^- \neq 0$, generically containing
  several sectors, will be called `blocks'.} In this way the above
mentioned diffeomorphism may serve as a gluing diffeomorphism, 
allowing to extend the generically just local solution \re{011h})
to one that applies wherever $X^+$ and $X^-$ do not vanish
simultaneously. Let us remark here, furthermore, that the two
representatives \re{inKarte}) are mapped into each other by \be
e^+ \longleftrightarrow e^- \; ,\quad \omega \longleftrightarrow
-\omega \; ,\quad X^+ \longleftrightarrow -X^- \; ,\quad \phi
\longleftrightarrow \phi \, . \el Strafo This transformation
reverses the sign of the action integral \I{grav}) only and
therefore does not affect the equations of motion. From \re{Strafo}) it
is obvious that the above gluing diffeomorphism (cf also equation
\re{glue}) below) maps one
set of null lines onto the respective other one, leaving the form
\re{011h}) of $g=2e^+e^-$ unchanged. It corresponds to a discrete
symmetry of \re{011h}) (called 'flip' in section \ref{Secmaxext}
below), which is independent from the continuous one generated by
$\6/\6_1$. Further details shall be provided in section
\ref{Secmaxext}.

\subsubsection{Solution in the neighborhood of saddle-points}
We are left with finding the local shape of the solutions in the
vicinity of points on $M$ that map to $X^+=X^-=0$. (This is true for
all the approaches applied so far.) Here we have to
distinguish between two qualitatively different cases: First,
$C=C_{crit}\equiv C(0,\phi_{crit})$, where $\phi_{crit}$ has been chosen
to denote the zeros of $W(0,\phi)$, and second, $C \neq C_{crit}$.
The special role of the critical values of $C$ comes about as
the Poisson structure $\CP$ vanishes  precisely at $X^+=X^-=0$ on the
two-surfaces $C=C_{crit}$,  cf equation  \ry crit . 

We start treating the noncritical case $C \neq C_{crit}$: For this,
one could attempt to construct a CD coordinate system valid in a
neighborhood of a (noncritical) point $X^+=X^-=0$.  At least in the
torsion--free case this may be done in an explicit way, but the
formulas are somewhat cumbersome. We therefore follow a somewhat less
systematic, but simpler root here: {}From the three-component of \rz
eoma it is straightforward to infer that points with $X^+=X^-=0$ are
saddle points of $\phi$. This suggests to replace the gauge conditions
\re{x0}), (\ref{x1}) by an ansatz of the form \ba \int^{\phi}
C,_u[X^2(z,C),z] dz &:=& x y + a \, , \pl{x0neu} \\ f &:=& b \ln x \,
,\pl{x1neu} \ea where $a$ and $b$ are constants to be determined
below. As a first justification of \re{x0neu}), \re{x1neu}) we find
$C,_u d\phi \wedge df$ to be finite and nonvanishing on $(x,y) \in
\dR^2$. Implementing the above conditions in \ry gX3 , the metric
becomes \be g=2b \, dx dy + b {2xy + b \, h(xy + a) \over x^2 } \,
(dx)^2 \, , \el saddle where $h$ is the function defined in equation
(\ref{hbar}) or, in the torsion--free case, in \ry h .
For generic values of $a$ and $b$ \rz saddle is singular at $x=0$.
However, the choice \be a := h^{-1}(0) \qquad b:= -
2/h'\left(h^{-1}(0)\right) \el ab is seen readily to yield a smooth
$g$!

The singularity of the gauge choice \rz x1neu at $x=0$ was devised
such that it compensated precisely the singularity of the CD
coordinate system at $X^\pm=0$, used to derive \ry gX3 .
The charts \rz saddle provide a simple alternative to a generalized
Kruskal extension (cf section \ref{Secmaxext}). For Schwarzschild,
$h(r)=1-2m/r$, it is a global chart (as $h^{-1}(0)$ is single-valued),
in the more complicated Reissner-Nordstr\"om case,
$h(r)=1-2m/r+q^2/r^2$, the constant $a$ may take one of the two values
$r_{\pm}=m \pm \sqrt{m^2-q^2}$ and \rz saddle provides a local chart
in the vicinity of the respective value of $r=x^0$.  A generalization
of the right-hand sides of \rz x0neu and \rz x1neu to $F(x) y + G(x)$
and $\int^x dz/F(z)$, respectively, with appropriate functions $F$ and
$G$, allows even for global charts of (two-dimensional) spacetimes of
the form $\dR\, \times$ `null-lines' (cf also the considerations in
section \ref{Secattain} above!).

Such global charts may be obtained quite systematically within the
gauge fixing approach of section \ref{Secgauge}. We will come back to
this issue in section \ref{SecRN} below.

For \rz saddle to exist, it is decisive that the corresponding zero $a$
of $h$ is simple, cf the second equation \ry ab . For noncritical
values of $C$, e.g., all zeros of $h$ are simple.  This is
particularly obvious for torsion--free theories, where $h=2X^+X^-$ and
$h'(\phi)=V(\phi)$, but holds also in general.  If $C \in \{C_{crit}\}$,
on the other hand, there exist zeros of $h$ of higher degree.  Then
the spacetime manifolds $M$ with varying $X^i$ do not contain the
critical points $X^i=(0,0,\phi_{crit})$.  This may be seen in two
different ways: First, studying extremals running towards such a
point, one finds the point to be infinitely far away.
Second, from the field equations point of view: Taking successive
derivatives $\6/\6 x^\m$ of the equations \re{eoma}) and evaluating them
at the critical points, we find \be X^- \equiv 0 \, ,\,\, X^+ \equiv 0
\, ,\,\, \phi \equiv \phi_{crit}=\hbox{const.} \el momdeSitter

\rz momdeSitter corresponds to additional, separate solutions of
the field equations. These are the generalizations of the
solutions discussed at the end of section \ref{Secmetric} for the
case of torsion--free theories.

From the point of view of PSMs, such solutions come as no surprise.
More or less by definition the critical points \rz crit of the target
space constitute zero dimensional symplectic leaves. It is a general
feature of PSMs, verified here explicitly in \rz momdeSitter and the
first equation of \re{Loesung}), that the image $X(x)$ of the map from
the worldsheet or spacetime $M$ into the target space $N$ has to lie
entirely within a symplectic leaf $\CS \subset N$ (cf section
\ref{SecglobalPSM} below for further details).

The remaining field equations \re{eomb}), which are, more
explicitly, \be De^a =0 \, ,\,\, d\omega = -W,_v(0,\phi_{crit})
\e\,, \el deSitter show that the solutions \rz momdeSitter have
vanishing torsion and constant curvature all over $M$. The metric
for such a solution can be brought into the form \re{011h}), too,
with $h(x^0)=W,_v(0,\phi_{crit}) \cdot [(x^0)^2 +d] \equiv
{\partial W \over \partial \phi}(0,\phi_{crit}) \cdot [(x^0)^2 +d],
d=const$. This in turn determines the zweibein and spin connection
up to Lorentz transformations.

 As remarked in section \ref{Secattain},  the coordinate
system underlying \re{011h}) is adapted to a Killing direction. In
the case of deSitter space, the Killing fields form a {\em
three\/}-dimensional vector space, in which the vector fields
$\6_1$ corresponding to the different choices  $d=-1,0,1$ in $h$
are pair-wise independent (for a basis take, e.g., two of them and
their Lie bracket).
\footnote{As we will find below, the universal covering solutions
of 2d deSitter or anti-deSitter space may be described by a
ribbon-like conformal diagram.  Fig.~\ref{fig:fig9} displays such
Carter-Penrose diagrams for deSitter space where the Killing lines
$x^0 = const$ have been drawn; {\bf J1} corresponds to $d<0$, {\bf
J2} to $d=0$, and {\bf J3} to $d>0$.}

\subsubsection{Inclusion of Yang--Mills fields \pl{SecYMlocal}}
In section \ref{Secinclumat} a 2d gravity--YM action was seen to allow
a description in terms of a PSM with the bracket \re{PoiGYM}) in an
$(n+3)$-dimensional target space $N$. We now need to analyze this
bracket further only; the local solution is then determined easily.

As any Poisson tensor, also \rz PoiGYM permits CD coordinates in the
neighborhood of generic points. If the rank of the chosen Lie algebra
is $r$, the rank of the Poisson tensor is $n-r+2$ and there will
be $r+1$ such Casimir coordinates.  Correspondingly there will be
$r+1$ field equations \rz eom2a and $n-r+2$ field equations \ry eom2b
. In the CD--coordinates the symmetries \re{syma}) and \re{symb}) take
a very simple form, and it is a triviality to realize that again the
local solutions are parameterized by  integration constants only, the
number of 
which coincides with the number of independent Casimir functions.
Note that in the present context the symmetries \re{syma}, (\ref{symb})
entail diffeomorphisms, Lorentz transformations, and nonabelian gauge
transformations all at once. So, more or less without performing any
calculation, we obtain the result that the local solutions are
parameterized by $r+1$ constants now. In the the particular case of a
$U(1)$-field considered in \cite{KunstU(1)}: $n=1, r=1$, and $r+1=2$.

The Poisson structure \rz PoiGYM has a very particular form: First the
$E^j$ span an $n$-dimensional Poisson submanifold of $N=\dR^{n+3}$.
Second, the Poisson brackets between the gravity coordinates
$(\varphi^a,\phi)$ and the coordinates $E^i$ of this submanifold
vanish. And, last but not least, the Poisson brackets between the
`gravity'-coordinates close among themselves up to a {\em Casimir}
function of the $E$-submanifold. With this observation it is a
triviality to infer the local form of $g$ of $L^{gravYM}$ from the
results of previous subsections: On-shell $E^jE_j$ is some constant
$C^{E}$. So in all of the formulas of, e.g., section \ref{Hui} we
merely have to replace $W$ by $W+C^{E} /\a$. Thus the metric $g$ again
takes the form \rz 011h locally, where now $h$ is parameterized by
{\em two\/} constants $C^{E}$ and $C^{grav}$ (`generalized Birkhoff
theorem for 2D gravity--YM-systems').  For torsion--free theories,
e.g., \be h(x^0)=\int^{x^0}_c V(z) dz + \2 C^{E} \int^{x^0}_c
\frac{dz}{\alpha(z)} + C^{grav} \, , \ee where $c$ is some fixed
constant (and again $2W(\cdot,\cdot)=V(\cdot)$).

(Here we ignored additional exceptional solutions of the type
\re{momdeSitter}) for simplicity, corresponding to maps into lower
dimensional symplectic leaves.)

\subsubsection{The model \I{gravtriv}) \pl{Sectriv}}
For the sake of completeness we finally turn to the field equations of
an action $F(R)$, where, however, torsion is {\em not\/} required to
be zero. This theory may be described by an action of the form
\I{gravtriv}). From its field equations we learn that
$\phi=\phi_{crit}=const$, where again $W(\phi_{crit})=0$. The
curvature is determined by \be d\o=-2 W'(\phi_{crit}) \e \pl{Rtriv}
\ee or by $R=2 W'(\phi_{crit})$. So this is a space of constant
curvature.  However, torsion need not be zero now.

It seems that one is left with an {\em arbitrary}\/ metric
$g=2e^+e^-$. As we learned in section \ref{Secattain}, by an appropriate
choice of coordinates, any 2d metric may be brought into the form
\re{lcgauge}). Then, up to gauge transformations $\o \to \o + d \a$,
$\o$ is determined by equation \re{Rtriv}): $\o=W'(\phi_{crit}) \, x^0
dx^1$. Now all field equations are satisfied. Diffeomorphism
invariance has been used to obtain \re{lcgauge}). The local symmetry
$e^+ \to \exp{[\l(x)]} \:  e^+$, $e^- \to \exp{[-\l(x)]} \: e^-$, which is
    independent of the symmetry $\o \to \o + d \a$ in the specific
    action \I{gravtriv}), does not affect $g$, furthermore. At least
    if there are no additional local symmetries hidden in \I{gravtriv}),
    there thus remain propagating modes. However, in a sense, the
    theory seems trivial, as the metric $g$ is not restricted by the
    field equations at all.

\subsection{Supergravity \pl{Secsugraloes}}
We finally solve the field equations for the supergravity extensions
of \I{gdil}) presented in section \ref{Secsugra}. Here the formulation
as (graded) PSM, presented in section \ref{Secextsugra}, will turn out
most powerful.  Beside its
notational compactness, the main advantage of the formulation \I{PSMsus})
as opposed to \I{Izq}) is its inherent target space covariance. Thus
we again may change coordinates in the target space of the theory so as to
simplify the tensor $\CP$, while the field equations in the new
variables still will take the form \I{eomsus}) (but then with the
transformed, simplified Poisson matrix).

We will first use this
method to show that locally the space of solutions to the field
equations of \I{Izq}) modulo gauge symmetries is just
one--dimensional.  Thereafter we will provide a representative of this
one--parameter family in terms of the original variables used in
\I{Izq}). As a byproduct we will find the somewhat surprising result
that locally the space of
solutions is {\em identical}\/ to the one of the bosonic theory; all the
fermionic fields may be put to zero by gauge transformations.

As mentioned already in section \ref{Secpoisson},
locally (more precisely, in the neighborhood of a generic point) any
(bosonic) Poisson manifold permits Casimir--Darboux (CD)
coordinates $(C^I, Q^A, P_B)$ \cite{Weinstein}. According to, e.g.,
\cite{Henneaux}, Darboux coordinates exist also for supersymplectic
manifolds (manifolds with a graded, nondegenerate Poisson bracket). We
thus assume that CD coordinates exist in the case of general, graded
Poisson manifolds. (At least in the case of the bracket \I{PBsus}) they
do: Although we did not succeed to find such coordinates explicitly as
of today, we will provide a Casimir function below. On its
level surfaces the Poisson bracket is (almost everywhere)
nondegenerate and the result on supersymplectic manifolds may be
applied.)

We now are in the position to show that locally there is only a
one-parameter family of gauge inequivalent solutions to the field
equations of \I{Izq}). For this purpose we only need to know about
the local {\em existence}\/ of CD coordinates $\wt X^i$.  (These
coordinates, to repeat, are coordinates on the target
  space, not on the worldsheet or spacetime. A  change of coordinates
$X^i \to \wt X^i$, which 
  induces the change $A_i \to \wt A_i \equiv A_j (\dl
  X^j/\partial \wt X^i)$, corresponds to a (local) change of field
  variables from the point of view of the 2d field theory.)
In these CD coordinates,
the second set of field equations \I{eomsus}) reduce to
  $d \wt A_i =0$. Thus locally $\wt A_i = d f_i$ for some functions
$f_i(x)$. However, the local symmetries \I{symmsus}) also simplify
dramatically in these new field variables: $\d \wt A_i = d \wt \e_i$.
This infinitesimal formula may be integrated easily showing that all
the functions $f_i$ may be put to zero identically.  But then we learn
from the first set of the field equations \I{eomsus}) that all the
functions $\wt X^i$ are constant. All of the constant values of $Q^I$
and $P_I$ may be put to an arbitrary value by means of the residual
gauge freedom in \I{symmsus}) (constant $\ep_i$). What remains as gauge
invariant information is only the constant values of the Casimir
functions $C^I$ again. Since the Poisson tensor \I{PBsus}) has rank four
(almost everywhere),\footnote{To determine the rank of the supermatrix
 $\CP^{ij}$, we may concentrate on the rank of the the two by two
  matrix $\CP^{\a\b}$ and the three by three matrix in the purely
  bosonic sector; $\CP^{a \a}$, being linear in Grassmann variables,
  cannot contribute to the rank of the matrix, cf \cite{Henneaux}.}
the model defined by equation \I{Izq}) has just one Casimir function and
its space of local solutions is thus indeed one--dimensional only.

The local solution obtained above in terms of CD--coordinates may be
transformed back easily to any choice of target space coordinates. We
find that also in the original variables: $A_i \equiv 0$ and $X ^i =
(const)^i$, where the latter constants may be chosen at will as long as
they are compatible with the constant values of the Casimir(s) $C^I$,
which characterize the (local) solution. As it stands, this solution
corresponds to a solution with vanishing zweibein and metric. In a
gravitational theory, the metric (and zweibein) is required to be a
nondegenerate matrix, however. The vanishing zweibein is a result of
using the symmetries \I{symmsus}), which, in contrast to
diffeomorphisms, connect the degenerate with the nondegenerate sector
of the theory. (We will come back to this problem in more detail and
from another point of view in chapter 
\ref{Sectowards}. However, we need to discuss it
already here somewhat, in order to depict the solution in a
physically acceptable gauge.) A similar phenomenon occurs, e.g., also
within the Chern--Simons formulation of $(2+1)$--gravity 
\cite{Wit}. The problem may be cured by applying a gauge
transformation \I{symmsus}) to the local solution $A_i \equiv 0$ so as
to obtain a solution with nondegenerate zweibein. However, in contrast
to Chern--Simons theory, where the behavior of $A$ under {\em
  finite}\/ (nonabelian) gauge transformations is known, the
infinitesimal gauge symmetries \I{symmsus}) cannot be integrated in
general (except for the case where the Poisson tensor is (at most)
linear in the fields $X^i$ and the theory reduces to a (non)abelian
gauge theory). We thus need to introduce one further step.

For the Poisson brackets \I{PBsus}) a possible choice for a Casimir
function $C$, $\{C, \cdot \} \equiv 0$, is \be C = \2 X_a X^a + 2 u^2
- \frac{i}{8} u' \chi_\a \chi^\a \, . \label{Csus} \ee This expression
basically coincides with the Casimir found in the bosonic theory: as
there the last two terms are the integral of
$\CP^{{\underline{0}}{\underline{1}}}$ with respect to $\phi$,
$\chi_\a \chi^\a$ Poisson commuting with all bosonic variables. We now
choose new coordinates on (patches of) $N$ (where $X^a \neq 0$)
according to \be \bar X^i := (C,\ln |X^+|, \chi^\a, \phi) \, .
\label{new} \ee
(An argumentation similar to the following one may be applied also if
$\ln |X^+|$ is replaced by $X^+$ in \I{new})). $(C,\ln |X^+|, \phi)$
provides a CD coordinate system in the purely bosonic sector, cf
equation \re{Xtilde}); however, in the five--dimensional target space $N$,
these coordinates are far from being CD, several brackets still
containing the potential $u(\phi)$. It thus seems rather difficult to
solve the field equations for field variables \I{new}). However, from
our considerations above, we know that up to PSM gauge transformations
the local solution always takes the from $\bar A_i \equiv 0$, $C =
const$, $\ln |X^+|= \chi^\a = \phi =0$. In these field variables,
induced by the coordinates \I{new}), it is now possible to gauge
transform explicitly to a solution with nondegenerate zweibein and,
{\em simultaneously}, the resulting solution may be transformed back
to the original variables used in \I{Izq}) --- this is possible since,
in contrast to the CD coordinates $\wt X^i$ used above, the
coordinates $\bar X^i$ are known explicitly in terms of the original
variables.

It is straightforward to verify that {\em on the above solutions}\/
the infinitesimal gauge transformations \I{symmsus}) with $(\bar \ep_i)
:= (\ep_C, \bar \ep_+, 0, 0, 0)$ are simply: \be \delta A_C = \delta
\ep_C \, , \; \delta \bar A_+ = d \bar \ep_+ \, , \; \delta \phi =
\bar \ep_+ \, , \label{symm2} \ee with {\em all}\/ other variations
vanishing. Note that, e.g., in the second relation we dropped terms
proportional to $\bar A_\a$, since $\bar A_\a$ can be {\em kept}\/
zero consistently by the above transformations due to $\delta \bar
A_\a = 0$. It is thus possible to {\em integrate}\/ the gauge
symmetries \re{symm2}): $A_C \to A_C + df_1$, $\phi \to \phi + f_2$,
$\bar A_+ \to \bar A_+ + df_2$ ({\em all}\/ other fields remaining
unaltered) where $f_{1,2}$ is an arbitrary pair of functions on the 2d
spacetime. The degenerate solution is then transformed into $A_C =
df_1$, $\bar A_+ = df_2$, $\bar A_\alpha= \bar A_\phi = 0$, $C =
const$, $\ln |X^+|= \chi^\a =0$, and $\phi = f_2$. Using $f_1$ and
$f_2$ as coordinates $x^1$ and $x^0$ on the worldsheet, respectively,
and transforming these solutions back to the original variables
\I{multisus}) (using $A_i = \bar A_j (\dl \bar X^j/\partial
X^i)$), we obtain: \ba &(e^+,e^-,\o) = (dx^1 , dx^0 + \2 h(x^0) dx^1,-
h'(x^0) dx^1) \, ,& \; \nn &(X^+,X^-,\phi) = (1 , \2 h(x^0) , x^0) \,
, & \ea where $h(x^0) \equiv C - 2 u^2(x^0)$, $C$ being the constant
value of the Casimir \re{Csus}).  {\em All}\/ the fermionic variables
vanish identically.

Thus, up to gauge transformations, the local
solution agrees completely with the one found in the purely bosonic
dilaton theory \I{gdil}). This applies at least to those patches
where the above coordinate systems are applicable. Since, e.g., all
the fixed points of the supersymmetric bracket \I{PBsus}) lie entirely
within the bosonic sector of the target space, we expect that there
are also no exceptional solutions, containing (necessarily)
nonvanishing fermionic fields.

So the characterization of the dynamics of the general supersymmetric
extension of \I{gdil}) turns out to be less difficult than it appeared
at the time when \cite{Parker} was written (cf the remarks
following equation (50) of that paper). Rather, it appears that the
supersymmetric extension of \I{gdil}) is trivial (on-shell), at least
at the classical level.

It would be interesting to check this result by some other method and
to possibly establish it in a less indirect way. It is to be expected,
moreover, that a similar result holds also on the quantum level.

Let us finally remark that the supersymmetric extension may still be
of some value even on the purely classical level: As remarked already
in section \ref{Secsugra}, in \cite{Parker} it was used to establish
some kind of positive-mass theorem under quite general circumstances
(and including nonsupersymmetric theories).  Further investigations of
2d dilatonic supergravity theories, including generalizations to
theories with nontrivial torsion and a comparison to the existing
literature \cite{Susyandere,Rivelles,Ertl,Susyrelated}, is in
preparation \cite{prep}.

\section{Dilaton theories with matter \pl{Secmatsol}}
In this section we will present several models with propagating modes
allowing an exact solution of their field equations. In the first part
we will consider (solvable) generalizations of the CGHS model. To
achieve solvability, we need to restrict the potentials in the
generalized dilaton part \re{gdil}). In the second part, the
potentials are left arbitrary. Solvability is achieved then only upon
restriction to chiral fields.  Some further related and in part
complementary work is \cite{Rest}.

\subsection{Solvable generalizations of the CGHS model}
In this section we consider the coupled gravity matter system \re{I}).
The field equations were derived in section \ref{Secfield} already.
This coupled system of field equations is not solvable in full
generality (cf already the, in comparison, relatively simple, but
still highly nontrivial system of spherically reduced gravity with
scalar fields, mentioned at the end of section \ref{Secspher}).
As remarked already in section \ref{Sectwo}, the main reason for the
possibility to solve the field equations of the CGHS model may be
found in the fact that the redefined metric $g$, cf equation \re{gneu}),
is {\em flat}.

This idea will be generalized here: We require the potential
$\V(\Phi)$ in \re{gdil}), and, if nonzero, also $\alpha(\Phi)$, to be
determined by the freely chosen functions $\U(\Phi)$ and $\Z(\Phi)$ up
to the choice of two resp.\ three (real) constants $a$, $b$, and
$\a_0$ by means of the relation(s) \ba
\V(\Phi)&=&\exp\(-\int^{\Phi}\frac{Z(u)}{\U'(u)}du \)
\, \(-4a \U + 2b\)  \pl{potrelation} \, , \\
\a(\Phi) &=& \a_0 \exp\(-\int^{\Phi}\frac{Z(u)}{\U'(u)}du \) \pl{a1}
\, , \ea {\em and\/} we require $\b(\Phi)$ to be constant (minimal
coupling).

The restrictions are devised such that the 
 auxiliary metric $g$ is of constant curvature, 
$R(g)=-4a$. The solution to the matter field equations is
straightforward then, as, in the appropriately redefined variables,
they reduce to likewise equations on a spacetime of constant
curvature. Still, this does not provide the full solution to the field
equations yet. The field equations for the (redefined) dilaton remain
to be solved. For $a \neq 0$ this turns out to be quite a nontrivial
problem, while still it may be resolved. Since the dilaton enters the
(true) metric $\bfg$, the latter is seen to respond to the matter fields.

The treatment constitutes a reasonable generalization of the CGHS
model. The CGHS
model satisfies equation (\ref{potrelation}) with $a=0$ and
$b=-\L/8$, the coupling to the scalar fields is minimal, $\beta \equiv
1$, and in that model there are no fermions or gauge fields
($\gamma=\alpha=0$). Also e.g.\ the one--parameter generalization of
the CGHS model considered in \cite{Fabbri} (with the {\em same}\/ $a$
and $b$ as above) is covered. Similarly, we generalize the recent
(independent) considerations of Cavagli\`a et al.\ \cite{Cava1}, where
they consider fermions coupled to generalized dilaton gravity; for
$a=0$ they provide the general solution to the fermion--gravity
system, while for the more complicated case $a\neq 0$ they find the
stationary solutions only (which are not exhaustive here).

Note that as we allow also for $U(1)$ gauge fields, the classically
solvable models considered here incorporate generalizations of
the massless Schwinger model to a nontrivial gravitational sector.
For reasons of better interpretation, all the matter dependence of the
metric will be expressed in terms of the energy momentum tensor, furthermore.

In the new variables, the metric $g$ decouples
completely from the matter sector. Up to a choice of coordinates,
$R(g)= -4 a$ forces $g$ to take the form \be g = \frac{2 dx^+
  dx^-}{(1+a x^+x^-)^2} \equiv : 2 \exp (2\xi) dx^+ dx^- \pl{const} \,
\, ,\ee where the function $\xi$ is defined by this equation. Note
that this does not imply that the gravitational and matter sectors of
the original model decouple. 

The situation is quite analogous to a
system of coupled harmonic oscillators: There the introduction of
appropriate variables (`normal coordinates') leads to a system of
decoupled harmonic oscillators.  Also here the field equations in the
new variables $g$ etc.\ take qualitatively the same form as the one of
the original variables $\gtrue$ etc., just that, upon restriction to
the class of Lagrangians considered in this section, in the new
variables the equations of motion simplify greatly and (in part)
decouple.  (Certainly, if one counts the dilaton to the gravitational
sector, this decoupling is not complete. The equations for the dilaton
depend on the energy momentum tensor, cf equations \re{eomg2}) or equations
\re{dil1x}), \re{dil2x}) below.  It is also in this way that the matter
enters the metric $\gtrue$, cf equation \re{gneu}) above).

Due to $\b = \wt \b= 1$, we find that $f$ is a superposition of left--
and right--movers, \be f = f_+(x^+) + f_-(x^-) \pl{Lsgphi} \,\, , \ee
just as in flat Minkowski space. This is the case since due to the
conformal invariance of \re{KG}) with $\b' =0$ and $f$ of 
conformal weight zero, $\Box f = 0$ reduces to just $\6_+ \6_- f=0$
for any metric in the conformal gauge.  So here $\gtrue \to g$ does
not make any difference.

\vskip5mm

\subsubsection{No gauge fields ($\a_0 = 0$):}

\vskip2mm

Next we turn to the field equations for the redefined fermion fields
$\psi$.  For simplicity we restrict ourselves to the case of no $U(1)$
gauge field first ($\a_0=0$ in equation \re{a1})).  As a consequence of the
reformulation, we merely have to solve the massless Dirac equation with
the background metric $g$ (even despite the fact that $\gamma' \neq
0$). Here it is not so much decisive that $g$ is a space of constant
curvature, the main point is that we know its conformal factor {\em
explicitly}, cf equation \re{const}). In such a case the solution of the
Dirac equation is the one of Minkowski space, i.e.\ again consisting
of right-- and left--movers, $\chi_R(x^+)$ and $\chi_L(x^-)$,
conformally  {\em transformed}\/ to the space with metric $g$ as a
field with conformal weight $-\frac{1}{2}$: \be \psi = \exp (-\xi/2)
\, \left(
\begin{array}{c} \chi_R(x^+) \\  \chi_L(x^-)
\end{array} \) \pl{RLLsg} \,\, , \ee where, in the case under
consideration, $\xi$ is given by equation \re{const}). (Note that the
result is no more a superposition of a left-- and a right--mover,
$\xi$ depending on $x^+$ {\em and}\/ $x^-$, but only a conformal
transform thereof).

Up to now the solution of the field equations was immediate. Now we
come to solving the equations \re{eomg2}), however, which is a less
trivial task (for $a \neq 0$). Due to the introduction of $g$ and
$\phi$ as new variables, there is no $\Z$-term and $\U$ is linear. The
essential restriction of this section, equation \re{potrelation}), moreover,
ensures that the whole system becomes linear in $\phi$.  So we are
left with the following three equations \ba \nabla_\pm \nabla_\pm \phi
&=& {\bf T}_{\pm \pm} \pl{dil1x} \\ e^{-2\xi} \6_+ \6_- \phi + 2a \phi
&=& -b \, , \pl{dil2x} \ea where $\xi_{,\pm} \equiv \6_\pm \xi$ and where we
used ${\bf T}_{+-}=0$ (as a consequence of $\CA=0$). The left-hand side
of the first two equations may be rewritten more explicitly as: $\(
\6_\pm - 2 \xi_{,\pm} \) \6_\pm \phi \equiv e^{2\xi} \6_\pm e^{-2\xi}
\6_\pm \phi$. (The only nonvanishing components of the Christoffel
connection are $\Gamma^+{}_{++}= 2\xi_{,+}$ and $\Gamma^-{}_{--}=
2\xi_{,-}$.) The respective righthand sides of equations \re{dil1x}) are
given by \ba {\bf T}_{++} &=& {\bf T}_{++}(x^+) \equiv (f_+')^2 + i
\chi_R^\ast \dlr_+ \chi_R \, ,\pl{T+} \\{\bf T}_{--} &=& {\bf
  T}_{--}(x^-) \equiv (f_-')^2 + i \chi_L^\ast \dlr_- \chi_L \,
.\pl{T-} \ea Here the indices $+$ and $-$ are world sheet indices
($\m$, $\n$, $\ldots$), not to be mixed up with the frame bundle
indices ($a$, $b$, $\ldots$).

Note that with both indices lowered (and only then!) the energy
momentum tensor ${\bf T}_{\m\n}$ is invariant with respect to the
conformal field redefinition \re{gneu}): ${\bf T}_{\m\n} =T_{\m\n}$.
Here the bold faced quantity is the energy momentum tensor of the {\em
  original}\/ theory \re{I}), given in equations
\re{Tscal}), \re{TU1}), \re{Tferm}), and $T_{\m\n}$ the energy momentum
tensor following from \re{J}) upon variation with respect to the
auxilary metric $g$.  This holds also for the $U(1)$ gauge field; the
fact that it is not conformally invariant in two dimensions, in
contrast to scalar fields, e.g., is reflected in that the
redefinition of $\a$ to $\wt \a$ contains the conformal factor, while
the latter is absent in the analogous transition $\b\to \wt \b$ (cf
equation \re{wtbeta})).

As ${\bf T}_{\pm \pm} = {\bf T}_{\pm \pm}(x^\pm)$ the solution of
\re{dil1x}), \re{dil2x}) is immediate for $a=0$, since then also $\xi
\equiv 0$ (cf equation \re{const})); the complication arises when
permitting $a \neq 0$ in equation \re{potrelation}).

The general local solution to the equations \re{dil1x}, \ref{dil2x}) is
of the form: \be \phi = \CT_+(x^+) + 2 \xi_{,+} \, \int^{x^+}\CT_+(z)
\, dz \, + \; (+ \lra -) \; - \, \frac{bx^+x^-}{1+a x^+x^-}
\pl{solution} \ee with \be \CT_\pm (u) = \int^udv\int^v {\bf T}_{\pm
  \pm}(z) \, dz + \2 K + k_\pm x^\pm \, , \pl{Tpm} \ee where in the
last line we displayed constants of integration $K$ and $k_\pm$
expli\-cit\-ly. There was no need to introduce two different constants
of integration instead of $K$, as within \re{solution}) they anyway
would contribute the same (i.e.\ there difference drops out and
without loss of generality the two constants may be set equal ---
assuming that the lower boundaries in the integrations are fixed in
some arbitrary way). The constants of integration $C_\pm$ from the
integral over $\CT_\pm$, on the other hand, may be absorbed into a
redefinition of $k_\pm$ for $a \neq 0$, $k_\pm \to k_\pm + C_\pm/a$,
while they disappear for $a=0$ due to $\xi_{,\pm} =0$; so we did not
display them.

That equation \re{solution}) is a solution of the coupled system
\re{dil1x}, \ref{dil2x}) may be established readily using two relations,
following from the definition of $\xi$ in equation \re{const}):
$\6_\pm \6_\pm \xi = (\xi_{,\pm})^2$ and $\6_+ \6_- \xi = - a \exp (2
\xi)$, the latter of which is equivalent to the statement that
$g=2\exp(2\xi) dx^+dx^-$ describes a space of constant curvature
$R(g)=-4a$.  These two equations are {\em essential}\/ for
\re{solution}) to solve equations \re{dil1x}, \ref{dil2x}). Thus despite the
simplicity and the apparent generality of our solution \re{solution}),
it solves the field equations only for the specific function $\xi$
defined above!  Having found a particular integral of the linear
equations \re{dil1x}, \ref{dil2x}), one is left with finding the general
solution of the homogeneous system. As we will see right below, the
homogeneous solutions $\phi_{hom}$ are indeed incorporated in
\re{solution}) if one takes the freedom in choosing the
constants $K$ and $k_\pm$ into account.  This then concludes the proof that
\re{solution}) is the general solution of equations \re{dil1x},
\ref{dil2x}).

To our mind the homogeneous system is most transparent in a different
coordinate system, namely the one in which $g=2dx^0dx^1 + 2a \(x^0
dx^1\)^2$, cf equation \re{011h}), 
resulting from \re{const}) by means of $x^-=x^1$ and $x^+ =
x^0/(1-ax^0x^1)$. These coordinates have
 the advantage that the zero--zero component
of the homogeneous system $\nabla_\m\6_\n \phi_{\hom} + g_{\m\n} \, (2a
\phi_{\hom} + b)=0$ reduces to $\6_0\6_0 \phi_{\hom} =0$, so that
$\phi_{\hom}$ is found to be at most linear in $x^0$! The $x^1$
dependence is then restricted by the remaining two equations, which
are $(\6_1 -2a x^0) \6_0 \phi_{\hom} +2a \phi_{\hom} =0$ and
$(1-x^0\6_0)\6_1\6_1 \phi_{\hom}=0$, where we have made use of the
former equation to simplify the one--one component of the field
equations to reduce to the latter equation. This then leads to \ba
\phi_{hom}&=& k_+ x^0 + k_- x^1 (1 - a x^0 x^1) + K (1 - 2a x^0 x^1)
\nonumber \\ &=& \frac{k_+ x^+ + k_- x^- + K (1 - a x^+ x^-)}{1 + a x^+x^-}
\pl{hom} \ea for three free constants $k_\pm$ and $K$. As the
notation already suggests, they indeed coincide with the three
constants found in \re{solution}).

As remarked already in section \ref{Secpurelocal}, in the absence of
matter fields and with a spacetime--topology $\dR^2$, the general
solution of the field equations modulo gauge symmetries is
parameterized by {\em one}\/ real quantity only. As the above three
constants $K$ and $k_\pm$ remain in the matterless (`vacuum')
solution, not all of them can describe physically (or geometrically)
different spacetimes.  Indeed, equation \re{const}) has a residual
gauge freedom: \be x^\pm \ra \l^{\pm 1} \, \frac{x^\pm + s^\pm}{1 -
  s^\mp a (x^\pm + s^\pm)} \, , \pl{resi} \ee parameterized by three
real constants $\l \neq 0$ and $s^\pm$. For $s^\pm :=0$, e.g., this is
nothing but the rescaling $x^+ \ra \l x^+$, $x^- \ra x^-/\l$, leaving
\re{const}) invariant at first sight; by means of equation \re{hom})
it leads to the identification $(k_+,k_-) \sim ((k_+/\l), \l k_- )$. A
more detailed analysis shows that the factor space of $K$ and $k_\pm$
modulo the action induced on them by equation \re{resi}) is indeed a
one parameter space. (E.g., for $a=0$, $b \neq 0$, and no matter
fields, $T_{\m\n}=0$, one easily establishes that $k_\pm$ may be put
to zero by shifts in $x^\pm \to x^\pm + s^\pm$; thereafter rescalings
with $\l$ have no effect anymore, so that $K$ remains as a physical
parameter. --- If, on the other hand $a=b=0$ and there are no matter
fields, then one can achieve, e.g., $k_+ = 1$ and $K=0$, with $k_-$
remaining.) In many cases the remaining parameter may be given the
interpretation of the `mass' $m$ of the spacetime described by
$\gtrue$. Strictly speaking, the above consideration applies to the
matterless case (reconsidered in the present gauge). However, we think
that the present {\em counting}\/ of gauge--invariant parameters will
be unmodified when adding matter, so that also in the general case
only one of the three parameters $k_\pm$ and $K$ will survive.

In summary we get the following results: Combining the solution
\re{solution}) with equations \re{gneu}), (\ref{const}), and
(\ref{relation}), the metric is brought into the form \be \gtrue =
\frac{4(2a\phi +b) \, dx^+\,dx^-}{(1+ax ^+x^-)^2 \, \V(\U^{-1}(\phi))}
\, , \label{hihi} \ee while $\Phi=\U^{-1}(\phi)$. The scalar field is
given by equation \re{Lsgphi}), the fermionic field $\bfPs$ implicitly
by equations \re{psi}) and \re{RLLsg}). The local solutions are
parameterized by the choice of the one--argument functions $f_\pm$ and
$\chi_\pm$ as well as by the one constant of the vacuum theory (cf the
discussion above). In terms of these data the energy momentum tensor,
as defined in equations \re{Tscal}), \re{Tferm}), has the form
\re{T+}), \re{T-}) (while ${\bf T}_{+-} =0$).



\subsubsection{Inclusion of gauge fields (arbitrary $\a_0$)}


We now turn to the system with an additional $U(1)$ gauge symmetry.
Assuming $\a$ in the action \re{U1}) to be subject to the constraint
\re{a1}), we obtain a constant coupling $\wt \a =\a_0$ in the
reformulated action \re{J}). The matter sector of the theory still
poses no problem, as it is mapped to a corresponding system on
a space of constant curvature. So we are left with the dilatonic
equations \re{eomg2}), or, equivalently, with: \ba e^{2\xi} \6_\pm
e^{-2\xi} \6_\pm \phi &=& {\bf T}_{\pm \pm} \pl{dil1y} \\ \6_+ \6_-
\phi - 2 (\6_+ \6_- \xi) \phi &=& -b e^{2\xi} - {\bf T}_{+-} \, .
\pl{dil2y} \ea For ${\bf T}_{+-}=0$ this  takes the form of
equations \re{dil1x}), \re{dil2x}). However, the present ${\bf T}_{\m\n}$ has
pronounced differences to the previous considerations: First, ${\bf
  T}_{++}$ and ${\bf T}_{--}$ depend on {\em both}\/ coordinates $x^+$
and $x^-$ now (otherwise we would be forced to $\CF=0$ and,
consequently, also to $\bfPs=0$, cf the remarks following equation
\re{eomA})!). Second, also ${\bf T}_{+-}=-(\a_0/2) \, \exp(-2\xi) \, \[
\CF_{+-}\]^2 \neq 0$ for $\CF \neq 0$; i.e., due to the presence of
the $U(1)$ field the energy momentum tensor is no more tracefree
except for $\bfPs = 0 = \CF$.

The form of ${\bf T}_{\m\n}$ is still restricted by some
decisive relations. These may be expressed most elegantly when using
the trace $T$ of the energy momentum tensor with respect to the
auxiliary metric, $T =g^{\m\n}T_{\m\n}=g^{\m\n}{\bf T}_{\m\n}$. (We
remind the reader that, in view of equation \re{gneu}), ${\bf T} = \O^2 \, T
\neq T$ in general). The energy momentum tensor satisfies:
\ba \6_+ \6_- \sqrt{|T|} &=&0
\pl{xv1} \\ \6_\pm T&=& - 2\exp(-2\xi) \, \6_\mp {\bf T}_{\pm\pm}
\pl{xv2} \ea The first of these relations follows from equation \re{eomA}).
It allows to write $T=-\a_0 \, \[ \exp(-2\xi) \,
\CF_{+-}\]^2$ in the form $-[F_+(x^+)+F_-(x^-)]^2$, where, again
through equation \re{eomA}), the functions $F_\pm$ are determined by the
fermion fields up to an additive constant.  The second relation, equation
\re{xv2}), is equivalent to $\nabla^\m {\bf T}_{\m\n} =0$, where this
equation is understood entirely with respect to the  {\em auxiliary}\/
metric $g$; it follows from the equations fulfilled by
the matter fields, and, simultanously, it is the integrability condition
of the dilatonic system of differential equations. Note that the
second relation allows to express ${\bf T}_{++}$ and ${\bf T}_{--}$ in
terms of $T$ and $\xi$ up to a function of the single
variable $x^+$ and $x^-$, respectively:
\be {\bf T}_{++}=-\2 \int_0^{x^-} \,
e^{\left(2\xi( {x}^+,{\wt x}^-) \right)} \,
(\6_+ T)( {x}^+, {\wt x}^-) \, d {\wt x}^-
+ {\bf T}_{++}(x^+,0) \, \ee
with a similar equation for ${\bf T}_{--}$. The functions
${\bf T}_{++}(x^+,0)$ and ${\bf T}_{--}(0,x^-)$
may then be eliminated from the right--hand side of the
dilatonic system of linear differential equations by a particular
integral of the type \re{solution}) found already above. One then is
left with finding a particular integral of the remaining system, with
inhomogeneities determined solely by  $T=-[F_+(x^+)+F_-(x^-)]^2$.

This still turns out to be quite a hard problem. After the dust
clears, however, it is possible to put the solution to the system
\re{dil1y}), \re{dil2y}) into the following form: \ba \phi&=& \int_0^{x^+}\! d\wt
x^+\int_0^{x^-} \! d\wt x^- \; \left(\frac{a(x^- -\wt x^-)(1+a\wt x^+
    x^-)}{1+a\wt x^+\wt x^-} -1\right)\; {\bf T}_{+-}(\wt x^+,\wt x^-)
\, - \nn && -\, \frac{bx^+x^-}{1+a x^+x^-}\, + \, \int^{x^+} \! d\wt
x^+ \int^{\wt x^+} \! du \; \frac{\(1+a\,ux^-\)^2}{\(1+a\, \wt x^+
  x^-\)^2} \; {\bf T}_{++}(u,0)\, +\,\nn &&+\, \int^{x^-} \! d\wt x^-
\int^{\wt x^-} \! dv \; \frac{\(1+a\,x^+ v\)^2}{\(1+a\, x^+ \wt
  x^-\)^2} \; {\bf T}_{--}(0,v) \,+\, \phi_{hom}\; . \pl{phiU1} \ea
Here $\phi_{hom}$ is the homogeneous solution \re{hom}), containing the
one gauge--invariant parameter of the vacuum theory that is left over when
taking into account the residual gauge freedom \re{resi}) discussed
above.

It is a somewhat cumbersome calculation to show that \re{phiU1})
indeed solves the system \re{dil1y}), \re{dil2y}) and we will not provide any
further intermediary steps here so as to not become too technical.
Actually, it is even not completely obvious to see how the above $\phi$
reduces to the previous form \re{solution}) for ${\bf T}_{+-}=0$ and
${\bf T}_{\pm\pm}={\bf T}_{\pm\pm}(x^\pm)$. We recommend this
consistency check as an exercise to the reader.

Up to gauge transformations, the general local solutions are
parameterized by the choice of the initial data for the matter fields
$f$ and $\bfPs$, as well as by the gauge invariant constant contained
in $\phi_{hom}$ and one further constant of integration in the
field strength $\CF$.

\vskip7mm

In this section we analyzed 2d gravity--matter models in which the
{\em auxiliary}\/ metric $g$, defined in equation \re{gneu}), has
three Killing vectors (cf equation \re{const})). Basically this was
the defining restriction for the class of models considered here. We
remark that for a completely general model \re{I}) it is also possible
to find {\em those}\/ solutions, in which $g$ has one Killing field
(stationary/homogeneous auxiliary spacetimes); for this special {\em
  subclass}\/ of solutions of the general model, one can reduce the
field equations to one ordinary differential equation of third order.
We refer to \cite{Diplom} for details.

\subsection{Generalized dilaton gravity with chiral matter}
As noted by W.\ Kummer \cite{Kummerferm}, the KV--model \re{KV}) may
still be solved, if chiral fermions are added.\footnote{This was noted
  also independently, but later, in \cite{Solodukhinferm}. In
  \cite{Solodukhinferm} it was claimed also that the solution for both
  chiralities can be found along similar lines; this was erroneous,
  however (cf footnote in \cite{Heiko}).} The solution becomes
drastically simplified and transparent again, when using the PSM
formalism. The KV--model is general enough such that the solvability
can be extended to the whole class \re{grav}) of gravity models.
Moreover, it is also possible to extend the calculation to a multiplet
of chiral fermions and Yang--Mills fields. Similarly, the chiral
sector of scalar fields may be included. We only sketch the idea for
the simplest, torsion--free case and refer to \cite{Heiko} for further
details and the more general scenarios.

In terms of CD coordinates, the action \re{PSJ}) takes the form
\be I=\int_{\CM} \wt
A_i \w (d \wt X^{i}+\wt J^{i})+\wt A_2\w \wt A_3 \pl{ICD} \, . \ee
Note, however, that now the originally simple currents \re{Jpar})
may have become complicated within the transition   $J^i
\ra \wt J^i \equiv (\6 \wt{X^i}/\6 X^a)\, J^a$.
A possible choice of CD coordinates for \re{P}) is provided by equation
\re{Xtildeneu}). Redefining $\wt X^1$ by a factor of two, the
transformed
currents become:
\be \wt J^i = (2X^+J^-+ 2X^-J^+\, , \; J^+/X^+,0) \,\, , \pl{wJ} \ee
where $X^\pm$ are to be understood as functions of the new variables
$\wt X^i$.  While in the case of pure gravity, $J^i \equiv 0$, also
the last of the initially three potentials, $V(\phi)$, could be
eliminated from the action by means of the above change of variables,
it now creeps in again through $\wt J^i$. {\em However}, from equation
\re{wJ}) it is obvious that $V(\phi)$
can be eliminated in the case of {\em chiral}\/ fermions:
$\psi_L=0$ implies $J^+=0$ and $V$ drops out of \re{ICD})! With
the further (Lorentz invariant) field redefinition \be \wt r := 2X^+
\, r^2 \pl{wr} \ee the transform of the chiral current $J^i =
(J^-,0,0)$ is $\wt J^i = (\wt r d\rho,0,0)$ and the action \re{ICD})
takes the trivial form \be I_{chiral}=\int_{\CM} \ti A_{i}\w d\ti
X^{i}+\ti r\ti A_1\w d\r+ \ti A_2\w \ti A_3 \pl{DA.3.81} \; .  \ee We
are thus left with solving the field equations of this very simple
action.

Variation with respect to $\wt A_2$, $\wt A_3$ yields \be \wt A_2=d\wt
X^3\; ,\quad \wt A_3=-d\wt X^2\; , \pl{DA.3.82} \ee which may be used
to eliminate these fields together with $\wt X^2$, $\wt X^3$. This
simplifies \re{DA.3.81}) further to \be I_{chiral}=\int_{\CM} \ti A_1\w
\(d\ti X^1 + \ti r d\r \)\; .  \pl{DA.3.84} \ee According to
\re{DA.3.84}), $\wt A_1$ is exact. Thus, with an appropriate choice of
coordinates, $\wt A_1 = dx^1/2$. This may be used to establish that all
the remaining fields $\rho$, $\wt r$, and $\wt X^1$ are functions of
$x^1$ only and the latter is determined up to a constant of
integration $c$ by the former two via \be \wt X^1= \wt X^1(x^1)=-\int \wt
r(x^1)\rho'(x^1) dx^1 + c \,\, \pl{solX1} . \ee The gauge freedom of our
gravity theory may be fixed completely by choosing $\wt X^3$ as the
second coordinate $x^0$ and by using a Lorentz frame such that $X^+ =1
\Rightarrow \wt X^2=0$.

Thus in the present framework the field equations
could be trivialized. The local solutions are parameterized by the
choice of initial data for $\wt r(x^1)$, $\rho(x^1)$, and the
integration constant $c$, furthermore. The latter is the only
parameter remaining in the matter-less case, in coincidence with
section \ref{Secpurelocal}.

We are left with reexpressing this solution in terms of the original
variables $\bfg$, $\Phi$, and $\Psi$. After a series of elementary
steps \cite{Heiko}, one may put the metric into the form \be \gtrue =
2 dx^0 dx^1 + [h_0(x^0) + k_0(x^0) h_1(x^1)] \, (dx^1)^2 \, \, ,
\pl{CEF} \ee where $h_0$ is the same function as the one in the
pure gravity case, equation \re{h}) above, and the dilaton field
$\Phi=\Phi(x^0)$ remains unchanged, too. The respective matter field
depends on the null coordinate $x^1$ only.  The function $k_0$ in equation
\re{CEF}) is determined completely by $\U,\V$, and $\Z$, furthermore.
Thus all the matter dependence in \re{CEF}) may be put into the
function $h_1$. More explicitly one finds \be h_1 = -2\int^{x^1} {\bf
  T}_{11}(u)du \, , \pl{h1} \ee where ${\bf T}_{11}={\bf T}_{11}(x^1)$
is the only nonvanishing component of the energy momentum tensor of
the matter fields. It is possible to prove \cite{Heiko}, finally,
that these systems do not show a Choptuik effect (cf the discussion
at the end of section \ref{Secspher}).

\chapter{Classical solutions on a global level}
\pl{Chapglobal}
In this chapter we will deal with gravity--YM-systems. Very little is
known on a global level for models with propagating modes, generically,
and we will not discuss them within this chapter.

Although locally the solutions to the field equations all look quite
alike for the pure gravity--YM theories, cf section \ref{Secpurelocal},
globally there may be decisive differences between different models.
Locally all solutions are parameterized by one constant $C$ (resp.\
$r+1$ in the gravity YM-case, cf section \ref{SecYMlocal}) and the
metric always could be put into the form \re{011h}) (with the dilaton
being a function of $x^0$, furthermore). Globally, on the other hand,
solutions characterized by a function $h$ in \re{011h}) (resp.\ $\bfh$
in \re{011href})) with at most two zeros show a qualitatively very
different behavior from solutions with more than two zeros. Indeed,
in the latter case there exist smooth maximally extended solutions on
any punctured 2d Riemann surface (in the topological sense:
`punctures' or `holes' may well be `extended' in a pictorial
sense, cf, e.g., Figs.~\ref{fig:fig16} and \ref{fig:fig17} in
section \ref{Secall} below, and
inextendible as, e.g., bounded by curvature singularities).  Such
solutions do {\em not}\/ exist, if $h$ has at most two zeros (as for
Schwarzschild (SS) or Reissner-Nordstr\"om (RN)).

Within a given (fixed) 2d gravity model, the number of zeros of the
one-parameter family of functions $h$ depends crucially on the number
of zeros of the potential(s) characterizing the model, cf, e.g., equation
\re{h})! For a potential $V(\phi)$ with, say, one zero only, there can
be at most two zeros in $h(x^0)$, irrespective of the choice of $C$.
This follows because the zeros of $V$ correspond to maxima of $h$,
which in turn is a continuous function.

It will be possible, moreover, to find {\em all}\/ global, maximally
extended solutions for a given model and to parameterize the gauge
inequivalent ones. The dimension of the resulting {\em moduli space}\/
of gauge inequivalent solutions will then depend on the chosen
topology of the underlying spacetime manifold $M$. E.g.\ for a
gravity--YM model giving rise to functions $h$ with more than two
zeros, the dimension of the moduli space will turn out to be
$(\pi_1(M) + 1)(r+1)$, where $\pi_1(M)$ is the fundamental group
of $M$ and $r$ is the rank of the YM gauge group.

It is remarkable, that within these two-dimensional models there
is so much control on all smooth and complete solutions for a wide
class of models. Nothing close to this is conceivable for 4d Einstein
gravity, where, in comparison to all existing solutions, only very few
are known. Certainly, in Einstein gravity the moduli space of
classical, gauge inequivalent solutions is by no means finite
dimensional (except if one restricts attention to assymptotically
flat, stationary black hole solutions, however, cf the Introduction). Still,
the finite dimensionality of the moduli space in the 2d models does
not imply that the solutions are more or less trivial from the point
of view of their spacetimes. The solutions will include
multi black hole spacetimes of almost arbitrary complexity and variety.

It may be hoped that the classification presented below finds
applications, e.g., in a path integral approach to quantization,
where, in contrast to Hamiltonian methods (cf section \ref{Sectowards}), the
topology of spacetime is not restricted a priori, and, on the other
hand, the classical solutions should give the dominating contribution
(at least in an appropriate limit).

The classical considerations below are, however, also interesting in
their own right. The machinery for constructing Carter-Penrose
diagrams as well as the ensuing construction of global solutions on
nontrivial topologies may be used for any (effectively)
two-dimensional metric with at least one Killing vector, and thus also
for many known solutions of 4d Einstein gravity (like RN-deSitter
spacetimes, etc).

We will in the following, first start with the construction of all
{\em simply connected}\/ maximally extended spacetimes. Solutions for
arbitrary topology are then considered in a subsequent section,
as all of them may be obtained as appropriate factor spaces of the
simply connected solutions.

\section{Maximal Extensions 
for all Gravity--Yang--Mills systems \pl{Secmaxext} }
After some remarks on (non)uniqueness of maximal
extensions and on general properties of the metric \re{011h}), the
symmetries of the metric will be exploited to solve the equation of
extermals. Then, we will derive a simple building block principle for
obtaining the universal covering solutions of any metric with local
form \re{011h}) \cite{TKII}, streamlining and generalizing a method
suggested by M.\ Walker \cite{Walker} (cf also \cite{Brill}) for a
rather restricted class of metrics. The outcome are so-called
(Carter-)Penrose diagrams, which permit a global and efficient view
on the causal structure of the covering solutions. The method will be
illustrated by means of examples at the end of this section.

\subsection{Generalities on maximal extensions \pl{Secgeneral} }
As emphasized already before (cf, e.g., section
\ref{Secattain}), in general a chart \re{011h}) will cover parts of
spacetime only.  For physical reasons, however, one is interested in
maximally extended spacetimes: All extremals (which, if timelike,
describe the motion of test particles) should be either complete (have
infinite length) or run into a true curvature singularity.

Thus, we have to construct the maximal extension of the local charts
\re{011h}). Certainly, in general such an extension is not unique.
Already the (analytic) Minkowski solution, $h \equiv 0$, permits
inextendible manifolds of planar, cylindrical, or toroidal topology
(the latter two even carrying further continuous parameters).  The
Minkowski-plane, however, is the unique universal covering space of
the cylinders and tori, and the latter, in turn, are obtained by
factoring out a discrete symmetry group from this covering manifold.
This generalizes: Any multiply connected solution, however complicated
it may be, can be obtained from the universal coverings (i.e.\ the
maximally extended {\em simply connected\/} manifolds) by factoring
out a discrete symmetry group (cf, e.g., \cite{Wolf}). Therefore, in
the present section we will restrict ourselves to the simply connected
extensions. The global solutions with nontrivial spacetime topology
will be dealt with then in the following section, section
\ref{Secall}.

Still, even if we restrict ourselves to simply connected spacetimes, the
extension of a local solution of the form \re{011h}) is not unique without
additional specifications. Take, e.g., the case that the function $h$ is
given in the interval $[0,1]$ and known to vanish there. Obviously already
within a chart of the type \re{011h}) there is an infinity of smooth
extensions of the function $h$ to values of $x^0$ outside the given
interval. Even if the function $h$ is known on a maximal domain of
$x^0$, having a boundary only if there $h'' = R(g) \to \pm \infty$, there
are similar ambiguities in a smooth extension of the (generically) local
chart \re{011h}) into regions not covered by that chart. One way to avoid
such ambiguities is to require that $g$ should be analytic everywhere.
Alternatively, working in the framework of a given gravity--Yang--Mills model
considered in chapter 
\ref{Chapgen}, we may require the field equations of the given
Lagrangian to restrict the extension. In this case the universal covering
solutions turn out to be unique also for nonanalytic functions $h$.
Either of these two philosophies underlies the simple extension rules
which we will derive in this section.

\subsection{Some remarks on Carter-Penrose diagrams \pl{Seccarter} }
In the following we will make extensive use of Carter-Penrose or
conformal diagrams. Following standard literature (cf, e.g.,
\cite{Waldbuch}), they are constructed roughly as follows (we phrase
it for the 2d case): First, one chooses null-lines as local
coordinates and extends local charts \re{confgauge}) as far as
possible, i.e.\ until one either runs into a true curvature
singularity or `ad infinitum', where the latter has to be checked by
means of an affine parameter of the null-lines. Spacetime then, in the
case of a simply connected spacetime is represented by an {\em open}\/
subset of $\dR^2$ (in fact, this is true only in the simpler cases
like for SS, RN or RN-deSitter; for the more general case cf below).
Next, one regards an {\em auxiliary}\/ metric $\wt g$, conformally
related to the true spacetime metric $g$, i.e.\ $\wt g = f^2 \, g$
(for some nonvanishing function $f$). This amounts to dropping part of
the prefactor in \re{confgauge}); note that in the coordinates chosen,
the direction of the null-lines does not change in this process.  Now
$f$ is chosen in such a way that, first, the `boundary' of the
spacetime is brought into a finite distance {\em measured in terms
  of}\/ $\wt g$, and, second, on the `boundary', on which $g$
certainly is no more well-defined, $\wt g$ remains smooth.  (For
further details cf \cite{Waldbuch} or \cite{HawkingEllis}.)  This
allows to {\em include}\/ the boundary into the mani\-fold to be
regarded, which in a general spacetime dimension is a nonunique
procedure from the purely topological point of view. Moreover,
asymptotic conditions for fields defined on the spacetime manifold may
be translated into ordinary boundary conditions on the extended
spacetime. And, last but not least, the (in part) finite extension of
the resulting pictures of spacetime, which result upon using Cartesian
null-coordinates for $\wt g$ and which are the above mentioned
Carter-Penrose diagrams, permit  a concise grasp of the global
causal structure.

In the two-dimensional setting and at least for many purposes (but
also for some purposes in higher dimensions), the use of an auxiliary
metric $\wt g$ is not necessary. The {\em interior}\/ of a
Carter-Penrose diagram is diffeomorphic to the spacetime under
discussion and a change of coordinates will bring the `boundaries'
to finite {\em coordinate}\/ distance (loosly speaking; in the
pictorial sense). Moreover, regarding the resulting charts of
spacetime as open subsets of $\dR^2$, there is also no ambiguity in
defining the boundary of the spacetime canonically. Thus, in
this way, the (interior of) Carter-Penrose diagrams are regarded as
{\em charts}\/ of {\em all of}\/ spacetime, with the null-rays running
through the diagram under 45 degrees. One only has to keep in mind
that distances are always to be measured by means of the metric.
(Distances in the picture are fictitious and correspond to the choice
of coordinates; {\em however}, the (true) distance between two points
in the diagram which, in the picture, are separated by a finite
distance, will {\em always}\/ be finite, too, {\em except}\/ possibly,
if one of the points lies on the boundary.)

\begin{figure}[ht]
\begin{center}
\unitlength 1cm
\begin{picture}(13,6)
\put(0.4,5){\mbox{${\cal J}^+_L$}}
\put(0.4,0.8){\mbox{${\cal J}^-_L$}}
\put(5.2,5){\mbox{${\cal J}^+_R$}}
\put(5.2,0.8){\mbox{${\cal J}^-_R$}}
\put(12.,5){\mbox{${\cal J}^+$}}
\put(12.,0.8){\mbox{${\cal J}^-$}}
\put(8,2.7){\mbox{$r=0$}}
\put(0,0){\psfig{file=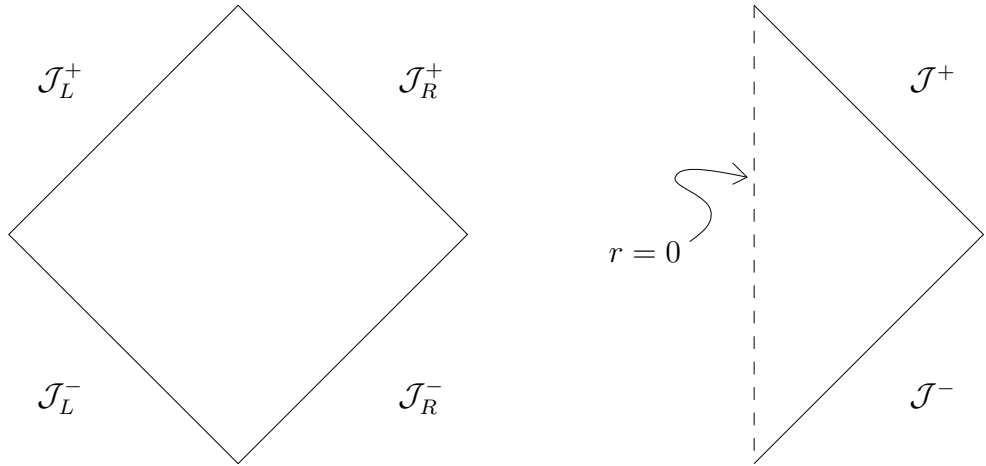,width=13cm}}
\end{picture}
\end{center}
\renewcommand{\baselinestretch}{.9}
\caption{ 
  \captionstyle \label{fig:neu}
  Carter-Penrose diagrams of Minkowski space in
  two (left) and four (right) spacetime dimensions.}
\end{figure}

Let us illustrate these considerations by means of 2d Minkowski
spacetime $g=(dt)^2-(dx)^2$.  Using null coordinates $x^\pm := (t \pm
x)/\sqrt{2}$, $g$ takes the form $g=2dx^+\, dx^-$. Maximal extension
implies that $x^+$ and $x^-$ each run over {\em all}\/ of $\dR$ (they
are affine parameters of the null-lines $x^-=const$ and $x^+=const$,
respectively). To bring the `boundary', i.e.\ infinity in this case,
into a finite coordinate distance --- without, at the same time,
bending any null-lines ---, we may apply a {\em conformal
  diffeomorphism}\/ $x^\pm \to f_\pm(x^\pm) =: y^\pm$.  These are the
{\em only}\/ diffeomorphisms, mapping straight null-lines $x^\pm
=const$ into likewise straight null-lines $y^\pm =const$. Choosing,
e.g., $f_+=f_-=tanh$, the Minkowski metric takes the form \be g
=\frac{2dy^+ \, dy^-}{(1-(y^+)^2)\, (1-(y^-)^2)} \, . \pl{Mink} \ee By
means of these coordinates, (null) infinity $x^\pm \to (\pm) \infty$
is mapped onto the lines $y^\pm = (\pm) 1$, bounding a square of
finite (coordinate) extension. Tilting this square by 45 degrees
and using $y^\pm$ as Cartesian coordinates, one arrives at the left part
of Fig.~\ref{fig:neu}.

Note that we did {\em not}\/ drop the conformal factor (the denominator)
in \re{Mink});
this would correspond to a transition to the above mentioned auxiliary
metric $\wt g$ (while simultaneously keeping track of the range of
variables $y^\pm$, not maximally extending $\wt g$,
certainly).\footnote{Conformal diffeomorphisms should be sharply
  distinguished from Weyl transformations of the metric.
  Unfortunately, there is no agreement on the use of the term
  `conformal transformations' in the literature. Mostly it is used
  for Weyl transformations of the metric, sometimes, however, for
  conformal diffeomorphisms instead.}
\par
Let us at this point add a comparison of some features of 2d and
4d spacetimes at the example of Minkowski space. Dropping the
angular part of the 4d Minkowski metric when represented in
spherical coordinates, it takes again the standard 2d Minkowski 
form, where, however,
$x=r >0$ by definition of the radial coordinate $r$. Thus, we may
repeat the above steps leading to \re{Mink}) and Fig.~\ref{fig:neu}, keeping
track, however, of the restricted range of the new coordinates.
Obviously we then end up with the right picture in Fig.~\ref{fig:neu}, the
vertical boundary of the triangle corresponding to $r=0$. This is
the Carter-Penrose diagram for 4d Minkowski space.

Any internal point $(r,t)=const$ of this diagram represents a
two-sphere now, with the 2d coordinate $r$, which now becomes a
curvilinear coordinate in the diagram, characterizing the 
volume of the respective invariant two-surface. At the left boundary, however,
$r=0$ and the two-surfaces
 are degenerate, i.e.\ they are points only.
A careful analysis of the 4d boundary as defined by means of $\wt g$,
on the other hand, shows that future (past) null infinity ${\cal J}^+$
(${\cal J}^-$) has topology $S^2 \times \dR$, i.e.\ any point on {\em those\/}
boundaries of the diagram corresponds to a two-sphere again.  The
three corner points (future, past, and spacelike infinity,
respectively), on the other hand, are just true points.

In the two-dimensional case, the situation is simpler: Any point in
the diagram is a point and nothing else. However, note that in the
process of the compactification, spacelike (or timelike) infinity of
2d Minkowski space, which is {\em not}\/ part of Minkowski space, was
mapped to just two points.

There are further important differences between already these
simplest spacetimes in two and four dimensions.\footnote{In section
\ref{Recipe}, we will compare (anti)deSitter spaces for two and
higher dimensions briefly, too.} First, only in $D=2$ spacetime
dimensions, future (or past) null infinity consists of two
disconnected parts, denoted by $L$ and $R$ for `left' and
`right' in the diagram. This is the case because possible
(oriented) directions of null-rays span an $S^{D-2}$, defined by
$\sum_{i=1}^{D-1} x^i \, x^i = 1$, which, only for $D=2$ becomes
disconnected. Correspondingly, an initially left-moving light ray
in the {\em four\/}--dimensional Carter-Penrose diagram necessarily ends up
at the same ${\cal J}^+$ (in the diagram) as an initially
right-moving one. If a 45 degrees null-line meets $r=0$ in the
diagram, it is just reflected.

One may question, if this simple observation, and the
corresponding separation of left-- and right--movers, does not
constitute a qualitatively important difference to the 4d analog when
discussing Hawking radiation. (Cf also related remarks in
\cite{VaraCGHS} as well as the use of `mirrors' in 2d models as,
e.g., in \cite{Verlindemirror}, so as to better simulate the 4d
situation.)

Moreover, one has to be aware that, in contrast to a 2d diagram, in a
4d Carter-Penrose diagram not all null rays are straight lines.
Clearly there are (say, originally left moving) null rays in 4d
Minkowski space which do not pass the arbitrarily chosen origin $r=0$
in the diagram.

Still, the {\em boundaries}\/ of the causal past or future of a point
(or region) will consist of straight, 45 degree pieces in the diagram.
This is one of the main strengths of the use of Carter-Penrose
diagrams, allowing, e.g.\ to quickly recognize if the spacetime
contains black holes or not. (For a careful definition of black holes
cf, e.g., \cite{Waldbuch}.)

In this context it may be worthwhile to mention that in $D=2$ the
boundary of the causal past of a region necessarily is smooth. There
is no room for cusps as in higher dimensions (cf, e.g., \cite{cusps}).

\subsubsection{Further preliminaries \pl{General}}
To find the maximal extension of the local solutions \re{011h}),  we want to
exploit the symmetries of this metric. As pointed out already
in section \ref{Secattain},  \re{011h}) permits a Killing field, namely
$\partial\over\partial x^1$. The function $h(x^0)$ measures the norm squared
of this Killing vector, furthermore. In particular a zero of $h$ indicates
that the line $x^0=const$ is a light-like (null) Killing trajectory. We
will adopt the customary term {\em Killing horizon\/} for such lines and
{\em degenerate Killing horizon\/} if the corresponding zero of $h$ is of
higher order (i.e., $h'|_{h=0}=0$). Also we will call the regions between
two zeros (or beyond the first/last zero) {\em sectors\/}. Finally, we call
a sector {\em stationary\/} if the Killing field is timelike there ($h>0$)
and {\em homogeneous\/} if it is spacelike ($h<0$).

The metric \re{011h}) has another symmetry, seen best in
the $(r,t)$--coordinates introduced to put the metric into generalized
SS form \re{gSS}). Equation \rz gSS has the disadvantage of becoming singular
at zeros of $h$, whereas \re{011h}) behaves perfectly well there.
However, it displays a further independent (discrete) symmetry, namely
one under an inversion of the Killing parameter $t$, $t
\leftrightarrow -t$.  If $h>0$, such that the transformation amounts
to a time--reversal, then the metric is called {\em static\/}, and thus
we see that in two dimensions stationary implies static. This comes,
however, as no surprise since in two dimensions any
vector field is hypersurface-orthogonal.
\footnote{A static metric can also be characterized by the condition
  that the timelike Killing field be hypersurface-orthogonal (cf,
  e.g., \cite{Waldbuch}), a
  nontrivial requirement in higher dimensions.} In connection with the
coordinates \re{011h}), this time-- (or space--)reversal symmetry
will prove to be a powerful tool in constructing the maximal
extension. Therefore we write it out in full also in the
original $(x^0,x^1)$--coordinates: \be {\widetilde x^0} =x^0 \quad ,
\qquad \widetilde x^1= -x^1-2f(x^0) \el glue with $f$ as in
\re{fun}). Of course, this symmetry works only within one sector
at a time, because on its edges (zero of $h$) $f$ diverges.  We
will call this transformation {\em flip}.  Note that changing the
constant in $f$ yields a different transformation (shifting the
origin of the Killing-parameter $t$ and thus the reflection axis),
so actually one has a one-parameter family of
flip-transformations.  The effect of this flip transformation on
the dynamical fields is exactly that of the gauge changing
transformation equation \re{Strafo}). Thus the flip is in fact a symmetry
transformation for the full solution to the models \I{grav}) or
\I{gdil}).

\subsection{Extremals}
The standard method for constructing the maximal extension involves
the determination of the extremals and the study of their completeness
properties.  Since we will directly apply the transformation
\re{glue}) to extend our local solutions, this is not really necessary
in the present case.  Still let us solve their equation for
illustrative purposes; after all, extremals, i.e.\ curves extremizing
the action $m\!\int\!ds^2$, are supposed to describe the motion of a
test particle in the curved background and are used to study the
completeness properties of the solutions.

Using the Christoffel symbols
$\G_{\m\n\r} \equiv (g_{\m\n,\r} + g_{\r\m,\n} - g_{\n\r,\m})/2 \,$,
the equation for the extremals reads
\be \ddot x^\m  +  \G^\m{}_{\n\r} \dot x^\n \dot x^\r =0\,.
  \el extrem1
For the metric  \re{011h}) this reduces to
\be \ddot x^0=-h'\dot x^1 (\dot x^0+ {h\over2} \dot x^1)
  \quad, \qquad
  \ddot x^1 =   {h'\over2}(\dot x^1)^2 \, , \el extrem2
with $h'\equiv dh(x^0)/ dx^0$.
Up to affine transformations, equation \re{extrem1}) yields a unique
parameterization of the solutions $x(\t)$. For nonnull extremals
the arc length $s$ itself is such an {\em affine parameter\/}.
Let us remark that in the torsion-free case extremals
coincide with auto-parallels, satisfying $\nabla_{\dot x} \dot x = 0$, while
in the case of nonvanishing torsion they are auto-parallel only
with respect to the Levi-Civita connection.

Clearly in two dimensions any null-line $g(\dot x, \dot x)=0$
solves \re{extrem1}), \re{extrem2}). (It is again a peculiarity of
$D=2$ that {\em any\/} null--line is also a null--extremal, as follows
by uniqueness of the latter.) In the chart
(\ref{011h}), these {\it null extremals} are:
\ba   x^1&=&\hbox{const} \, , \pl{null1} \\
{dx^1\over dx^0}&=&-{2\over h} \quad,\quad \mbox{wherever }h(x^0) \neq
0,
\pl{null2a} \\
x^0&=&\hbox{const} \, ,\quad \mbox{if} \,\, h(x^0)=0 \, .\pl{null2b}
\ea The extremals \re{null2b}) are exactly the Killing horizons
mentioned above. Plugging these solutions back into (\ref{extrem2}),
we see that for \re{null1}) and \re{null2a}) the affine parameter is
$\t=ax^0+b$, ($a,b=const$). Thus these null extremals are complete at
the boundaries, iff the coordinate $x^0$ extends to infinity into both
directions of the charts \re{011h}).  For the models under study this
is the case e.g.\ for all torsion--free theories, whereas the KV--model
\I{KV}) provides an example of incomplete extremals \re{null1},
\ref{null2a}) at $x^0=0$ (which is still a true singularity as $R$
blows up there, cf equation \re{RKV})). The affine parameter for the
extremals \re{null2b}) depends on the kind of zero of $h(x^0)$: For
nondegenerate horizons ($h'(x^0)\ne0$) we get $\t=a\exp(-{h'\over
  2}x^1)+b$, so they are incomplete on one side, whereas degenerate
horizons ($h'(x^0)=0$) are always complete, as then $\t=ax^1+b$.

The nonnull extremals are found most easily by making use of the constant
of motion associated with the Killing field $\partial\over\partial x^1$
(cf, e.g., \cite{Thi}): $g({\partial\over\partial x^1},\dot x)
\equiv \dot x^0+h\dot x^1=const$ along extremals. Knowing that for nonnull
extremals we may choose the length as an affine parameter,
this equation may be rewritten in the form
\be (dx^0+hdx^1)=\hbox{const}\cdot ds = \hbox{const} \cdot
  \sqrt{\Big|2dx^0dx^1+h(dx^1)^2\Big|} \,. \ee
The resulting  quadratic equation has the solutions
\ba {dx^1\over dx^0}&=&{-1\pm\sqrt{\displaystyle{c\over c-h(x^0)}}\over
  h(x^0)} \quad,\quad c=\mbox{const}\, , \pl{exsol1} \\
  \noalign{\vspace{-0.3 cm}}\nonumber\\
  x^0&=&\hbox{const},\quad \mbox{if} \,\, h'(x^0)=0 \, . \pl{exsol2} \ea
\re{exsol1}) is meant to hold  only  when  meaningful; the
condition in \re{exsol2}) is immediate from \re{extrem2}).
The null-extremals \re{null1}), \re{null2a}) can be obtained from
\re{exsol1}) as the limiting case $c\rightarrow\infty$.
Equation \re{exsol1}), even if it cannot be integrated explicitly, permits
a comprehensive
qualitative discussion of the extremals. For brevity we do not go into
details here.  Let us just point out that under the flip transformation
\re{glue}) each extremal \re{exsol1}) is mapped onto another one with the same
value of $c$, but with the opposite sign of the square root. Similarly the
two types of null extremals \re{null1}) and \re{null2a}) get interchanged.
It is straightforward to see that the line element for the extremals
(\ref{exsol1})
is
\be ds={1\over\sqrt{|c-h(x^0)|}}dx^0 \, ,
  \el 11
which will be used to determine their completeness properties.
The extremals \re{exsol2}), on the other hand, are obviously always complete,
since $x^1 \propto s + const$ ranges over all of $\dR$.

\subsection{Construction of the Building Blocks, Carter-Penrose
  Diagrams \pl{Blocks} }
In this and the following subsection we will derive
the general rules of how to find the Carter-Penrose diagrams
starting from any given metric of the form \re{011h}). Thereafter
then we will summarize the resulting simple building block
principle and apply it to the specific models \I{JT}), \I{R2}),
and \I{KV}) for illustration.

Let us first assume that $h$ has no zeros. Then the diffeomorphism
\be x^+=x^1 + 2f(x^0),\quad x^- =x^1
  \el conf
with $f(x^0)$ as before \re{fun})
brings the solution into conformally flat form, i.e.,
the metric in the new coordinates reads $g=h(x^0)dx^+dx^-$.
The Killing field $\partial\over\partial x^1$ then becomes
${\partial\over\partial x^+}+{\partial\over\partial x^-}$ and the flip
transformations \re{glue}) are simply the reflections at any of the
lines $x^++x^-=const$.

Now, according to the asymptotic behavior of $h$ there are three cases
to be distinguished: First, the range of $f$ may be all of $\dR$
(e.g. for $h\sim{(x^0)}^{k\leq1}\hbox{\ as $x^0 \to \pm\infty$}$). Then the
image of the diffeomorphism \re{conf}) is the whole $(x^+,x^-)$--coordinate
plane. Second, the range of $f$ may be bounded from one side (e.g.\ for
$h\sim{(x^0)}^{k>1}\hbox{\ as $x^0 \to \pm\infty$}$). In this case the image is
only the half plane
\footnote{This reflects the fact that then in the original
  $(x^0,x^1)$--coordinates any of the null extremals \re{null2a}) has one of
  type \re{null1}), i.e.\ $x^1=const$, as an asymptotic and thus the
  null-lines \re{null2a}) do not intersect all of the null-lines \re{null1}).}
bounded by the line $x^+-x^-=\lim f$.
Finally, $f$ may be bounded from both sides, in which case the image is a
ribbon $a<x^+-x^-<b$.
\par
%
%
\begin{figure}[ht]
\leavevmode
\begin{center}
\epsfig{file=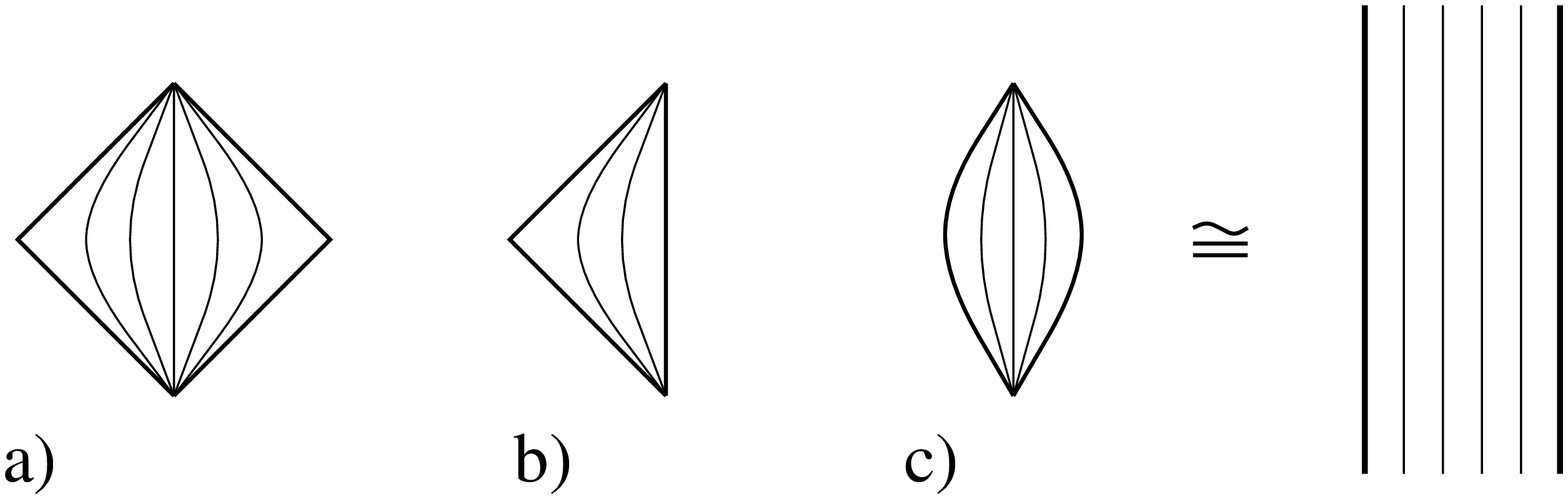,width=12cm}
\renewcommand{\baselinestretch}{.9}
\caption{
  \captionstyle \label{fig:fig1}
  Carter-Penrose diagrams
  for $h>0$ (thin lines represent Killing-trajectories).}
\end{center}
\end{figure}
In order to get the Carter-Penrose diagrams we further apply a
conformal diffeomorphism like $x^\pm\rightarrow\tan x^\pm$ which
maps the solutions into a finite region, and finally we turn the
patch 45 degrees (counter-)clockwise for $h\,(<)\!>0$, given our
convention that a positive (negative) $ds^2$ corresponds to a
timelike (spacelike) distance.
\footnote{Remember that $h$ measures the norm squared of the Killing field.
  Thus $h\!<\!(>)\,\,0$ corresponds to homogeneous (stationary) spacetimes.}
The three cases then correspond to a square, triangle or a lens-shaped region,
resp.\ (cf Fig.~\ref{fig:fig1}; the diagrams for $h<0$ are similar but
turned by 90 degrees). In the last case, however, we will sometimes prefer the
(uncompressed) infinite ribbon form.

\medskip

\begin{figure}[hpt]
\begin{center}
\leavevmode
\epsfig{file=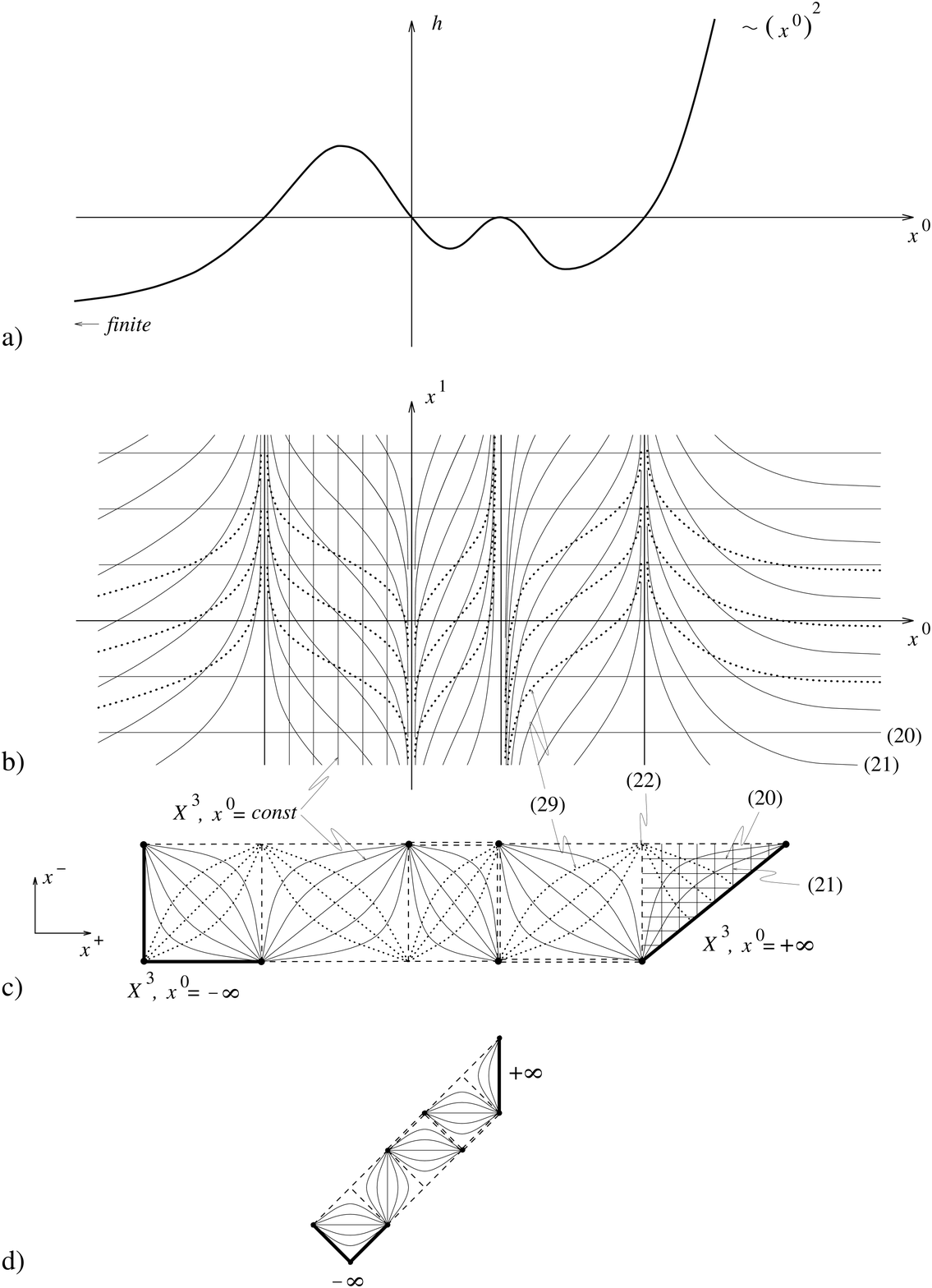,width=12cm}
\end{center}
\renewcommand{\baselinestretch}{.9}
\caption{
  \captionstyle \label{fig:fig2}
  Construction of the
  fundamental building block for a fictitious function $h(x^0)$. In
  (b) and part of (c) the null-extremals \re{null1}), \re{null2a})
  are drawn; in the final building block (d) they would run under
  $\pm 45$ degrees, but have been omitted. In (c),(d) the horizons
  \re{null2b}) have been drawn as dashed lines (degenerate ones as
  multiply dashed lines). There and in the Carter-Penrose diagrams
  we have also included the Killing trajectories (= lines of
  constant $\phi\equiv X^3$), which in (b) have been the straight
  lines $x^0= const$, as thin solid lines. The function $\phi$, and
  simultaneously $x^0$, increases monotonically throughout the block
  and the `boundary' values of $\phi$ are written to the
  corresponding boundary segments. Finally, in (b),(c) we have drawn
  the special extremals \re{13}), which are simultaneously the
  possible flip-axes for the extension. }
\end{figure}

Let us now come to the cases where $h$ has zeros.
These shall be treated by means of a generic example, the
function $h$ of which is drawn in Fig.~\ref{fig:fig2}a.
In Fig.~\ref{fig:fig2}b we qualitatively depicted  representatives of the null extremals
\re{null1}, \ref{null2a}, \ref{null2b}) corresponding to the metric of
Fig.~\ref{fig:fig2}a.
Now, within each sector the diffeomorphism \re{conf}) could be applied (again
exhibiting the flip symmetry of the sector);
it, however, breaks down at $h(x^0)=0$.
It may be cumbersome to write down explicitly the diffeomorphism that
brings $g$ into conformal form on all of the chart underlying Fig.~\ref{fig:fig2}b.

Fortunately, the explicit form of such a diffeomorphism need not be
constructed; we can proceed with a simple geometric argumentation:
By an $x^1$-dependent distortion of the $x^0$--coordinate one can
straighten the null extremals \re{null2a}), leaving the horizons
\re{null2b})  as well as the null extremals \re{null1}),
i.e.\ $x^1=x^- = const$, unmodified.
Note that in our example the $(x^+, x^-)$--chart
cannot be all of $\dR^2$ any more; rather on the right-hand side there will be
some boundary, because due to the asymptotic behavior of $h$ the null lines
of type
(\ref{null2a})
do not intersect all
null lines $x^- = const$. By means of a subsequent conformal diffeomorphism
(like $x^- \to \tan x^-$; cf, however, the following paragraph)
and a similar one for $x^+$
the new coordinate chart covers only a finite region in $\dR^2$; the
result is drawn qualitatively in Fig.~\ref{fig:fig2}c.
The boundary on the right-hand side can be made straight by a conformal
transformation in $x^+$, by means of which one can also transform
all rectangles into squares. The final building block for the
Carter-Penrose diagram is obtained by turning the patch 45 degrees
counter-clockwise, as before, and is depicted in Fig.~\ref{fig:fig2}d.

In this example all the lines $\phi \to \pm \infty$ are complete:
The null extremals running there are complete since the coordinate
$x^0$ ranges over all of $\dR$ in Fig.~\ref{fig:fig2}a (cf the discussion
following equation \re{null2b})); for the nonnull extremals this
follows from \re{11}), since $h(x^0)=O((x^0)^2)$. We will draw
complete boundary lines boldfaced and incomplete ones as thin
solid lines. Horizons are drawn as dashed lines (degenerate
horizons, i.e.\ at higher order zeros of $h$, as multiply dashed
lines); the other null-extremals, which run through the
Carter-Penrose diagrams under $\pm 45$ degrees, are omitted.
Finally, massive dots indicate points at infinite distance.

As pointed out before, each sector exhibits the flip symmetry. In the
example Fig.~\ref{fig:fig2}c,d we took this into account by drawing
the sectors symmetrically; especially those segments of the boundary depicted
as dashed lines are horizons, too, similar to the interior horizons.
The solution will thus be extendible at these boundary lines
and we must pay
attention that the conformal diffeomorphism transforming the $x^-$--coordinate
has the proper asymptotic behavior to allow a smooth extension.
Due to the flip symmetry there is of course a distinguished choice for the
$x^-$-transition function, namely the one making the
sector symmetric (this amounts essentially to taking $x^-$ equal to
some affine
parameter of an $x^+=const$ null-extremal). Unfortunately, different sectors
may yield contradictory functions, so in general one cannot simultaneously
achieve symmetry for all the sectors of the building block.

\newpage 

%
%

\newpage

%
%
\begin{figure}[ht]
\begin{center}
\leavevmode
\epsfig{file=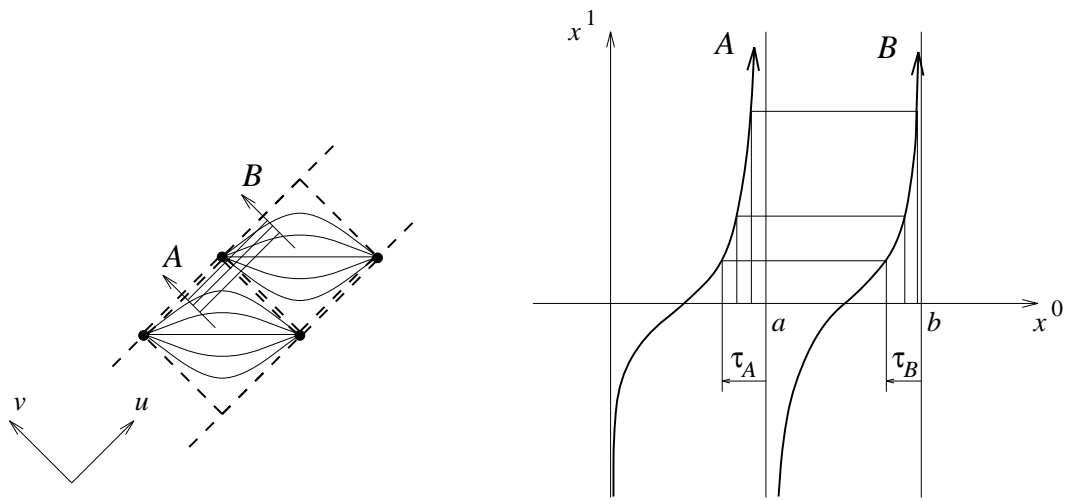,width=14cm}
\end{center}
\renewcommand{\baselinestretch}{.9}
\caption{
  \captionstyle \label{fig:fig3}
  Illustration of relative blue-shift.}
\end{figure}
The situation becomes particularly drastic
if $h(x^0)$ has zeros of different order as, e.g., in the example chosen in
Fig.~\ref{fig:fig3} (cf also Fig.~\ref{fig:fig2}  and the examples
{\bf R3,4, 7,10} in
section \ref{Recipe}):
Let $A$ and $B$ be null-extremals leaving the building block across a
degenerate (double zero of $h$) resp.\ nondegenerate horizon as shown in
Fig.~\ref{fig:fig3}. In the chart \re{011h}) they are described by equation \re{null2a})
and as their affine distance to the horizon we may choose $\t_A:=a-x^0$
and $\t_B:=b-x^0$. The conformity of the building block requires that a chosen
spacing along the null-extremal $A$ transfers to the extremal $B$ via the
family of perpendicular null-extremals \re{null1}), i.e.\ the lines
$x^1$ or $v=const$,
(cf Fig.~\ref{fig:fig3}). Now, near the horizon the extremal $A$ may be described
asymptotically by $x^1\sim1/\t_A$ and $B$ by $x^1\sim-\ln\t_B$, hence
$\t_B\sim e^{-1/\t_A}$. When approaching the horizons $\t_A$ and $\t_B$
vanish simultaneously, but also ${d\t_B\over d\t_A}\rightarrow 0$.
This already impedes a smooth diagram, which is seen as follows:
Choose a conformal gauge \re{confgauge}) and let $A$ and $B$ be represented
by $u=const=u_A$ or $u_B$, resp. The affine parameter along these
null-extremals then satisfies $d\t_A/dv \propto \exp(\r(u_A,v))$ and
likewise for $B$, as is easily shown. But from the above we know that
at the horizons
\be {d\t_B\over d\t_A} = {d\t_B/dv\over d\t_A/dv} \propto
  {\exp(\r(u_B,v))\over\exp(\r(u_A,v))} \rightarrow 0 \, .\nonumber \ee
Thus the conformal factor $\exp(\r(u,v))$ will either diverge
along $A$ when approaching the horizon or, if rendered finite
along $A$, vanish along $B$. It is thus {\em impossible\/} to draw
a smooth Carter-Penrose diagram in this case! A physical
consequence of this is that an observer $B$ approaching the
horizon, watching $A$ approaching the degenerate horizon (both
observers may also be timelike), will notice
that $A$ becomes infinitely blue-shifted.
\footnote{More precisely, if $A$ is sending out
  signals towards $B$ at a constant rate with respect to its affine
  parameter (proper time), $B$ will receive them with an ever increasing and
  finally diverging frequency (with respect to its own affine parameter).}
Similar problems occur for any combination of different horizons, only
if all horizons are of equal degree the diagram will be smooth.

Nevertheless, it is still possible and instructive to use such
`nonsmooth Carter-Penrose diagrams' for book-keeping when
constructing the extension and for studying the causal structure.
In the illustrations Fig.~\ref{fig:fig11}, \ref{fig:fig13} we have marked those
nonsmooth diagrams with the sign \attn. At any rate, the extended
solutions themselves will be smooth.

But even for only nondegenerate horizons there may occur problems, if two
triangular sectors meet. While for {\em one\/} sector the boundary can always
be made straight by a conformal diffeomorphism, this may be impossible for
{\em two\/} adjacent sectors simultaneously:
Any conformal diffeomorphism (which is a reparameterization of each of the two
lightcone-coordinates $x^+,x^-$) alters the angles between boundary and
horizon on both sides of the horizon in the same sense; more precisely, the
ratio of the tangents of the respective angles is invariant.
\begin{figure}[pht]
\begin{center}
\leavevmode
\epsfig{file=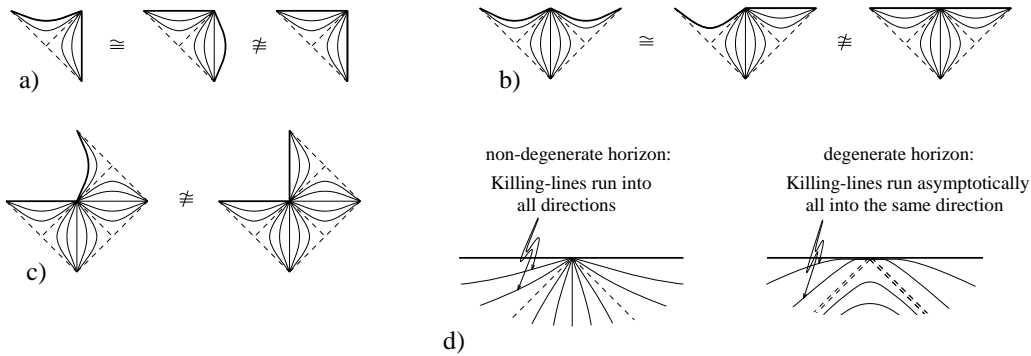,width=14cm}
\end{center}
\renewcommand{\baselinestretch}{.9}
\caption{
  \captionstyle \label{fig:fig4}
  Distorted Carter-Penrose-diagrams.}
\end{figure}
For instance, in the situation of Fig.~\ref{fig:fig4}a, a conformal diffeomorphism which
makes the upper boundary straight must necessarily bulge out the right
boundary (at least near the corner point), so the angles of the boundaries
against the horizon can never be made equal. This argument is also valid for
the (already `extended') diagram Fig.~\ref{fig:fig4}c, and for similar reasons the angle
in Fig.~\ref{fig:fig4}b cannot be smoothed. At the end of section \ref{Saddle}
(cf Fig.~\ref{fig:fig7}) we will show by an example that the distorted
diagram is in fact the generic case, if two `triangular' sectors meet.

Finally, for degenerate horizons of equal degree these distortions
do not occur; one can always obtain right angles or straight
lines. But also the Killing lines behave differently then: As can
be shown easily using \re{011h}), \re{null1}), \re{null2a}) they no
longer run from the corner point of the sector in all directions;
instead they leave the corner point in only one direction
asymptotically (tangential to the boundary, if there is one; cf
Fig.~\ref{fig:fig4}d). This feature applies also to (4d) extremal
Reissner-Nordstr\"om with its degenerate horizons, which is in
this respect drawn incorrectly in most textbooks (e.g.
\cite{HawkingEllis}). But also the other issues illustrated by
Figs.~\ref{fig:fig4} and \ref{fig:fig5} are usually ignored in the literature.

\subsection{Maximal Extension and Saddle-Point Charts \pl{Saddle}}
In the previous section we have compressed the
coordinate patches \re{011h}) into finite building blocks. Now we
will investigate whether these blocks have to be extended and how
to do so.

Let us again examine the cases without zeros of $h$ first. Their
building blocks are drawn in Fig.~\ref{fig:fig1} and they are already
inextendible, which is seen as follows: If an extension were
possible at all, then it must be possible (also) along the
null-infinities (i.e., the 45-degree boundaries of Fig.~\ref{fig:fig1}a,b)
or along the timelike (for $h>0$) singularities
\footnote{We will use this term even if the `singularity' turns out to be at
  an infinite distance.}
(the vertical resp.\ curved
boundaries in Fig.~\ref{fig:fig1}b,c). The extension cannot be possible over a corner
point {\em alone\/}, since this point then cannot be an interior point
of a new local chart.
%
%

Now, if $x^0$ ranges over all of $\dR$, then these boundary lines
lie at infinite distance since $x^0$ is an affine parameter for
the null-extremals \re{null1}, \ref{null2a}). Thus they cannot be
regular interior points and the solution is inextendible. Note,
however, that in general null- and nonnull extremals may have
different completeness properties! The completeness of the
null-extremals depends on the domain of $x^0$, whereas the
completeness of the other extremals hinges largely on the
(asymptotic) behavior of the function $h$ (cf equation \re{11})).
E.g., the boundaries for the $R^2$--model \I{R2}) are
`null-complete' (as $x^0 \in \dR$) but `nonnull incomplete'
(as $\lim_{x^0 \to \pm \infty} s =$ finite, cf \re{11})).
Usually a boundary point is called complete only if {\em all\/}
extremals running into it are complete. On the other hand, {\em
one\/} complete extremal is sufficient to make the metric
inextendible through that boundary point.

If $x^0$ does not range over all of $\dR$, then the null-extremals
are incomplete and perhaps also the other extremals. However, even
then the solution is inextendible, since on these boundaries some
physical field ($\phi$ or some dilaton field) diverges. Of course,
if these fields are not taken seriously (or if one drops the
restriction $\U'(\Phi) \neq 0$, requiring it locally only), then one
might  try to extend
the function $h(x^0)$ smoothly or analytically beyond its original
domain and repeat the analysis with this new $h$ to obtain an
extension.

\begin{figure}[pht]
\begin{center}
\leavevmode
\epsfig{file=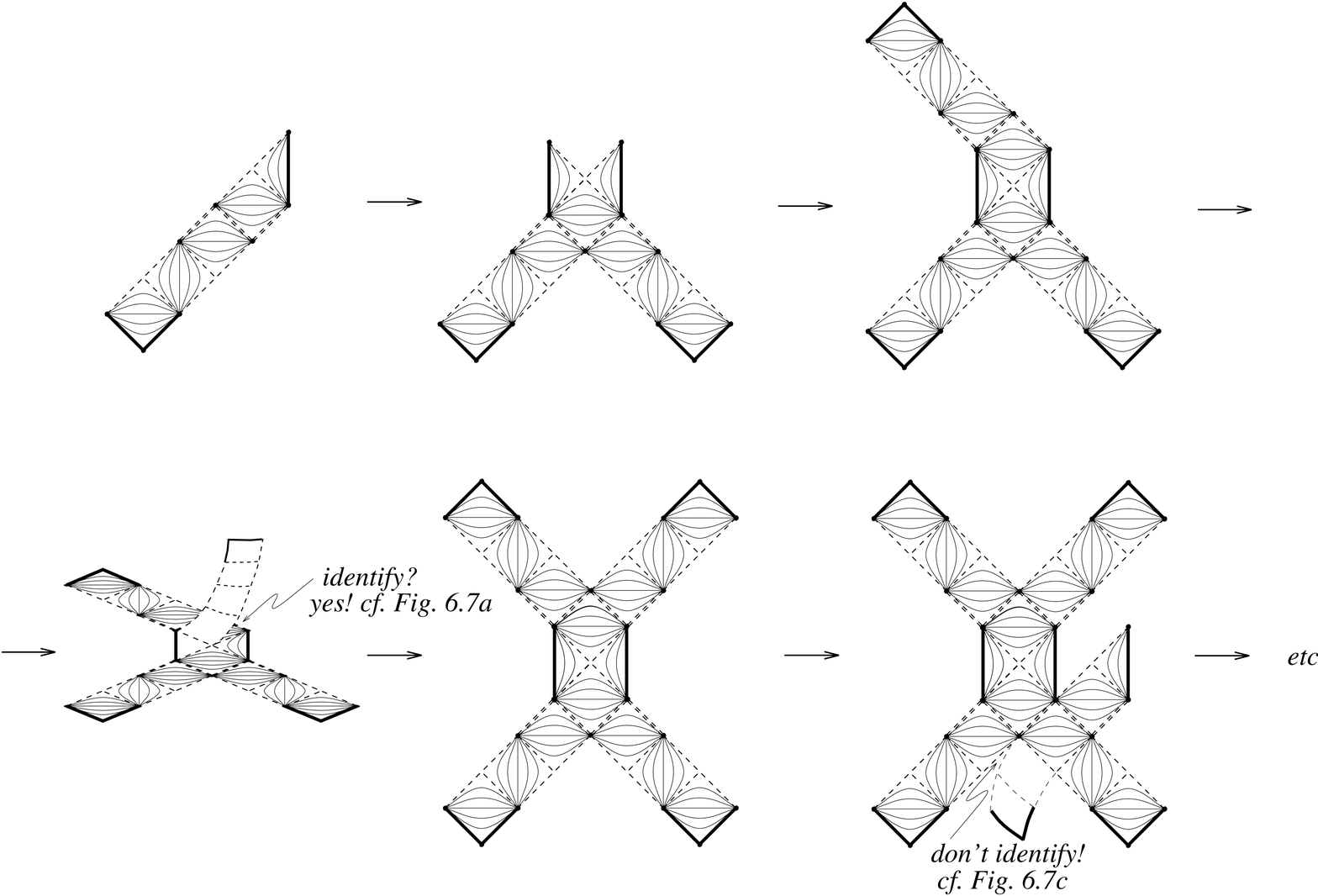,width=8cm}
\end{center}
\renewcommand{\baselinestretch}{.9}
\caption{
  \captionstyle \label{fig:fig5}
  The gluing procedure.}
\end{figure}

In the case where $h$ has zeros, the discussion above is still
valid for the first and the last sector, i.e., an extension is not
possible beyond those boundary lines. However, at the other,
interior, boundary lines the solutions are extendible and this can
be done as follows: Already in section \ref{General} we have
introduced the flip transformation \re{glue}). In the
Carter-Penrose diagrams this flip shows as a reflexion of the
sector in question. The reflexion axis, running diagonally through
the sector, has to be {\em transversal\/} to the Killing lines
(i.e.\ horizontal for stationary sectors and vertical for
homogeneous ones; cf e.g.\  the dotted lines in Fig.~\ref{fig:fig2}c).
The flip transformation breaks down at the horizons bounding the
sector, because it maps the interior horizons onto exterior ones.
We can, however, flip (i.e.\ reflect) the whole building block and thus obtain
an extension beyond the previously `exterior' horizons.
The gluing thus amounts to taking the mirror image of the block and patching
the corresponding sectors together.
This procedure has to be performed at each sector of the first building block
and after that also at each sector of the new building blocks and so on
(cf Fig.~\ref{fig:fig5} for the first steps).
\par
This gluing process is essentially unique; the only free parameter
is the choice of the symmetry axis of the flip transformation
(i.e.\ the choice of the constant in \re{fun})), but it only
results in a coordinate change (shifting the $x^1$-origin of the
charts \re{011h})). As long as only the universal covering is
pursued this does not affect the solution. However, when further
identifications of sectors are made, such that the resulting
solution is not simply connected, then we can have an effect of
this parameter (this will be discussed in section \ref{Secall}).

\medskip

\begin{figure}[pht]
\begin{center}
\leavevmode
\epsfig{file=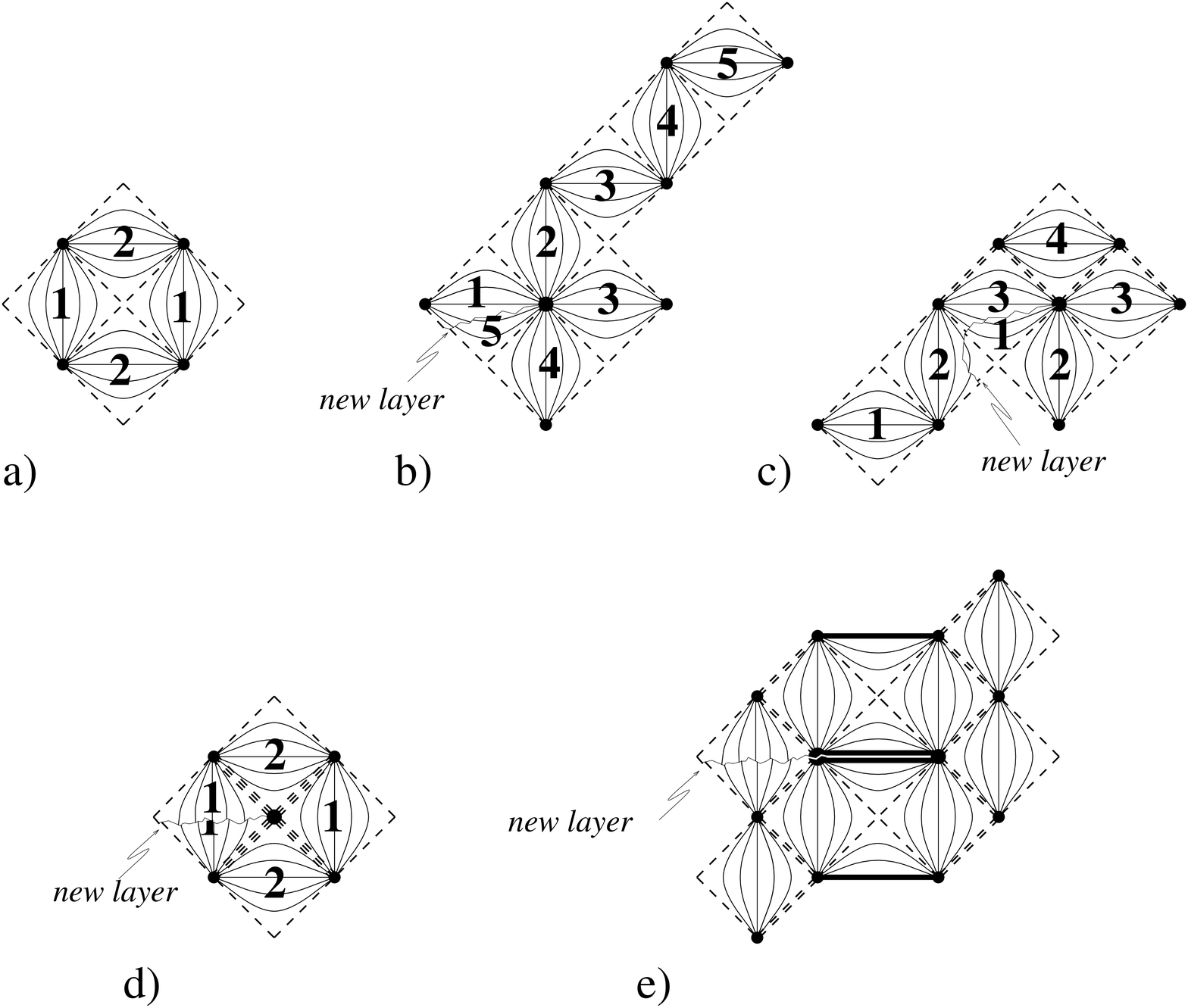,width=14cm}
\end{center}
\renewcommand{\baselinestretch}{.9}
\caption{
  \captionstyle \label{fig:fig6}
  Gluing around a vertex point. Only in (a) the four
  sectors make up a single sheet around the vertex point. In all other cases
  (b--d) the vertex point is not an interior point and the solutions must be
  extended into further new layers covering the original sectors. The same is
  true for the general situation of e.g.\ (e), showing part of the extended
  diagram {\bf G7} (cf Fig.~\ref{fig:fig13}).}
\end{figure}
We have yet to investigate the case (see Figs.~\ref{fig:fig5}) and 
\re{fig:fig6})
that after surrounding the  point at the vertex of four blocks
the overlapping sectors of the first and the fourth block
match. Shall they be identified?

This depends: There is exactly one case where they must be identified
(Fig.~\ref{fig:fig6}a) and there are a couple of cases where they could (but shall not).
A necessary requirement is of course that the sectors are equal
(i.e.\ isometric, fields like $\phi$ coincide). This is the case in (a),(d)
of Fig.~\ref{fig:fig6}, but generically not in (b),(c), because there different sectors
overlap (e.g.\ in (b) the sectors 1 and 5, if numbered consecutively in the
building block); only under rare circumstances (e.g.\ periodic functions $h$
and $\phi$) they could happen to fit onto one another.

Now to the question whether they shall: If the solution {\em
can\/} be extended smoothly into the vertex point, then of course
this must be done, and in this case the overlapping sectors have
to be identified to make a single sheet around this vertex point
(as shown below this will be the case precisely in the situation
Fig.~\ref{fig:fig6}a). If, on the other hand, the vertex point turns out to
lie at infinite affine distance, then it cannot possibly be an
interior point. If the sectors were identified, then one would
have a noncontractible loop around this vertex point, which
cannot be accepted since we intend to construct the (simply
connected) universal covering. Thus we must not identify the
overlapping sectors but start a `new layer" and continue the
gluing, giving rise to the winding-staircase like structure
outlined in Fig.~\ref{fig:fig6}b--d (cf also Fig.~\ref{fig:fig11}/{\bf R2} or
Fig.~\ref{fig:fig13} {\bf G4}). This applies also to the situation (Fig.~\ref{fig:fig6}e) that a whole slit is surrounded before overlapping, or even
more generally to any occurence of overlapping sectors different
from the arrangement Fig.~\ref{fig:fig6}a. Certainly, even if we identify the
sectors we still have a (multiply connected) solution of the
equations of motion. However, this gluing process is not
unambiguous but introduces a (geometrically meaningful) parameter.
We will analyze those solutions in all detail in section
\ref{Secall}.

Our considerations on the completeness of the horizons (cf the paragraph
below \re{null2b})) show already that Fig.~\ref{fig:fig6}a is the only candidate for a
regular vertex point: Degenerate horizons are complete in both directions,
whereas nondeg.\ horizons are complete on one side only, which is easily
recognized as that side of the building block where the Killing lines converge
(in Fig.~\ref{fig:fig6} all such complete points have been marked by
massive dots). But even the general extremals running towards those vertex
points show the same behavior: Those running along with converging
Killing-lines into the vertex point are either oscillating ones
(between two fixed $\phi$-- resp.\ $x^0$--values, cf \re{exsol1})) or of the
kind \re{exsol2}),
which are both complete. And those running transversally to the
Killing lines are precisely those of \re{exsol1}) with $c=0$, i.e.,
\be {dx^1\over dx^0}=-{1\over h},
  \el 13
and coincide, furthermore, with the possible symmetry axes for the flip
transformations (in Fig.~\ref{fig:fig2}b,c we have drawn some of them as dotted lines).
Their length follows from \re{11}) where $a$ is the zero of $h(x^0)$ in
question:
\be s=\int^a {dx^0\over\sqrt{|h|}} \sim \int^a
  {dx^0\over{(x^0-a)^{n\over2}}} \to \left\{
                         \begin{array}{r@{\quad}l}
                                <\infty & n=1 \\
                                 \infty & n\ge 2
                         \end{array} \right. \quad. \el length
It is finite only at simple zeros.
Thus, whenever a degenerate horizon runs into the vertex point or Killing lines
focus there, then this vertex point is at infinite distance
and has to be taken out of consideration. In these cases, to obtain the
universal covering we must not identify the overlapping sectors but continue
the gluing in a new layer.

In the case of Fig.~\ref{fig:fig6}a (only nondeg.\ horizons and the Killing lines
avoiding the vertex point), however, all extremals are incomplete, and it is
thus to be expected that the vertex point is a regular interior point of the
mani\-fold. This is indeed true:
equation \re{saddle}) represents a smooth nondegenerate
metric for the neighbourhood of such points,
which reveals the vertex point as regular interior point
(a saddle point of $\phi$, with $X^a=0$).
The four adjacent sectors then constitute one single sheet (cf also
diagrams 3--5 of Fig.~\ref{fig:fig5}).

Keeping in mind the Schwarzschild-like form of \re{gSS}), there is
also an alternative way of obtaining such a saddle-point chart,
imitating the Kruskal-Szekeres procedure. It was initially
proposed by M. Walker \cite{Walker} for a special rational form
of $h$. We want to show here that it works for {\em any\/}
sufficiently smooth $h$ (say $C^n$) with simple zero at $x^0=a$.
The transformation reads (with $f$ as in \re{fun}))
\be
u=\hbox{sgn}(h)\exp{h'(a)\left[f(x^0)+\frac{x^1}{\scriptstyle2}\right]}
  =\hbox{sgn}(h)\exp{\int\limits^{\,\,\,\,x^0}\!\!\frac{\scriptstyle
  h'(a)}{\scriptstyle h(x)}dx
  +\frac{\scriptstyle h'(a)}{\scriptstyle2}x^1 }
  \el Krtrfu
\be v=\exp{-\frac{\scriptstyle h'(a)}{\scriptstyle2}x^1}  \, \, .
  \el Krtrfv
It brings the metric \re{011h}) into the form
\be ds^2=-\frac{4\wt h(uv)}{h'(a)^2}\frac{dudv}{uv}\, ,
  \el Kruskal
which is evidently nonsingular and nondegenerate around $u,v=0$.
Here $\wt h(uv)$ is determined  implicitly
via $\wt h(uv)=h(x^0(u,v))$.
The integration constant hidden in $u$ has to be chosen
such that $u$ is continuous at $x^0=a$.

To show that the transformation \re{Krtrfu}), \re{Krtrfv}) is really a
(local) diffeomorphism
consider the following decomposition of the integrand in \re{Krtrfu})
(separating the singular term):
\be \frac{h'(a)}{h(x)}=\frac1{x-a}-\frac{\frac{h(x)}{x-a}-
  h'(a)}{h(x)}\,\, . \ee
The first term yields (after integration and exponentiation) a factor
$(x-a)$,
\footnote{Note that the coefficient $h'(a)$ in \re{Krtrfu}) is absolutely
  necessary: Otherwise we would have a factor $k\ne1$ in the singular term,
  which then integrates to $(x-a)^k$, and the transformation would no longer
  be a diffeomorphism. This seems to have been ignored in the literature
  sometimes.}
and also the second term
(being $C^{n-2}$, if $h\in C^n$) behaves
perfectly well. It is thus not at all necessary that $h$ is of the special
rational form given in \cite{Walker,Brill}.

\medskip

We are now finally in the position of proving the statement on the
distorted boundary of the $R^2$--gravity solutions made at the end
of section \ref{Blocks}. Let e.g.\ $\L=-{1\over3},C={4\over3}$,
hence (cf equation (\I{R2h}))
$h={2\over3}\left[2-x^0-(x^0)^3\right]$. Equations
\re{Krtrfu}), \re{Krtrfv}) can be integrated easily to \ba
uv&=&\mbox{sgn}(h)\exp\int\limits^{\,\,\,\,x^0}{{h'(1)}\over{h(x)}}dx=
                                                               \nonumber \\
      &=&\mbox{sgn}(h)\sqrt[8]{(x^0)^2-2x^0+1\over(x^0)^2+x^0+2}
\exp\left({{-3\over4\sqrt{7}}\arctan{1+2x^0\over\sqrt{7}}}\right)\,\, .  \ea
The singularities lie at $x^0=\pm\infty$, i.e.\ at
$uv=\exp\left(-{3\p\over8\sqrt{7}}\right)\approx1.561$ and
$uv=-\exp\left(+{3\p\over8\sqrt{7}}\right)\approx-0.641$,
respectively.
\begin{figure}[ht]
\begin{center}
\leavevmode
\unitlength 1cm
\begin{picture}(14,3)
\put(0,0){\epsfig{file=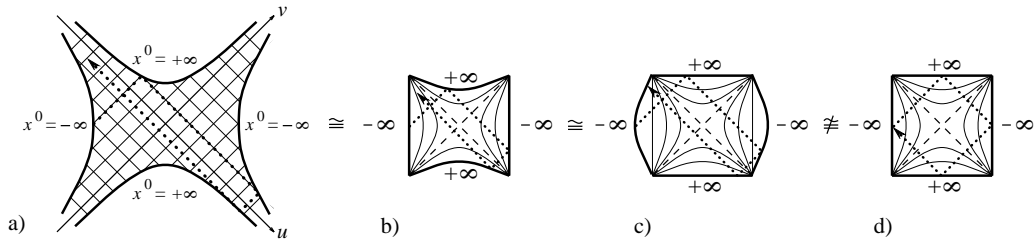,width=14cm}}
\end{picture}
\end{center}
\renewcommand{\baselinestretch}{.9}
\caption{
  \captionstyle \label{fig:fig7}
  Distorted Carter-Penrose-diagrams for the $R^2$--model.}
\end{figure}
The familiar conformal diffeomorphism $u \rightarrow \tan \hat u,
v \rightarrow \tan \hat v$ shows immediately that the correct
shape of the Carter-Penrose diagram must be like in Fig.~\ref{fig:fig7}b,c (it
is of course possible to straighten the one boundary line at the
cost of the other by means of conformal diffeomorphisms). Only if
the two hyperbolae $x^0=\pm\infty$ happen to lie at equal values
of $|uv|$ then we can obtain a square. Thus for nonsymmetric
functions $h$ (with one zero) squares are rather the exception to
the rule. This remark applies also to the square shaped
Carter-Penrose diagrams obtained, e.g., in \cite{Brill,Lemos} for
other special models covered by the present treatment. Finally we
want to mention some physical difference between a world with a
square-shaped Carter-Penrose diagram and another of the generic
form Fig.~\ref{fig:fig7}b,c: Only in the former case a series of null-lines
`bouncing off' the boundaries closes to a rectangle (cf dotted
lines in Fig.~\ref{fig:fig7}).

\subsection{Recipe and examples \pl{Recipe}}

Let us summarize briefly the principle of how to construct the
maximally extended Carter-Penrose diagram corresponding to any
function $h$ in
(\ref{011h}):\\ $\bullet$ The number and kind of
zeros of the function $h$ determines the number of sectors and
their `orientation' in a fundamental building block ($h>0$
corresponds to stationary sectors, $h<0$ to homogeneous ones).
More explicitly: For $n$ zeros of $h$ there are $n+1$ sectors in
the building block. If a zero of $h$ is of odd order, then the
corresponding Killing horizon separates a stationary from a 
homogeneous sector; otherwise the two neighboring sectors are
clearly of the same type (as $h$ does not change its sign).
\\
$\bullet$ The end sectors of the building block are either a square or a
triangle, depending on the asymptotic behavior of $h$. It is a triangle,
if $f(x^0)=\int^{x^0} du/h(u)$ remains finite at the boundary (e.g., for
$h(x^0)\sim (x^0)^{k>1}$ as $x^0\to\pm\infty$), and it is a square, if
$f(x^0)$ diverges (e.g., for $h(x^0)\sim (x^0)^{k\le1}$
as $x^0\to\pm\infty$).
\\
$\bullet$ Choose any sector and establish the symmetry axis, running
diagonally through the sector, transversal to the Killing lines
(i.e.\ horizontal for stationary sectors and vertical for homogeneous ones).
Reflect the whole block at this symmetry axis and identify the corresponding
sectors.
\\
$\bullet$ Proceed in this way with all sectors until you come to an end, or
ad infinitum.
\\
$\bullet$ If after surrounding a vertex point sectors overlap, the Killing
horizons running into that point being nondegenerate (simple
zero of $h$), and if the Killing lines do not focus in that point (Fig.~\ref{fig:fig6}a),
then identify the overlapping sectors such as to make a
single sheet around this vertex point. In all other cases (e.g.\ higher degree
zero of $h$, Killing-lines running into that point, etc.) do not identify the
sectors, but continue gluing in a new layer to get the universal covering.
\\
$\bullet$ Any boundary  is null-complete, iff $x^0 \to \pm \infty$ there.
A boundary point is complete with respect to all other extremals
\re{exsol1}), iff $\int^{x^0} du/\sqrt{|c-h(u)|}$ diverges there,
(cf equation (\ref{11})).
Complete boundaries have to satisfy both conditions (except that only
extremals of one kind run there) and are depicted  boldfaced.

As pointed out repeatedly before, the Carter-Penrose diagrams
obtained in this way are to be understood as {\em schematic\/}
ones only. However, if the zeros of the respective function $h$
are all of the same order, then these diagrams are also smooth. By
this we mean that there exists a (smooth) diffeomorphism from the
universal covering solution to the respective diagram, so that the
diagram may indeed be regarded as a chart to the spacetime (cf
our discussion in section \ref{Seccarter}). Still, if the zeros are
all simple (nondegenerate horizons) and two nonnull boundary
lines meet in a point, then it will in general be necessary to
deform the boundary lines somewhat (cf Figs.~\ref{fig:fig4}a--c, \ref{fig:fig7}).
For a null and a nonnull boundary line meeting
\footnote{If, say, a spacelike boundary meets a null boundary at a corner
  point (e.g.\ in the diagrams {\bf G1-4, 8,9,11} of Fig.~\ref{fig:fig13}), then
  there is obviously no problem to change the angle of the spacelike boundary
  at will, since the null boundary always remains null under conformal
  transformations.}
or for higher order zeros these complications do not occur.

If, on the other hand, $h$ has zeros of different degree, then
there is {\em no} smooth diffeomorphism from the universal
covering solution to the Carter-Penrose diagram; the latter has to
be regarded as purely schematic then. An explanation from the
physical point of view has been
provided by means of a relative red/blue-shift between observers
approaching Killing horizons of a different type (cf Fig.~\ref{fig:fig4}).
Certainly the universal covering solutions themselves are still
smooth everywhere  in these cases, too: The gluing diffeomorphism
\re{glue}) may be used to patch together charts \re{011h}) (this
poses no problem since only {\em one} kind of Killing horizon at a
time is involved) and an atlas is obtained when completing  these
charts by those of type \re{saddle}) or \re{Kruskal}). Obviously,
also in these cases it is instructive to keep track of the
patching by a (schematic Carter-Penrose) diagram; only  it may not
be taken as a true smooth image of the respective spacetime then.

\vspace{.3cm}

\begin{figure}[pht]
\begin{center}
\leavevmode
\epsfig{file=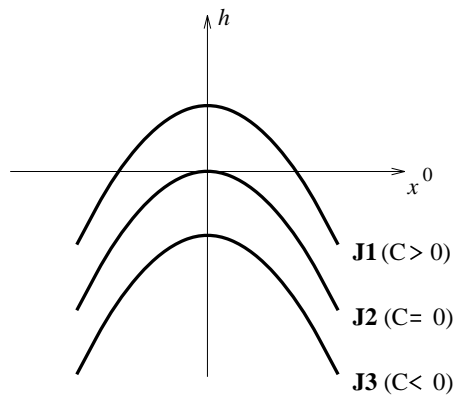,width=6cm}
\end{center}
\renewcommand{\baselinestretch}{.9}
\caption{\captionstyle \label{fig:fig8}
  Functions $h(x^0)$ for the JT--model.}
\end{figure}

We now come to the announced examples, starting with
(anti-)deSitter gravity \I{JT}). Since all values $\L \neq 0$
yield (basically) equivalent Carter-Penrose diagrams, we will set
$\L:=-2$ in the following. This describes deSitter gravity. For
positive values of $\L$ (anti-deSitter gravity) the Carter-Penrose
diagrams obtained below have to be turned by an angle of 90
degrees, while $\L=0$ yields the diamond like Minkowski space
diagrams, of course. We then get a one-parameter family of
functions $h$ [cf equation (\ref{h})], $h^{JT}(x^0)=C - (x^0)^2$,
parameterized by the Casimir constant $C = X_aX^a + (\phi)^2$ (cf
Fig.~\ref{fig:fig8}).

\par
\begin{figure}[pht]
\begin{center}
\leavevmode
\epsfig{file=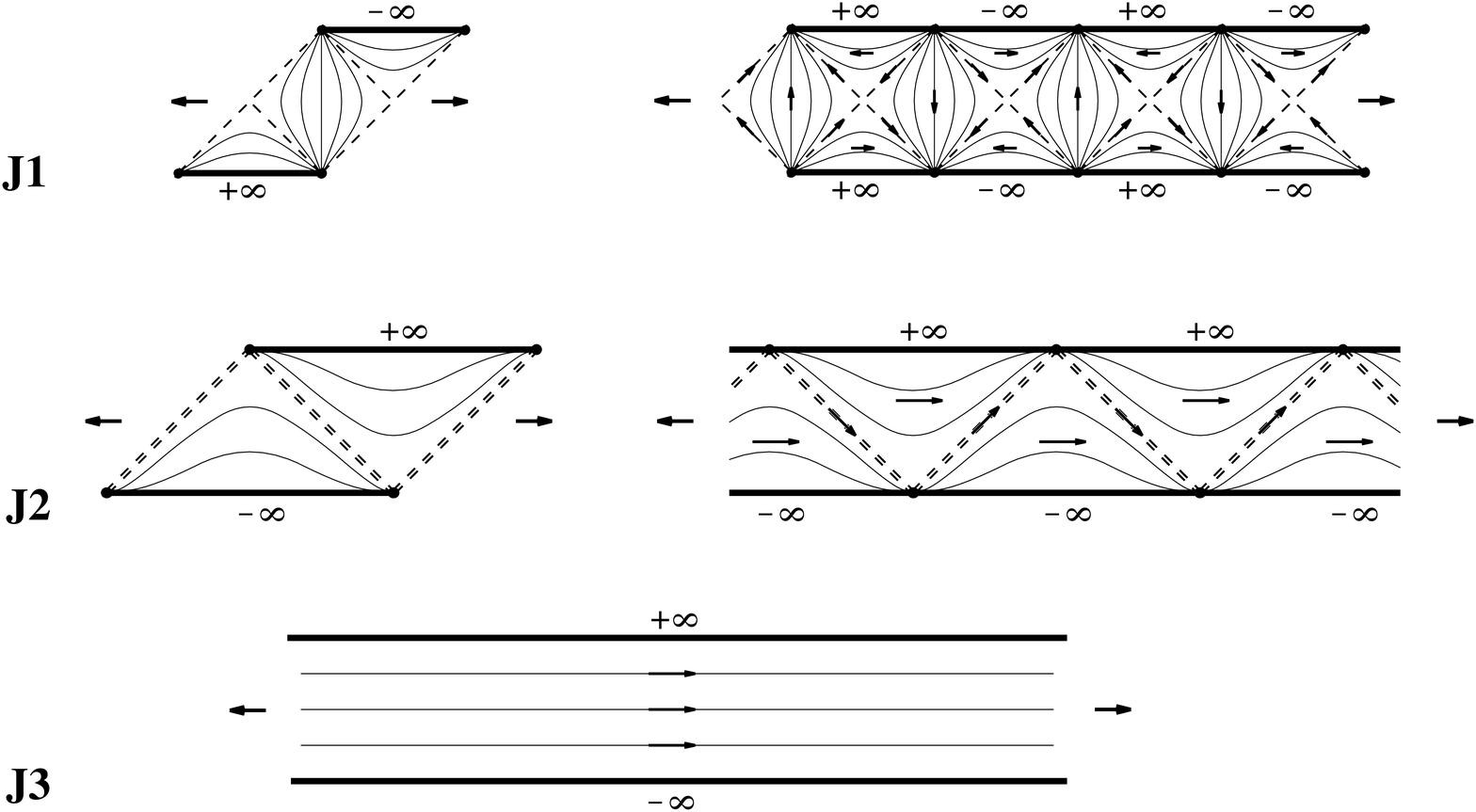,width=13cm}
\end{center}
\renewcommand{\baselinestretch}{.9}
\caption{
  \captionstyle \label{fig:fig9} Carter-Penrose diagrams
  for the JT--model; left the building blocks, right the extended
  diagrams. Arrows inside the diagrams indicate Killing fields for
  those solutions.}
\end{figure}

{\bf J1:} For any $C>0$ this curve $h(x^0)$  has two simple zeros, leading to
one square within the fundamental building block.
Asymptotically we have $h\sim-{(x^0)}^2$, so that
adjacent to the square there will be a triangle at each side, the boundary
of which, $\phi = \pm\infty$ resp., is spacelike and
complete. Gluing leads to the ribbon-like diagram shown in Fig.~\ref{fig:fig9}.
(For the other sign of the cosmological constant, e.g.\ $\L=+2$, we get the
same diagram for $C<0$, rotated, however,  by 90 degrees, as the infinity
is timelike then).

{\bf J2:} For $C=0$ we get no square, but only two triangles. The
corresponding Carter-Penrose diagram is again plotted in Fig.~\ref{fig:fig9}.

{\bf J3:} For $C<0$ the function $h$ has no zeros (Fig.~\ref{fig:fig8}). We therefore may
apply directly the diffeomorphism (\ref{conf})  with the function
(\ref{fun}), which in the present case can be written in terms of
elementary functions:
\be f(x^0)=-{2 \over \sqrt{-C}} \arctan \left( {x^0 \over \sqrt{-C}} \right) \, .
  \el JTfun
The resulting region $(x^+,x^-)$ is again a ribbon (Fig.~\ref{fig:fig9});
but this time without such a periodic internal structure as before,
since the Killing lines $\phi=x^0=const$ become the parallels $x^+-x^-=const$
in the present case.

Clearly, as manifolds the solutions {\bf J1-3} are all the same, namely
the unique (simply connected, maximally extended) manifold with zero
torsion and constant curvature ($R\equiv\L$, cf equation \re{JT})).
%
%
The difference between the spacetimes {\bf J1-3} arises only from the
function $\phi$ defined on them.  Preservation of $\phi$ reduces the
originally three independent Killing-fields to only one symmetry
direction (indicated by arrows in Fig.~\ref{fig:fig9}).

Note that while the Carter-Penrose diagrams for anti-deSitter
space coincide for all dimensions (vertical ribbon), they
apparently differ for deSitter space: D-dimensional deSitter space
can be obtained from restricting a $D+1$-dimensional Minkowski
metric to the one-sheet hyperboloid, the topology of which is
$\dR\times S^{D-1}$. In particular, this hyperboloid is simply
connected for $D>2$ and thus yields already the universal covering
space. Also, upon spherical reduction, its Carter-Penrose diagram
is the familiar square \cite{HawkingEllis}.
For $D=2$, on the other hand, this hyperboloid is topologically a
cylinder $\dR\times S^1$ and the universal covering, obtained by
unwrapping the $S^1$ to $\dR$, is clearly the horizontal ribbon of
Fig.~\ref{fig:fig9}. Thus for $D=2$ (and only there) deSitter and
anti-deSitter space are related by a simple signature change and
the corresponding Carter-Penrose diagrams are just rotated by 90
degrees against one another.

\medskip

Let us now turn to the second example: $R^2$--gravity \I{R2}). The
corresponding function $h(x^0)$ is in this case [cf equation
\I{R2h})] $h^{R^2}=-\frac23(x^0)^3+2\L x^0+C$. Since $x^0$ ranges
over all of $\dR$ and at infinity $h\sim (x^0)^3$ we get
incomplete (null-complete but nonnull-incomplete) triangular
sectors at both ends of the building block. The corner points are,
however, complete, since only the null extremals \re{null2b}) run
into them. A more detailed analysis yields five  qualitatively
different cases depending on the parameters $C$ and $\Lambda$ (see
Fig.~\ref{fig:fig10}):

\begin{figure}[pht]
\begin{center}
\leavevmode
\epsfig{file=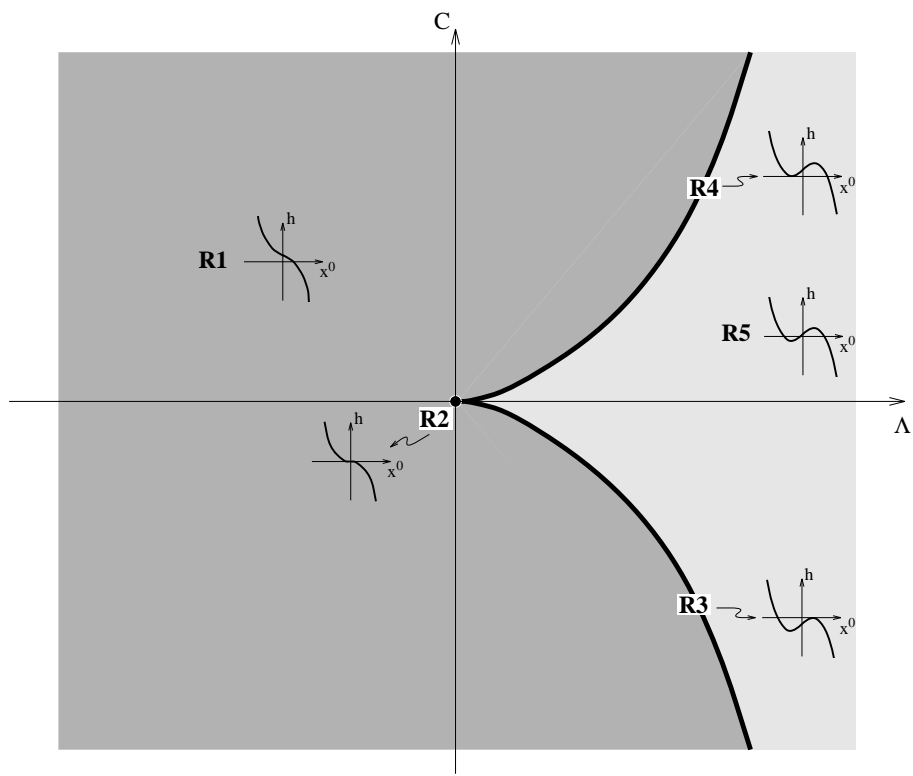,width=12cm}
\end{center}
\renewcommand{\baselinestretch}{.9}
\caption{
  \captionstyle \label{fig:fig10}
  $R^2$--gravity, overview.}
\end{figure}

\smallskip
\halign{{\bf R#}:\quad&#\hfill\cr
      1 &one simple zero of $h$ at $x^0=B$ \cr
2 &one triple zero
at 0            \cr
3 &one
simple zero at $B_1$ and one double zero at $+\sqrt\Lambda$ \cr
4 &one double
zero at $-\sqrt\Lambda$ and one simple zero at $B_3$   \cr
5 &three simple zeros at $B_1$, $B_2$, and $B_3$,\cr}

\begin{figure}[pht]
\begin{center}
\leavevmode
\epsfig{file=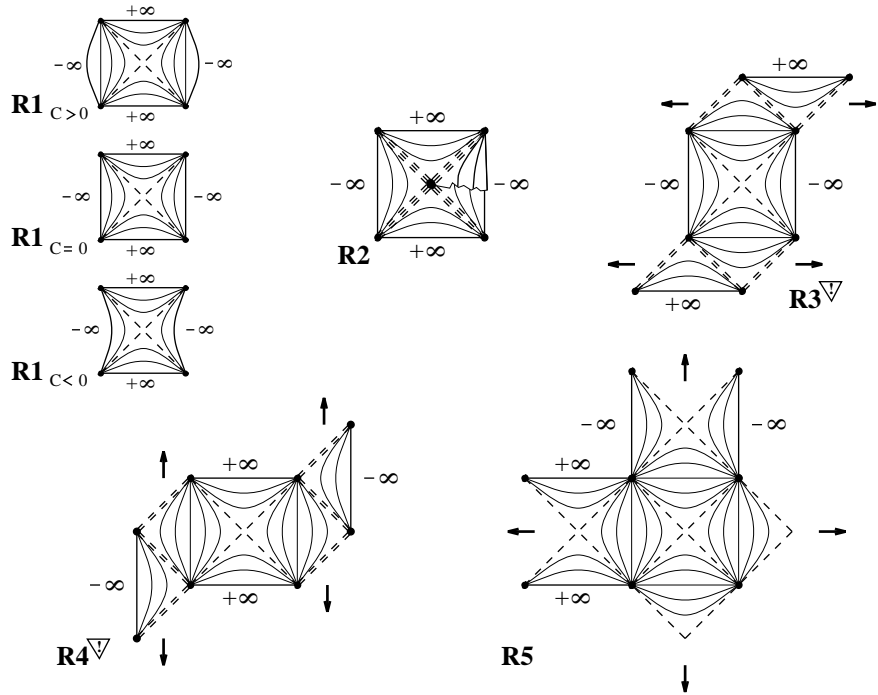,width=11.5cm}
\end{center}
\renewcommand{\baselinestretch}{.9}
\caption{
  \captionstyle \label{fig:fig11}
  Carter-Penrose
  diagrams for the $R^2$--model. Those marked \attn \ \,are
  nonsmooth; furthermore, the singularities of {\bf R5} might be
  distorted in the same way as sketched in Fig.~\ref{fig:fig4}c. (For more details
  cf caption of Fig.~\ref{fig:fig13}).}
\end{figure}

\smallskip
\noindent where $B_1<-\sqrt\Lambda<B_2<+\sqrt\Lambda<B_3$ and
$-\infty<B<+\infty$. Obviously ${\phi}_{crit}=\pm \sqrt{\L}$ (by
its definition as zero of the potential $W(0,\phi)$) and the curve
along {\bf R4,2,3} in Fig.~\ref{fig:fig10} corresponds to the critical values
$C_{crit} \equiv \pm (4/3)\Lambda^{(3/2)}$ of $C$. It is
completely straightforward to construct the  Carter-Penrose
diagrams according to the above rules. The result is depicted in
Fig.~\ref{fig:fig11}. Note that not only {\bf R2} but also the extended solutions
{\bf R3-5} will be multi-layered; for instance, the copies appended to
{\bf R4} at the upper left and the upper right horizon, resp., lie in
different layers of the universal covering (cf Fig.~\ref{fig:fig6}e for a similar
situation).
For nonnegative $\L$ there are, in addition to those diagrams, also
infinite ribbons for the  constant curvature solutions $d\o = \mp 2
\sqrt{\L} \e$
(cf equation \re{deSitter})); in the diagram Fig.~\ref{fig:fig10} they are located at the
curve  {\bf R4,2,3}, i.e.\ at $C=C_{crit}(\L)$.

So, for a negative cosmological constant $\L$, which in contrast
to $C$ is fixed by the choice of the action \I{R2}), there is only
one schematic Carter-Penrose diagram (cf Fig.~\ref{fig:fig10}). The choice
of the one free parameter $C$, however, influences the causal
structure of {\bf R1} somewhat, giving rise to the distorted
boundaries (cf also Fig.~\ref{fig:fig7}). For positive values of $\L$ there
are mainly two kinds of (again schematic) Carter-Penrose diagrams:
{\bf R1}  and the a bit more complicated diagram {\bf R5}. Again,
to have them smooth, some of the boundaries will be curved for
generic values of $C$ (also for {\bf R5}, although this has not
been indicated in Fig.~\ref{fig:fig10}!). The separation of the solution space
into solutions of the type {\bf R1} and {\bf R5} occurs at the
values $\pm (4/3)\Lambda^{(3/2)}$ of $C$. At these values we have
the spacetimes corresponding to the (inherently nonsmooth)
Penrose diagrams {\bf R3,4} {\em and\/} those of deSitter type
with $R \equiv \pm 4 \sqrt{\L}$ (but now with no internal
structure as the functions $X^i$ are constant all over $M$ here).
So at $C=C_{crit}$ the Casimir constant does no longer classify
the universal coverings uniquely. This is related to the fact that
there is more than one symplectic leaf for a critical value of
$C$: Besides a two-dimensional leaf, corresponding to {\bf R4}
resp.\ {\bf R3}, there is the pointlike one $X^+=X^-=0$,
$\phi=\phi_{crit}$, corresponding to a deSitter solution of
positive resp.\  negative constant curvature.

\medbreak The third example is the Katanaev-Volovich model
\I{KV}). Its function is $h^{KV}(x^0) = {1 \over \a}  \left\{ C
x^0 - 2 (x^0)^2 [(\ln x^0-1)^2+1-\Lambda] \right\} \,,$ where $x^0
\in \dR^+$. The function $h^{KV}$ always has a  zero at $x^0 =0
\LRA \a \phi = - \infty$, which is at least of order one. Thus the
function $f=\int^{x^0}du/h(u)$ (cf equation \re{fun})) has infinite
range as $x^0\to 0$, which shows that the Carter-Penrose diagrams
have a square-shaped sector at this end, the null-infinities,
however, being incomplete. On the other boundary of the coordinate
patch, $\a \phi = + \infty$, we find $\a h \sim (x^0)^2 \ln^2
x^0$, corresponding to a complete triangular sector at this end.

\begin{figure}[ht]
\begin{center}
\leavevmode
\epsfig{file=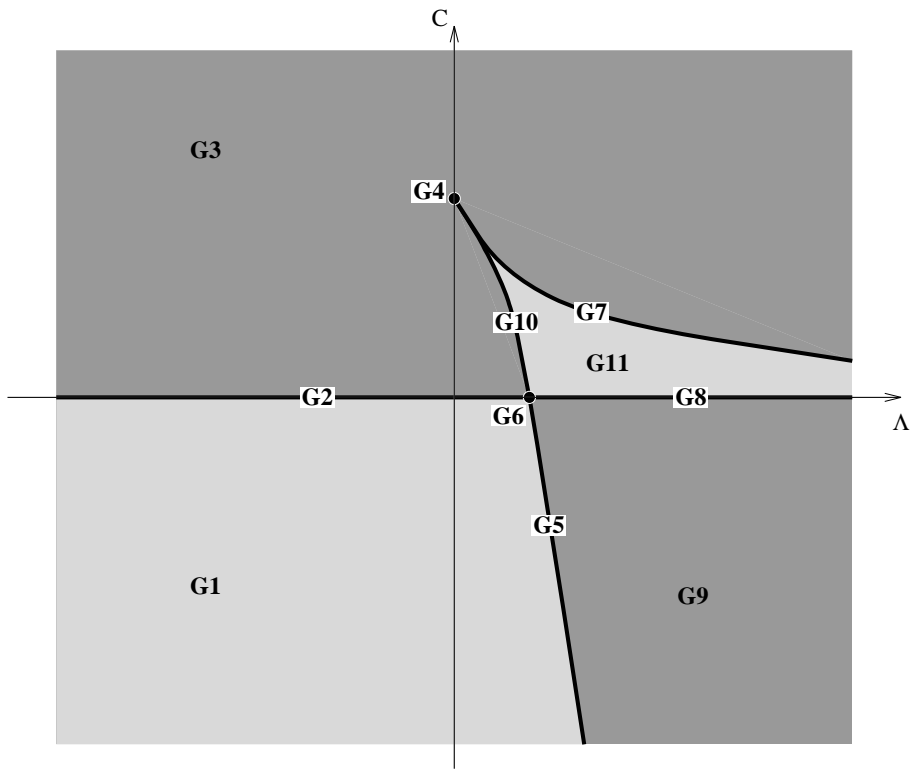,width=12cm}
\end{center}
\renewcommand{\baselinestretch}{.9}
\caption{
  \captionstyle \label{fig:fig12}
  KV--model, overview. The different regions
  correspond to qualitatively different functions $h$ (number and
  degree of zeros, asymptotic behavior at 0 and $+\infty$).}
\end{figure}

This time we get 11 qualitatively different cases (in addition to two deSitter
solutions) depending on the parameters
$C$ and $\Lambda$:

\smallskip
\halign{{\bf G#}:\quad&#\hfill\cr
     1,2 &no zeros of $h$ \cr
      3  &one simple zero at $x^0=B$                         \cr
      4  &one triple zero at $x^0=1$                      \cr
5,6 &one double zero at $x^0=e^{\sqrt\Lambda}$                \cr
7  &one double zero at $e^{-\sqrt\Lambda}$
and one simple zero at $B_1$       \cr
8,9 &two simple zeros at $B_2$ and $B_1$               \cr
10 &one simple zero at $B_3$ and one double zero at $e^{\sqrt\Lambda}$ \cr
11 &three simple zeros at $B_3$, $B_2$, and $B_1$,\cr}

\begin{figure}[ht]
\begin{center}
\leavevmode
\epsfig{file=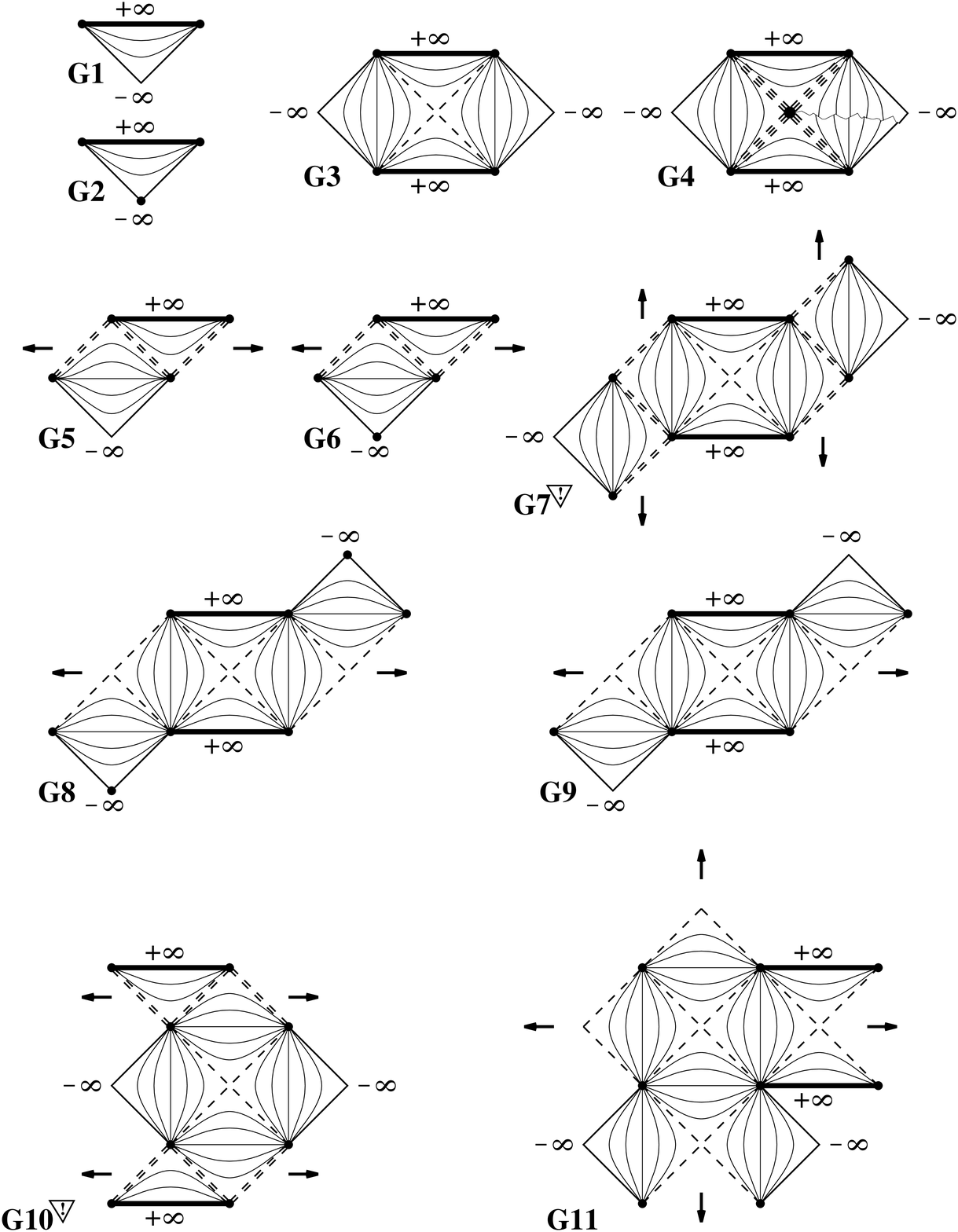,width=10cm}
\end{center}
\renewcommand{\baselinestretch}{.9}
\caption{
  \captionstyle \label{fig:fig13}
  Carter-Penrose diagrams for the KV--model (those marked \attn \
  \,are nonsmooth!). Complete boundary lines
  are indicated by boldfaced lines, complete points (points at infinity)
  by massive dots, incomplete boundary lines by thin solid lines,
  and horizons by dashed lines (degenerate horizons, i.e.\ at higher order
  zeros of $h$, by multiply dashed lines).
  The null extremals \re{null1}, \ref{null2a})
  which run through the diagrams under $\pm 45$ degrees are omitted.
  Arrows outside a diagram ({\bf G5-11}) indicate that the solution
  should be extended by appending similar copies at the corresponding
  boundaries (but cf Fig.~\ref{fig:fig6}e).}
\end{figure}

\smallskip
\noindent where $0<B_3<-\sqrt\Lambda<B_2<+\sqrt\Lambda<B_1$
and $B \in \dR^+$. An overview is provided by Fig.~\ref{fig:fig12}.
In the above list we took $x^0 \in \dR^+$.
The cases {\bf G2,6,8} and {\bf G1,5,9}, respectively, differ only in the
asymptotic behavior of $h$ at $x^0\rightarrow 0$ ($h^{KV}\sim x^0$ for
$C\ne 0$, but $h^{KV}\sim (x^0\ln x^0)^2$ for $C=0$), which influences the
completeness properties (see below).
The critical values of $\phi$ are easily
determined to be $\pm \sqrt{\L}/\a$; the corresponding value of the
Casimir function $C$ is
\be C_{crit} = -4 \( \pm \sqrt\L -1 \)
  \exp\(\pm \sqrt{\L}\)  \, , \el CdeS
which marks the curve {\bf G5,6,10,4,7} of Fig.~\ref{fig:fig12} and
simultaneously the deSitter solutions
$De^a=0, \, d\o = \pm {2\over\a}\sqrt{\L} \e$ (cf equation \re{deSitter})).

\begin{figure}[ht]
\begin{center}
\leavevmode
\epsfig{file=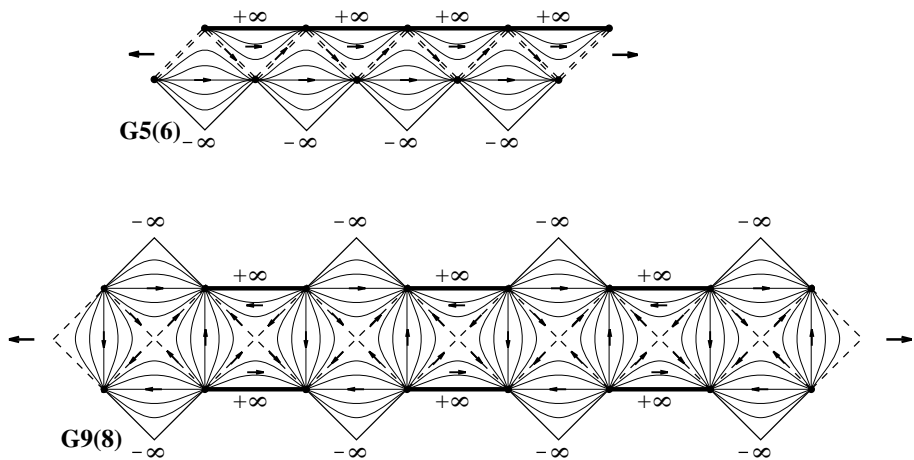,width=12cm}
\end{center}
\renewcommand{\baselinestretch}{.9}
\caption{
  \captionstyle \label{fig:fig14}
  Some extended Carter-Penrose diagrams for the KV--model. Arrows inside
  the diagrams indicate Killing fields.}
\end{figure}

It is now straightforward to draw the Carter-Penrose diagrams of
the KV--model. The result for $\a>0$ is depicted in Fig.~\ref{fig:fig13}
(extended versions of {\bf G5,9} are given in Fig.~\ref{fig:fig14}); the diagrams for $\a <0$ are obtained by rotating these by
$90$ degrees. According to the different behavior of $h$ at the
boundary $x^0=0$ the Carter-Penrose diagrams {\bf G2,6,8} ($C=0$)
differ from  {\bf G1,5,9}, respectively, only by having a complete
time--(space--)infinity, $\a>\!(<)\,0$, at that boundary (use
\re{11})). Let us stress that all Carter-Penrose diagrams in
Fig.~\ref{fig:fig13} except for {\bf G7,10} are smooth now, the boundary
lines being actually straight (cf previous footnote). Again,
however, a representation of the extended diagrams for {\bf
G4,7,10,11} will require infinitely many overlapping layers (cf
Fig.~\ref{fig:fig6}e for the case {\bf G7}).

The numbering {\bf G1-11} has been chosen as in \cite{Kat}, where
the Carter-Penrose diagrams Fig.~\ref{fig:fig13} have been constructed
first. It should be noted, however, that our procedure to obtain
these diagrams is incomparably faster than the one of \cite{Kat}.
The main reason is that the local solutions used there (resulting
also from ours through the diffeomorphism \re{conf})) are valid
only in coordinate patches which are part of ours (the sectors
$h(x^0)\neq 0$); they had to be glued along their border, which
entailed lengthy considerations of the asymptotic behavior. In
the chiral gauge \re{011h}), instead, the charts overlap and
simply have to be matched together.  As a consequence we also
could prove that all the solutions of \I{KV}), (and in fact also
of \I{grav}) with an, e.g., analytic potential $W$) are analytic.
As pointed out repeatedly, the Carter-Penrose diagrams {\bf G7,10}
are, however, not smooth but have to be regarded as schematic
diagrams only (of course, the universal covering solutions
themselves {\em are\/} smooth).

\medskip

As a further example for the application of the simple rules
outlined at the beginning of this subsection we could consider,
e.g., a model \re{Proto3}) where $V$ is chosen to be the
derivative of our generic function $h$ in Fig.~\ref{fig:fig2}a. In this
torsion-free case with $g=2e^+e^-$, the function $h$ in \re{011h})
will  coincide with the function in Fig.~\ref{fig:fig2}a up to a `vertical'
shift  according to \re{h}).  It is now not difficult to see that
there will be 11 qualitatively different schematic Carter-Penrose
diagrams in this case, as well as further four deSitter solutions
(corresponding to the four distinct extrema of the function $h$ in
Fig.~\ref{fig:fig2}a or, equivalently, to the four zeros of the potential
$V$). We leave it as an exercise for the interested reader to
sketch these Carter-Penrose diagrams explicitly. Note that some of
them will be  quite ramified already (cf, e.g., Fig.~\ref{fig:fig5}), a
feature that becomes more and more pronounced with an increasing
number of zeros of $V$ (or $h$).

This completes our analysis of the universal coverings of a
general pure gravity (or gravity--YM) model. Starting from any of
these Lagrangians, the determination of the corresponding
Carter-Penrose diagrams is straightforward now. One just has to
determine the function $h$ as, e.g., in \re{h}). Thereafter the
coverings follow from simple rules where only the zeros of $h$ and
its asymptotic behavior enters. This was illustrated by several
examples above. As remarked already before, the procedure is
applicable also to the {\em analytic}\/ extension of {\em any}\/
metric with a Killing field (cf the remarks in section
\ref{Secgeneral}).

\subsection{Remarks on global coordinates \pl{SecRN} }
%
For many of the Carter-Penrose diagrams, such
as e.g.\ for {\bf R1,G3,5,9}, it is possible to find also global
coordinates, displaying explicitly the analyticity of the metric.  E.g.,
for {\bf R1,G3} we have already found the Kruskal-like coordinates
\re{Kruskal}) or the more explicit ones \re{saddle}), and instead
of for {\bf G5,9} we provided a global, explicitly analytic chart
in a chiral gauge \re{lcgauge}) for the (mathematically) quite
similar (four-dimensional) Reissner-Nordstr\"om solution in
\cite{rnletter}. The method presented there may be adapted easily
to cases such as {\bf G5,9} or in fact to any two-dimensional
spacetime with a Killing field that permits a foliation by
null-lines.

The main idea of \cite{rnletter} is that in such cases there is no
Gribov obstruction \cite{Gribov} to picking the light cone gauge
\re{lcgauge}) or \re{gauge2}). Thus there {\em must\/} exist functions
$F$ and $G$ in \re{Loes}) yielding a globally well--defined solution.
The residual gauge freedom of \re{gauge2}) may then be used to
simplify these two functions. The resulting problem for $F$ may be
mapped to the problem of classifying vector fields on a line modulo
diffeomorphisms; the number of zeros of $F$ is thus gauge invariant,
among others.  The field equations pose restrictions on the slope of
$F$ at its zeros. For the case of SS or RN, e.g., the residual gauge
freedom may be used to bring $F$ (and also $G$) into a form containing
elementary functions only. E.g., for SS $F \propto x^1$ and $G=const$
may be achieved, while for RN $F$ and $G$ may be fixed by using $\cos$
and $\sin$. (Cf \cite{rnletter} for the details).

For SS the resulting coordinates provide a hardly known, explicit
global chart (found first by \cite{Israel}) as an to interesting
alternative to the much better known Kruskal coordinates \cite{Krus}.
The method provides also an explicit analytical global chart for the
RN spacetime. Both of these charts should be at least of pedagogical
interest.

\section{All global solutions of (all 2d) Gravity--Yang--Mills
  models \pl{Secall}}

\subsection{Introductory remarks \pl{Secintrorem}}
In this section we want to classify all global, diffeomorphism
inequivalent classical solutions of a general pure gravity model.
In particular the analysis covers the generalized dilaton gravity
models \I{gdil}). The classification shall be done without
any restriction on the topology of the spacetime $M$.

For some of the popular, but specific choices of the potentials
$\U,\V,\Z$, in \I{gdil}) such as those of string inspired dilaton
gravity (section \ref{Secstring}), of deSitter gravity (equation
\I{JT})), or of spherically reduced gravity (section \ref{Secspher}),
the possible topologies of the maximally extended solutions turn out
to be restricted considerably through the field equations. In
particular their first homotopy is either trivial or (at most) ${\Z}$.
(Permitting e.g.\ also conical singularities, cf section
\ref{Kinks}, the fundamental group might become more involved.)

For any `sufficiently generic' (as specified below) smooth/analytic
choice of $\U,\V,\Z$, on the other hand, the field equations of  $L$ 
permit maximally extended, globally smooth/\hspace{0cm}analytic 
solutions on all noncompact two-surfaces with an arbitrary number of
handles (genus) and holes ($\ge1$).
\footnote{That
   there are no solutions on compact manifolds (except in the flat case,
   cf section \ref{Const})
 can be seen by inspection of the possible fundamental groups;
 however, it may in many cases also be deduced from the fact that the range
 of the field $\Phi$ in \re{gdil}) is not compact. Let us note in this context
 that according to, e.g., \cite{Thm} there are {\em no\/} compact two-manifolds
 without boundary (closed surfaces) that may be endowed with a metric of
 Lorentzian signature, except for the torus and the Klein bottle.}
This shall be one of the main results of the present section.
These solutions are smooth and maximally
extended, more precisely, the boundaries are either
at an infinite distance (geodesically complete) or they correspond to true
singularities (of the curvature $R$ and/or the dilaton field $\Phi$).
We will call such solutions {\em global\/}, as there are other kinds of
inextendible solutions (cf below, in particular section \ref{Kinks}).

\subsubsection{Motivation}
The existence of solutions on such nontrivial spacetimes is a
qualitatively new challenge for any program of quantizing a gravity
theory. Take, e.g., a Hamiltonian approach to quantization: In any
dimension $D+1$ of spacetime the Hamiltonian formulation necessarily
is restricted to topologies of the form $\S \times \R$ where $\S$ is
some (usually spacelike) $D$-manifold. In our two-dimensional setting,
$\S$ may be $\dR$ or $S^1$ only. Thus $\pi_1(M)$ can be ${\Z}$ at most.
According to our discussion above, this is far from exhaustive in most
of the models \I{gdil}).

Let us compare this to the
case of full four-dimensional Einstein gravity. Clearly, there the
space of solutions will include spacetimes of complicated
topologies.  Therefore, a restriction to topologies of the
form $M =\S \times \R$ seems hardly satisfactory in the 4d scenario as well.
A path integral approach to quantum gravity, on the other hand, does
not place an a priori restriction on the topology of the base manifold
$M$. However, also in the path integral approach for 
$M \neq \S \times \R$, the
definition of an integration measure is plagued by additional
ambiguities and problems, cf \cite{PSMDubna}. 
The class of models \I{gdil}) (and also the
theories with torsion, equation  \I{grav})) may serve as a
good laboratory to improve on that situation and to gain new insights
in such directions.

For spacetimes of topology $\S \times \R$, furthermore, we are
interested in an explicit comparison of the solution space or moduli space
of a given action (space of all solutions to the field equations
modulo diffeomorphisms) with the reduced phase space (RPS) in a
Hamiltonian formulation of the theory. In the simply connected case
($\S = \R$) we already classified all global diffeomorphism
inequivalent solutions in the previous section. The solution space was
found to be {\em one\/}-dimensional, parameterized by a real number $C
\in \R$.  As the result of a symplectic reduction must lead to an {\em
  even\/}-dimensional RPS, we may conclude that in the case of an
`open universe' ($\S = \R$) the proper definition of a Hamiltonian
system, describing the same physics as \re{gdil}) (or \re{grav})), is
in need of some additional external input. This seems to creep in
implicitly when defining boundary/fall-off conditions for the
canonical phase space fields and/or by declaring some particular gauge
transformation of the Lagrangian formulation of \re{gdil}) as ({\em
  nongauge}) symmetry transformation (cf also section \ref{Secham}
below).  Periodic boundary conditions, on the other hand, lead to a
Hamiltonian formulation which is perfectly well-defined without any
further input besides that of periodicity (with respect to some
arbitrarily fixed coordinate period).  Effectively they describe the
case of a `closed universe' $\S = S^1$ and we conclude that for
cylindrical topologies of $M$ the solution space of \re{gdil}) must be
even-dimensional. Indeed it will turn out to be two-dimensional, a
second parameter `conjugate to $C$' arising from the
`compactification' (in one coordinate).

For generic 2d gravity theories the solution space for $M \sim S^1
\times \R$, and thus the corresponding RPS, will have a highly
nontrivial topology. This is the second challenge which has to be
faced in a quantization scheme: One has to cope with this
nontriviality of the orbit space, as for sure the RPS of
four--dimensional gravity will be even more intricate.

We will come back to these issues in chapter \ref{Sectowards}. We
mentioned them here already as they also serve as part of the interest
for the classification performed in the present section.




\subsubsection{Organization}
To classify the global solutions, we will employ three different,
alternative methods. First, at the example of the KV model \I{KV}),
solutions of nontrivial topology are obtained from a `cut-and-paste'
technique: Starting from the Carter-Penrose diagrams, we cut out some
fundamental region and glue it together appropriately. In this
process, we keep track of different parameters labeling inequivalent
gluings. Together with the Casimir parameter $C$ labeling the
Carter-Penrose diagrams, these parameters are coordinates on the
moduli space.

Second, we will employ the $\s$-model approach of chapter \ref{SecPSM}
for a direct classification. In the case of pure gravity models
it is possible to solve the field equations globally within this
framework, for arbitrary topology of $M$. The respective calculation
may again be performed within a few lines only. However, per se the
PSM formalism does not care about the physical restriction that the
metric should be nondegenerate globally. Thus, the solutions obtained
will include solutions which have to be excluded on these grounds in a
second step. We did not manage to incorporate this restriction in a
sufficiently practical way. Therefore, as of present state, the PSM
formalism gives an upper bound to the dimension of the moduli space
only.

Last but not least a systematic, group theoretical approach will be
applied.  For that purpose we will determine the full isometry group
${\cal G}$ of the universal covering solutions first.  All global
solutions may be obtained as factor solutions of the universal
coverings by appropriate discrete subgroups of ${\cal G}$. This will
yield a complete classification for a  general pure gravity model in
terms of conjugacy classes of discrete groups. The simple machinery
developed will fail, however,  for those solutions with three
Killing vectors, which describe spaces of constant curvature; they are
thus discussed separately (and, although apparently simpler, they
are not fully classified yet). In the subsection on the group
theoretical approach  we will also provide a
geometrical interpretation of the additional parameters arising.

In a final subsection, furthermore, we will briefly discuss
`nonglobal inextendible' solutions (some remarks on them are found
already in section \ref{Seccut}, too). They are connected to so-called
kinks (of the metric). For a cylindrical spacetime, for example, the
(discrete) kink number of a metric may be defined as follows: It is
the {\em minimal}\/ number of disconnected null-like parts (or points)
of any noncontractable line. Loosely speaking, the light cone tilts
around nontrivially along the noncontractable loop. The moduli space
then splits into separate, disconnected parts, characterized
precisely by the kink number.\footnote{This splitting is absent in the
  PSM approach or also in the Hamiltonian framework presented in
  chapter \ref{Sectowards}. This is connected to subtleties concerning
  degenerate metrics,
  and will be discussed in some detail in chapter \ref{Sectowards}.}

\subsubsection{Anticipation of some of the results \pl{Secanticipate}}

Also the classification of all global solutions
will be found to depend solely on the number and order
of the zeros of the function $h(r)$ in (\ref{011h}) (as well as there
asymptotic behavior). To provide an overview on the results, we restrict
ourselves to simple zeros $h$ (zeros of higher order occur only for specific
values of the parameter $C$).

Besides the uniquely defined universal covering, one obtains the following
global space-- and time--orientable solutions ($n=\mbox{\# simple zeros
  of h}$):
\begin{list}{}{\leftmargin1.2cm\labelwidth1cm\itemindent.0cm\labelsep.2cm}
\item[$n=0$:] Cylinders labelled by their circumference (real number).
\item[$n\ge1$:] The above cylinders are still available, but
  now incomplete in a pathological manner (Taub--NUT space).
\item[$n=2$:] Complete cylinders labelled by a discrete parameter
  (patch number) and a further real parameter.
\item[$n\ge3$:] Noncompact surfaces of arbitrary genus with
  an arbitrary number ($\ge1$) of holes. The number of continuous parameters
  equals the rank of the fundamental group
  $\pi_1(\mbox{solution})\equiv \pi_1(M)$, and
  there are also further discrete parameters.
\end{list}
The solution space for a fixed model (\ref{grav}) or \re{gdil}) and
for a fixed topology is labelled by the above parameters {\em plus\/}
the (mass) parameter $C$. For the case of additional YM--fields, its
dimension generalizes to $(\mbox{rank }\pi_1(M)+1) \, (\mbox{rank (gauge
  group)}+1)$.

Thus by a `sufficiently generic' generalized dilaton theory at the
beginning of the present section \ref{Secintrorem}, we meant an action
characterized by potentials $\U,\V,\Z$ which, at least for some values
of $C$, give rise to a function $h$ with three or more zeros (cf,
also, equation \re{h})).

\subsection{Cut-and-paste technique for the KV model \pl{Seccut}}

\begin{figure}[t]
\begin{center}
\leavevmode
\epsfig{file=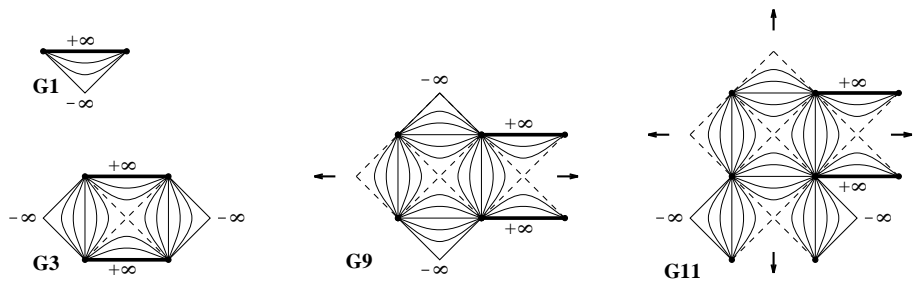,width=12cm}
\end{center}
\caption{
\captionstyle \label{fig:fig14a}
Some Penrose-diagrams for the KV--model.
The thin lines denote Killing trajectories, the broken lines Killing
horizons. The style of the boundary lines indicates their completeness
properties; however, since they are irrelevant for our topological
considerations, we will treat them rather sloppily.
The arrows in {\bf G9} indicate that the patch should be extended horizontally
by appending similar copies, and likewise {\bf G11} should be extended
vertically and horizontally.}
\end{figure}
In this section we discuss possible global solutions for the
Katanaev-Volovich (KV) model \I{KV}). Its universal covering
solutions, resulting from the analysis of the function $h$ in
\re{hKV}), were obtained already in section \ref{Recipe}, cf
Figs.~\ref{fig:fig12} and \ref{fig:fig13}.  For reasons of brevity we skip the
solutions where $h$ has higher order zeros in this section; also we
postpone a discussion of the deSitter solutions (section \ref{Const}).
They both occur only for positive $\L$ {\em and}\/ only for those two
particular values $C_\pm(\L)$ of $C$ on the boundary lines between
{\bf G1/G9} or {\bf G3/G11} (i.e.\ at {\bf G4-7,10} in
Fig.~\ref{fig:fig12})! The remaining, `generic' diagrams are collected in
Fig.~\ref{fig:fig14a}.

The KV model is adequate to illustrate much of what has been said in
the introductory remarks to this section as well as to get a grasp on
the general case. In particular, for a negative cosmological constant $\L$
there are only the simple Carter-Penrose diagrams {\bf G1} and {\bf
  G3}. Despite some differences, in the present context, the situation for
negative $\L$ reminds much of spherically symmetric vacuum gravity: Also
there a horizon is present only for positive Schwarzschild mass $m
\sim C$ and none for negative values of $M$. Moreover, like
the spherical model,the KV--model with $\L<0$
belongs to the class of particular models where the possible
topologies of spacetimes are restricted severely.

As mentioned above one obtains all noncompact two-surfaces, if the
one-parameter family $h=h_C(x^0)$ contains functions with three or
more zeros. In the KV--model this is the case for positive $\L$, where
the additional diagrams {\bf G11} if $0<\L<1$ as well as {\bf G9} if
$\L >1$, cf Figs.~\ref{fig:fig12}, \ref{fig:fig14a} arise.

\begin{figure}[ht]
\begin{center}
\leavevmode
\epsfig{file=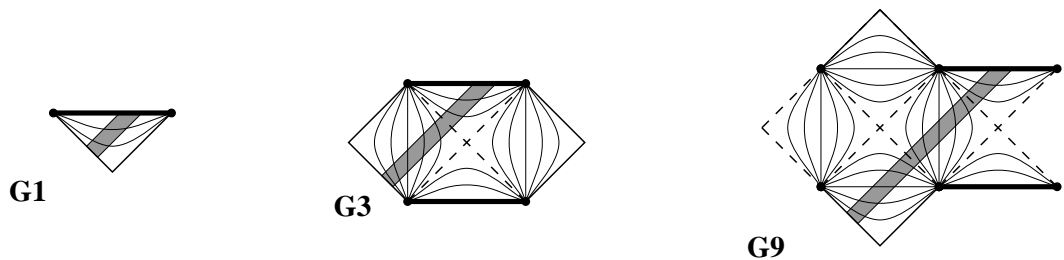,width=14cm}
\end{center}
\renewcommand{\baselinestretch}{.9}
\caption{
  \captionstyle \label{fig:fig15}
  Cylinders from {\bf G1,3,9}. The opposite
  sides of the shaded strip have to be identified along the Killing lines
  (cf also Fig.~\ref{fig:fig21}). Note that in {\bf G3,9} there occur closed Killing
  horizons (broken lines), which leads to pathologies of the Taub--NUT type.}
\end{figure}

We start with {\bf G1}: Obviously this is a spatially homogeneous
spacetime and \re{011h}) provides a global chart for it. Identifying
$x^1$ and $x^1 + \o$, $\o=const$, evidently we obtain an everywhere
smooth solution on a cylindrical spacetime $M \sim S^1 \times \R$. It
results from the Carter-Penrose diagram {\bf G1} by cutting out a
(fundamental) region, e.g.\ the strip between two null-lines in Fig.~\ref{fig:fig15}, and gluing both sides together in such a way that the
values of the curvature scalar $R$, constant along the Killing lines,
coincide at the identified ends.  The constant $\o$ (or a function of
it) becomes the variable conjugate to $C$ here. It may be
characterized in an inherently diffeomorphism invariant manner as the
(metric induced) distance between two identified points on a line of
an arbitrarily fixed value of $R$ (e.g.\ $R=0$). Thus $\o$ is a
measure for the `size of the compact (spacelike) universe'.

Next {\bf G3}: Clearly also in this case we can identify $x^1$ with $x^1+\o$
in a chart \re{011h}); obviously the resulting metric is completely smooth
on the cylinder obtained, the fundamental region of which is drawn in
Fig.~\ref{fig:fig15}.  However, this cylinder has some pronounced
deficiencies:
Not only does it contain closed timelike curves as well as one closed
null-line (the horizon); this spacetime, although smooth, is geodesically
incomplete. There are, e.g., null-lines which wind around the cylinder
infinitely often, asymptotically approaching the horizon while having only
finite affine length (Taub--NUT spaces, see section \ref{Factor}). So, from a
purely gravitational point of view such solutions would be excluded. Having
the quantum theory in mind, one might want to regard also such
solutions. Being perfectly smooth solutions on a cylinder,
certainly they will  be contained in the RPS of the Hamiltonian theory.
We leave it to the reader to exclude such solutions
by hand or not.

The above Taub--NUT solutions are not the only dubious ones.
Take any maximally extended spacetime, remove a point from it, and
consider the $n$-fold covering of this manifold: clearly the
resulting spacetime is incomplete. However, for $n \neq 1$ it is
impossible to extend this manifold so as to regain the
original spacetime. This is a very trivial example of how
to obtain an in some sense maximally extended $2n$-kink solution
from any spacetime. In the presence of a Killing vector there
are, however, more intricate possibilities of constructing kinky
spacetimes (resulting, e.g., also in inextendible
2-kink solutions, even flat ones);
we postpone their discussion to section \ref{Kinks}.
\footnote{
  One justification for this separation (besides technical issues)
  is that the RPS of the
  Hamiltonian formulation introduced in chapter 
\ref{Sectowards} below is in
  some sense insensitive to these solutions (so effectively one may ignore
  them to find the
  RPS). This may be different in other Hamiltonian treatments. We
  will come back to this issue in chapter \ref{Sectowards}.}
These solutions are certainly not global in the sense pointed out before.
Excluding them as well as the Taub--NUT type spaces from {\bf G3},
the above solutions are {\em all\/} global solutions for the KV--model
with negative cosmological constant $\L$. In particular we see that
the topology of spacetime is planar or cylindrical only.  Also
the RPS ($\S = S^1$) is found to have a simple structure: It is
a plane, parameterized by $C$ and $\o$.

\begin{figure}[hbt]
\begin{center}
\leavevmode
\epsfig{file=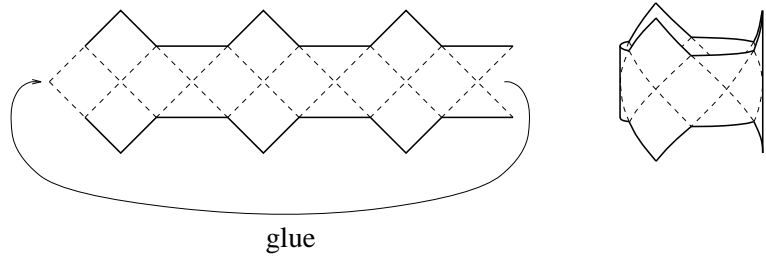,width=10cm}
\end{center}
\renewcommand{\baselinestretch}{.9}
\caption{
  \captionstyle \label{fig:fig16} Non--Taub--NUT cylinder from {\bf G9}.}
\end{figure}

Now the KV--model with $\L > 0$: The discussion of {\bf G1,3} is as
above. Also for {\bf G9,11} an identification $x^1 \sim x^1 + \o$ in a
chart \re{011h}) leads to a smooth (but incomplete) cylinder again.
However, for {\bf G9,11} there are also cylindrical solutions without
any deficiencies. Take, e.g., {\bf G9}: Instead of
extending the patch from Fig.~\ref{fig:fig14a} infinitely by adding further copies
one could take only a finite number of them and glue the faces on the left and
the right 
side together (cf Fig.~\ref{fig:fig16}). Clearly the result will
be an everywhere smooth, inextendible
spacetime. Also it permits a global foliation into
$\S \times {\R}$ with a spacelike $\S \sim S^1$. Now, however,
the `size' of the closed universe is `quantized' (the circumference being
fully determined by the (integer) number of blocks involved).
Still, there is some further ambiguity in the gluing process that leads to a
one-parameter
family of diffeomorphism inequivalent cylinders for any fixed value of
$C$ and fixed block number. This second quantity, conjugate to
$C$ in a Hamiltonian formulation, and its geometric interpretation
shall be provided in section \ref{Secgroup} (cf Fig.~\ref{fig:fig25}) below.

Thus in the case of {\bf G9} (and similarly of {\bf G11}) we find the
solution space for cylindrical spacetimes to be parameterized by $C$
(within the respective range, cf Fig.~\ref{fig:fig12}), by an additional
real gluing parameter, {\em and\/} by a further discrete label (block
number).
For {\bf G9} there is also the possibility of solutions on a M\"obius
strip: We only have to twist the ends of the horizontal ribbon prior to
the identification. It will be shown that these nonorientable
solutions are determined uniquely already by fixing $C$ and the block
number; there is now no ambiguity in the gluing!

\begin{figure}[ht]
\begin{center}
\leavevmode
\epsfig{file=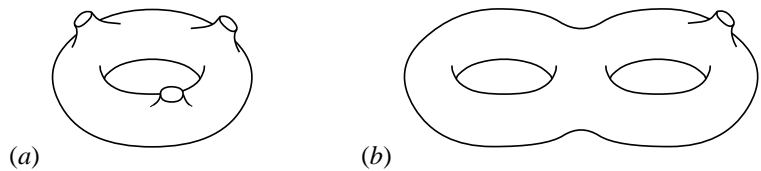,width=10cm}
\end{center}
\renewcommand{\baselinestretch}{.9}
\caption{
  \captionstyle \label{fig:fig17}
  Factor spaces from {\bf G11}. The
  corresponding fundamental regions are similar to those given in
  Fig.~\ref{fig:fig28} for {\bf R5}.}
\end{figure}

By far more possibilities arise for {\bf G11}. Again there are
cylinders of the above kind, with an analogous
parameterization of these solutions. However, now we can also identify
faces in vertical direction (cf Fig.~\ref{fig:fig14a}).
For instance, gluing together the upper and lower ends as
well as the right and left ends of the displayed region, one obtains a
global solution with the topology of a torus with hole. (It has
closed timelike curves (CTCs), but no further defects; also there are tori
without CTCs.)
The solution space for this topology is three-dimensional now, the two
continuous parameters besides $C$ resulting from inequivalent gluings
again. In addition, {\bf G11}
permits much more complicated global solutions. In fact, it
is one of the examples for which solutions on all (reasonable)
noncompact topologies exist. Fig.~\ref{fig:fig17} displays two further
examples: A torus with three holes and a genus two surface with one
hole. The respective fundamental regions (for the topologically similar
solution {\bf R5}) are displayed in Fig.~\ref{fig:fig28} below. As a general fact the dimension of the solution
space exceeds the rank of the respective fundamental group by one, and thus
it coincides with twice the genus plus the number of holes.  Also,
there occur further discrete labels.

\subsubsection{The moduli space for $M \sim \S \times \dR$}
In conclusion, let us consider the RPS (= the solution space for
topology $S^1 \times \R$) in the case of $\L>1$ (simultaneous
existence of {\bf G9} and {\bf G11}, cf Fig.~\ref{fig:fig12}).  It is
two-dimensional, being parameterized locally by $C$ and the respective
conjugate `gluing' variable. However, for $C$ taken from the open
interval $]C_-,C_+[$ (where $C_\pm(\L)= -4 \( \pm \sqrt\L -1 \)
\exp\(\pm \sqrt{\L}\)$, cf Fig.~\ref{fig:fig12} and equation \re{CdeS}))
there are infinitely many such two-dimensional parts of the RPS,
labelled by their `block number'. More precisely, for sufficiently
large negative numbers of the (canonical) variable $C$ ($C < C_-$) the
RPS consists of one two-dimensional sheet. At $C=C_-$ this sheet
splits into infinitely many two-dimensional sheets. At $C=0$,
furthermore, any of these sheets splits again into infinitely many,
all of which are reunified finally into just one sheet for $C \ge
C_+$.  Furthermore, at $C_\pm$ and $0$ the RPS is nonHausdorff.

So, while for $\L <0$, where the RPS was found to be planar globally,
a RPS quantization is straightforward, yielding wavefunctions
$\Psi(C)$ with the standard inner product (cf \ref{Sectowards} for
further details), an RPS quantization is not even well-defined for
$\L>0$ (due to the topological deficiencies of the RPS). In a Dirac
approach to quantization (cf also \ref{Sectowards} below), on the
other hand, related problems are encountered when coming to the issue
of an inner product between the physical wave functionals: For $\L>1$,
e.g., the states are found to depend on $C$, again, but for $C \in \;
]C_-,0[ \,$ there is one further discrete label, and for $C \in \;
]0,C_+[ \,$ there are even further labels. When no discrete indices
occur, as is the case for $\L < 0$, an inner product may be defined by
requiring that the Dirac observable $\o$ conjugate to $C$ becomes a
hermitian operator when acting on physical states \cite{Ashbuch}; this
again leads to the Lebesgue measure $dC$ then (cf also section
\ref{Secinner}).  Such a simple strategy seems to fail for $\L>0$ (and
also any generic case of \I{gdil})!).

Certainly one could {\em require\/} that only cylinders of the kind
Fig.~\ref{fig:fig15} (states with block number `zero') contribute to the
inner product. However, at least in view of the corresponding
classical solutions, this seems hardly satisfactory: For {\bf G9},
e.g., i.e.\ for $C \in \, ]C_-,0[$, we found precisely the above
solutions to be somewhat pathological, while those of nonzero block
number (Fig.~\ref{fig:fig16}) were perfectly admissible!

This is thus found as one of the points where an improvement of
quantization schemes may set in.

\subsection{A global $\s$-model approach\pl{SecglobalPSM}} In section
\ref{PSMlocalsolution} we determined the (local) solutions to the
field equations of the Poisson $\s$-models describing 2d gravity
theories by using Casimir-Darboux (CD) adapted field variables. As CD
coordinates exist locally on the target space $N$ only, we cannot just
apply the procedure directly in the global context without adapting it
appropriately.

In this section we will provide a classification of the global
solution for the pure gravity models.\footnote{We exclude additional
  YM-fields here. They may, however, be incorporated easily by
  standard argumentations used for classifying YM-fields on a fixed
  background. Alternatively, it may be possible to generalize Theorem
  2 below (maybe just by dropping its first assumption) to provide the
  general global solution of a general PSM; then, gravity--YM systems
  would be included automatically.} As remarked already in section
\ref{Secintrorem}, the classification within (the
present status of) this approach will provide only an {\em upper
  bound} to the dimension of the moduli space. The additional
requirement for globally nondegenerate metrics is not yet taken care
of.  It should be possible, however, to improve on this.

Pure 2d gravity models are described by means of a PSM with target
space $N=\dR^3$ and Poisson tensor \re{P}). The symplectic leaves are
either two- or zero-dimensional.

We first focus on the map $\CX$ from the spacetime (worldsheet) $M$
into the target space $N$. In the neighborhood of any (generic) point
we may introduce CD coordinates. In such coordinates, the field
equations yield $dC=0$. From this we may follow, that the image of
$\CX$ has to lie entirely within a symplectic leaf. This is already a
global result, obtained by patching together different charts with CD
coordinates. (As may be shown, it also holds for `nongeneric' points
in $N$, i.e.\ for zero dimensional leaves.)

Moreover, as becomes obvious upon inspection of the symmetries
\re{syma}), any {\em smooth}\/ deformation of the map $\CX$ is a gauge
symmetry. If $\CX$ is contractable, in particular, a gauge
representative may be chosen, such that $M$ is mapped into a single
point only. For a generic leaf this corresponds to a degenerate ($=$
noninvertible) metric (cf e.g.\ equation \re{eom2b}) or \re{gX3})). This
is not so dramatic {\em locally}, however. {\em Locally\/} one
always may use a gauge transformation such that the image of a small
region in $M$ is two-dimensional and then a nondegenerate metric is
available. Providing merely an upper bound to the dimension of the
moduli space, we do not need to check, if globally a nondegenerate
metric can be obtained by means of the PSM symmetries out of a
particular gauge representative. It is sufficient to determine the
dimension of the solutions identified by means of PSM symmetries.
(Recall that these symmetries coincide with the gravitational
symmetries for nondegenerate metrics. They are only somewhat stronger,
as they allow to gauge transform a nondegenerate metric into a
degenerate one. The very same mechanism works also in 2+1 gravity
\cite{WittenCS}. We will further discuss this issue in chapter 
\ref{Sectowards}.

In general there will, however, be maps $\CX$ also, with an image that
may not be contracted to a point. We thus obtain a {\em discrete\/}
label for the classical solutions, which characterize the homotopy
classes of maps $M \to N$. These labels are in one-to-one
correspondence with the block number(s) mentioned in the other two
approaches within the present chapter (or the conjugacy classes of the
discrete groups).

This completes the discussion of the fields $X^i(x)$. The more
difficult part is now the determination of gauge inequivalence classes
of the fields $A_i$. More precisely, this part is nontrivial only,
provided $\CX$ is noncontractable (otherwise we can use CD again,
inducing globally well-defined fields on $M$.)

For this purpose we return to the more abstract language, in which
$A=A_idX^i$ is regarded as a one-one-form on $M \times N$. Then we
have

\noindent {\em Theorem 1}. Any $A_i \equiv \CX^* (\6_i \ins A)$ of the form
\be A_i = \CX^*(\6_i \ins \bar \O ) \pl{thm1} \ee solves the field equations
\re{eoma}) and \re{eomb}) (globally), provided only $\bar \O$ is a
closed 
pseudo-inverse of $\CP$.

In the above $\ins$ denotes the insertion of the respective vector
field into the respectively subsequent differential form.

We know that $\CP$ has no inverse, as it has a (generically
one-dimensional) kernel. A pseudo-inverse $\bar \O$ is a matrix
(two-form) coming as close to an inverse as possible: its
defining property is $\bar \O \CP \bar \O = \bar \O$. Clearly, the
pullback of $\bar \O$ to a symplectic leaf is determined uniquely and
equals the symplectic form $\O$ introduced already before. There
certainly is some ambiguity in defining $\bar \O$; we will come back
to this shortly.

Theorem 1 follows from noting that it is formulated in a target space
covariant way and that, on the other hand, it is in agreement with
\re{eom2a}) and \re{eom2b}) upon specialization to (local) CD
coordinates. (A similar argumentation is often applied in general
relativity after using Riemann normal coordinates to establish a
result.) Note that within the theorem we may also replace $\6_i$ by
some arbitrary vector field $v$ on the target space.

More important in the present context, is, however, the following theorem:

\noindent {\em Theorem 2}. In the case that the orthonormal bundle of
the symplectic leaf $\CS \supset \mbox{Im}(\CX)$ may be trivialized, the
general local solution of $A$ is provided by \be A_v :\equiv \CX^* (v
\ins A) = \CX^*(v \ins \bar \O ) + \a \CX^*(v \ins dC) \, , \pl{thm2}
\ee where $C$ is some Casimir function and $\bar \O$ a closed 
pseudo-inverse
of $\CP$ (both defined in a neighborhood of $\CS$), and $\a$ is a closed
one-form on $M$. The gauge inequivalence classes of $A$, moreover, are
in one-to-one correspondence with elements of $\mbox{H}^1(M)$.

We will not provide the details of the proof here. Sufficiency is
demonstrated along the same lines as for Theorem 1. For the converse
direction of the proof, we used the fact that the orthonormal bundle of
Im$(\CX)$ (i.e., of the symplectic leaf under consideration) may be
trivialized.  This ensures that a vector field $\6/\6 C$ nowhere
parallel to Im$(\CX)$ may be found. It may be possible to relax or
even drop this assumption. (This is suggested, e.g., by the known
correspondence of gauge-inequivalent solutions of YM gauge theories to
elements of $\mbox{H}^1(M)$ for each independent Casimir element of
the gauge group.)

Note also that the choice of the Casimir function $C$ is irrelevant in
\re{thm2}), as a change $C\to f(C)$ only rescales $\a$ by a constant.

Let us remark, moreover, that a change in the choice of $\bar \O$
produces a term like the second one in \re{thm2}). So at least part of
the freedom in choosing elements of the first cohomology of $M$ may be
transferred also to the freedom of selecting a pseudo-inverse.
(Possibly this permits a simplification of the statement.)
However, within the present formulation of the theorem, $\bar \O$ is
fixed (and any fixed choice will do).

It remains to note that in the case of pure 2d gravity, the (generic)
symplectic leaves have codimension one and thus their orthonormal
bundle is trivial always.

We now have achieved the classification we strived for:

\noindent {\em Theorem 3}. Any point in the moduli space of a PSM with bracket
\re{P}) is in one-to-one correspondence to a homotopy class of maps
from $M$ into a symplectic leaf $\CS$ and an element from $H^1(M)$.  Thus,
$\mbox{dim}({\mbox{moduli space}}) = \mbox{dim}(\pi_1(M))+1$.

Here, by moduli space we mean {\em all\/} globally smooth solutions to
the field equations {\em modulo\/} the PSM symmetries \re{syma}),
\re{symb})!

Since some of those solutions will not permit PSM gauge
transformations such that the metric $g = 2A_+ A_-$ (symmetric tensor
product) is nondegenerate everywhere on $M$, Theorem 3 provides an
upper bound for the dimension of the gravitational moduli space only.
(Furthermore, from the gravitational point of view, an additional {\em
  discrete\/} label has to be introduced when also permitting kinky
spacetimes, cf sections \ref{Kinks} and \ref{Secmissingkink} below.)

As the results of the detailed group theoretical approach, collected
in section \ref{Secanticipate} above, prove, the existence of globally
nondegenerate representatives within a given PSM equivalence class
depends only on topological features of the respective symplectic leaf
(encoded in the parameter $n$ of section \ref{Secanticipate}) and the
topology of the spacetime (worldsheet). It thus may be expected that
there is also a simple direct approach taking care of global
nondegeneracy of the metric within the PSM language.

We add some preliminary considerations in this direction. The
nondegeneracy of the metric may be formulated in an elegant way in the
PSM language under the following assumptions: For solutions of the
form \re{thm1}) with a spacetime $M$ that is a covering of (or that is
covered by) Im$(\CX)\subset \CS$,\footnote{One these two scenarios is
  found to be realized always within the group theoretical approach
  below.}  the degenerate points of $g$ correspond to points on $\CS$
where the kernel of $\bar \O$ (which necessarily points out of $\CS$,
certainly) lies in an $X^3 \equiv \phi = const$ surface. Further work
along these lines should be pursued.

\subsection{Systematic group theoretical approach \pl{Secgroup}}

\subsubsection{Preliminaries \pl{Secprelim}}

\begin{figure}[ht]
\begin{center}
\leavevmode
\epsfig{file=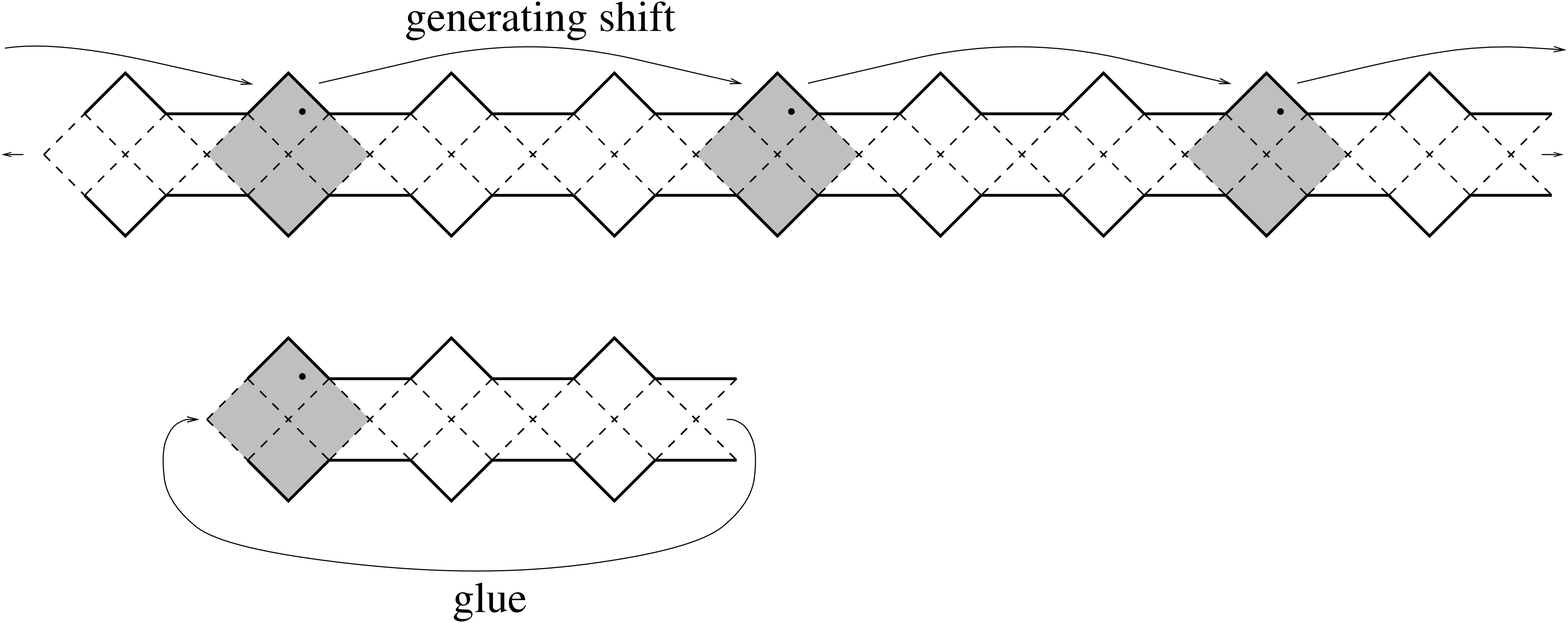,width=11cm}
\end{center}
\renewcommand{\baselinestretch}{.9}
\caption{
  \captionstyle \label{fig:fig18}
  Cut-and-paste approach versus factorization.
  In the upper figure we indicated the action of a transformation group
  generated by a shift three copies to the right. To obtain the orbit space
  one identifies all sectors which are a multiple of three copies apart
  (e.g.\ all shaded patches). This space may be described equivalently by
  cutting out a fundamental region (lower figure) and gluing together the
  corresponding faces.}
\end{figure}

The method employed for finding the multiply connected solutions will be to
factor the universal covering solutions
by a properly acting transformation group.  Let ${\cal G}:=
\mbox{\it Sym}({\cal M})$ be the symmetry group of the manifold
${\cal M}$. For any subgroup ${\cal H}\le{\cal G}$ we can construct the
factor- (or orbit-)space ${\cal M}/{\cal H}$ which consists of the
orbits ${\cal H}x$ ($x\in{\cal M}$) endowed with the quotient topology
(e.g.\ \cite{Massey}).
To pass from this approach to a cut-and-paste description
choose a fundamental region, i.e.\ a subset
${\cal F}\subseteq{\cal M}$ such that each orbit ${\cal H}x$ intersects
${\cal F}$ exactly once. The group action then dictates how the points of
the boundary of ${\cal F}$ have to be glued together (cf Fig.~\ref{fig:fig18}).

A priori orbit spaces may be topologically rather unpleasant,
they need e.g.\ not even be Hausdorff.
However, iff the action of this subgroup ${\cal H}$ is
free and properly discontinuous,
\footnote{`Free action' in this context means fixed-point-free (not
  to be confused with `free group', which means that there are no
  relations between the generators of the group).  For the definition
  of properly discontinuous see e.g.\ \cite{HawkingEllis,Wolf,Massey}.  These
  two conditions on the action of $\cal H$ certainly imply that $\cal
  H$ is discrete with respect to any reasonable topology on $\cal G$.
  (The converse is, however, not true: a finite rotation group, e.g.,
  is discrete but has a fixed point).}
then the orbit space is again
locally ${\R}^n$ and Hausdorff, i.e.\ a manifold. If, furthermore,
${\cal H}$ preserves some (smooth, metric, etc.)  structure or fields
(e.g.\ $\Phi$), then the orbit space inherits such a structure in a
unique way, i.e.\ the metric and the other fields are well-defined on
${\cal M}/{\cal H}$ (they `factor through') and still fulfill the
equations of motion (e.o.m.).

In this way one can obtain new `factor'-solutions of the e.o.m. We
now want to sketch shortly that when starting in this way from the
universal covering ${\cal M}$,  one obtains {\em all\/} multiply
connected global solutions:
Given a multiply connected manifold $M$, one can always
construct the (unique) simply connected universal covering space $\wt
M$ and, furthermore, lift all the structure (metric, fields) to it.
Certainly, the lifted fields again satisfy the e.o.m.\ (since these
equations are purely local).  Also, the lifted geodesics are geodesics
on the covering space, and they have the same completeness properties.
Thus the universal covering of a global solution is again a global
solution of the e.o.m.\ and coincides with ${\cal M}$ (which was found
to be determined uniquely, cf the previous chapter). Conversely, the
original multiply connected solution $M$ can be recovered from the
universal covering $\wt{M}={\cal M}$ by factoring out the group of
deck-transformations.  Let us note in passing that the fundamental
group of a factor space is isomorphic to the group factored out,
$\p_1(M)\equiv\p_1({\cal M}/{\cal H})\cong{\cal H}$ (more on this in
\cite{Massey}).

The solutions obtained by this approach are all smooth,
maximally extended, and Hausdorff. Of course, if one is less
demanding and admits e.g.\ boundaries (nonmaximal extension),
conical singularities (failures of the differentiable
structure), or violation of the Hausdorff-property, then there
are many more solutions. We will shortly touch such
possibilities in section \ref{Factor} (Taub--NUT spaces) and section
\ref{Kinks}.  On the other hand,
from the point of view of classical general relativity even
the globally smooth solutions may still have unpleasant
properties such as closed timelike curves or the lack of global
hyperbolicity (cf the previous section). In any case, our strategy will be
to describe all of them;
if necessary they may be thrown away afterwards by hand.

\medskip

The metric \re{011h}) displays two symmetries, namely the
Killing field $\6\over\6x^1$, generating the transformations
\begin{equation}
  \mytilde x^0 =x^0 \quad , \qquad \mytilde x^1= x^1 + \omega\,,
  \pl{Kill}
\end{equation}
valid within one building block, and the (local) {\em flip\/}
transformation provided by equation \re{glue}),
valid within a sector, which has been used as gluing
diffeomorphism for the maximal extension.
The special family of nonnull extremals \re{13}) provide
possible symmetry axes for the flip-transformations
\re{glue}) (cf dotted lines in Fig.~\ref{fig:fig2}).

The building block is usually incomplete (unless there is only one
sector) and has thus to be extended.  This process (described in
detail in the previous chapter) consisted of taking at each sector the
mirror image of the block and pasting the corresponding sectors
together (using the gluing diffeomorphism \re{glue})).  Usually,
overlapping sectors should not be identified, giving rise to a
multi-layered structure, cf e.g.\ the spiral-staircase appearance of
{\bf G4} (Fig.~\ref{fig:fig22} below). Only where nondegenerate horizons
meet in the manner of {\bf G3} (Figs.~\ref{fig:fig14a}, \ref{fig:fig22}, called {\em
  bifurcate\/} Killing horizons), the enclosed vertex point is an
interior point (saddle-point for $\Phi$, called {\em bifurcation
  point\/}), and the overlapping sectors have to be glued together,
yielding one sheet.


\medskip

Any symmetry-transformation of a solution to the model \re{gdil}) must
of course preserve the function $\Phi$ (more generally, for the model
(\ref{grav}) the functions $\phi$ and $X^a$) and also scalar curvature
(and, if nontrivial, also torsion).  However, by means of the e.o.m.\
of \re{gdil}) the scalar curvature may be expressed in terms of
$\Phi$ (cf also chapter \ref{Chapgen}).
Similarly, for (\ref{grav}) curvature and torsion can by the e.o.m.\
be expressed in terms of the functions $\phi$ and $X^a$ (cf chapter
\ref{Chap2dgeom}).
Moreover, since $X^a$ carries a Lorentz index, one only has to
preserve $X^2 \equiv X^a X_a$, which in turn may be expressed in
terms of $\phi$ via another field equation (Casimir function
$C\left[X^2,X^3\right]= const$).  Hence, in order
to preserve all the functions above, it is sufficient to preserve
$\phi$
only.
\footnote{One could consider also to neglect preservation of $\Phi$
  resp.\ $X^i$ and regard isometries only, e.g.\ when being interested merely
  in a classification of all global $1+1$ metrics with one (local) Killing
  field. In cases where $R(\phi)$ is not
  one-to-one this may lead to further discrete symmetries. We will not
  discuss the additional factor spaces that can arise as a
  consequence.}
(Recall that, in its specialization to vanishing
torsion,  (\re{grav}) describes \re{gdil}) upon the identification
$\Phi=\U^{-1}(\phi)$; 
so $\Phi$ is preserved, iff $\phi$ is in this case (as common
throughout the literature, $\U$ is assumed to be a diffeomorphism).
Thus, also for notational simplicity, we shall speak of $\phi$ only;
readers interested merely in \re{gdil}) may, however, well replace
`$\phi$' by `$\Phi$' in everything that follows.]

\medskip

Finally, we shortly summarize some facts concerning free groups
(details can be found in \cite{Massey, CombGr}). A free group is a
group generated by a number of elements $g_i$ among which there are no
relations. The elements of the group are the words
${g_{i_1}}^{k_1}\ldots{g_{i_l}}^{k_l}$, subject to the relations
(necessitated by the group axioms) $g_i{g_i}^{-1}=1$ and
$g_i1=1g_i=g_i$ (the unit element 1 is the empty word). A word is
called {\em reduced\/}, if these relations have been applied in order to
shorten it wherever possible. Multiplication of group elements is
performed simply by concatenating
the corresponding words and reducing if necessary. While for a given
free group there is no unique choice of the free generators, their
number is fixed and is called the {\em rank\/} of the group.  Free
groups are {\em not\/} abelian, except for the one-generator group; if
the commutation relations $ab=ba$ are added, then one speaks of a
{\em free abelian\/} group.

Subgroups of free groups are again free. However, quite contrary to
what is known from free abelian groups and vector spaces, the
rank of a subgroup of a free group may be larger than that of the
original group.
The number of the cosets (elements of ${\cal G}/{\cal H}$)
of a subgroup ${\cal H}\le {\cal G}$ is called
the {\em index\/} of ${\cal H}$ in ${\cal G}$. If this index is finite,
then there is a formula for the rank of the subgroup $\cal H$:
\begin{equation}
  \mbox{index $\cal H$}={\mbox{rank $\cal H$}-1 \over 
\mbox{rank ${\cal G}$}-1}\,,
  \pl{index}
\end{equation}
(cf \cite{Massey,CombGr}). Especially, subgroups of finite index
have {\em never\/} a smaller rank than the original group.  On the
other hand, $\mbox{rank $\cal H$}-1 = n \cdot (\mbox{rank ${\cal G}$}-1)$
for some $n$ does not guarantee that the index of the subgroup
$\cal H$ is finite;
\footnote{For instance, there are a lot of proper subgroups $\cal H$
  with the same rank as ${\cal G}$ (e.g.\ those generated by powers of
  the original free generators). However, none of them can be of
  finite index: If they were, then according to \re{index}) they
  should have index 1; but this means that there is no coset besides
  $\cal H$, thus ${\cal H}={\cal G}$, contrary to the assumption.}
still, the question whether a given subgroup has finite index is
decidable (cf \cite{CombGr}), but the algorithm is rather
cumbersome.

\subsubsection{The symmetry group \pl{Sym}}
As pointed out above, any
symmetry transformation must preserve the function $\phi$; thus sectors
must be mapped as a whole onto corresponding ones (i.e.\ with the same
range of $\phi$). Since $\phi(x^0)$ is
always monotonic (except for the deSitter solutions, which are
therefore discussed separately in section \ref{Const}), this has also the nice
consequence that within one building-block a sector cannot be mapped onto
another one. So each
transformation gives rise to a certain permutation of the sectors and
we can thus split it into ({\it i\/}) a sector--permutation and ({\it ii\/})
an isometry of a sector onto itself.  Furthermore, it is evident that the
whole transformation is already fully determined by the image of only
one sector (the transformation can then be extended to the other
sectors by applying the gluing diffeomorphism \re{glue})).

Let us start with ({\it ii\/}), i.e.\ determine all isometries of one sector
onto itself. Again $\phi$ must be preserved, so in the chart
(\ref{011h}) the map must preserve the lines $\phi=const
\Leftrightarrow x^0=const$. But also null-extremals must be mapped
onto null-extremals. This leaves two possibilities: If the
null-extremals \re{null1}) (i.e.\ $x^1=const$) are mapped onto
themselves, then the only possibility is an overall shift of the
$x^1$--coordinate, $x^1 \to x^1+\omega$, i.e.\ a Killing-transformation
\re{Kill}).  The gluing diffeomorphism (\ref{glue}) shows that such a
transformation extends uniquely onto the whole universal covering, and
that it is in all charts \re{011h}) represented as an $x^1$-shift of the same
amount (but on the `flipped' ones in the opposite direction!).
In the neighbourhood of bifurcation points (simple zero of $h(x^0)$) we can
also use the
local Kruskal-coordinates (\ref{Kruskal}), where the same transformation reads
$u \to u\exp(\frac{h'(a)}2\omega)$,
$v \to v\exp(-\frac{h'(a)}2\omega)$,
which 
looks in this case very much like a Lorentz-boost (cf e.g.\ the
arrows around the bifurcation points in Figs.~\ref{fig:fig25}, \ref{fig:fig31} below). We will (thus) call these Killing-transformations shortly {\em
  boosts\/} and denote them by $b_\o$.
The composition law is clearly $b_\o b_\n=b_{\o+\n}$, so the boosts
form a group isomorphic to $\dR$, which shall henceforth be denoted by
$\Rboost$.

\begin{figure}[ht]
\begin{center}
\leavevmode
\epsfig{file=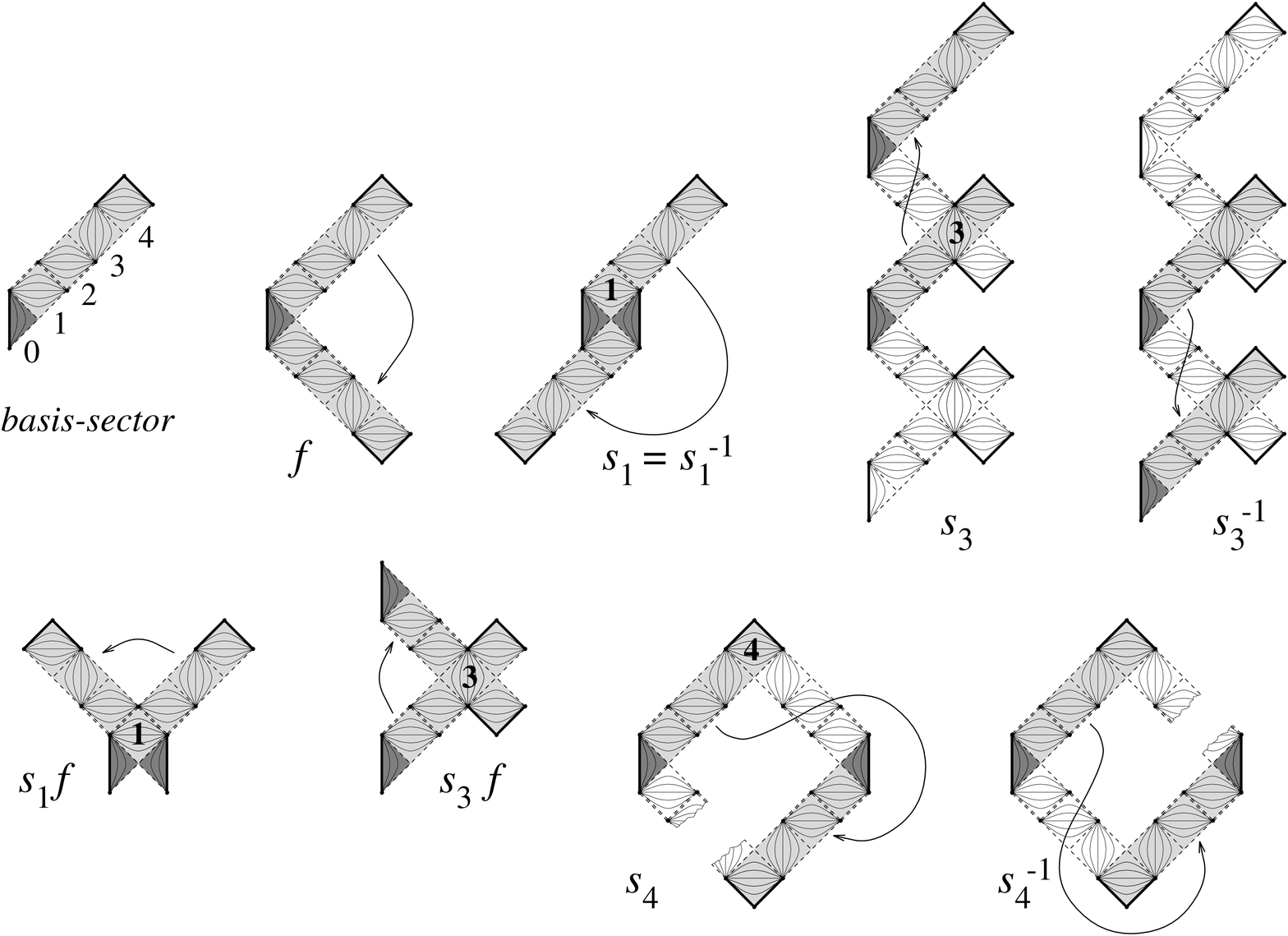,width=14cm}
\end{center}
\renewcommand{\baselinestretch}{.9}
\caption{
  \captionstyle \label{fig:fig19}
  Some basic sector-moves, their
  inverses, and flips. The light shaded block is the basis-block
  resp.\ its image under the move, the dark shaded sector is the
  basis-sector (or its image). Only a few sectors of the universal covering
  are displayed. Note that the basic move $s_4$ and its inverse
  ${s_4}^{-1}$ transport the basis-block into different (but
  overlapping) layers of the universal covering.
  And note also that $s_1$, $s_4$, and their inverses turn the
  basis-block upside down!
  }
\end{figure}

If, on the other hand, the two families of null-extremals,
\re{null1}) (i.e.\ $x^1=const$) and \re{null2a}) (i.e.\
$dx^1/dx^0=-2/h$), are interchanged, then we obtain precisely the
flip transformations. These transformations are the gluing
diffeomorphisms \re{glue}), but this time considered as active
transformations. Due to the constant in \re{glue}) the flips come
as a one-parameter family; however, they differ only by a boost,
i.e.\ given a fixed flip transformation $f$, any other flip $f'$
can be obtained as $f'=f b_\o \equiv b_{-\o}f \equiv
{b_{\o/2}}^{-1}fb_{\o/2}$ for some $\o$.  We will thus consider
only {\em one\/} flip and denote it by $f$.  In the Carter-Penrose
diagrams such a flip is essentially a reflexion at some axis
(horizontal or vertical, according to sgn$\,h$; cf Fig.~\ref{fig:fig19}) and it is of course involutive (i.e.\ self-inverse,
$f^2=1$). In the `Kruskal'-coordinates \re{Kruskal}) it
amounts essentially to an exchange $u\leftrightarrow \pm v$.
Let us finally point out that while under a pure boost each sector is
mapped onto itself (the corresponding sector--permutation is the
identity), a flip transformation always (unless there is only one
sector) entails a nontrivial sector--permutation (which is clearly
self-inverse, since $f$ is).

\medskip

We now turn to task ({\it i\/}), the description of the
sector--permutations: As pointed out above, each transformation is
fully determined by its action on only one sector. Let us thus choose
such a `basis-sector'. If the transformation does not preserve
orientation,
\footnote{By this we mean the orientation of the spacetime considered as
  a 2-manifold, not the {\em spatial\/} orientation.}
we may apply a flip at this basis-sector and we are left
with an orientation--preserving transformation.  For these, however,
the corresponding sector--permutation is already uniquely determined by
the image of the basis-sector, i.e., for each copy of the basis-sector
somewhere in the universal covering there is exactly one
sector--permutation moving the basis-sector onto that copy.  All
 these
orientation--preserving sector--permutations thus make up a discrete
combinatorial group (henceforth called \smovs), which  now shall be
described in more detail.

Also choose a `basis-building-block' and within this basis-block fix
the first (or better: zeroth) sector as basis-sector (we label the
sectors by $0,\ldots,n$ and the horizons between by $1,\ldots,n$).  By
a `basic move' across sector $i$ we mean the following (cf Fig.~\ref{fig:fig19}): Go from the basis-sector to the $i$th sector of the
basis-block 
and from there to the zeroth sector of the
perpendicular (i.e.\ flipped)
block. 
The corresponding sector-move, mapping the basis-sector onto this
other sector, shall be denoted by $s_i$ (of course $s_0$ is the identity,
$s_0=1$).  Also the inverse basic moves can be easily described: One
has to perform the same procedure at the {\em flipped\/} basis-block
only (cf Fig.~\ref{fig:fig19}).  An inverse basic move is thus the
conjugate of the original move by a flip, $fs_if={s_i}^{-1}$ (note
$f=f^{-1}$).  Here $f$ has been supposed to be a flip at the
basis-sector. Flips at other sectors can be obtained by composition
with sector-moves: $s_if=f{s_i}^{-1}$ is a flip at sector $i$ of the
basis-block (Fig.~\ref{fig:fig19}).

Evidently there are two qualitatively different cases: If basis-sector
and `flip'-sector are both stationary resp.\ spatially homogeneous, then the
basic move is essentially a translation (e.g.\ $s_3$ in Fig.~\ref{fig:fig19}).
However, if the sectors are of a different kind, then the move ($s_1$,
$s_2$, $s_4$ in the example of Fig.~\ref{fig:fig19}) turns the whole
solution upside down, inverting space {\em and\/} time (thus still
preserving the orientation, as required for elements of \smovs{} --- in
contrast to, e.g., $s_1 f$, which inverts space only, cf Fig.~\ref{fig:fig19}).

\begin{figure}[ht]
\begin{center}
\leavevmode
\epsfig{file=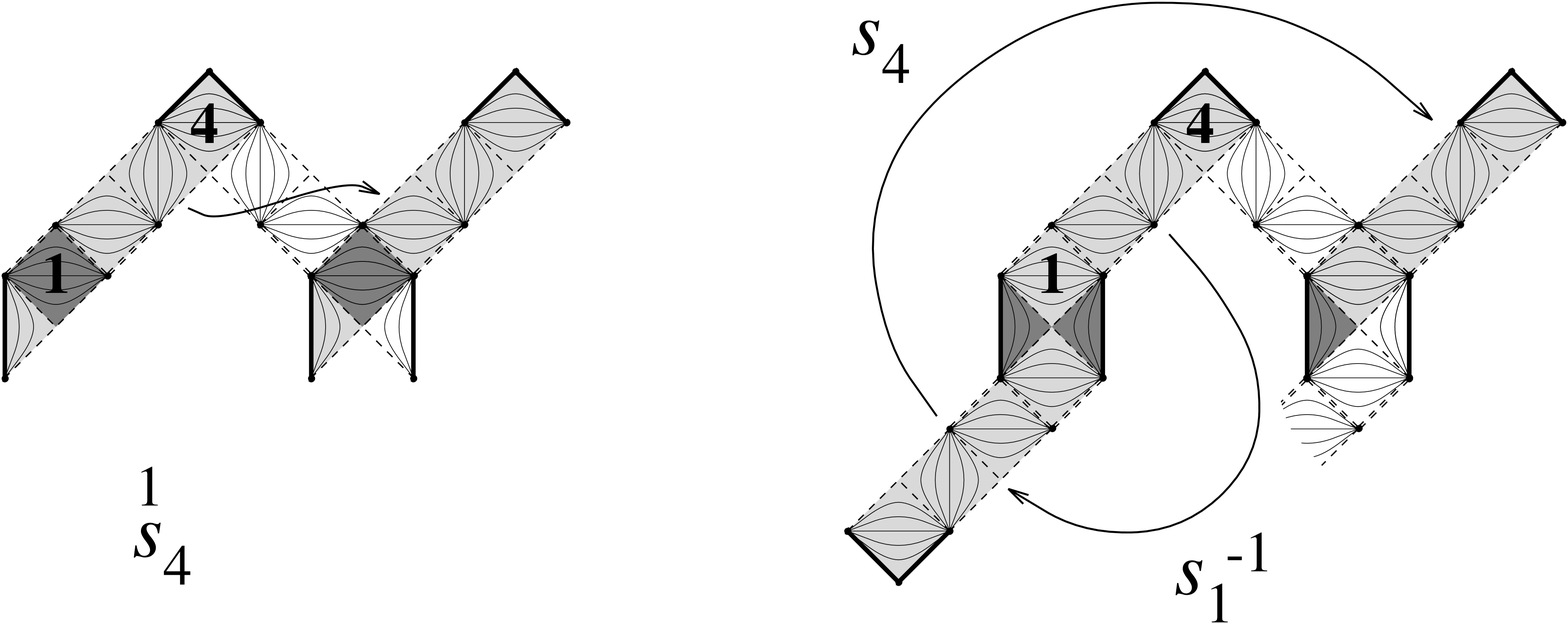,width=10cm}
\end{center}
\renewcommand{\baselinestretch}{.9}
\renewcommand{\baselinestretch}{.9}
\caption{
  \captionstyle \label{fig:fig20}
  Basic sector-move with different
  basis-sector. As is seen from this figure the move $\protect\scoaf$ can be composed
  of two moves with basis-sector zero, $\protect\scoaf=s_4{s_1}^{-1}$. Although in
  this case ${s_1}^{-1}=s_1$ (relation at a bifurcation point), in general the
  right element has to be an inverse move!}
\end{figure}

(Of course it is not necessary to choose the zeroth sector of the block as
basis-sector. Let us denote the basic move across sector $i$ with basis-sector
$k$ by $\sckai$ (thus $\sczai \equiv s_i$).
They can, however, be expressed in terms of the old moves:
As may be seen from Fig.~\ref{fig:fig20} we have $\sckai=s_i{s_k}^{-1}$,
and consequently even $\sckai=\sclai\sclakinv$ for an arbitrary sector $l$.
Obviously always $\sckak=1$ (generalizing $s_0=1$) and $\sckai=\sciakinv$.
Thus, there is no loss of generality in choosing the basis-sector zero.)

The above basic moves $s_i$ form already a complete set of generators for
\smovs:
By the extension process each location in the universal covering can be
reached from the basis-sector by (repeated) application of the following
step: Move through the basis-block or the flipped basis-block to some sector,
then continue along the perpendicular block (i.e., applying a flip at this
sector). But this step {\em is\/} exactly a basic move.

There may yet be relations between the generators. If, however, all
horizons are degenerate, then there are no relations, and the
resulting group is the free group with generators $s_i$. The reason is
that in this case the vertex points between sectors are at an infinite
distance (cf \re{length})) and thus the overlapping sectors (after
surrounding such points) must not be identified, yielding a
multilayered structure (cf e.g.\ $s_4$ in Fig.~\ref{fig:fig19}, or the
`winding staircase'-like {\bf G4} in Fig.~\ref{fig:fig22} below).  A
nontrivial word composed of basic moves $s_i$ describes such a move,
which must therefore necessarily lead into a different `layer' of the
universal covering, since this manifold is simply connected.

The situation is different if there are nondegenerate horizons: In this case
there are (interior!) bifurcation points at which four sectors meet, which
then constitute a single sheet. Moving once around such a bifurcation point
leads back to the original sector. Consequently there emerges a relation:
If e.g.\ the first horizon is nondegenerate then the basic move $s_1$ turns
the solution $180^\circ$ around with the bifurcation point as centre
(cf Fig.~\ref{fig:fig19}). A second
application of $s_1$ yields the original configuration again, so we
have ${s_1}^2=1$. To find the analogous relations for a nondegenerate horizon
say between the $(k-1)$th and $k$th sector it is wise to
temporarily
shift the basis-sector to the $k$th sector. Then evidently $\sckakmo$ turns
the solution $180^\circ$ around that bifurcation point and we have the
relation $(\sckakmo)^2=1$ which
translates back to $(s_{k-1}{s_k}^{-1})^2=1$.

\medskip

Summarizing, we have found that the group \smovs\ has the following
presentation in terms of generators and relations ($n$ being the number of
horizons):
\begin{equation}
\smovs = \left\<\; s_1,\ldots,s_n\;
     \Big|\; (s_{k-1}{s_k}^{-1})^2=1
        \mbox{\ \ for each nondeg.\ horizon $k$} \;\right\>
        \,\, .
\pl{smovs}
\end{equation}

Any symmetry transformation can thus be written as a product of possibly
a flip $f$ (if it is orientation-reversing), a sector-move from the group
\re{smovs}), and a boost $b_\o$ ($\o\in{\R}$).
Furthermore, this representation is unique, provided the elements are in
this order.
We have yet to describe the group product:
Evidently boosts and sector-moves commute, since, as pointed out before,
a boost is in all equally oriented charts (and sector-moves preserve the
orientation) presented as a shift $x^1\rightarrow x^1+\o$, equation \re{Kill}).
Furthermore, the conjugate
of a boost by a flip is the inverse boost, $fb_\o f=b_{-\o}={b_\o}^{-1}$,
and the conjugate of a basic move by a flip is the inverse basic move,
$fs_if={s_i}^{-1}$, which also
defines the conjugate of a general (composite) sector-move.
The group product is thus completely determined, since
in the formal product of two elements the factors can be
interchanged to yield the above format.
Thus the structure of the full isometry group is a semi-direct product
\begin{equation}
  {\cal G}=\Zflip\ltimes \Big(\Rboost\times\smovs\Big),
  \pl{symmgroup}
\end{equation}
where the right factor (in round brackets) is the normal subgroup,
$\Zflip$ denotes the group $\{1,f\}$,
and $\Rboost$ is the group of boosts described previously.
Especially, if we restrict ourselves to orientable factor spaces and
thus to orientation--preserving isometries, flips must be omitted and
we are left with a {\em direct\/} product of the combinatorial group
\smovs\ with $\Rboost$.

Still, this description is not always optimal, for two reasons: First,
one is often interested in orientable {\em and\/} time--orientable
solutions.  Second, while for only degenerate horizons the group
\smovs\ is a free group, this is not true if there are nondegenerate
horizons (cf equation \re{smovs})).  Interestingly, both problems can be
dealt with simultaneously.
Clearly the time--orientation--preserving sector-moves constitute a
subgroup of \smovs, which we shall denote by \osmovs\ (in analogy to the
notation used frequently for the orthochronous Lorentz group). If all
horizons are degenerate and of even degree, then all sectors are
equally `oriented' (stationary or spatially homogeneous) and consequently all
sector-moves are automatically time--orientation--preserving; thus
$\osmovs=\smovs$, and its rank equals the number of horizons.

This is no longer the case if there are horizons of odd degree.  Then
some sectors will have an orientation contrary to that of the
basis-sector and a sector-move at such a sector will reverse the
time--orientation (cf, e.g., $s_1$ in Fig.~\ref{fig:fig19}).  The group
\osmovs\ is then a proper subgroup of \smovs\ (with index 2), which
consists of all elements containing an even number of time--orientation
reversing sector-moves.

Fortunately, this group can be described quite explicitly.  Let us
start with the case that there are nondegenerate horizons.  To
simplify notation we assume for the moment the first horizon to be
nondegenerate (below we will drop this requirement).  The sector-move
$s_1$ is then an reflection at a bifurcation point and thus inverts the
time--orientation (cf Fig.~\ref{fig:fig19}).  But also the sector-moves
around all other sectors `oriented' contrary to the basis-sector
(sectors 2 and 4 in Fig.~\ref{fig:fig19}) will reverse the
time--orientation.  The idea is to extract the space--time--inversion
$s_1$ from the group \smovs: For each sector $i$ introduce the new
generators $s_i$ and $s_1s_is_1$, if the $i$th sector has the same
`orientation' as the basis-sector, and  $s_is_1$ and $s_1s_i$, if it
has the opposite `orientation'.  These new generators are all
time--orientation--preserving. Together with $s_1$ they obviously still span
the whole group \smovs\ (relations between the new generators will
be discussed presently). Conjugation by $s_1$ only permutes them among
themselves (use ${s_1}^2=1$): $s_is_1\leftrightarrow s_1s_i$ and
$s_i\leftrightarrow s_1s_is_1$.
Also, any element of \smovs\ can be expressed as a word composed of the new
generators with or without a leading $s_1$. Thus the group is a
semidirect product
\begin{equation}
  \smovs=\ZPT\ltimes\osmovs,
  \pl{smovsplit}
\end{equation}
where $\ZPT=\left\{1,s_1\right\}\,$ (the `PT' stands for parity and
time--inversion) and \osmovs\ is the
group generated by the new elements given above (with $i \ge 2$).
There may still be some relations among those generators:
{}From \re{smovs}) we know that any nondegenerate horizon $k$ adds a relation
$(s_{k-1}{s_k}^{-1})^2=1$ or equivalently
$s_{k-1}{s_k}^{-1}=s_k{s_{k-1}}^{-1}$. This yields a relation between the
new generators, e.g.\ $(s_1s_{k-1}s_1)(s_ks_1)^{-1}=(s_1s_k){s_{k-1}}^{-1}$,
by means of which either of the (four) generators involved can be
expressed in terms of the remaining ones.
Apart from those relations there are no
further ones, so if we eliminate the redundant generators we are left
with a {\em free\/} group!  To determine the rank of this group note
that each sector $>1$ contributes two generators and each
nondegenerate horizon (except for the first, which was taken into
account already in \ZPT) yields a relation which in turn kills one
generator. Thus
\begin{eqnarray}
\mbox{rank \osmovs} & = & 2 \;(\mbox{number of degenerate horizons})  +
                                                               \nonumber\\
            &   & \mbox{}+ \mbox{number of nondegenerate horizons} - 1 \,\,.
\pl{rank}
\end{eqnarray}
Finally, if the first horizon is degenerate but the $k$th is
not, then one can replace $s_1$ above by $\sckakmo$ and proceed
in an analogous way. The result is \re{smovsplit}), \re{rank}) again.%
\footnote{Of course there may be other possibilities to split the isometry
  group into a product. For instance, if there are bifurcate horizons, then
  one can choose a basis-bifurcation-point and instead of tracking the motion
  of the basis-sector (which leads to \smovs) follow 
  the bifurcation point. The resulting group of space-- and
  time--orientation--preserving bifurcation point moves coincides exactly with
  \osmovs, which can thus also be interpreted as \bpmovs\
  (cf e.g.\ Figs.~\ref{fig:fig24},
  \ref{fig:fig26} ({\it b\/}), \ref{fig:fig27}). The 
remaining bifurcation point
  preserving isometries make up the 1+1 dimensional Lorentz group
  $O(1,1)$, which contains the boosts, flips (i.e.\ space--inversions and
  time--inversions), and reflection at the bifurcation point
  (i.e.\ space--time--inversion). Hence the full isometry group can also be
  written as ${\cal G}=O(1,1)\ltimes \bpmovs\,$.
  Furthermore, when restricting to
  space-- {\em and\/} time--orientable solutions we may use
  $
    \oG=SO^\uparrow(1,1)\times\bpmovs 
  $
  where the proper orthochronous Lorentz group
  $ SO^\uparrow(1,1) \cong \Rboost $ contains only the boosts.
  Clearly, this is just the same as equation \re{osymmgroup}).} 


The remaining case to consider is the one in which all the horizons are
degenerate. Then \smovs\ \re{smovs}) 
is a free group and correspondingly its
subgroup \osmovs\ is  free, too (cf section \ref{Secprelim}). Let us determine
its rank: As noted already above, if all the horizons are of
even degree, then $\osmovs\ = \smovs$, and the rank equals the total
number of (degenerate) horizons (thus \re{rank}) does not generalize to this
case). Finally, assume that there is an odd-degree horizon and let it again
be the first one (if it is not the first but the $k$th, just replace
$s_1$ by $\sckakmo$ in what follows): A set of generators can be
found in the same way as above (introducing $s_1s_k$, $s_ks_1$,
or $s_l$, $s_1s_ls_1$, respectively, $k,l
\ge 2$); however, now the element ${s_1}^2$, which no longer is the unit
element,  has to be added as a further generator. If $d$ denotes
the number of (degenerate) horizons, we thus find $2d -1$
generators for \osmovs\ (two for each degenerate horizon except
for the first horizon, which adds one only).  Here we also could
have used formula \re{index}), since
\osmovs\ has index 2 in \smovs.
\footnote{Note, however, that in this case \smovs\ does not split into a
  semi-direct product like in \re{smovsplit}), since now there is no subgroup
  \ZPT (\smovs\ is free!); it is only a nontrivial extension of
  ${\Z}_2$ with \osmovs.}

Let us finally summarize the above results:
\begin{quote}
 \noindent{\bf Theorem:}\quad The group of space-- and
 time--orientation--preserving symmetry 
transformations \oG\ is a direct product
 of $\dR$ with a free group,
 \begin{equation}
   \oG=\Rboost\times\osmovs.
   \pl{osymmgroup}
 \end{equation}
 Let furthermore $n$ denote the number of nondegenerate horizons and $d$
 the number of degenerate horizons.
 Then the rank of the free group \osmovs\ is
 \begin{eqnarray}
      d   & \qquad\mbox{if all horizons are of even degree,}\nonumber\\
   2d+n-1 & \mbox{in all other cases.}
   \pl{oGrank}
 \end{eqnarray}
\end{quote}
We will mainly work with this subgroup, but nevertheless discuss at a few
examples how (time--)\-orientation-violating transformations can be
taken into account.

\subsubsection{Subgroups and factor spaces \pl{Factor}}

We now come to the classification of all factor solutions.
As pointed out in section \ref{Secprelim}, 
they are obtained by factoring out a
freely and properly discontinuously acting
(from now on called shorthand {\em properly acting\/})
symmetry group from the
universal covering solutions. Thus we have first to find all
properly acting subgroups of the full
transformation group. However, not all different subgroups give rise to
different (i.e.\ nonisometric) factor spaces. If, for instance, two
subgroups $\cal H$ and ${\cal H}'$ are conjugate (i.e., there is a symmetry
transformation
$h\in{\cal G}$ such that ${\cal H}'=h{\cal H} h^{-1}$) then the factor spaces
${\cal M}/{\cal H}$ and ${\cal M}/{\cal H'}$ are 
isometric. (Roughly speaking, this conjugation can be interpreted as a
coordinate change).
But also the converse is true:

\begin{quote}
 \noindent{\bf Theorem:} \quad Two factor spaces are isometric iff the
 corresponding subgroups are conjugate.
\end{quote}
(for a proof cf e.g.\ \cite{Wolf}, Lemma 2.5.6).\\
The possible factor spaces are thus in one-to-one correspondence with the
conjugacy classes of properly acting subgroups.
Still, apart from this abstract characterization it would be nice to have
some `physical observable', capable of discerning between the different factor
spaces. This concerns mainly the boost-components of the subgroup elements,
since information about the sector-moves is in a rather obvious way
encoded in the global causal structure (number and arrangement of the
sectors and singularities) of the factor space.
Indeed we will in general be able to find such observables.
[Here `physical observable' means some quantity that remains unchanged under
the group of diffeomorphisms (= gauge symmetry); so it will
be a {\em geometrical invariant\/}, which not necessarily can
be `measured' also by a `physical observer', strolling along his
timelike worldline].

The above theorem is also valid for the case of a restricted symmetry group.
For instance, a spacetime is often supposed to have --- apart from its metric
structure --- also an orientation and/or a time--orientation. If the universal
covering $\cal M$ carries a $\mbox{(time--)}$orientation and $\cal H$
is (time--)\-orientation--preserving, then also the factor space inherits a
(time--)\-orientation. Now, if two subgroups are conjugate,
${\cal H}'=h{\cal H} h^{-1}$, but the intertwining transformation $h$
does not preserve the (time--)\-orientation, then the corresponding factor
spaces will
have different (time--)\-orientations (while still being isometric, of course),
and should thus be regarded as different. Thus, in this case the conjugacy
classes of subgroups should be taken with respect to the restricted symmetry
group (e.g.\ \oG).

\medskip

The requirement that the subgroup acts freely
already rules out some transformations. First of all the subgroup must not
contain pure flips at any sector: A flip has a whole line of fixed points,
namely an extremal of the kind $dx^1/dx^0=-1/h$ (cf \re{13}) and
Fig.~\ref{fig:fig2}) as
symmetry axis. This symmetry axis would become a boundary line of the
factor space, which then would no longer be maximally extended.
Consequently not only $f$, but also $f{s_i}^n$ and their conjugates have
to be omitted.
\footnote{As pointed out before $s_if$ is a flip at sector
  $i$. By the group product one has further $f{s_i}^{2k}={s_i}^{-k}f{s_i}^k$
  and $f{s_i}^{2k-1}={s_i}^{-k}(s_if){s_i}^k$, so they are conjugate to pure
  flips and thus also pure flips at displaced sectors.}

But also reflections at a bifurcation point (turning the solution $180^\circ$
around this bifurcation point, e.g.\ $s_1$ in Fig.~\ref{fig:fig19}) must
be avoided: Not only does the factor space fail to be
time--orientable in this case, but also the bifurcation point is a
fixed point, which upon factorization develops into a singular
`conical tip' making the solution nonsmooth.  Thus, if the
$k$th horizon is nondegenerate, then the elements $(\sckakmo)=
s_{k-1}{s_k}^{-1}$ and conjugates thereof must not occur in the
subgroup $\cal H$.

\begin{figure}[ht]
\begin{center}
\leavevmode
\epsfig{file=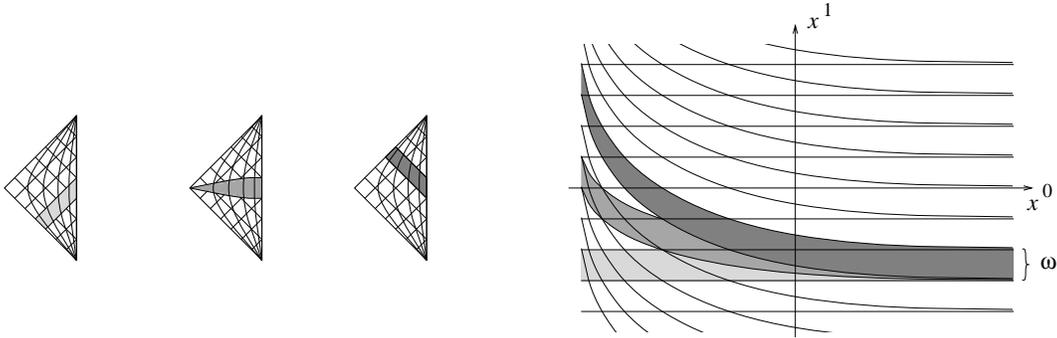,width=14cm}
\end{center}
\renewcommand{\baselinestretch}{.9}
\vspace*{-1cm}
\caption{\captionstyle \label{fig:fig21}
  A cylinder resulting from pure boosts
  and possible fundamental regions $\cal F$; left in the
  Carter-Penrose diagrams, right in the EF-chart \re{011h}).
  Boundaries have to be glued along Killing-trajectories, i.e.\
  along the curved lines in the left, resp.\ along vertical fibres
  in the right diagram.}
\end{figure}

Let us next assume pure boosts, which form a group isomorphic to $\dR$
(\Rboost).
The only discrete subgroups of boosts are the infinite cyclic groups
generated by one boost, ${\cal H}_\o := \left\{{b_\o}^n,\, n\in{\Z}
\right\},\: \o>0$.
In the coordinates (\ref{011h}) such a boost $b_\o$ is a shift of length $\o$
in $x^1$-direction. The factor space is then clearly a cylinder.
Also, since boosts commute with sector-moves and `anticommute' with flips
($fb_\o=b_{-\o}f$), the group ${\cal H}_\o$ is invariant under conjugation
and the parameter $\o$ cannot be changed. Thus the
cylinders are labelled by one positive real parameter $\o$. In order to find
some geometrical meaning of this parameter it is useful to adopt a
cut-and-paste approach to the factorization procedure
(used already intuitively in section \ref{Seccut}): In the above example a
fundamental region $\cal F$ can be obtained from the patch
\re{011h}) by
cutting out a strip of width $\o$ parallel to the $x^0$-axis and gluing it
together along the frontiers (preserving $x^0$, i.e.\ vertical fibres).
Of course there are several choices for
${\cal F}$; it need not even be a horizontal strip, but any strip
with vertical cross-section $\o$ will work (cf Fig.~\ref{fig:fig21}).
The width $\o$ of this strip (i.e.\ of the generating shift) is proportional to
the length of an $\phi=const$-path (resp.\ constant curvature or $\Phi$)
running once around the cylinder.
This is the desired geometrical observable: for any metric $g$
(i.e.\ function $h$ in \re{011h})) we get a set of distinct solutions
parameterized by their `size' or circumference (any positive real
number).
\footnote{In section \ref{Seccut} we took the second continuous parameter
  besides $C$ without restriction on its sign. For a comparison with the RPS
  this is more appropriate, since $\o \sim -\o$  is induced
  by the {\em large\/} diffeomorphism $x^1 \to - x^1$, which is not connected
  to the identity and thus cannot be generated by the flow of the constraints.}
Let us finally point out that this parameter $\o$ has nothing
to do with the Casimir-value $C$, present in the function $h$ of the metric.
It is a new, {\em additional\/} parameter resulting from the factorization.

\begin{figure}[htb]
\begin{center}
\leavevmode
\epsfig{file=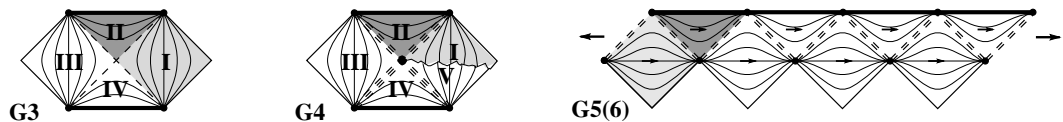,width=14cm}
\end{center}
\renewcommand{\baselinestretch}{.9}
\caption{
  \captionstyle \label{fig:fig22}
  Carter-Penrose diagrams for one
  horizon ({\bf G3-6}). For a nondegenerate horizon ({\bf G3}) the
  vertex point is an (interior) bifurcation point and the manifold a
  single sheet. For degenerate horizons of odd degree (e.g.\ {\bf
    G4}), on the other hand, this point is at an infinite distance,
  and thus the manifold has infinitely many layers. Finally, for
  horizons of even degree it is an infinite ribbon ({\bf G5(6)},
  interior arrows indicate the Killing field).}
\end{figure}

This all works perfectly well as long as there is only one sector, i.e.,
$h$ has no zeros, but at horizons the boosts do not act properly
discontinuous. As a consequence the factor space will not be Hausdorff
there. 
Furthermore, at bifurcation points the action is not free, so the factor
space is not even locally homeomorphic to $\dR^2$ there.
At the first glance this might seem surprising since when starting from an
EF-coordinate patch \rz 011h the construction of Fig.~\ref{fig:fig21} should yield
regular cylinders. They are not global, however:
`half' of the extremals approach the (closed) horizon asymptotically,
winding around the cylinder infinitely often while having only
finite length.
This phenomenon is well-known from the Taub--NUT space or
its two-dimensional analogue as described by Misner
\cite{Mis} (cf also \cite{HawkingEllis}).
A detailed description with illustrations can also be found in \cite{Geroch}.
[In {\bf G3}, Fig.~\ref{fig:fig22}, for instance, let the EF-coordinates cover
the sectors I and II. The above class of incomplete extremals then comprises
those which run across the lower horizon from I to IV, leaving
the EF-patch. Thus to obtain an extension a second half-cylinder
corresponding to IV has to be attached to I via a second copy of the closed
horizon and likewise for the sector III, but of course this violates the
Hausdorff property.]
Similar results hold in general for solutions with zeros of $h(x^0)$:
whenever the group factored out contains a pure boost we obtain a
Taub--NUT like solution (a cylinder-`bundle', where at each horizon, two sheets
meet in a non--Hausdorff manner) labelled by its `size' (metric-induced
circumference along a closed $\phi=const$ line). If there are
further sector--moves in the group $\cal H$, they have only the effect of
identifying different sheets of this cylinder-bundle.


\medskip

\begin{figure}[htb]
\begin{center}
\leavevmode
\epsfig{file=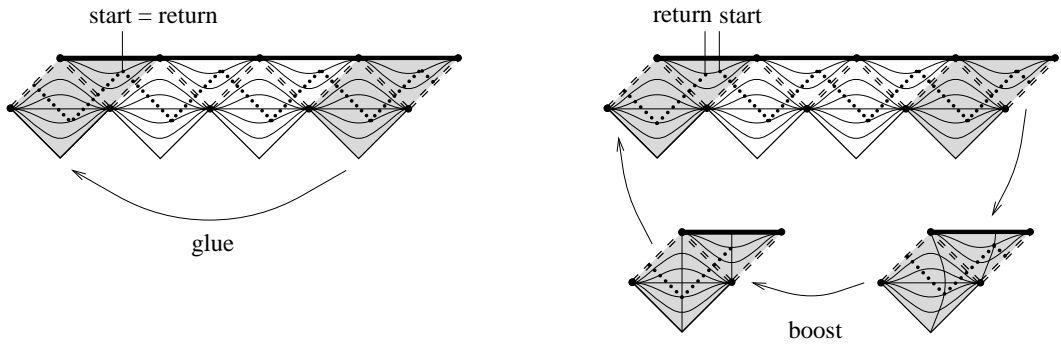,width=14cm}
\end{center}
\renewcommand{\baselinestretch}{.9}
\caption{
  \captionstyle \label{fig:fig23}
  Boost-parameter for {\bf G5(6)}. In both
  cases the generating shift ${s_1}^3$ (or its inverse) is used to glue
  the right shaded patch onto the left one. If this generator has a nontrivial
  boost-component, then one has to apply a boost before gluing the patches
  (right figure).
  To illustrate the effect we have drawn a polygon of null-lines (pointed lines).
  Due to the boost the endpoint of this null-polygon will be shifted against the
  start, and this
  deviation may serve as a measure for the boost-parameter. Of course the
  same construction can also be applied to the other cases
  (e.g.\ {\bf G4}, Fig.~\ref{fig:fig22}).}
\end{figure}

\begin{figure}[htb]
\begin{center}
\leavevmode
\epsfig{file=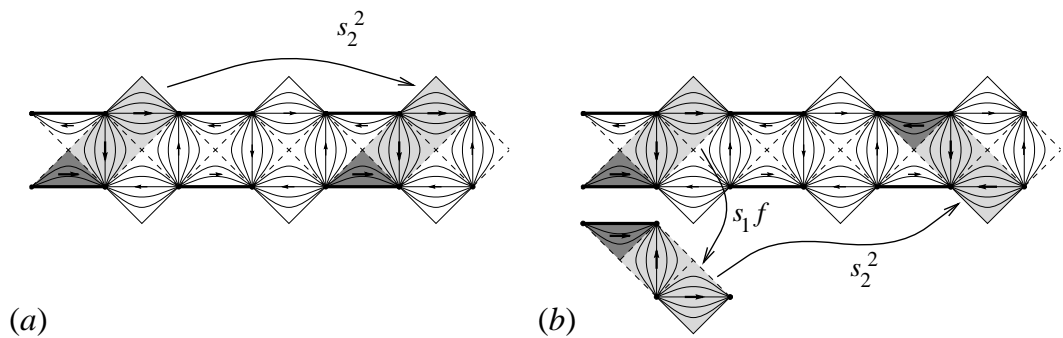,width=14cm}
\end{center}
\renewcommand{\baselinestretch}{.9}
\caption{
  \captionstyle \label{fig:fig24}
  Sector-moves for {\bf G8,9}; ${s_2}^2$ gives
  rise to a cylinder ({\it a\/}), while ${s_2}^2s_1f$ yields a M\"obius-strip
  ({\it b\/}). Note that in ({\it b\/}) the (Killing-)arrows in the blocks
  identified by ${s_2}^2$ do not match.}
\end{figure}

\begin{figure}[htb]
\begin{center}
\leavevmode
\epsfig{file=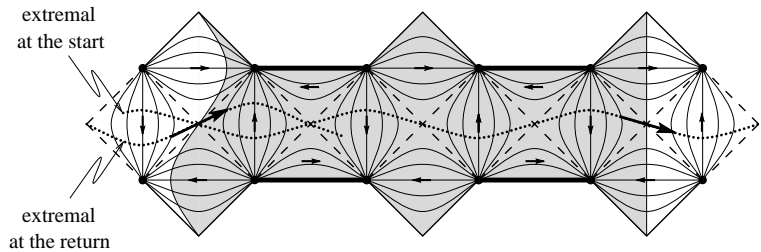,width=10cm}
\end{center}
\renewcommand{\baselinestretch}{.9}
\caption{
  \captionstyle \label{fig:fig25}
  Fundamental region (shaded) and interpretation
  of the boost-parameter for {\bf G8,9}. The right straight boundary has to be
  glued to the curved left one, and also the extremal \re{13})
  (dotted line) returns tilted (i.e.\ boosted).}
\end{figure}

After these preliminaries we will now start a systematic
classification of the factor solutions. We do this in order by number
and type of horizons and illustrate it by the solutions of the JT--,
$R^2$--, and KV--model (examples {\bf J1-3}, {\bf R1-5}, and {\bf
  G1-11}, respectively; cf Figs.\ in the previous chapter, but also
Figs.~\ref{fig:fig14a}--\ref{fig:fig16}, \ref{fig:fig22}--\ref{fig:fig25}).

\medskip
\noindent{\bf No horizons (e.g.\ G1,2, J3):}\\
\noindent This case has already been covered completely by the above
discussion: There is only one sector, so \smovs\ is trivial. Furthermore,
flips are not allowed, thus only boosts remain and they yield
cylinders labelled by their circumference (positive real number).

\medskip

In the following we will continue studying all factorizations possible
for solutions {\em with\/} horizons.
As pointed out before, there exist pathological Taub--NUT-like solutions for
all of these cases (resulting from pure boosts).
We will from now on exclude them, so
pure boosts must not occur in the subgroup $\cal H$. But this also means
that no sector--permutation can occur twice with different
boost-parameters,
since otherwise they could be combined to a nontrivial pure boost.
This suggests the following strategy:
factor out $\Rboost$ from $\cal G$, i.e.\ project
the whole symmetry group onto ${\cal G}/\Rboost=\Zflip\ltimes\smovs$,
respectively \oG\ onto $\oG/\Rboost=\osmovs$. All above (non--Taub--NUT)
subgroups $\cal H$ are mapped one-to-one under this projection.
One can thus `forget' the
boost-components, and first solve the combinatorial problem of finding the
projected subgroups $\overline{\cal H}$ (which we will nevertheless simply
denote by $\cal H$).
Only afterwards we then deal with
re-providing the sector-moves with their boost-parameters.

\medskip
\noindent{\bf One horizon (e.g.\ G3-6, R1,2, J2):} (Fig.~\ref{fig:fig22})\\
\noindent%
The group \smovs\ is in these cases still rather simple:
$\<s_1|{s_1}^2=1\>\cong{\Z}_2$ for a nondeg.\ horizon
({\bf G3, R1}), and $\<s_1|-\>\cong{\Z}$ for the others.
First of all we see that flips have to be omitted altogether:
The most general form in which they can occur is $f{s_1}^kb_\o$, but this is
by the group product always
conjugate to $f$ or $s_1f$ and thus a pure flip.

The action of \smovs\ on the manifold is most evident for a zero of even
degree, in which case we have the ribbon-structure of {\bf G5(6)}.
Then $s_1$ is a shift of one block to the right (in the situation of
Fig.~\ref{fig:fig22}). The nontrivial subgroups are the cyclic groups generated
by ${s_1}^n$, for some $n\ge 1$, and the corresponding factor space is obtained
by identifying blocks which are a multiple of $n$ blocks apart, i.e., gluing
the $n$th block onto the zeroth. This yields a cylinder of a `circumference'
of $n$ blocks (see Fig.~\ref{fig:fig23} for $n=3$). For degenerate horizons of
odd degree (e.g.\ {\bf G4}) the
situation is rather similar, only the basic move is now a
$180^\circ$-rotation around the (infinitely distant) central point,
`screwing the surface up or down' (i.e.\ mapping I $\rightarrow$ III,
II $\rightarrow$ IV, III $\rightarrow$ V, etc.). Thus we get again cylinders
of $n$ blocks circumference.
When passing once around such a cylinder, however, the lightcone tilts upside
down $n$ times, so we
have got an {\it$n$-kink\/}-solution. Especially, the solutions with $n$ odd
are not time--orientable. Finally, for nondeg.\ horizons (e.g.\ {\bf G3})
the only nontrivial sector-move is $s_1$ (${s_1}^2$ is already the
identity again), but this is the reflexion at the bifurcation point
and thus not a free action. So we get no smooth factor solutions in this case.

If we had restricted ourselves a priori to orientable and time--orientable
solutions, then we would have had to start from \osmovs. For {\bf G5(6)}
this makes no difference since then $\osmovs=\smovs$, and indeed all cylinders
were time--orientable in this case. For {\bf G4}, on the other hand, \osmovs\
is a proper subgroup, $\osmovs\cong2{\Z}<{\Z}\cong\smovs$,
generated by ${s_1}^2$. This reflects the
fact that only solutions of even kink-number are time--orientable. Finally,
for {\bf G3} \osmovs\ is trivial and there are no factor solutions at all.

We have so far neglected the boost-component of the generator. After all
the full-fledged generator of the subgroup will be ${s_1}^nb_\o$!
Does this have any consequences on the factor space?
According to the theorem at the beginning of this section we must check
whether the corresponding subgroups are conjugate.
This is not the case here:
The generator ${s_1}^n b_\o$ of the subgroup commutes with everything
except flips, and even a flip only transforms the group elements to their
inverse, $f{s_1}^nb_\o f={s_1}^{-n}b_{-\o}=({s_1}^nb_\o)^{-1}$.
Thus the group remains the same, and
one cannot change the boost-parameter $\o$ by conjugation.
Let us again look for some geometrical meaning of this parameter.
In the case of pure boosts discussed above (which also lead to cylinders)
we have found the metric-induced
circumference as such an observable. This cannot be transferred to the
present case, however: there is no closed $\phi=const$-line along which a
circumference could be measured, only the number of blocks $n$ `survives'.
So we have to be  more inventive: one could, e.g.,
take a series of null extremals, zigzagging
around the cylinder between two fixed
values of $\phi$ (see pointed lines in Fig.~\ref{fig:fig19})
and interpret the deviation from being closed, i.e.\ the distance between
starting- and endpoint on this $\phi=const$-line, as measure for the boost.
This distance is of course independent of the choice of the starting point on
the $\phi=const$-line, but it depends on the two $\phi$-values; especially,
these $\phi$-values can always be chosen such that the deviation is zero
(closed polygon). This interpretation of the boost-parameter may be
somewhat technical (we will find a much nicer one for e.g.\ {\bf G8,9} below),
but there is certainly no doubt that the parameter is geometrically
significant.

\medskip

\noindent{\bf Two nondegenerate horizons (e.g.\ G8,9, J1):}

\noindent%
Here $\smovs=\langle s_1,s_2 \,|\, {s_1}^2=1, (s_1{s_2}^{-1})^2=1 \rangle$,
where the second relation may be replaced by $s_1s_2s_1={s_2}^{-1}$.
With the help of these two relations any element can be expressed in
the form ${s_2}^n$ or ${s_2}^ns_1$, and the group
can thus (in coincidence with \re{smovsplit})) be written as a semi-direct
product, $\ZPT\ltimes{\Z}$,
where $\ZPT:=\{1,s_1\}$ and ${\Z}:=\{{s_2}^n,n\in{\Z}\}=\osmovs$.
However, not all those elements can be permitted: $s_1$ is a reflections at one
bifurcation point, $s_2s_1=\scoat$ is a reflection at the other bifurcation
point, and the general element ${s_2}^ns_1$ is
conjugate to one of them, thus also a bifurcation point reflection.
The only freely acting group elements are
thus the ${s_2}^n$ and they shift the manifold horizontally (in
Figs.~\ref{fig:fig24}, \ref{fig:fig25}) a number of copies side-wards. The factor space is then a
cylinder. Each cylinder carries again a further real boost-parameter, but
let us postpone this discussion and before admit flips
(i.e.\ nonorientable solutions).
The most general transformations involving flips (still omitting boosts) are
${s_2}^nf$ and ${s_2}^ns_1f$, but ${s_2}^nf$
and $s_1f$ are pure flips and cannot be used. A small calculation shows
further that the only remaining admissible groups are those generated by {\em
one\/} element ${s_2}^ns_1f$, $n\ne 0$. Since $s_1f$ is a flip at the middle
sector (a reflection at the horizontal axis in Fig.~\ref{fig:fig24}) and
${s_2}^n$ a shift along that axis, the whole transformation is a
glide-reflection and the factor space a M\"obius-strip of $n$ copies
circumference.

Now concerning the boost-parameter:
The complete cylinder-generator is ${s_2}^nb_\o$. So we must check
whether subgroups with different parameters $\o$ are conjugate.
Of course conjugation with $s_2$, $f$, and boosts does not change the
parameter. However, conjugation with $s_1$ or $s_1f$ changes $\o$ to its
negative, $(s_1f)({s_2}^n b_\o)(s_1f)^{-1}={s_2}^n b_{-\o}$.
Hence the factor spaces
corresponding to $\o$ and $-\o$ are isometric and one could consider to
restrict the boost-parameter to nonnegative values, $\o\in{{\R}_0}^+$.
However, the transformation $s_1f$ is a flip at the central sector, i.e., a
reflection at the longitudinal axis (horizontal in
Figs.~\ref{fig:fig24}, \ref{fig:fig25})
and thus inverts the time.
Hence, if the spacetime is supposed to have a time--orientation,
then the generators with boost-parameter $\o$ and $-\o$ are no
longer conjugate in the restricted symmetry group and the parameter has to
range over all of $\dR$.

In order to give this parameter a geometrical meaning we could of course
employ the `zigzagging' null-polygon again, but in this case there
is a much nicer description:
In Fig.~\ref{fig:fig25} part of the infinitely
extended solution {\bf G8,9} is drawn. A generating sector-move
shifting the solution two copies (in this example) to the right dictates
that the rightmost large patch (four sectors) is pasted
onto the corresponding left patch. The boost-parameter then describes that
before the pasting a boost has to be applied to the patch. As shown
in Fig.~\ref{fig:fig25} its effect is, e.g., that the thin timelike line crossing
the right patch vertically has to be glued to the curved line of the left
patch. The (shaded) region between these two lines is a possible fundamental
region for this factor space! Also, due to the boost the spacelike tangent
vector (arrow) to the dotted curve is tilted.

In section \ref{Secmaxext} it was shown that the bifurcation points
are conjugate points and the extremals running between them are those
of \re{13}) (in Fig.~\ref{fig:fig25}, they have been drawn as dotted
lines). They run through the bifurcation points into all directions
between the two null-directions.  A boost bends them side-wards,
altering the angle of their tangent.  One can now start from a
bifurcation point in a certain direction along a spacelike extremal.
This extremal will eventually return to the original point, but due to
a boost its tangent (cf arrow in Fig.~\ref{fig:fig25}) at the return
may be tilted (boosted) against that at the start.  This boost is of
course independent of the chosen extremal and is thus a true
`observable'; in particular, there is one solution without boost.
Thus,
the cylindrical solutions are
parameterized by a positive integer (number of patches) and a real constant
parameterizing the boost.
\footnote{If we had chosen $\a <0$ in (\ref{KV}), then the whole
Carter-Penrose
  diagram would have to be rotated by $90^\circ$.  The above extremals
  would then be timelike and the boost at the return could be
  interpreted nicely as acceleration during one journey around the
  cylinder.}
As discussed before, if there is no time--orientation, then the
boost-parameter has to be restricted to ${{\R}_0}^+$.
In the above interpretation of the boost-parameter this restriction
arises if one cannot distinguish
between boosts to the past or future. Of course, if a time--orientation is
given, then such a distinction is possible.
Note that in this case the
(time--)\- direction of the boost is independent of the sense
in which the extremal runs through the diagram.

It may seem that for the M\"obius-strip one would also have such a continuous
parameter due to different flips. However, this is {\em not\/} the case:
Any two flips are conjugate
(via a boost, $f'\equiv fb_\o=b_{-\o/2}fb_{\o/2}\sim f$), hence the
corresponding subgroups are conjugate and all
M\"obius-strips are equivalent.
(There is always one extremal \re{13}) which returns unboosted).
Also, in contrast to the former examples,
a boost cannot be defined consistently on the M\"obius-strip;
the boost transformation does not `factor through' the canonical projection
onto the factor space. This is also seen immediately from
Fig.~\ref{fig:fig24} ({\it b\/}), where in the M\"obius-case the sectors
occasionally have to be identified with their mirror images and thus the
arrows indicating the boost-direction do not match.
However, locally this Killing symmetry is still
present.
Hence, the M\"obius-strip solution is only parameterized by a
positive integer (number of copies).

\subsubsection{More than one generator \pl{Two}}

So far we have treated the cases where $\osmovs\cong{\Z}$ (one
generator) or trivial.
In those cases also all
subgroups have been one-generator groups $\Z$, and the possible
topologies have thus been restricted to cylinders and M\"obius strips
(remember that the subgroup $\cal H$ factored out equals the fundamental group
of the factor space, $\p_1({\cal M}/{\cal H})$).
This situation changes drastically when there is more than one generator,
as there are then subgroups of arbitrarily high rank (even infinite).

Ultimately we want to know the conjugacy classes of subgroups of \osmovs\
(in this section we restrict ourselves to space-- and time--orientable
solutions; then all those subgroups are properly
acting). Subgroups of free groups are again free, so in principle any
solution can be obtained by the choice of a free set of
generators. But this is only the easier part of the job:
\begin{itemize}
 \item Given a subgroup (say, in terms of generators) it may be hard to find a
  {\em free\/} set generating this group.
 \item Also the free generators are by no means unique (one can, e.g., replace
  $g_1$, $g_2$, \ldots by $g_1$, $g_2g_1$, \ldots). Only the number of free
  generators (the {\em rank\/} of the subgroup) is fixed. So there is the
  problem to decide whether two sets of generators describe the same group
  or not.
 \item We have to combine the subgroups into conjugacy classes.
\end{itemize}
Since the group is free, these three issues can be solved explicitly (at
least for finitely generated subgroups);
\footnote{Surprisingly, for nonfree groups this is in general impossible.
  For instance, there is no (general) way to tell whether two given words
  represent the same group-element (or conjugate elements); and it may also
  be undecidable whether two presentations describe isomorphic groups
  (word-, conjugation-, and isomorphism-problem for
  combinatorial groups, cf \cite{CombGr}).}
however, the algorithms are rather cumbersome and thus we will not extend on
this here (details in \cite{CombGr}).

\medskip

Due to the more complicated fundamental groups it is to be expected
that one gets interesting topologies.
As already mentioned in section \ref{Secintrorem}, all the solutions will be
noncompact (this is also clear, since there is no compact manifold without
boundary with a free fundamental group!).
Thus it would be nice to have a classification of noncompact
surfaces at hand. Unfortunately, however, there is no really satisfactory
classification which could be used here (cf \cite{Massey}).
Let us shortly point out the wealth of different possibilities:
A lot of noncompact surfaces can be obtained by cutting holes into compact
ones. Of course the number of holes may be infinite, even uncountable
(e.g.\ a Cantor set). A more involved example is that of surfaces of
countably infinite genus (number of handles). Finally, there need not even be
a countable basis of the topology (this does not happen here, though, since by
construction there are only countably many building blocks involved).
For an important subcase (finite index), however, the resulting topologies
are always of the simple form `surface of finite genus with finitely many
holes'.

Abstractly the index of a subgroup $\cal H$ is the number of cosets of
$\cal H$. But it has also a nice geometrical meaning:
Since \smovs\ acts freely and transitively on the sectors of the same type
(and thus on the building blocks), the index of $\cal H$ in \smovs\
counts the number of building blocks in the fundamental region. Actually,
since we started from \osmovs, it would be more convenient to use
the index of $\cal H$ in that group. If all horizons are of even degree,
then $\smovs=\osmovs$ and there is thus no difference.
On the other hand, if there are horizons of odd degree, then $\osmovs$ is a
subgroup of index 2 of $\smovs$ and consequently the number of building
blocks is {\em twice\/} the index of $\cal H$ in \osmovs. This is also obvious
geometrically, since the fundamental regions in this case are built from
patches consisting of two building blocks (e.g.\ those situated
around a saddle-point in the examples of  Figs.~\ref{fig:fig26}
({\it b\/}), \ref{fig:fig27}  below).
The index counts the number of these basic patches in the
fundamental region then. For finite index, furthermore, it is correlated
directly to the rank of $\cal H$ via formula \re{index}) (with
$\cal G$ replaced by \osmovs), thus
\begin{equation}
   \mbox{index}\, {\cal H}=
   \mbox{number of basic patches} = \frac{\mbox{rank}\,{\cal H}-1}{n-1}\,
  \pl{indexH}
\end{equation}
($n$ being the number of generators of \osmovs).

We will now provide the announced examples, starting with a
discussion of the respective combinatorial part of the
(orientation and time--orientation--preserving) symmetry group and
followed by a discussion of possible factor spaces. Figures
\ref{fig:fig26} and \ref{fig:fig27} contain basic patches as well
as the generators of \osmovs.  Although this group is the same
in all these cases (rank 2), its action for {\bf R5} is
different from the others, and correspondingly will be found to
give rise to different factor solutions.

\begin{figure}[htb]
\begin{center}
\leavevmode
\epsfig{file=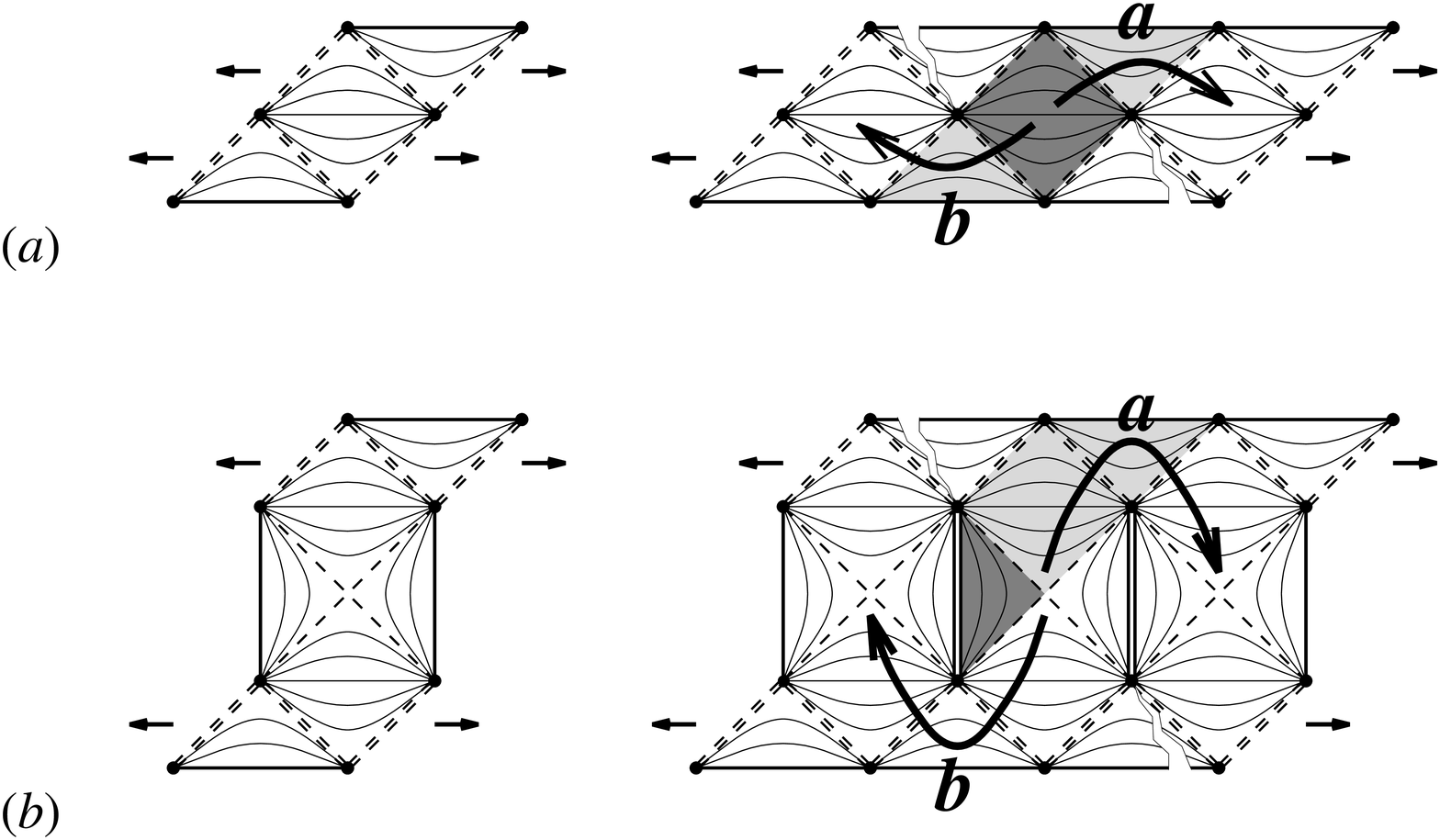,width=10cm}
\end{center}
\renewcommand{\baselinestretch}{.9}
\caption{
  \captionstyle \label{fig:fig26}
  Basic patches and generators of
  \osmovs\ for a fictitious example with two doubly degenerate
  horizons ({\it a\/}) and for {\bf R3} ({\it b\/}). Note that in
  the upper example the points in half height are at infinite
  distance, and that in the lower example the vertical singularities
  meet, yielding a slit (double lines). Going once around this
  point/slit leads into a different layer of the universal covering,
  as indicated by the jagged lines (multi-layered Carter-Penrose
  diagrams). Consequently $ab \neq 1$.}
\end{figure}

In the example Fig.~\ref{fig:fig26} ({\it a\/}) (two doubly degenerate horizons)
the group \smovs\ ($=\osmovs$) is free already, with the
two generators $s_1$ and $s_2$. Geometrically,
however, the moves with basis-sector 1 have a nicer representation.
Let $a:=\scoat \equiv s_2{s_1}^{-1}$ and $b:=\scoaz \equiv {s_1}^{-1}$.
Clearly $a$ is a move one block to the right {\em above\/} the singularity
and $b$ a move to the left {\em below\/} the singularity. Note that
since we are in the universal covering their composition, $ab$, is {\em
not\/} the identity but leads into another layer of the covering;
if an identification shall be enforced,
then the element $ab$ must occur in the subgroup factored out.

In Fig.~\ref{fig:fig26} ({\it b\/}) only the second horizon is degenerate
{\bf (G7,10, R3,4)}. Thus the basic patch consists of two building blocks.
Here \osmovs\ is a proper subgroup of \smovs\ and has the two free
generators $s_1s_2$ and $s_2s_1$ (since the
second sector is spatially homogeneous but the basis-sector stationary).
Again, $a:=s_2s_1$ is a move one patch to the right {\em above\/} the
singularity and $b:=s_1s_2$ a move to the left {\em below\/} the singularity.
Thus the action is similar to that in Fig.~\ref{fig:fig26} ({\it a\/}).
However, if (time--)\-orientation--preservation is not
required, then here one has the additional symmetries $f$ and $s_1$, i.e.\
reflection at the horizontal axis resp.\ at the saddle-point.

Our last example is {\bf R5, G11} (three nondegenerate horizons,
cf Figs.~\ref{fig:fig27}, \ref{fig:fig14a}). Also in
this case there are two free generators: $a:=s_2$, which is a move one
patch upwards, and $b:=s_3s_1$, a move one patch to the right. (The other
two potential generators can be expressed in terms of $a$ and $b$ by means
of the saddle-point relations: $s_1s_2s_1=a^{-1}$ and
$s_1s_3=a^{-1}b^{-1}a$). Again  going once around the
singularities leads into a new layer of the universal covering
($\LRA [a,b]:=aba^{-1}b^{-1}\ne 1$).
\vspace*{-0.5cm}
\begin{figure}[htb]
\begin{center}
\leavevmode
\epsfig{file=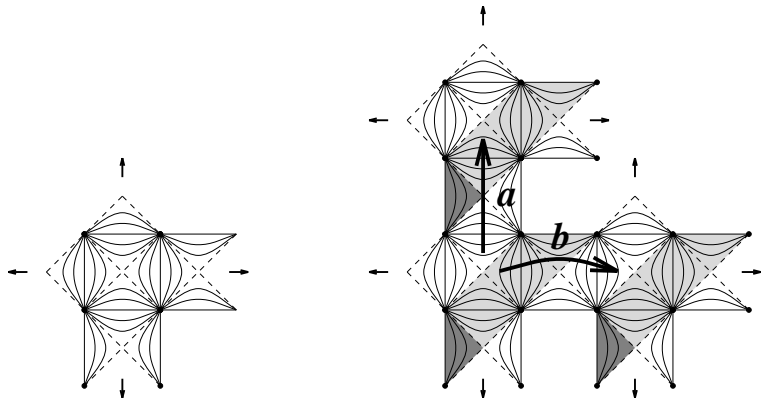,width=10cm}
\end{center}
\renewcommand{\baselinestretch}{.9}
\caption{
  \captionstyle \label{fig:fig27}
  Possible basic patch and generators of
  \osmovs\ for {\bf R5}. As is seen here, the basic patch need not
  consist of two {\em entire\/} blocks, but the involved sectors may be
  rearranged somewhat.}
\end{figure}

\begin{figure}[htb]
\begin{center}
\leavevmode
\epsfig{file=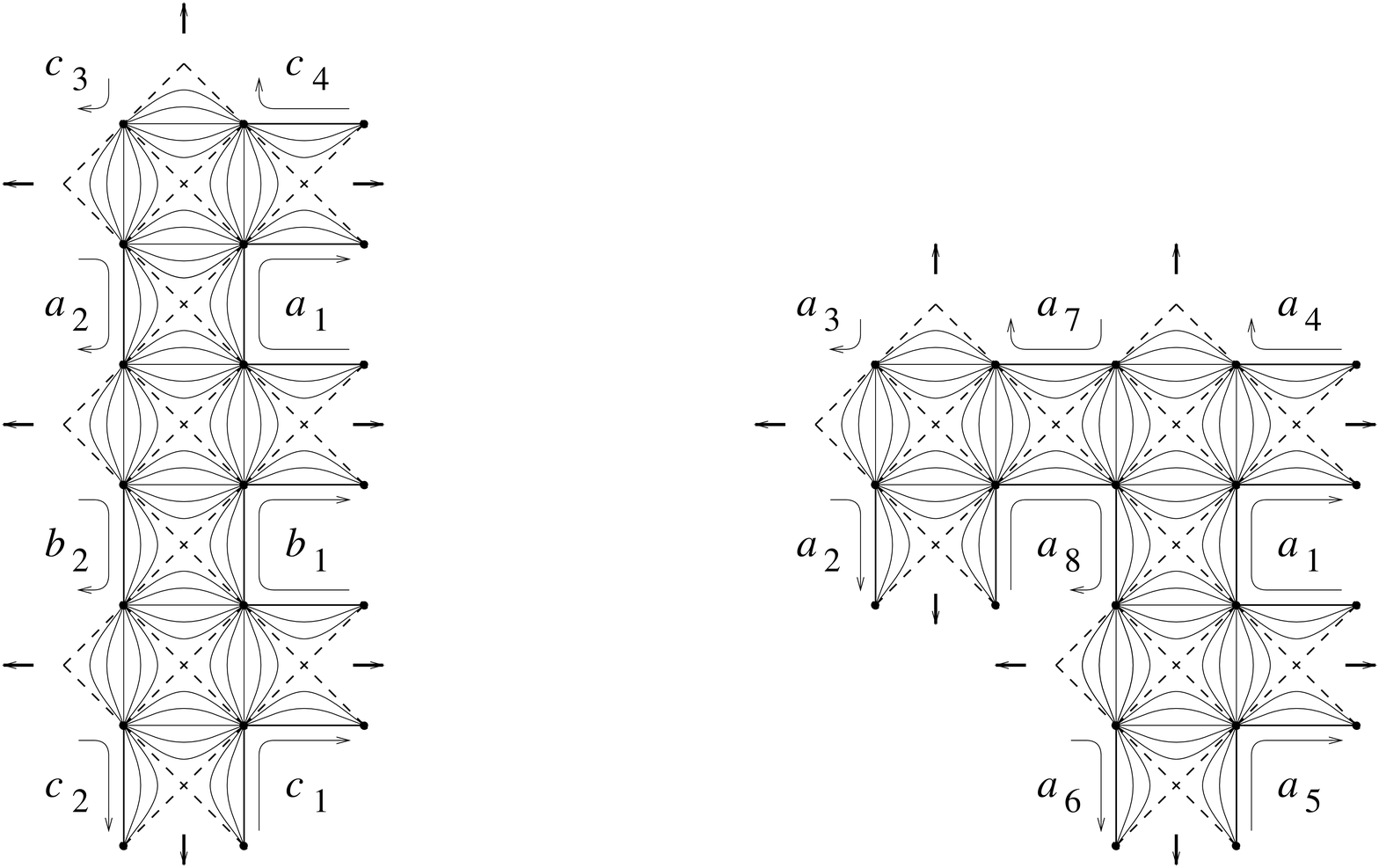,width=10cm}
\end{center}
\renewcommand{\baselinestretch}{.9}
\caption{
  \captionstyle \label{fig:fig28}
  Counting the boundary components
  (opposite faces have to be glued together). In both cases there are three
  basic patches (cf left part of Fig.~\ref{fig:fig27})
  and thus (use  \re{index}) and $\mbox{rank}\,\osmovs = 2$ !)
  four generators for $\cal H$, given in the text.
  However, due to the different number of boundary components (holes) the
  topologies differ, cf equation \re{genus}): In the
  left example there are three components ($a_{1,2}$, $b_{1,2}$,
  $c_{1\!-4}$), thus the topology
  is that of a torus with three holes. In the right example there is only
  one component ($a_{1\!-8}$); therefore this solution is a
  genus-2-surface with one hole. The resulting manifolds are shown in
  Fig.~\ref{fig:fig17}.}
\end{figure}

Let us now determine the topology for the solutions with finite index.
It is clear that then there are also only finitely many boundary segments.
[This is a slightly informal terminology, since these `boundary
segments' (singularities, null infinities, points at an infinite distance)
do {\em not\/} belong to the manifold. Still, this can be made precise and
such boundaries are called `ideal boundaries' or `ends'; we will
thus simply use the notion boundary.]
The generators of the subgroup $\cal H$ determine
how the faces of the fundamental region have to be glued and thus also how
the boundary segments are put together to form boundary components.
This is shown at two examples in Fig.~\ref{fig:fig28}. There opposite faces
should be glued together, which can be achieved by using the following
generators: $b$, $a^{-1}ba$, $a^{-2}ba^2$, $a^3$ for the left case,
and  $b$, $a^{-1}b^2a$, $a^{-1}bab^{-1}a$, $a^2$ in the right case, provided
one starts from the lowest basic patch. When starting from another
patch, the subgroups and their generators will be conjugates of the above
ones, but clearly this does not change the factor solution.

Now, topologically to each boundary component (which is clearly an $S^1$)
a disk can be glued. This yields a compact orientable surface,
which is completely determined by its genus. The original manifold is then
simply this surface with as many holes as disks had been inserted (each
boundary component represents a hole).
The genus in turn depends on
the rank of the fundamental group $\p_1({\cal M}/{\cal H})\cong{\cal H}$
and on the number of holes:
\begin{equation}
   \mbox{rank}\, {\cal H}=
   \mbox{rank}\,\,\p_1({\cal M}/{\cal H}) = 2\,\, \mbox{genus} +
                                             \mbox{(number of holes)}-1\,.
  \pl{rankH}
\end{equation}
Expressing the rank by means of \re{indexH}), this yields
\begin{equation}
 \mbox{genus}\,=\,\frac{(\mbox{number of basic patches})\cdot(n-1)-
                     (\mbox{number of holes})}2+1 \,,
  \pl{genus}
\end{equation}
where $n$ is the number of generators of \osmovs.

\begin{figure}[htb]
\begin{center}
\leavevmode
\epsfig{file=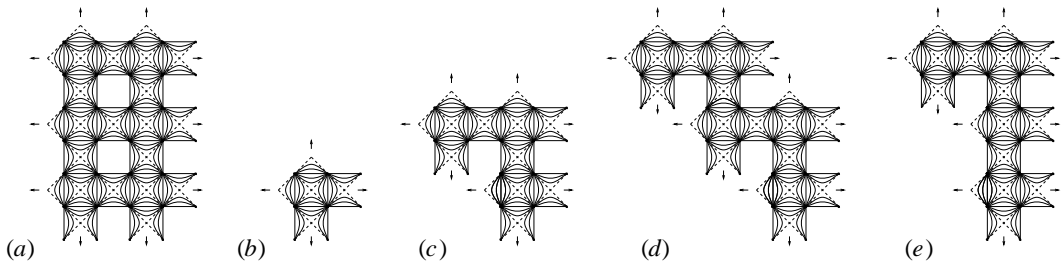,width=14cm}
\end{center}
\renewcommand{\baselinestretch}{.9}
\caption{
  \captionstyle \label{fig:fig29}
  Fundamental regions of factor spaces for
  {\bf R5}. Opposite faces have to be glued together.
  ({\it a\/}) torus with six holes,
  ({\it b\/}) torus with one hole,
  ({\it c\/}) genus-2-surface with hole,
  ({\it d\/}) genus-3-surface with hole,
  ({\it e\/}) genus-2-surface with two holes.}
\end{figure}

The general procedure to determine the topologies of the factor spaces can
thus be summarized as follows:
Draw a fundamental region for your chosen subgroup and
determine which faces have to be glued together. Then count the connected
boundary components (= number of holes) and calculate the genus from
\re{genus}). We illustrate this procedure at two examples in Fig.~\ref{fig:fig28}.
Some further examples for {\bf R5} are given in Fig.~\ref{fig:fig29}.
Actually, they show that surfaces of any genus ($\ge1$) and with any number
($\ge1$) of holes can be
obtained: continuing the series Fig.~\ref{fig:fig29} ({\it b\/}),
({\it c\/}), ({\it d\/}) one can increase the number of handles arbitrarily,
while attaching single patches from `below' like in ({\it e\/})
allows one to add arbitrarily many holes.

For the cases of Fig.~\ref{fig:fig26} (e.g.\ {\bf R3}) the same analysis can
be applied. For instance, it is obvious that cylinders with an arbitrary
number of holes ($\ge 1$) can be obtained. Also surfaces of higher genus are
possible; however, now the number of holes is always $\ge3$ (past and future
singularity, and at least one hole in `middle height'). [Note that this is
no contradiction to the cylinder-with-hole case, since a cylinder with one
hole is a sphere (genus-0-surface) with three holes].
In contrast to the {\bf R5}-examples these factor solutions
do not have closed timelike curves.


%

\medskip

We turn to the cases with infinite index and thus infinite
fundamental regions. All subgroups of infinite rank belong to this
category, but also many subgroups of finite rank (see below). One topological
reason for an infinite rank of the subgroup (and thus also of the fundamental
group $\p_1$) is the occurrence of infinitely many holes. For instance, it was
pointed out that in the solution {\bf R3} (Fig.~\ref{fig:fig26}) the move $ab$ is
not the identity but leads into a new layer of the universal covering. One
can of course enforce the identification of overlapping layers by imposing
the relation $ab\stackrel!=1$ and its consequences. This is tantamount to
factoring out the group generated by $ab$ and all its conjugates
(the elements $a^kaba^{-k}$, $k\in{\Z}$,  form already a free set of
generators). The result is a ribbon with infinitely many holes (slits).
Clearly the parameter space of such solutions is
infinite dimensional now (cf also remarks at the end of this section).
[If  in addition one imposed the relation
$a^n\stackrel!=1$, then the previously infinite set of
generators would boil down to $n+1$ generators ($a^n$, and
$a^kaba^{-k}$ for $0\le k<n$), and the resulting factor space
(finite index again) would be a cylinder with $n$ holes.]

Likewise, in the example {\bf R5} (Fig.~\ref{fig:fig27}) the identification of
overlapping layers is obtained by factoring out the infinitely generated
commutator subgroup (generated freely e.g.\ by $a^mb^nab^{-n}a^{-m-1}$,
$(m,n)\ne(0,0)$; cf \cite{Massey}); the factor space is a planar,
double-periodic `carpet' then.
Adding, furthermore, the generator $a^n$ (or $b^n$) yields a cylinder with
infinitely many holes (e.g.\ Fig.~\ref{fig:fig29} ({\it a\/}) extended infinitely
in vertical direction); and adding both $a^n$ and $b^k$ yields a torus with
$nk$ holes, which is again of finite index.

Another possible reason for an infinite rank is an infinite genus (number of
handles); such a solution is obtained for instance by  continuing
the series Fig.~\ref{fig:fig29} ({\it b\/}), ({\it c\/}), ({\it d\/})
infinitely. Of course, both cases can occur
simultaneously (infinite number of holes {\em and\/} infinite genus).

\begin{figure}[htb]
\begin{center}
\leavevmode
\epsfig{file=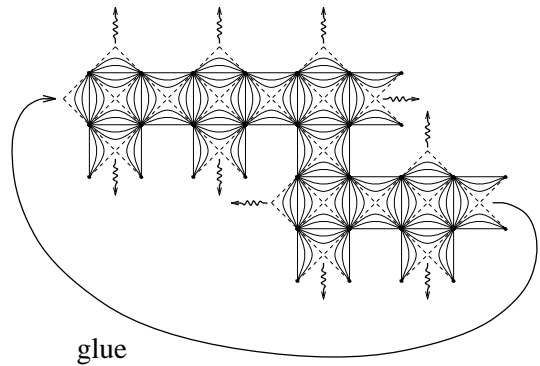,width=7cm}
\end{center}
\renewcommand{\baselinestretch}{.9}
\caption{
  \captionstyle \label{fig:fig30}
  Frazzled cylinder from {\bf R5}. Only the
  utmost left and right faces have to be glued together, such as to make a
  closed `ribbon' of five patches (a possible generator for this gluing is
  e.g.\ $b^2a^{-1}b^2$). At all other faces (indicated
  $\rightsquigarrow$) the solution has to be extended without further
  identifications, similarly to the universal covering.}
\end{figure}

Let us now discuss the groups of finite rank and infinite index.
Already the universal covering itself, being  topologically an open
disk (or $\dR^2$),
\footnote{The reader who has difficulties to imagine that such an
  infinitely branching patch is really homeomorphic to a disk may
  recall the famous Riemann mapping theorem, which states that
  any simply connected (proper) open subset of $\dR^2$, however
  fractal its frontier might be, is not only homeomorphic but even
  biholomorphically equivalent to the open unit disk
  (e.g.\ \cite{Burckel}). [Clearly, the universal covering is a priori
  not a subset of the plane (due to the overlapping layers), but
  by a simple homeomorphism it may be brought into this form.]}
provides such an example
(with the trivial subgroup factored out).
But also one-generator subgroups can by \re{index}) never be of finite
index (if $\mbox{rank}\osmovs\ge2$).
These subgroups yield proper (yet slightly pathological) cylinders without
holes: Let ${\cal H}_g=\langle g \rangle \equiv\{g^n,n\in{\Z}\}$ for arbitrary
nontrivial $g \in \osmovs$.
To speak in pictures: The generator $g$ of this infinitely cyclic
subgroup
\footnote{or rather the corresponding cyclically reduced element
  (remember that conjugate groups yield isomorphic factor spaces).}
defines a path in the universal covering. Now the end-sectors of the path
(i.e.\ of the corresponding ribbon) are identified and at all other junctions
the solution is extended infinitely without further identifications (cf
Fig.~\ref{fig:fig30}). Thus a topological cylinder, although with a
terribly frazzled boundary, is obtained (as before it is possible to
smooth out the boundary by a homeomorphism). [The cylinders obtained in
this way may have kinks of the lightcone or not. For instance, in the example
Fig.~\ref{fig:fig30} the lightcone tilts by (about) $90^\circ$ and then tilts
{\em back\/} again. So, this is not a kink in the usual sense of the word;
still, there is not one purely
spacelike or purely timelike loop on such a cylinder.]

And even for higher ranks of $\cal H$ there are solutions of infinite
index.  The topologies obtained in this way are again of the simple
form compact surface with hole(s). There is, however, a much greater
flexibility in the rank (which is no longer restricted by formula
\re{indexH})) as well as in the number of boundary-components
(remember that in the examples Fig.~\ref{fig:fig26} all solutions of
finite index had at least three boundary-components). Indeed, one can
obtain any genus and any nonzero number of holes in this way, as
sketched in the following paragraph. Equation \re{rankH}) is still
valid, however, $\mbox{rank}\, {\cal H}$ can no longer be expressed by
\re{indexH}) but has to be determined directly from the gluings.

As already mentioned we may abuse the classification of compact two-manifold
with boundary for our purpose; one has just to replace the true boundaries
by `ideal boundaries' which do not belong to the manifold.
Note that the periphery of the basic patch consists of a couple of faces
which have to be glued together pairwise, separated by (`ideal') boundary
components. [It is topologically immaterial whether these boundary components
are pointlike or extended singularities, since they do {\em not\/} belong to
the manifold; each boundary point can be stretched to an extended segment by a
homeomorphism and vice-versa.]
The faces come in pairs and the number of pairs equals the rank of the group
\osmovs. Thus in the present case there are at least two such pairs, which is
sufficient to produce fundamental regions with arbitrarily many faces.
Furthermore, by virtue of the infinitely branching extensions one can get
rid of redundant faces: just extend the solution infinitely at this face
so as to obtain a new (`frazzled') boundary segment which connects the two
adjacent ones, yielding one larger boundary segment. Thus it is possible to
produce polygons, the faces of which are to be glued in an arbitrary order.
According to \cite{Massey} this already suffices to produce all topologies
announced above.

\medskip

Finally, we have to discuss the boost-parameters.
Since each free generator of $\cal H$ carries a boost-parameter, their total
number equals the rank $r$ of this subgroup.
Also an interpretation can be given in analogy to the cases dealt with
before (zigzagging null-polygon and/or boosted saddle-point extremals).
However, not all such choices of
an $r$-tuple of real numbers are inequivalent.
We show this at the example of the torus with three holes (Fig.~\ref{fig:fig28},
left part): There are four generators and thus also four
boost-parameters, three of them describing the freedom in the
horizontal gluing ($b$, $a^{-1}ba$, and $a^{-2}ba^2$) and one for
the vertical gluing ($a^3$); let us denote them by $(\o_1,\o_2,\o_3;\o_4)$.
Now, since conjugate subgroups lead to equivalent factor spaces, we can
e.g.\ conjugate all generators with $a$. This leads to new generators,
but since $a$ lies in the normalizer
\footnote{The {\em normalizer\/} of a subgroup $\cal H$ contains all
  elements $g$ for which $g^{-1}{\cal H}g = {\cal H}$.}
$\cal NH$ of $\cal H$ in \osmovs\
they still span the same (projected) subgroup  $\cal H$. Thus it is possible
to express the old generators in terms of the new ones.
However, during this procedure the boost-parameters change: For instance,
the (full)
\footnote{Here we denoted the boost 
 by $\o_i$ instead of $b_{\o_i}$ in order to avoid confusion with the
 sector-move $b$.}
first generator $\o_1b$ is mapped to $a^{-1}\o_1ba=\o_1(a^{-1}ba)$, i.e.\
the parameter $\o_1$ is shifted from the first to the second
generator. Altogether, the three `horizontal' boost-parameters $\o_{1\!-3}$
are cyclically permuted and thus we get an equivalence relation among the
4-tuples, $(\o_1,\o_2,\o_3;\o_4) \sim (\o_3,\o_1,\o_2;\o_4)$.
This is of course also geometrically evident: While the timelike loop
corresponding to $\o_4$ is uniquely characterized, the three spacelike loops
corresponding to $\o_{1\!-3}$ are indistinguishable (there is no `first
one').

In general, we have a (not necessarily effective) action of
the group ${\cal NH}/{\cal H}$ on the space of boost-parameters
${\R}^r$.  Here $\cal NH$ is the normalizer
of $\cal H$ in \osmovs\ (or, if no \mbox{(time--)}orientation
is present, also in
\smovs\ or $\Zflip\ltimes\smovs$, respectively).  The true
parameter space is the factor space under this action,
${\R}^{r}\big/({\cal NH}/{\cal H})$. Locally, it is still $r$-dimensional;
however, since the action may have fixed points (e.g. in the above example
the whole plane $(\o,\o,\o;\o_4)$), it is an {\em orbifold\/} only.

\subsubsection{Remarks on the Constant Curvature Case \pl{Const}}

So far we have only dealt with those solutions where the metric (or
$\phi$--preservation) restricted us to only one Killing field.
For reasons of completeness one should
treat also the (anti-)\-deSitter solutions \re{deSitter}) of the
general model, corresponding to the critical values $\phi={\phi}_{crit}$ and
$X^a=0$, which have constant curvature (and zero torsion).
Constant curvature manifolds have already occurred as
solutions of the Jackiw-Teitelboim (JT) model \cite{JTmodel}, equation
(\re{JT}),  where, however, the symmetry group was still restricted
to only one Killing field since
$\phi$ had to be preserved. Here, on the other hand, these fields are constant
all over the spacetime
manifold, and thus the solutions have a much higher symmetry.

Already the flat case offers numerous possibilities: The symmetry group
(1+1 dimensional Poincar\'e group) is generated by translations, boosts, and
if space-- and/or time--orientation need not be preserved, also by spatial
and/or time inversion.
As before pure reflexions would yield a boundary line (the reflexion axis)
or a conical singularity (at the reflexion centre) and boosts a
Taub--NUT space. The only fixed-point-free transformations are thus
translations and glide-reflexions
\footnote{A {\em glide-reflexion\/} is a translation followed by a reflexion
  at the (nonnull) translation axis. Note that this axis must not be null,
  if the (orthogonal) reflexion shall be well-defined.}
We have thus the following generators and corresponding factor spaces:
\begin{itemize}
\item One translation: Cylinders; parameters = length squared of the generating
translation = circumference (squared) of the resulting cylinder (especially,
there is only {\em one\/} cylinder with null-circumference).
\item One glide-reflexion:  M\"obius-strips, again parameterized by their
circumference.
\item Two translations: Torus, labelled by three parameters: The lengths
(squared) of
the two generators $\vec{a}$ and $\vec{b}$ and their inner product.
Globally, however, this is an overparameterization: Replacing, for instance,
the translation vector $\vec{b}$ by $\vec{b}+n\vec{a}$ changes one length and
the inner product, but still yields the same torus (just the
original longitude is now twisted $n$ times around the torus).
\item One translation and one glide-reflexion: Klein bottle.  Here
  only two parameters survive (the inner product of the two generators
  can be conjugated away always). [This is somewhat similar to the
  situation cylinders versus M\"obius strips in the case of {\bf G8,9} in
  section \ref{Factor}, where also a potential continuous parameter
  did not occur due to the nonorientability.]
\end{itemize}
Since two glide-reflexions combine to a point-reflexion or boost,
and three translations generically yield a nondiscrete orbit, this
exhausts all cases.

\begin{figure}[htb]
\begin{center}
  \leavevmode
  \epsfig{file=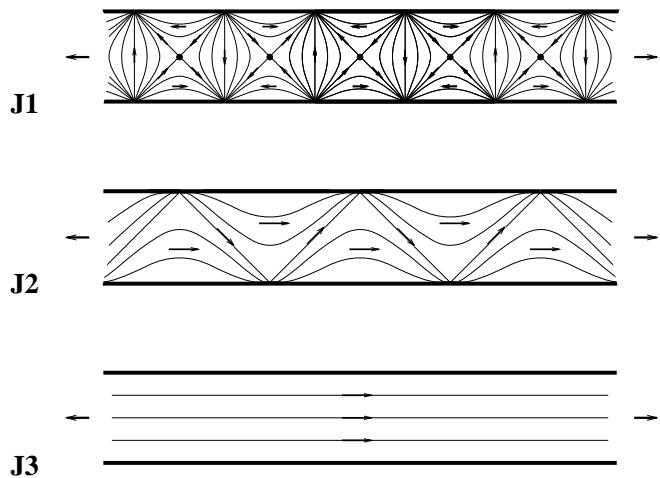,width=9cm}
\end{center}
\renewcommand{\baselinestretch}{.9}
\caption{
  \captionstyle \label{fig:fig31}
  Killing fields for the JT--model
  and deSitter.}
\end{figure}

Now concerning the `proper' ($R\ne0$) (anti-)\-deSitter
solutions resp.\ their universal coverings:
Here the situation is slightly more involved.
Of course all factor solutions of the JT--model (\JTref)
are also available for the (anti--)\-deSitter case, as both have
constant curvature and zero torsion.
For instance, the Killing fields corresponding to the solutions {\bf J1,2}
give rise to cylinders labelled by block number and a boost-parameter,
while {\bf J3} yields cylinders labelled by their (real number)
circumference. Now, however, there are three independent Killing fields
(e.g.\ {\bf J1,3} and their Lie-bracket)
\footnote{Note that this is the {\em only\/} case with more than one (local)
  Killing field: Whenever curvature is not constant, the Killing trajectories
  are restricted to the lines of constant curvature, which leaves at most one
  independent field. There is thus no 2D-metric with only two local
  Killing fields.}
and thus one would expect further
factor solutions. As an example, one has now in addition to cylinders also
M\"obius-strips of {\em arbitrary\/} circumference and not only of an
integer number of blocks as in the case of {\bf J1 (G8,9)}.
Unfortunately, the full isometry group of the space is $\widetilde O(2,1)$,
whose connected component equals $\widetilde{SL}(2,\R)$, and this group is
famous for having no faithful matrix representation and is thus
rather difficult to handle. A partial classification has been
accomplished by Wolf \cite{Wolf}, who deals with homogeneous spaces
\footnote{A space is called {\em homogeneous\/}, if its group of
  isometries
  acts transitively on it (the space then looks `the same' from every point).}
only and obtains a discrete series of cylinders and M\"obius-strips for them.
There are strong hints that even in the general case these are the
only possible topologies
\footnote{It is clear that no compact topologies can occur: According to
  \cite{Thm} compact Lorentz-manifolds should have Euler characteristic zero
  (i.e., torus or Klein bottle), but by the Gauss-Bonnet theorem this is
  impossible for nonvanishing constant curvature.}
(for instance, it would suffice to show that all
properly acting subgroups of $\widetilde{SL}(2,\R)$ are isomorphic to $\Z$.)
However, a proof requires a different approach.

\subsection{Remarks on kinks \pl{Kinks}}

The solutions obtained so far have all been geodesically complete, or, if
not, the curvature or some physical field blew up at the boundary, rendering
a further extension impossible. However, these global spacetimes are not all
inextendible ones: it is, for instance, possible that the extremals are all
incomplete, the fields and the curvature scalar remain all finite, yet when
attempting to extend the solution one runs into problems, because the
extension would no longer be smooth or Hausdorff or the like
(cf e.g.\ the Taub--NUT cylinders of section \ref{Factor}).
The purpose of this section shall be to give some further
examples, to outline some general features of such solutions, and
to discuss to which extent a classification is possible.

A familiar example for the above scenario is the metric \cite{Dunn}
\begin{equation}
  g=e^{-2t}\left(-\cos2x\,dt^2-2\sin2x\,dt\,dx+\cos2x\,dx^2\right)\,,
  \pl{flatkink}
\end{equation}
where the coordinate $x$ is supposed to be periodically wrapped up,
$x \sim x + n\pi$. These are $n$-kink solutions, which means that,
loosely speaking, the lightcone tilts upside-down $n$ times
when going along a noncontractable nonself\/intersecting loop on the
cylindrical spacetime. For $n=2$, however, \re{flatkink}) is nothing but
flat Minkowski space, the origin being removed, as is easily seen by
introducing polar coordinates
\begin{equation}
  \mytilde x=e^{-t}\cos x\,,\quad \mytilde t=e^{-t}\sin x
  \pl{polar}
\end{equation}
into the metric
\begin{equation}
  g=d\mytilde t^2-d\mytilde x^2 \,,
\end{equation}
a fact that seems to have been missed in most of the literature.
Consequently, the metric is incomplete at the origin; it has a hole which
can easily be filled by inserting a point, leaving ordinary Minkowski space
without any kink.
For $n \ne 2$, on the other hand, (which are covering solutions of the above
punctured Minkowski plane, perhaps factored by a point-reflection)
this insertion can no longer be done, because it would yield a `branching
point' (conical singularity) at which the extension could not be
smooth. Thus these manifolds are inextendible but nevertheless incomplete
and certainly the curvature does not diverge anywhere ($R\equiv0$).

Such a construction is of course possible for {\em any\/} spacetime,
leading to inextendible $n\ne2$-kinks.
However, if there is a Killing symmetry present, then even in the
2-kink situation inextendible metrics can be obtained.
To see this let us try to adapt the factorization approach of
the previous section to these kink-solutions. First of all, since a point
has been removed the manifold is
no longer simply connected, so one must pass to its universal covering,
which now winds around the removed point infinitely often in new layers
(cf {\bf G3} versus {\bf G4} in Fig.~\ref{fig:fig22}).
All above kink-solutions can then be obtained by factoring out a `rotation'
of a multiple of $2\p$ (or $\p$, if there is a point-reflection symmetry)
around the hole.
But according to the previous sections there should also
occur a kind of boost-parameter. Is it meaningful in this context?

\begin{figure}[htb]
\begin{center}
\leavevmode
\epsfig{file=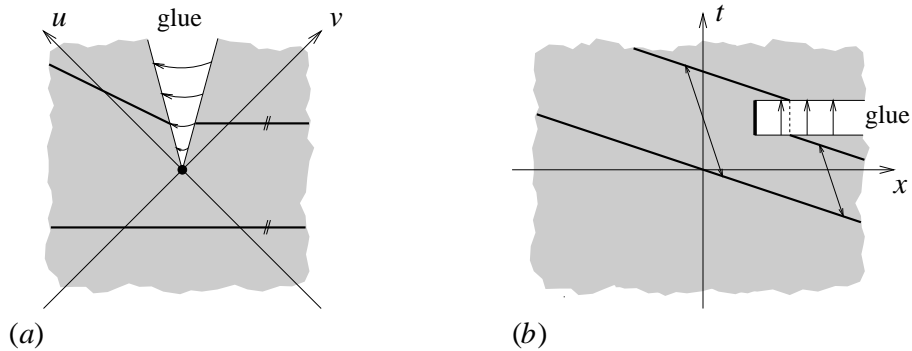,width=12cm}
\end{center}
\renewcommand{\baselinestretch}{.9}
\caption{
  \captionstyle \label{fig:fig32}
  ({\it a\/}) Minkowski kink with nontrivial
  holonomy. This space can be obtained by removing a wedge from flat Minkowski
  space and gluing together the corresponding boundary lines by a boost.
  Due to this construction two extremals which are parallel on one side of the
  origin are mutually boosted on the other side (cf bold lines).
  Thus the holonomy is nontrivial (surrounding the origin yields a boosted
  frame), and at the origin there would occur a conical singularity.
  ({\it b\/}) Another inextendible Minkowski kink; it has trivial holonomy but
  the distance of parallels passing the hole changes.}
\end{figure}

The answer to this question is yes, and this is perhaps best seen at the
flat 2-kink example, the Killing field chosen to describe boosts around the
(removed) origin. As long as the origin was supposed to belong to the manifold,
smoothness singled out one specific boost value for the gluing of the
overlapping sectors, leading to Minkowski space; otherwise there would have
occurred a conical singularity at the origin. However, if this point is
removed, then there is no longer any restriction on the boost-parameter. Its
geometrical meaning is that after surrounding the origin a boost has to be
applied before gluing or, in terms of fundamental regions, that a wedge has
to be removed from the original (punctured) Minkowski space and the
resulting edges are glued by the boost (also the tangents must be mapped
with the tangential map of this boost, cf Fig.~\ref{fig:fig32} {\it a\/})).
Of course, it is also possible to insert a wedge, but this is
equivalent to removing a wedge from an adjacent (stationary) sector.

Clearly such a space is everywhere flat (except at the
origin, which is considered not to belong to the manifold) but has
nontrivial holonomy.
For instance, two timelike extremals which are parallel `before' passing
the origin at different sides will be mutually boosted afterwards
(bold lines in Fig.~\ref{fig:fig32} ({\it a\/})).

In the above example we chose as Killing symmetry the boosts centered at
the origin. However, Minkowski space also exhibits translation
symmetries. An analogous construction can be applied also in this
case with the following geometrical interpretation:
cut out a whole slit (in direction of the chosen translation), remove the
strip on one side of the slit, and glue together the resulting faces
(cf Fig.~\ref{fig:fig32} ({\it b\/})). This manifold has now trivial holonomy,
but the metric distance of two generic parallels passing the hole
changes. Thus the manifold is so badly distorted that it cannot be completed
to ordinary Minkowski space, either.
In contrast to the former case this space can still be smoothly extended
further: one can simply continue beyond the remaining left edge of the slit
(bold line)
into an overlapping layer whose upper and lower faces have to be glued together
(since the endpoints of the slit have to be identified).
This yields a maximally extended cylinder, where the (identified) endpoints of
the slit constitute a conical singularity and should be removed.

\begin{figure}[htb]
\begin{center}
\leavevmode
\epsfig{file=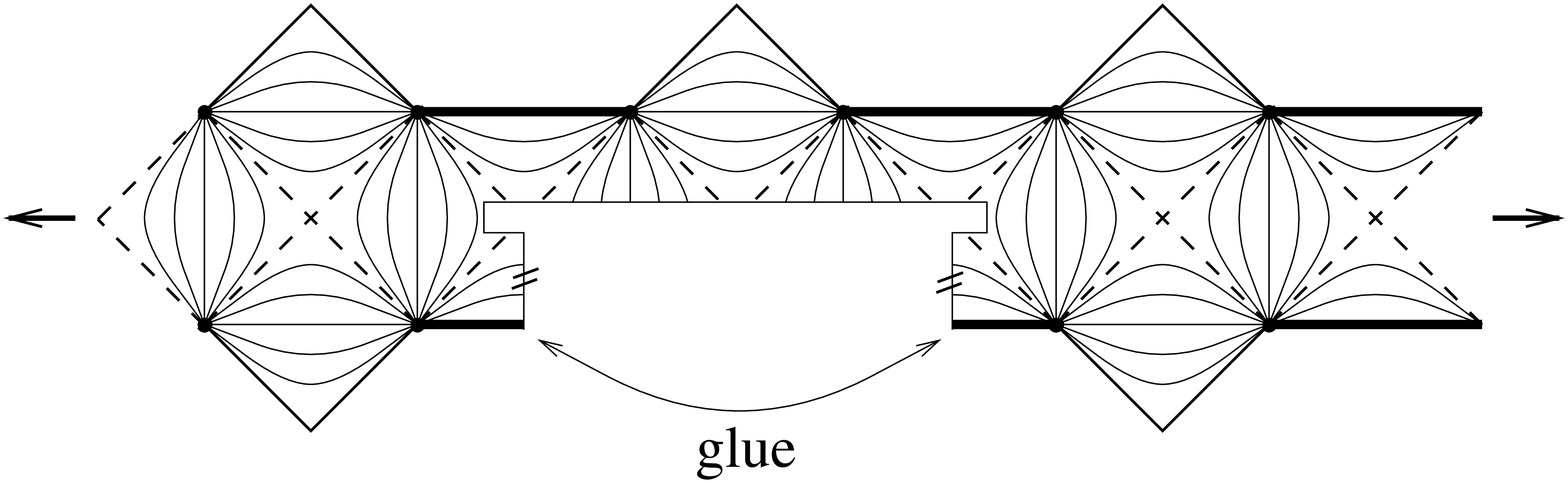,width=10cm}
\end{center}
\renewcommand{\baselinestretch}{.9}
\caption{
  \captionstyle \label{fig:fig33}
  Yet another kink for {\bf G8,9}.}
\end{figure}

It is relatively straightforward to write down the metric for the
above examples in a circular region around the hole (but not too close to the
hole), using smooth but nonanalytic functions. 
Analytic charts are more difficult to obtain, but at least for the
case Fig.~\ref{fig:fig32} ({\it b\/}) also this is possible
(cf \cite{KloeschStroblkinks}). This last example, furthermore, can easily be
generalized to an arbitrary metric with
Killing symmetry: one has just to introduce Eddington-Finkelstein
coordinates \re{011h}) in one patch. Then the analogous construction
with $(x,t)\rightarrow(x^0,x^1)$ yields a one-parameter family of
inextendible 2-kink solutions (resp.\ $2n$-kink). Explicit charts can be
found in \cite{KloeschStroblkinks}.
There is (sometimes) even the possibility to introduce discrete
parameters like a `block-number':
one could, e.g., take {\bf G9} and make a long horizontal slit
over a number of blocks, then remove a few blocks on the one side
and glue together again (cf Fig.~\ref{fig:fig33}).

\medskip

It is somewhat problematic to give a complete classification of the kink
solutions found above. Of course they could be described as factor spaces
of {\em limited\/} coverings of the original universal covering solution.
First, however, this would rather be a mere enumeration of the possible cases
than a classification. Second, we do not want to distinguish two solutions,
one of which is just an extension of the other. Thus we should only consider
maximally extended limited coverings. In the above examples they all
had only conical singularities, but it is not evident that this
should be the most general scenario.
Disregarding this question of the extension, the kinks are of
course characterized by their kink-number, a (real) boost-parameter, and
perhaps further discrete parameters (block-number or the like).

Finally we want to mention that such surgery is not restricted to cylindrical
solutions (i.e.\ one hole only), but, within any of the global solution
obtained in the previous sections, one can cut any number of holes, each
giving rise to one boost-, one kink-, and perhaps some further discrete
parameter. And as is well-known from complex analysis (Riemann surfaces),
one can even obtain surfaces of higher genus (e.g.\ genus 1 with four
branching points, etc.) in that way.

For further details on kinks we refer to
\cite{KloeschStroblkinks}. This completes also our discussion of
possible global solutions of the gravity models under discussion.



\chapter{Towards quantum gravity \pl{Sectowards}}
The main part of this chapter is devoted to one particular
quantization scheme applied to 2d gravity--YM theories, namely the
so--called Dirac quantization \cite{Dir}. The quantum treatment of
theories with additional matter fields (and thus with propagating
modes) is mentioned briefly only. In fact, even the Dirac quantization
of the topological theories is not fully satisfactory yet, as, in the
general case, the construction of an inner product between `physical
quantum states' is not available as of present day. We emphasize this
because the purpose of lower dimensional models is to help us develop
further current schemes of approaching the quantization of gravity
theories.

Still, we will be able to find and classify all the physical states.
The resulting space of states is found to crucially depend on the
details of the potentials chosen in the original Lagrangian (cf
equations \re{gdil}) and \re{grav})). Moreover, the classification of
states will be seen to nicely fit to the classification of classical
solutions for topology $\S \times \dR$ as determined in the previous
chapters.

Thus, there {\em are\/} some nontrivial results to be presented.
However, even in the merely `topological' case of pure 2d gravity-YM
theories, there are still open problems on the technical as well as on
the conceptual level, which we do not attempt to hide.

\section{Hamiltonian formulation \pl{Secham}}
According to chapter \ref{SecPSM}, the large variety of actions of
gravity--Yang--Mills models can all be cast into the framework of Poisson
$\sigma$-models (PSMs). In this section we will first provide the
Hamiltonian structure of a general PSM (the increased generality poses
no complication, rather notational simplicity is gained). Thereafter
several issues of interest for quantization are addressed,
like the construction of a reduced phase space (RPS).\footnote{As usual, the
  RPS is defined as the quotient of the restriction of the original
  phase space to the surface of constraints by the symmetries
  generated by the first class constraints.}  In the simplest cases,
which includes, however, spherically symmetric vacuum Einstein gravity
(SS) and string inspired dilaton gravity (SIDG) (cf sections 
\ref{Secspher} and \ref{Secstring}), the RPS is just an
$\dR^2$ and quantization is trivial. It is straightforward to
extend the Hamiltonian formalism to also include, e.g., massless
scalar fields. For this, however, we refer to the literature (cf.,
e.g., \cite{JackiwCGHS,KucharCGHS}).

In general, a Hamiltonian formulation needs the restriction to
spacetimes $M$ which, topologically, are of the form $M \sim \S
\times \dR$. For the case of a two--dimensional spacetime, $\S$ can
thus either be a circle or a line, only.

The former case is simpler and more transparent in two respects:
First, as $\S = S^1$ has no boundary, there is no room for boundary
contributions to the Hamiltonian. The Hamiltonian framework can be
made well--defined without bothering about (nontrivial) boundary
conditions and contributions, which, in contrast, is necessary in the
open case $\S \sim \dR$ (where, again, one can distinguish between
`bounded' regions and infinitely extended ones).

Second, in the previous chapter, the moduli space, i.e.\ the space of
 classical solutions modulo gauge symmetries, was found to be
of even dimension for cylindrical spacetime topologies, and of odd
dimension for planar topologies.  (E.g., in the case of pure SS this
moduli space is one-dimensional, being parameterized by the
Schwarzschild mass.) In the case of $\S \sim S^1$, the RPS and the
moduli space will be found to coincide, while for $\S \sim \dR$ they
are different.  (To be precise, there are also two possible
definitions of the moduli space on the Lagrangian level, depending on
which precise notion of symmetries is used. Only for one of these
choices, agreement with the RPS is obtained. We will come back to this
below.) 

We thus first start with the case of spacetimes of cylindrical
topology and only then discuss the modifications necessary for
the treatment of planar spacetimes.

\subsection{For a cylindrical spacetime \pl{Sechamcyl}}
The action of a PSM is already in first order form. It
is therefore possible to compare it with a Hamiltonian action right
away and read of Hamiltonians and constraints directly. For 
convenience of the reader we rewrite the action \re{PSM}) or
\re{actioncoor}) here in component form, where, for the
spacetime coordinates $(x^0,x^1)$, the more suggestive $(t,x)$ is used
and, for notational simplicity, $A_{ri} : = A_i$, $A_{ti} := \L_i$,
$\6_0 X^i \equiv \6_t X^i := \dot X^i$, and $\6_1 X^i \equiv \6_x X^i
:= \6 X^i$. (So within this section, $A_i$ does no more
  denote a one-form, but a function, or, better, $A_i \, dx$ a
  one--form on $\S$,  and, similarly, $x$ does no more
  denote a point on the two-manifold $M$, but just on $\S$. We hope,
  this does not lead to any confusion.)  Then \re{PSM}) is
rewritten identically as \be L^{PSM} = - \int dt \oint_{S^1} dx \,
[A_i \dot X^i - \L_i ( \6 X^i + \CP^{ij} A_{j} )] \, . \pl{PSMham} \ee
Our variables on phase space are $X^i(x)$ and $A_{j}(x)$, required to
be $2\pi$-periodic $C^\infty$-functions of $x$. They are canonically
conjugates; dropping the overall, and thus irrelevant, minus sign in
front of \re{PSMham}), their {\em field theoretic}\/ Poisson brackets
are given by \be \{ X^i(x), A_{j}(y) \} = \d^i_j \d(x-y) \, , \el
PBfield while $\{ X^i(x),X^j(y) \} = \{A_i(x),A_j(y) \}=0$. (The delta
function on the right--hand side of \re{PBfield}) is understood as
periodic delta function, certainly: $\sum_n \d(x-y+2 \pi n)$.) These
Poisson brackets are {\em sharply}\/ to be distinguished from the
Poisson brackets generated by the tensor $\CP$ on the {\em
  finite-dimensional}\/ target space spanned by the coordinates $X^i$ (cf
chapter \ref{SecPSM}): \be \CP^{ij} = \{X^i,X^j\} \, . \pl{PBtarget}\ee

The variables $\L_i$, on the other hand, are recognized as Lagrange
multipliers for the constraints \be G^i\equiv \6X^i + \CP^{ij} A_{j}
\approx 0 \, \, . \el cons As suggested already by the notation, they
are of first class. A straightforward calculation, making essential
use of the Jacobi property of \re{PBtarget}), yields \be
\{G^i(x),G^j(y)\}={\6 \CP^{ij}\over \6 X^k}(X(x)) \: G^k (x) \,
\d(x-y) \, . \el consalg

Due to the diffeomorphism invariance of the underlying theory, the
Hamiltonian is only a combination of the constraints: $H= -
\oint_0^{2\pi} dx \L_{i}G^i$. This comes as no surprise: After all,
$H$ generates translations in $t$ and they are part of the local
symmetries of the theory and thus should be generated by constraints.
`On-shell' then there is {\em no}\/ nontrivial Hamiltonian in the
theory, $H \approx 0$. In the context of the quantum theory
(apparently there is no nontrivial Schr\"odinger equation) this gives
rise to what is called the `problem of time' in the literature (cf
Introduction and section \ref{Secinner}).

Let us note that up to now we did not require $t \equiv x^0$ to be
timelike and $x\equiv x^1$ to be spacelike. The present formulation is
flexible enough to describe also spacetimes $M$ of, e.g., the Taub-NUT
type, where initially space-like circles $x^0 =const$ turn into light-
and timelike ones at later values of the evolution parameter $x^0$!

It is worthwhile to comment on the symmetries generated by the
constraints, i.e., in more abstract terms, to study the flow of the
Hamiltonian vector fields $\{ \cdot, G^i(x) \}$, or, smearing out the
distributional quantities by means of periodic test functions
$\ep_i(x)$, the flow of the vector fields $\CV_\ep := \{ \cdot, \oint
\ep_i(x) G^i(x) dx \}$. By means of \re{PBfield}) and \re{cons}) we
find: \ba \d_\ep X^i(x) \equiv \CV_\ep \, X^i(x)
&=& \ep_j(x) \CP^{ji}(X(x)) \pl{symaham} \\
\d_\ep A_i \equiv \CV_\ep \, A_i(x) &=& \6\ep_i(x) + \CP^{lm}{}_{,i}
A_l \ep_m \, .\pl{symbham} \, . \ea These are just the spatial
components of the PSM symmetries found in equations \re{syma}) and
\re{symb}). Allowing $\ep$ to be also $t$--dependent, we thus may
generate {\em all\/} the Lagrangian PSM symmetries (cf
\cite{AllsymmsStrobl} for more details on the correspondence).

How are these symmetries related to the standard gravitational
symmetries. Inspection of the Poisson tensor \re{P}), characterizing
the pure gravity models, shows that $G^3(x)$ generates the local
Lorentz boosts in the frame bundle. $G^\pm(x)$, on the other hand,
generate diffeomorphisms in null directions (up to Lorentz boosts).
There are several ways of seeing this. The probably simplest one is
the following \cite{p2}: By construction, the
Hamiltonian $H$ is the generator corresponding to $\6_t \equiv \6_0$
on spacetime. $H \sim G^+$, on the other hand, implies $\L_- \equiv
A_{0-} = 0$.  However, $A_{0-} = e_0{}^+ = 0$ yields zero spacetime
norm of $\6_0$, which proves the assertion.  (Eventual conformal
rescalings of the metric, cf section \ref{Secconf}, clearly do not
affect this argument).

The symmetries generated by the $G^i$ thus contain the diffeomorphism
symmetry and the local Lorentz symmetry present in the original
formulation (provided Einstein--Cartan variables are used). As will be
shown in a subsequent section, section \ref{Secmissingkink}, for invertible
metrics  also the converse holds: The diffeomorphisms
and local Lorentz boosts contain the transformations generated by the
constraints $G^i$; one thus establishes a bijection between the
symmetries for such metrics.  Since one usually restricts attention to
invertible metrics, the difference of the flow of two kind of
symmetries on degenerate metrics seems irrelevant. We will, however,
have to come back to this issue in sections \ref{Seccomparison} and
\ref{Secmissingkink}, when comparing the physical quantum states with
the moduli space of the classical theory found in chapter \ref{Chapglobal}.

One may also easily establish contact with lapse and shift. Recall,
e.g., that $\L_i \equiv A_{0i}$ and that the Hamiltonian vector field
of $H=\oint A_{0i} G^i dx$ is the generator $\6_0$ on spacetime. Thus,
$\oint A_{1i} G^i dx$ corresponds to $\6_1 \equiv \6_x$. So, the
diffeomorphism constraint ${\CH}_x$ is simply $A_{1i} G^i\equiv A_i \6
X^i$. As we will not need lapse and shift in what follows, we leave
the rest as an exercise to the reader.

\subsection{Modifications for open spacetimes \pl{Sechamopen}}
For $\S \simeq S^1$, periodic boundary conditions for
the variables in phase space are canonical, and there is no room for
boundary terms in, e.g., the Hamiltonian since $\6 S^1 = 0$. A
well--defined Hamiltonian formulation on $\S \simeq \dR$, on the other
hand, needs an additional external input. We are in the need to
specify boundary conditions or asymptotic fall-off conditions and,
simultaneously, have to take care for  well--defined Hamiltonian
vector fields \cite{ReggeTeitelboim} associated to functions on phase
space which are of (physical) interest to us (such as the
Hamiltonian $H$). The situation becomes technically involved easily
and, in lack of uniqueness, sometimes nontransparent. In particular,
we know from the classification obtained in section 6.1, that the
moduli space of classical solutions on $M \simeq \dR^2$ {\em modulo\/}
the gravitational (Lagrangian) symmetries is one-dimensional
only. However, the reduced phase space (RPS) in a properly defined
Hamiltonian framework necessarily is even-dimensional. Thus, during
the process of selecting boundary conditions, somewhere a second moduli
parameter will appear. We would like to see clearly, at which
point this occurs (and, in a second step, what alternatives arise). 

Again, also in the present context, the framework of $\s$-models will
provide a well-adapted tool for transparency within this procedure.
We focus on a Hamiltonian formulation on a bounded region $\S$
(which, topologically, still is $\dR$). An `infinitely extended' $\S$
may be visualized as a limit of the bounded case.

We will consider a region in spacetime bounded by two Killing lines.
Denote the values of $\phi$ at these boundaries by $\phi_R$ and
$\phi_L$, respectively. Fix the range of parameter $x$ to similarly
lie between $x_L$ and $x_R > x_L$ (so the parameter is increasing from
`left' to `right'). Furthermore, fix the Lorentz frame at the boundary
by requiring that $X^+$ equals 1 there. This is possible always,
except if the boundary point coincides with a bifurcation point of the
spacetime. (This exception does not bother us, if, at the end we are
interested in the limiting case $x_{R,L}\to \pm \infty$, with similar
conditions for $\phi_{L,R}$, depending, however, on the model
considered. If, on the other hand, we {\em want\/} one of the
boundaries to coincide with the bifurcation point (cf, e.g., 
\cite{Barvinsky}), we may require
$X^+=X^-=0$ on that side of $\S$, leaving $\phi$ unspecified there!)
Summarizing, we require: \be \phi(x_{L,R}) \stackrel{!}{=} \phi_{L,R}
= const \;\; , \quad X^+(x_{L,R}) \stackrel{!}{=} 1 \; . \pl{boundary1} \ee

To determine eventual boundary contributions to the Hamiltonian $H$,
we need to investigate the action functional \re{PSMham}) with the
integral over $S^1$ replaced by an integral over $\S$. We need to
check, if the Hamiltonian, as read off from this action, has a
well--defined Hamiltonian vector field (on the field theoretic phase
space) \cite{ReggeTeitelboim}. The potentially troublesome terms are
those with spatial derivatives acting on the canonical variables.
Variation of \re{PSMham}) (with $S^1$ replaced by $\S$) with respect
to $X^+$ and $X^3$ leads to boundary contributions that vanish on
behalf of \re{boundary1}). However, without further manipulations,
this does not apply also to $\d X^-$.

We now have two options: First, we may also require $X^-$ to be fixed
at the boundary. Then, $H$ is functionally differentiable and the
Hamiltonian formulation is well--defined. On the other hand we observe
that when fixing $X^-$ beside $X^+$ and $X^3$ at the boundary, the boundary
conditions also determine the symplectic leaf and the respective value
of the Casimir $C$.  As $C$ was the only local,
gauge--invariant parameter left in the analysis on the Lagrangian
level, it is plausible (and indeed the case) that such boundary
conditions will lead to a {\em zero\/}--dimensional RPS. This is not
very interesting, certainly.  However, the RPS then is seen to be
even--dimensional, as it should be, {\em and\/} this is also understood
from the Lagrangian point of view: Fixing $C$ leaves only a single 
gauge--inequivalence class of classical solutions.

Our second option is to introduce a boundary term to the Hamiltonian
action, the variation of which will cancel the undesired contribution.
This is of the form $(\L_- X^-)_R-(\L_- X^-)_L$, where the suffix
denotes evaluation at the respective boundary. It may appear that now
we have shifted the problem to the variation of $\L_-$ only (as the
addition of the above boundary term amounts to a partial integration
in the action, with the roles of $\L_-$ and $X^-$ interchanged).
However, $\L_-$ is {\em not\/} a variable on phase space.

Next, we need to properly define the constraints. Here one usually
requires the appropriate test functions $\e_i$ to become zero at the
boundaries. Therefore the  constraints are 
\be \int_\S \e_i(x) G^i(X(x)) \, dx \approx 0 \, \,  \quad
\mbox{with} \;\; \e_i(x_{L,R})=0 \, . \pl{consham} \ee (Note the
target space covariance of the expression.)

To make sure that the boundary conditions \re{boundary1}) are
conserved with respect to the flow of the Hamiltonian, we need to
require that some combinations of the $\L_i$ have to vanish at the
boundary. These restrictions are deteremined most easily by using CD
coordinates. Note that in the neighborhood of each of the boundary
lines we may introduce CD coordinates $\wt X^i = (\2 C, ln |X^+|,
\phi)$, cf equation \re{Xtilde}). In these coordinates, the boundary
conditions \re{boundary1}) read: $\wt X^2 (x_{L,R})=0$ and $\wt X^3
(x_{L,R})= \phi_{L,R}$. The constraint densities $G^i(x)$ take the
simple form: \be \wt G^1 = \6 {\wt X}^1 \; , \quad \wt G^2=\6 {\wt
  X}^2 + A_{\wt 3} \; , \quad \wt G^3 = \6 {\wt X}^3 - A_{\wt 2} \; .
\el neueconstraints Here and in what follows $A_{\wt i} \equiv \wt
A_i$, $\wt G^i \equiv G^{\wt i} = (\6 \wt X^i/\6 X^j) G^j$ etc. Thus,
it is obvious that our boundary condtions are preserved, iff $\L_{\wt
  2}$ and $\L_{\wt 3}$ vanish at the boundary.  (These requirements
are in no contradiction with $\det g \neq 0$, as one easily may
verify.)

Since the fields $X^i(x)$ are required to be smooth (in $x$), we may
follow from \re{consham}) that the value of the Hamiltonian $H$ is
nonzero now: \be H \approx (\L_{-} {X}^-)_R-(\L_{-} {X}^-)_L
\pl{Hamwert} \, . \ee Note that due to \re{boundary1}), the value of
$X^-$ at the boundary is a unique function of the Casimir $C$ at the
boundary. In particular in the torsion--free theories $2X^- = C +
const$ at the boundary (cf equation \re{Ctorless})). Then upon
appropriate conditions on $\L_{-}(x_{L,R})$ (or $\L_{\wt 1}(x_{L,R})$)
the (value of the) Hamiltonian is found to coincide with the Casimir
$C$. As shown in \cite{Lau}, up to an appropriate normalization
constant, this quantity may be identified with the ADM mass,
furthermore.

Now, with these definitions, the RPS will turn out to be {\em
  two\/}-dimensional (cf below for a simple proof). For the case of
Schwarzschild (SS), this coincides with the result found in
\cite{Kastrup,Kuchar}. 

Let us now shed some light on the appearance of the second parameter
by comparing explicitly the Hamiltonian with the Lagrangian
symmetries. By means of considerations completely analogous to the
case $\S \simeq S^1$, one may establish equivalence of the Lagrangian
and the Hamiltonian symmetries {\em inside}\/ of the region under
consideration. However, in \re{consham}) the test functions were
required to vanish at the boundary. Thus there the symmetries are
frozen. So, there are less (gauge) symmetries available at the
boundary in the Hamiltonian formulation as compared to the Lagrangian
one.

On the other hand, we, however, also imposed boundary conditions, not
present when determining the moduli space on the Lagrangian level.
From the Lagrangian point of view, the boundary conditions
\re{boundary1}) may be recognized as {\em gauge fixing conditions\/}
for local Lorentz transformations and diffeomorphisms nonparallel to
the Killing trajectories $\phi = const$. In other words, we lost
Lagrangian symmetries corresponding to $G^{\wt 3}$ and $G^{\wt 2}$,
but, simultaneously, we fixed boundary conditions which are cross
sections to the flow of these symmetries. So, by this step, clearly no
new gauge--invariant quantities are introduced. 
 
Freezing, on the other hand, the action of $G^{\wt 1}$ at the
boundaries is {\em not\/} accompanied by appropriate (gauge
fixing)/boundary conditions! Thus, the Lagrangian (gauge) symmetries
corresponding to the action of $G^{\wt 1}$ at its boundaries have been
dropped in formulating the constrained Hamiltonian system
above.\footnote{In the above considerations we found the use of $\wt
  G^i$ more advisable than the use of $G^i$ (although much the same
  may be said also by appropriate combinations of the $G^i$). This was
  admissible because the considerations were constrained to the
  neighborhood of the boundaries, where \re{boundary1}) was
  required. Below we will regard $G^{\wt 1}=\2 \6 C$ on all of $\S$;
  this poses no problem provided only that the respective leaf is
  noncritical (cf equation \re{crit})).}

Since there are two boundaries, it might appear that now we introduced
{\em two\/} new parameters. However, (providing first a rather
heuristic argument), $G^{\wt 1} \equiv \6 \wt X^1 \equiv \2 \6 C$ and
thus $\int G^{\wt 1} dx$ is {\em identically}\/ zero, if $\wt X^1$ on
the left boundary has the same value as on the right boundary. So,
there are not two independent constraints for each boundary, but, in
effect, just one. These considerations may be made more precise by
regarding the Hamiltonian flow generated by \be \int f(x) G^{\wt 1} dx
- \int \e_{\wt 1}(x) G^{\wt 1} dx \, , \el difference where $f(x)$ is
some unrestricted function and the term subtracted is a particular
constraint (thus $ \e_{\wt 1}$ vanishes at the boundaries). This flow
generates diffeomorphisms parallel to the Killing lines $\phi =
const$, which is part of the Lagrangian gauge symmetries, but, now, no
more an allowed gauge transformation on the Hamiltonian level. (There
is also a simple argument showing explicitly that the Casmir $C=2\wt
X^1$ generates transformations which are necessarily parallel to lines
of constant $\phi$: Since $C=C(X^i)$, it has vanishing (field
theoretic) Poisson brackets with $\phi = X^3$.)

Some remarks on a possible (in general only local) presentation of a second
parameter $p$ in the RPS will be provided below. 


\subsection{Abelianization and the reduced phase space}
\pl{Secabel} Before we turn to the Dirac quantization of this theory,
let us comment briefly on a simpler but, in the present general
context, less rigorous approach to quantization: it was used in
\cite{Kastrup,Kuchar} to treat the quantization of SS and
(independently) in \cite{p2} for the KV-model.  It makes use of the
fact that in these two--dimensional gravity theories it is possible to
define an {\em explicit\/} canonical transformation to new canonically
conjugate variables such that the constraints become part of the new
generalized coordinates. Such a transformation, found originally in
specific cases by more or less tiresome inspection and calculation,
pops out at once for the general case in the $\sigma$-model
formulation: Using, e.g., the CD parameterization \re{Xtilde}), it is
a {\em triviality\/} to see (cf \rz consalg and remember $P^{\wt 2 \wt
  3}=+1$ or cf \re{neueconstraints})) that \be \left(\int_\S \wt
  X^1(x) dx, G^{\wt i}(x); A_{\wt 1}(x), -\wt X^3(x) , \wt X^2(x)
\right) \el abel  where ${\wt i} \in \{ \wt 1,\wt 2, \wt 3 \}$, forms a set
of new canonical variables on the unconstrained phase space, or, more
precisely, on those parts of it where the field parameterization does
not become singular. To avoid misunderstandings about the notation
used in \re{abel}), let us recall that $G^{\wt 1} \equiv \6 \wt X^1$ so
that the first two entries of \re{abel}) may be replaced {\em
  equivalently\/} by $\wt X^1(x)$, which certainly is canonically
conjugate to $A_{\wt 1 }(x)$. We recommend the reader to check that
\re{abel}) is indeed a set of canonical variables.

{\em Ignoring\/} the global deficiencies of \re{abel}), any approach
to a quantization is equivalent now. One may apply an RPS quantization
(i.e.\ first reducing and then quantizing) or a Dirac quantization.
Obviously $\int_\S \wt X^1 dx$ and \be p := \int_\S A_{\wt 1} dx \,
\el Impuls (with $\S$ being either $\dR$ or $S^1$, depending on the
context) are the only two physical phase space variables ($\wt G^i(x)
\approx 0$). In this (somewhat simplified) picture, the physical wave
functions become functions of, e.g., the zero mode of $\wt X^1 =
\mbox{$\2$} C$, while $p$ becomes the corresponding derivative
operator.

\subsubsection{Remarks on the variable $p$}
Actually, in a CD coordinate system, $\int_\S A_{\wt i} dx$ commutes
with the constraints for all values $\wt i$. However, as obvious from
\re{neueconstraints}), only \re{Impuls}) is a nonconstant function on
phase space. 

More generally we may consider the change of $\int_\S A_i dx$
generated by the constraints (equations \re{consham}) in the case $\S
\sim \dR$). The Poisson brackets yield $\int_\S A_j(x)
\varepsilon_k(x) \CP^{jk}{}_{,i}(X(x)) dx$. In a CD coordinate system 
$\CP$ is constant and this expression vanishes. As observed above, only
one of the resulting three gauge invariant quantities yields an
independent observable. 

For $\S = \dR$, the quantity $p$ in equation \re{Impuls}) is not gauge
invariant on the Lagrangian level. $p$ changes with respect to the
flow of \re{difference}) with unrestricted $f$, which is part of the
local Lagrangian symmetries. For $\S = S^1$, on the other hand, $p$ is
easily seen to be gauge invariant also on the Lagrangian level. (It is
a kind of generalized parallel transporter in this case.)

The quantity $p$ is not defined globally in phase space in general
since CD coordinates are available locally only. (This may change 
in particular models, cf e.g.\ the following subsection.) This is a
relict of the fact that the RPS has a nonsmooth topology in general
(the nonsmooth regions are of lower dimension, however). 

Moreover, the variable $p$ is not unique. It depends on the choice of
the vector field $\6/\6 \wt X^1$. We illustrate this by the two
choices resulting from \re{Xtilde}), which also provides some insight
into the physical significance of $p$. (The applicability of this 
particular choice of CD coordinates has been discussed with care in
chapter 5; they are applicable {\em simultaneously}, if both $X^+ \neq 0$ and
$X^- \neq 0$, i.e.\ for regions of spacetime which are {\em sectors\/}
in the nomenclature chosen in chapters 5 and 6. Other choices of CD
coordinates may be applicable on larger regions of spacetime or on phase
space, cf e.g.\ the subsequent subsection for the example of string
inspired dilaton gravity as well as spherically reduced 4d gravity.)

The Hamiltonian field equations combined with the Hamiltonian
constraints are equivalent to the Lagrangian field equations. Thus, in
order to evaluate $p$ (as provided by the two CD charts and for a
spacetime region $\S \times \dR$, which does not exceed a `sector') on
classical solutions, it is most advisable to translate this quantity
into a Lagrangian one and to use the classical solutions obtained in
previous chapters. This is very simple: Obviously, in the
notation used within chapter 5, $p = \int_\S A_{\wt 1}$. Combining
equations \re{Loesung}) and \re{x1}), $p$ is found to coincide with
the difference of the parameters $x^1$ at the right and left boundary.
According to \re{011h}), $x^1$ is a Killing parameter of the lines
$\phi =const$. $x^1 = const$ is a null line, on the other hand. Thus,
in this version, $p$ measures the difference in Killing time 
of the boundary points of a given (initial time)
slice $\S$, transported from left to right along a {\em null\/}
line. According to the considerations around equation (5.48), however,  
these are two different null lines, ingoing  and outgoing,
respectively, depending on the choice of the CD system
\re{Xtilde}). (This may be compared with the analysis in
\cite{Kuchar}, where the analogous quantity to $p$ was obtained as the
difference in Killing time between left and right boundaries upon
orthogonal (instead of null) transport. In any case, we observe some
spacetime ambiguity in the presentation of a canonically conjugate
variable to $C$ for $\S = \dR$, all of which appear to be similar in
spirit, however.)

Let us remark, finally, that for $\S = S^1$, $\oint_\S A_{\wt 1}$ may be
brought into contact with the boost parameter found in section 
\ref{Secall} above. E.g., choosing $\S$ on a homogeneous spacetime
solution to coincide with a  closed Killing line, the above
consideration shows that the result for $p$ is  proportional to the
metric induced circumference. We intend to provide further
details on the precise relation in the various possible cases elsewhere.
(Cf also \cite{Wintergruendiplom} for similar considerations.)

\subsubsection{Global abelianization for Schwarzschild \pl{Secabelschwarz}}
In some cases, as, e.g., for SS, for string inspired dilaton gravity
(SIDG), or for the KV-model with negative cosmological constant $\L$,
this simple picture is correct. It is correct, iff the symplectic
leaves of the considered gravity model are of trivial topology. This
is the outcome of the exact operator quantization in section
\ref{Secdirac} below. The {\em argumentation\/} above is, however, on
shaky grounds as models with nontrivial symplectic leaves will show
drastically, in particular in some of the Euclidean theories.

In fact, for SIDG and SS, one may even provide a {\em globally}\/
well--defined set of CD coordinates. {\em Then\/} the abelianization above
becomes a stringent and correct argumentation. (Note that in the
previous literature \cite{p2,Kuchar,Kastrup} there always was some
problem at (at least one of) the horizon(s), such as is the case when
using CD coordinates \re{Xtilde}).)

For SIDG $\{ X^-,X^+\}=\L/2$. Thus, beside $C=X^+X^-+\L\phi/2$, one
may use just $X^+$ and a rescaled $X^-$ as a set of possible CD
coordinates. So, basically, in SIDG, $X^+$ and $X^-$ {\em are}\/
already Darboux coordinates. 

For SS, on the other hand, $\{ X^-,X^+\}=1/4\sqrt{\phi}$. The Casimir
\re{Ctorless}) becomes $C=2X^+X^-+\sqrt{\phi}/2$. To find global
Darboux coordinates, we first note that $\phi$ cannot be one of the
two, because its Hamiltonian vector field (on the target space)
vanishes at $X^+=X^-=0$. However, $X^+$ may be supplemented by an
appropriate canonically conjugate variable, namely by $4X^- \,
(X^+X^-+\sqrt{\phi})$!

We finally emphasize that only when the abelianization is performed on
a global level in the space of fields, it is a trustworthy starting
point for a quantization. Patching arguments alone, as those suggested
in \cite{p2}, or, smoothness arguments, as suggested in \cite{Kuchar}
(where the associated danger also was not left unnoticed) will not do.

\section{\sloppy Dirac quantization of gravity--Yang--Mills 
models \pl{Secdirac}}
\subsection{Physical states \pl{Secstatescyl}}
In this subsection we discuss the case $\S = S^1$. The considerations
for open boundaries are, however, very similar and basically lead to
the same results. The corresponding changes will be discussed briefly
in the subsequent subsection, section \ref{Secfurther}, where the
general quantization of the present subsection will be applied to
gravity theories with Euclidean signatures, too. 

According to Dirac \cite{Dir}, the first class constraints \re{cons})
are to be turned into an operator condition on wave functions. Since,
in general, the constraints are linear in $A_i$, but nonlinear in the
$X^i$, we go into an $X$-representation of the wave functions. Thus,
to start with, our `wave functions' $\Psi$ are functionals of
$X^i(x)$, i.e.\ $\Psi$ is a complex-valued functional of parameterized
smooth loops $\CX: x \ra X^i, \, S^1 \to \, N$. In the context of pure
gravity, always $N=\dR^3$. For gravity--YM models it is also a linear
space, but of some higher dimension. Much of our considerations below
will, however, be performed for completely general $N$, thus covering,
e.g., also the G/G WZW model \cite{Anton}. Only in our application of
the general results, we will then restrict ourselves to the purely
gravitational theories for simplicity.

According to Dirac, the {\em physical}\/ wave functions are those
functionals which satisfy \be \hat G^j(x) \Psi[\CX] \equiv \left( \6
  X^j(x) - i \CP^{jk}(X) {\d \over \d X^k(x)} \right) \Psi[\CX] =0 \,
, \el qcons where we have replaced $A_i(x)$ by the standard
(functional) derivative operator and $\hbar$ has been set to one.

Note that in this approach it is not necessary to equip the {\em
  original\/} space of wave functions with an inner product. We just
define it as a vector space and then determine the kernel of a set of
operators.  This is a mathematically well-posed problem. Only
thereafter we will deal with the question of an inner product (cf
section \ref{Secinner} below). Note that equipping the present space
of functionals with an inner product would not be of much help in this
context, since, generically, at least for some of the $\hat G^i$, zero
will be part of their continuous spectrum, and then the corresponding
eigenstate cannot be normalizable with respect to the original inner
product.

Returning to \re{qcons}), first one has to check if the integrability
conditions for \re{qcons}) are satisfied. (By this one usually means
{\em local\/} integrability conditions. We mention this here
explicitly, because later on we will also encounter global
integrability conditions.) We thus need to check, if the quantum
constraint algebra closes also on the quantum level, i.e.\ if the
commutators of the operators $\hat G^i$ close with all the operators
$\hat G^i$ to the {\em right\/} of the structure functions (cf
equation  \re{consalg})). Critical terms in this calculation can
arise only from terms containing {\em two\/} functional derivatives.
They, in principle, need regularization.  However, the derivative term
in the quantum constraint \re{qcons}) is, up to a proportionality
constant, recognized as nothing but the corresponding classical
Hamiltonian vector fields ($\CV_\ep$, when smeared out by test
functions). The calculation of the commutator of two Hamiltonian
vector fields, on the other hand, is purely classical and thus no
anomalies can arise from this source. (This calculation may, however,
be confirmed also by employing an explicit lattice regularization.
Some explicit calculations may be found also in
\cite{Wintergruendiplom}.)

Since the $G^i$ are the generators of the classical symmetries, having
checked the local integrability of the quantum constraints, we have
established that there are no anomalies.

To solve the constraints \re{qcons}), we make use of the fact that
they are covariant with respect to target space coordinate
transformations.  Let \be \O = \(\CP|_{T^*S}\)^{-1} \el Om denote the
symplectic form on each of the symplectic leaves $\CS \subset N$. Note
that the above restriction of $\CP$ is well--defined since, by
definition of the symplectic leaves $\CS$, all its Hamiltonian vector
fields are tangential to $\CS$. (Alternatively, we could also work with
a pseudo-inverse $\bar \O$ of $\CP$, cf section \ref{SecglobalPSM}.)
Denote, furthermore, by $(C^I)$ a set of independent Casimir functions
characterizing the symplectic leaves (locally). In the case of pure 2d
gravity, the label $I$ may be dropped, as there is just one Casimir.

Then evidently the set of constraints \re{cons}) implies: \be \6
C^I(x) \, \Psi[\CX] = 0 \, . \pl{Ccons} \ee This shows that the wave
functional $\Psi$ has its support only on loops $\CX$ with an image
that lies entirely within some symplectic leaf $\CS$. We thus may
restrict our attention to such wave functionals in the following. (We
do not work with distributional wave functionals here, although this
would be possible, too.)  Now the remaining constraints \re{qcons})
may be reformulated easily as \be \left[ { \d \over \d X^\a(x)} + i
  \O_{\a\b} \6 X^\b(x)\right] \Psi [\CX] = 0 \; ,\pl{Rest} \ee where
$X^\a$ are local coordinates on the respective leaf $\CS$ under
consideration.

Choosing the $X^\a$ as Darboux coordinates of $\O$, $\O_{\a\b}$
becomes independent of $X(x)$. An example for such a coordinate system
has been provided for the purely gravitational models in equation 
\re{CDLor}). However, as we intend to solve the constraints \rz qcons
on a global level and in general a single Casimir-Darboux (CD)
coordinate system is applicable to parts of $N$ only, we require
\re{Ccons}) and \re{Rest}) for {\em all\/} local CD-coordinate systems
in $N$. (The situation and the following steps remind 
somewhat of choosing Riemannian normal
coordinates in general relativity. Several formulas simplify
drastically in such a coordinate system. If the result of a
calculation is reformulated in a covariant manner at the end, we know
that it holds globally and with respect to any coordinate system.)

The two remaining equations \rz Rest may then be integrated easily
(locally in the space $\G_\CS$ of loops on $\CS$): \be \Psi = \Psi_0 \exp
\left({-i \over 2} \oint_0^{2\pi} X^\a \O_{\a\b} \6 X^\b dx^1 \right)
\pl{lokalLoesung} \, , \ee where $\Psi_0$ denotes a proportionality
constant, depending on the symplectic leaf under consideration (but
cf also below).

In a next step we need to restore the global information. Thus far we
found the local integral to the functional differential equations
\re{qcons}), i.e.\ we solved \re{Rest}) locally in the space of loops
$\G_\CS$. We now need to include global properties of $\G_\CS$.

The essential features of interest here are the set $\pi_0(\G_\CS)$ of
(disconnected) components of the space $\G_\CS$ of loops as well as its
first homotopy group $\pi_1(\G_\CS)$.

Clearly, if $\pi_0(\G_\CS)$ is nontrivial, then the proportionality
constant $\Psi_0$ in \re{lokalLoesung}) depends also on the chosen
component. Obviously, $\pi_0(\G_\CS)=\pi_1(\CS)$. So,
$\Psi_0=\Psi_0(C^I,a) \equiv \Psi_0(\CS,a)$, where $a$ is an element
of $\pi_1(\CS)$. If, e.g., $\pi_1(\CS) \sim \dZ$, $\pi_1(\CS)$ has a generator
$b$, such that $a=b^n$. Then $\Psi_0=\Psi_0(C^I,n)$, where $n$ is
a {\em winding number}\/ of the loop $\CX$ around the (single)
hole on the symplectic leaf.

Equation \re{lokalLoesung}) with $\Psi_0 = \Psi_0(\CS,a)$, $a \in \pi_1(\CS)$,
provides still only the {\em local\/} solution to \re{Rest}).  If
$\pi_1(\G_\CS)$ is nontrivial, {\em global obstructions\/} to the
solvability of the quantum constraints can arise. Let us remark on
this occasion that it is hardly conceivable to carefully check these
global issues in the space of fields, if the significance of the
leaves (and their topology) in the target space of the theory is not
realized and taken into account properly.

\subsubsection{The case of simply connected symplectic leaves}
Let us first focus on the case that $\pi_1(\CS)$ is trivial. In this
case we have the following identifications: $\pi_1(\G_\CS)= H_2(\CS)=
\pi_2(\CS)$.\footnote{Loosely speaking, nontrivial elements of $H_2(\CS)$,
  the second homology group of $\CS$, are (homotopy-equivalence classes
  of) arbitrary two-surfaces without boundary (within $\CS$) which are
  not the boundary of a three-surface. $\pi_2(\CS)$, the second homotopy
  group of $\CS$, similarly consists of homotopy classes of two-spheres
  in $\CS$. For trivial fundamental group $\pi_1(\CS)$, these two {\em
    groups\/} coincide.}  Then the global integrability condition may
be formulated in terms of a simple integrality condition on $\O$,
namely \cite{PSMold,PSM1} \be \frac{1}{2 \pi \hbar} \, \oint_\s \O \in
\dZ \qquad \forall \s \in \pi_2(\CS) \el integrality where we have
reintroduced Planck's constant $\hbar$ in this formula.

The origin of this condition becomes obvious, if we rewrite the phase
of \rz lokalLoesung in a (target space) coordinate independent way:
\be \2 \oint_0^{2\pi} X^\a \O_{\a\b} \6 X^\b dx^1 = \oint_{\CX} \Theta
\el Phase where $\Theta$ is a symplectic potential of $\O$ (in local
Darboux coordinates $\Theta = \mbox{$\2$} X^\a \Omega_{\a\b} dX^\b$), i.e. \be
\O = d \Theta \qquad \mbox{{\em locally} on $\CS$} \, . \el Poten

Now suppose that we start the integration of the (functional)
differential equations \re{Rest}) in $\G_\CS$ at the loop $\CX_0$.
Then, by means of Stoke's theorem, we may rewrite our local solution
in the form \be \Psi = \Psi_0 (C^I) \, \exp(-i \int_F \O ) \, \, ,
\el symplsol where $F$ is a path in $\G_\CS$ along which we integrate, or in
other words, it is a (possibly at some points degenerate)
two-surface in $\CS$. Here we only redefined $\Psi_0$ by a phase
factor depending on the reference loop $\CX_0$.  Integration of
\rz Rest along two paths in $\G_\CS$ will yield a unique result for
$\Psi$ precisely if \re{integrality}) is met for all possible
closed paths in $\G_\CS$, resp.\ for two-surfaces in $\CS$ generated
in this way. (Cf also \cite{PSM1,PSMDubna,Schladming} for
further details and illustration.)

In general the condition \re{integrality}) will be met only for a
subclass of all possible symplectic leaves. We will illustrate
this at examples below.

It is interesting to note that \re{integrality}), which arises here as
sufficient and necessary condition for the (exact) solvability of the
quantum constraints of the field theoretic model \re{PSMham}), is
equivalent to the semiclassical quantization condition (cf., e.g.,
\cite{Woodhouse}) of a fictitious point particle system with phase
space $\CS$ and symplectic form $\O$, i.e.\ with the Poisson brackets
provided by (the restriction of) \re{PBtarget}).\footnote{As remarked
  already in chapter 4, PSMs were rediscovered in \cite{Cat} in the
  context of quantizing finite dimensional Poisson manifolds. The
  above (nonperturbative) result, establishing precisely a correlation
  between the topological quantum field theory and the quantization of
  the brackets \re{PBtarget}), probably constitutes an interesting
  complement to the (perturbative) consideration of that paper.}

\subsubsection[Multiply connected leaves]{Multiply connected
leaves\footnote{The content 
of this subsection provides an improvement/correction of the
global integrability condition provided in
\cite{PSM1,PSMDubna,Schladming}. I am grateful to T.\ Kl\"osch
and P.\ Schaller for discussions leading to the improved resp.\ 
correct form of the restrictions below.}} In the case of a
nontrivial $\pi_1(\CS)$, we already found the additional dependence
of $\Psi_0$ on elements $a$ of this $\pi_1(\CS)$: $\Psi_0 =
\Psi_0(S,a)$. We now need to find the analog of
\re{integrality}). This has to be done for each fixed $a$
separately only, since $a$ also labels the disconnected parts of
$\G_\CS$.

We will find below that in the context of the pure 2d gravity
models, $H_2(\CS)$ and $\pi_1(\CS)$ cannot be nontrivial
simultaneously. This allows to establish that in the case of a
nontrivial fundamental group (of the leaf), there never is a
further global integrability condition in those cases. For
completeness we still provide the general integrability conditions
also for  multiply connected leaves; we will, however, not
motivate or prove them further.

For $a = 1$, the identity element of $\pi_1(\CS)$, the loop under
consideration is contractable. Then global integrability is
provided for those leaves which satisfy \re{integrality}).

For $a \neq 1$ one has to scrutinize $\pi_1(\CS)$ for elements $b
\in  \pi_1(\CS)$ satisfying $ab=ba$. If there are no such elements,
then the respective physical wave function exists, {\em
irrespective}\/ of $\CS$ fulfilling \re{integrality}) or not!

If, on the other hand, there are elements $b$ commuting with $a$,
then they generate a torus $T(a,b)$. $\CS$ now has to satisfy \be
\frac{1}{2 \pi \hbar} \,  \oint_{T(a,b)} \O \in \dZ \qquad \forall
b \in \pi_1(\CS) \quad \mbox{with} \;\; ab=ba \, ,  \el integrality2
i.e.\ the analog of \re{integrality}) with spheres being replaced
by tori.

These are the necessary and sufficient conditions for the
existence of a physical quantum state associated to an element of
the fundamental  group of a symplectic leaf.

Note that it is not necessary to require, e.g., that an
integrality condition is satisfied for all elements of $H_2(\CS)$.
Not all elements of $H_2(\CS)$ can be generated by loops of loops.
On the other hand, $H_2(\CS)$ may be trivial and still $\pi_1(\G_\CS)$
is nontrivial; in  fact if  $H_2(\CS)=0$, then
$\pi_1(\G_\CS)=\pi_1(\CS)$. Still, in the latter case,
\re{integrality2}) is satisfied always, because then there cannot
be any element $b$ commuting with $a$ (otherwise $H_2$ necessarily
is nontrivial).

Thus it is sufficient (although not necessary) to establish that
$H_2(\CS)=0$ ($\Ra \pi_2(\CS)=0$). Then any element $a \in \pi_1(\CS)$
gives rise to precisely one (independent) physical quantum state.

\subsubsection{Lorentzian signature of the theory \pl{Seclorstates}}
In the present subsection we specialize the results above to the case
of pure gravity models with Lorentzian signature. After all, our
spacetime metric is of Lorentz signature and the classical
considerations of the previous chapters dealt with this case only,
too.\footnote{Let us remark on this occasion that on the local level
  the PSM approach yields classical solutions for Euclidean signature
  in an equivalently straightforward way as in the Lorentzian case.
  In fact, for the local form of the metric, the result is the same as
  when one naively Wick rotates \re{gSS}) and, simultaneously,
  restricts the function $h$, (given again by \re{011h}) in the
  torsion--free case) to positive values. The global analysis may
  yield quite different results, however, since the framebundle
  generically is nontrivial in the Euclidean case.}  
In the subsequent two subsections, the resulting quantum states will 
then be compared with the classical moduli space. 

We will show below (at the end of this subsection) that for all the
pure gravity models of chapters 2 and 3 with Lorentzian signature
(i.e., in other words, for the Poisson tensor provided by equation
\re{P})) the symplectic leaves have trivial second homology.
$\pi_1(\CS)$, on the other hand, will be found to be nontrivial in
general. Thus, for the purely gravitational models, the general
solution $\Psi[\CX]$ to the quantum constraints \rz qcons may be
described as follows: It vanishes for all loops $\CX$ that do not lie
entirely within a symplectic leaf $S \subset N$ of $\CP$; otherwise we
may always represent it in the form \be \Psi[{\CX}] = \Psi_0(\CS,a)
\exp \left(-i \oint_{\CX} \Theta \right) \; , \qquad a \in \pi_1(\CS)
\, , \ee where, as above, $\Theta$ is a symplectic potential:
$\O=d\Theta$.  Note that such a symplectic potential exists {\em
  globally\/} on $\CS$ since, by de Rahm's theorem, $H_2(\CS)=0$
implies $H^2(\CS)=0$!

To establish contact with the literature on the subject, we first
choose a Casimir coordinate to label (generic) symplectic leaves.
Moreover, we display the phase factor by using coordinates
\re{Xtilde}). Then: \ba &\Psi[{\CX}] = \Psi_0(C,a) \exp \left(- i
  \oint_{\CX} \Theta \right) \; , \quad C \in \dR \, , \; a
\in \pi_1(\CS_C) \, ,& \pl{quantLoes} \\
&\Theta = \pm \ln |X^\pm| \, dX^3 \quad \Ra \quad \oint_{\CX} \Theta =
\mp \oint_0^{2\pi} \ln |X^\pm(x)| \, \6 X^3(x) dx \; ,& \pl{locphase} \ea
where $X^3(x) \equiv \phi(x)$.  Note that $\Theta$ needs to be defined
only on a surface $C=const$.  There the two expressions given for
$\Theta$ are easily seen to coincide on the overlap of coordinates.

In addition to these solutions we get one state for every critical
point of $\CP$, cf equation \re{crit}).

The above representation of the physical wave functionals coincides
with the one found in \cite{p2,IssueoftimeStrobl} for particular
cases.  A similar choice of CD-coordinates reproduces the phase factor
of \cite{Jacki}. Note that even if one recognizes the break-down of the
specific target space coordinates used to display the  phase factor, 
and if one introduces others so as to cover $N$ (as, e.g., 
$X^-, 2X^+X^-, X^3$ in our example), one hardly
will be able to {\em reintroduce\/} the {\em field theoretic\/} label $a$
without recognizing its origin as a winding number of $X^i(x)$, $(x=0
\ra x=2\pi)$, around symplectic leaves in the {\em target} space of the
theory.

Let us on this occasion comment also on the relationship of our
approach to the one in \cite{Gegenbergetal}, where the torsion--free
case \re{gdil}) respectively \re{Proto}) has been studied. They obtain
a wave function that might appear as a coordinate representation of
\re{quantLoes}) at first sight (interpreting ours as the respective
momentum representation). However, their analog of our Casimir
function $C$ is not a function of those phase space variables only
which they have chosen the initial (unconstrained) wave functionals to
depend on!  They obtain $C$ as an integration constant on the {\em
  classical\/} level; so what they actually employ is something like a
mixture between a reduced phase space quantization and a Dirac
quantization.  The discrete label $a$, on the other hand, was lost
when they reformulated one of the constraints by dividing it through
the derivative of the Dilaton field $\phi$ (which is very much like
that what happens in \re{abel})).

Nevertheless, although not rigorous, the paper \cite{Gegenbergetal}
showed that a unifying treatment of several gravity models may be
feasible. It was this paper, which lead me to start investigating
generalized dilaton theory, and, in the sequel, to consider the
generalization to PSMs (cf end of \cite{PRD}). 

\vskip5mm

Let us now give the prove that $H_2(\CS)$ is always trivial in the
Lorentzian gravity theories. The symplectic leaves may be
generated by the vector fields $V^i:= \CP^{ij} \6_j$, with
$i=+,-,3$. $V^3$ is nothing but the generator  of Lorentz boosts
in the target space. As a consequence of the noncompactness of the
Lorentz group also the symplectic leaves are noncompact. The
assertion follows now by noting that the second homology of a
two-dimensional noncompact manifold is necessarily trivial.  

The situation changes for Euclidean signature. There $V^3$ is the
generator of rotations in the coordinates $X^1$ and $X^2$. In a
generic theory, {\em some\/} of the symplectic leaves are then
diffeomorphic to a two--sphere for some values of the Casimir. In this
case, \re{integrality}) reduces to the condition that the `surface
volume' $\oint \O$ has to be a multiple of $2\pi$. This selects only a
{\em discrete\/} subset from the space of symplectic leaves (within the
corresponding range of $C$) on which the wave functions have support.
We will illustrate this further in section \ref{Secfurther} below.

Finally, we want to get hold of the first homotopy of the symplectic
leaves $\CS$: The leaves may be characterized by
$C(2X^+X^-,X^3)=const=:C_0$ in $N=\dR^3$.  Let $m$ be the number of
intersections of the curve $C(u,v)=C_0$ with the $v$-axis in a
$u,v$-diagram. Then, $m-1$ is the number of {\em free\/} generators of
$\pi_1(\CS)$.

\subsubsection{Comparison of classical and quantum results 
  \pl{Seccomparison}} We finally obtain a  
close relationship between the quantum
states \re{locphase}) and the classical solutions with cylindrical
topology as obtained in section \ref{Secall}: One of the two
continuous parameters characterizing such a solution was the Casimir
function $C$  (or $\sim$ ADM mass, \cite{Lau}). The second continuous
parameter (the metric induced circumference or the `boost' parameter,
respectively) becomes the (locally) conjugate variable in the RPS (cf
also equation \re{Impuls}) with $\S = S^1$ and section \ref{Secabel}
above); thus it will not appear as argument of the wave functions.
The `winding number' $n$ in $a=b^n$, on the other hand, corresponds
precisely to the minimal number of Carter--Penrose building blocks
covered by an initial (noncontractable) circle $x^0 = const$ on $M$.
(Remember that the number of horizons within a fundamental building
block was determined by the number of zeros of the function $h$ in
\re{h}).  The latter equals $m$, since the zeros of $h$ coincide with
those of $X^2$, cf \re{hbar}).)

This close relation between the physical quantum states and the
classical, gauge--inequivalent solutions is comforting. It gives
support to the applicability of the present quantization
scheme. Moreover, it is felt that also from the technical point of
view, something for the canonical quantization of higher dimensional 
gauge or gravity theories may be learnt.

\subsubsection{The missing kink number \pl{Secmissingkink}}
From the comparison of the physical wave functions with the
classical solutions on cylindrical $M$ we infer that the wave
functions captured the `block number' of section \ref{Secall}.
However, we do not find the kink number $k$ within them. This is
somewhat astonishing as we know that such an index appears, if we
study the space of smooth solutions to the field equations
\re{eoma}), (\ref{eomb}) on $S^1 \times \dR$,  dividing out
diffeomorphisms and Lorentz transformations in the frame bundle.
From the previous chapter we know that elements within 
{\em this\/} moduli space carry also an additional discrete label for
kinks. So, what went `wrong' within the Hamiltonian
formulation so that this index is missing within the wave functions?

The main point is that the RPS of the Hamiltonian formulation (still
we are speaking of cylindrical spacetimes) is equivalent to the space
of solutions to the field equations {\em modulo}\/ the PSM symmetries
\re{syma}) and \re{symb}). These symmetries are equivalent to the
above gravitational ones on-shell, {\em provided}\/ the metric
(zweibein) is nondegenerate. This may be seen easily by means of 
\re{diff}). 

In the classical context we focused only on spacetime solutions
with a nondegenerate metric. Thus the above observation may appear to
provide a contradiction. 

The resolution of the apparent paradox comes about by noting that the
PSM symmetries allow to connect solutions with different kink numbers
(but otherwise equal labels and moduli parameters). The situation may
be compared with the symmetries generated by means of $(d/dq)$ and $q
\, (d/dq)$ on a line $\dR \supset q$.  Even if one is interested in
gauge representatives with $q \neq 0$ only, the orbit spaces differ.
It may be shown that it is precisely this mechanism that is
responsible for the discrepancy (with $q$ being the analog of $\det
g$). We refer to \cite{SchallerStroblPhysletts} where, for the example
of the JT-model \re{JT}), this is shown in detail. (A similar
observation was made just recently in the context of the Chern--Simons
formulation of 2+1 gravity, cf \cite{Matschull}.)

Two final remarks: It would be interesting to see, if a careful
treatment within the lapse--and--shift approach will capture the kink
number. (On the other hand, in view of the somewhat strange character
of spacetimes with nontrivial kink-number, it may be seen also as a
blessing that the PSM or Einstein--Cartan formalism does not care about
this index.) Second, the Ashtekar formulation of 4d General
Relativity is equivalent to the metrical formulation also only for 
 $\det g \neq 0$ and the above considerations may be of interest in
 this context, too.

\subsection{Further remarks \pl{Secfurther}}
\subsubsection{Comparing the Lorentzian and Euclidean theories}
Next let us specify the general results of the previous subsection for
the example of {\em Euclidean\/} $R^2$--gravity, characterized by the
Poisson structure \ry PR2 .  As we found in section \ref{Secpoisson},
for any value of the Casimir $C_{R^2}$ there is one symplectic leaf
which is diffeomorphic to $\dR^2 \sim T^\ast \dR$ (for some values
there are additional ones; these will be dealt with below).  These
leaves are certainly always quantizable phase spaces, \rz integrality
is satisfied trivially. So the proportionality constant in \rz
symplsol becomes an arbitrary function of $C_{R^2}$ here.  This leads
to physical wave functions on a line effectively.  However, this is
not all of the story. We know that for $C_{R^2} \in [-1/6,1/6]$ there
are further symplectic leaves and we have to find out which of them
are quantizable.

For the point-like leaf at $C_{R^2}=- 1/6$ the integrality condition
\rz integrality is satisfied trivially. So, we obtain one additional
quantum state of $R^2$-gravity for this value of the Casimir. (It is
arguable, however, whether this state contributes to the physical
Hilbert space: Possibly it should not be taken into account because
the corresponding gauge orbit is of lower dimension than the generic
ones; cf also \cite{Anton}). Next we have to clarify which of the
spherical leaves are quantizable ($C_{R^2} \in (-1/6,1/6)$).  This is
not too difficult: Integrating the symplectic form $\O=dZ \wedge
d\Phi$ over the respective `ellipsoid', obviously one obtains an integer
multiple of $2\pi$, iff the `height' $Z_{max}-Z_{min}$ of the
this surface is an integer. An explicit calculation shows that this is
the case only for the ellipsoid--like surface at $C_{R^2}=1/9\sqrt{6}$ ($\ra
Z_{max}=1/2 -1/\sqrt{6}$, $Z_{min}=-1/2 -1/\sqrt{6}$). This definitely
provides an additional quantum state of $R^2$-gravity and the
spectrum of the Casimir becomes twofold degenerate at that specific
value.  At $C_{R^2}=1/6$ the situation is not that clear: There are
three leaves, one of which is not simply connected. The latter leave
would lead to an infinite number of states at that value of
$C_{R^2}$. However, the space of gauge orbits is not Hausdorff
precisely there. Some continuity requirement thus might lead
effectively to an identification of all these additional states.
Evidently, at this point there is some ambiguity in how to define a
quantum theory (and this question is crucially connected to the
construction of an inner product).

Studying $R^2$--gravity for Lorentzian signature metrics, one again
obtains a continuous spectrum for $C_{R^2}$. Now, however, this
spectrum is infinitely degenerate within all of the range $(-
1/6,1/6)$. The reason is that those leaves that had a nontrivial
second homology in the Euclidean formulation of the theory, turn into
(two--dimensional) noncompact ($\Ra$ condition \rz integrality
satisfied always) and multiply connected leaves.

The above results are quite generic: For Lorentzian type theories the
spectrum of the Casimir $C$ (respectively of the mass) is continuous
always. Generically there are also `winding numbers' (discrete labels)
present in the wave functions; however, the presence of these labels
generically changes noncontinuously with varying values of $C$.
Regarding the same theory from the Euclidean point of view, the
generic spectrum of $C$ is a continuous one with additional discrete
values (typically within a range, where in the Lorentzian theory the
leaves are no more simply connected). In some extreme cases, such as
in the JT--model, equation (\ref{JT}), for one choice of the sign of
the constant $\L$, the symplectic leaves are spheres {\em only}. Then the
spectrum of the mass operator becomes purely discrete.

The symplectic leaves of Euclidean theories are always simply
connected, moreover. 

For further examples and illustration we refer to
\cite{PhD,PSMold,Schladming}. It is in particular quite illustrative
to trace the global deficiencies of a local representation of the
phase factor of the physical wave functions on a (nonintegral)
spherical symplectic leaf explicitly \cite{Schladming,Amati}. Ignoring
global issues in such cases leads to a completely incorrect spectrum
of physical observables.

\subsubsection{Modifications for open spacetimes} 
In this case the wave functions in \re{qcons}) are functionals of open
(instead of closed) strings, with boundaries fixed to the one--branes
$X^+=1$, $\phi = \phi_L$ and $X^+=1$, $\phi=\phi_R$, respectively.
Upon restriction to a symplectic leaf as in equation \re{Ccons}), the
boundaries of the (parameterized) string are completely
fixed.\footnote{If one considers $\S = \dR$ as an infinitely extended
  open interval, the endpoints of the string are possibly tied to
  `infinity' also on the symplecitic leaves in the target space, with
  an asymptotically constrained parameterization of the string. Still,
  the ensuing considerations are unmodified in their essence.} It is
then easy to see that in the case of simply connected symplectic
leaves the remaining quantum constraints yield the {\em same\/}
integrability condition \re{integrality}) in this case, too. Moreover,
trivial $H^2$ of multiply connected leaves again ensures global
solvabibilty of the quantum constraints.

Thus, the spectrum of $C$ is the same as in the case of a closed
universe $\S=S^1$: For the Lorentzian theories, the spectrum of $C$
will remain continuous always, while for the Euclidean regime
the resulting spectrum again depends crucially on the topology of the
symplectic leaves; one can obtain either
scenario, a continuous spectrum of $C$ (like for SS or SIDG), a
discrete one (like for JT), or, more generically, a superposition of
both as in the previous example of $R^2$--gravity. 

Note that this result is in contradiction to \cite{Barvinsky}, where a
discrete spectrum for {\em any\/} generalized 2d dilaton theory with
Euclidean signature (and appropriately chosen $\S \sim \dR$) was
obtained. According to the present analysis, the situation is more
subtle.

\subsubsection{The inner product and the issue of time \pl{Secinner}}
One of the open issues within the present quantum theory 
is the construction of an inner product between
the physical wave functionals in the Lorentzian case, where nontrivial
labels $a \subset \pi_1(\CS)$ occur. The main reason is that
generically these labels arise only within some {\em part\/} of the
range of the continuous variable $C$ on which the physical states
depend on (cf.\ equation \re{quantLoes}) above). 

We are not aware of a similar problem in constrained quantization of
inherently mechanical models. These labels are a relic of the field
theoretical content of the theory. We propose that this issue should
be clarified for the JT theory first. The reason is that in this case
the PSM formulation is equivalent to the one of a $BF$ gauge theory,
with, however, noncompact gauge group in the Lorentzian case (the
universal covering group of $SL(2,\dR)$, cf
\cite{SchallerStroblPhysletts} 
and references therein).

Recently there has been a proposal for a `refined' Dirac quantization,
which is supposed to also provide an inner product between physical
states, cf \cite{Marolf} and references therein. The status of this
formalism has not yet been developed far enough to be directly
applicable to the present case.  It seems be an interesting task for
future investigations, however, to adapt `refined Dirac quantization'
so as to cover also the above 2d gravity models.

For SS or SIDG, however, (or any other model in which {\em all\/} the
symplectic leaves are simply connected), the observable $p$ is defined
globally on the (field theoretical) phase space. Then there are also
{\em no\/} discrete labels present in the physical wave functions. A
general inner product between physical states is then provided by
integration over $C$ with respect to some measure. Requiring the
observable $p$ to become a hermitian operator with respect to this
inner product \cite{Ashbuch,PhD},  $p=p^\dagger$,  fixes the inner
product. Up to unitary equivalence the measure for the integration 
for physical states is then just the Lebesgue measure $dC$. 

If, on the other hand, a theory has  a purely discrete spectrum of $C$
as, e.g., for $BF$ theories with compact gauge group, the inner
product also poses no problem (as remarked above, the Euclidean JT
model falls into this class; note, however, that generically a
Euclidean gravity theory does not have a purely continuous
spectrum). States corresponding to different values of $C$ are then
just orthonormal (since they are eigenstates of an operator, namely
$C$, which  we require to be hermitian). General $BF$ theories in the momentum
representation have been discussed first in \cite{Amati}. 

\vspace{2mm}

We briefly add a remark on the issue of time: For a closed universe
$\S = S^1$ the Schr\"odinger equation is apparently trivial.
Nontrivial dynamics may, however, still be extracted upon appropriate
choices of gauge conditions (determining with respect to what
parameter the states should be evolved). The main idea is that all the
dynamics is {\em contained\/} already within the equations \re{qcons})
(remember also, that $G^i$ generate all the local symmetries, which
include diffeomorphisms). This may be made quite explicit and, in
particular cases, unitarity of the time evolution may be shown.  A
detailed analysis would go too far here, however, and we refer the
reader to \cite{PhD,p2,IssueoftimeStrobl} for several examples. (Cf
also \cite{Isham,Kuchartime} for discussion of the issue in time in
General Relativity.)






\section{Brief outlook on Hawking radiation \pl{Sechawking}} 
There is a vast literature on this subject of Hawking radiation for
two-dimensional black holes (for some recent reviews cf 
\cite{Stromingerrev,Wipf}). Except for some scattered remarks used as
part for motivating the study of two-dimensional gravity models,
however, we did not yet come to discuss Hawking radiation. And it will
also not be possible to do this properly within some pages. Still we
want to at least mention briefly two main {\em directions}, both of which
are dealing with the CGHS-model (or some of its modifications).

First, there is the semiclassical approach. The scalar matter fields
are integrated out, resulting in the nonlocal Polyakov term
\cite{Polyakov} to be added to the classical gravity action. It is
argued that, in a first (semiclassical) approximation, this term
includes both Hawking radiation {\em and}\/ back reaction. The ensuing
discussion is purely classical then ($\hbar$ entering only as
parameter in the action under investigation). Only for some particular
actions the resulting field equations may be analyzed exactly (cf,
e.g., \cite{Thorl}). It is indeed found that the back reaction leads
to a decrease in the mass of the black hole due to the Hawking
radiation. However, one of the main problems is that usually the
approximation breaks down before a conclusion about the endpoint of
the Hawking radiation may be obtained. Also, in some cases
conservation of total energy seems to be violated. Cf the above
two reviews and references therein for further details.

The second approach is an exact canonical quantization of the CGHS
model \cite{JackiwCGHS,KucharCGHS} along the very same lines as those
in the previous section, however, now with propagating modes: A
controversy between the two groups arose which still does not seem to
have been settled fully satisfactorily.  While in \cite{JackiwCGHS}
the appearance of quantum anomalies is asserted, in \cite{KucharCGHS}
it is claimed that the anomaly is in fact removable or absent.

Related, and quite interesting, is the work of Mikovic \cite{Mikovic}:
Mikovic  does not apply a Dirac quantization but a reduced phase space
quantization. This approach does not allow to decide about the
presence or absence of quantum anomalies. Still, within his setting,
he does obtain an operator quantization of the CGHS model which seems
to correspond to a {\em unitary\/} time evolution. On the other hand, the
semiclassical results about Hawking radiation may still be obtained
within this framework by means of an approximate treatment of the
metric operator \cite{Mikovic}!


The peak of the activity within these attempts has passed by now. We
do not have the impression that this is due to the fact that all the
essential conceptual problems have been answered {\em nor\/} that one
found out that they cannot be answered within the 2d setting. Only,
the simpler problems have been solved so far. Many fundamental and
conceptual problems still wait for a resolution, even in the
simplified context of two-dimensional gravity theories. To our mind
there are still interesting and promising problems to be settled
within the realm of two-dimensional gravity.



\newpage


\begin{center}
\section*{Acknowledgement}
\end{center}
\vspace*{1cm}

It is hardly possible to name all the people who helped and supported
me within the tense final period of completion of this thesis of
habilitation. Particular thanks in this respect are due to: Norbert
D\"uchting, Horst Fischer, Marc Flesch, Peter Schaller, Konstantinos
Stergios, Martin Bojowald, and, in some other respect, my parents, my
children, and, last but not least, my wife Ini.

I am also very grateful to all the people from the institute for
constantly encouraging me and for the nice atmosphere. Special thanks
go certainly to Prof.\ H.\ Kastrup, in particular also for the freedom
and time he gave me for work on my habilitation.

Finally, I gratefully acknowledge the profit I gained from
collaborations with Thomas Kl\"osch, Heiko Pelzer and Peter
Schaller. Furthermore I also profited from discussions or
collaborations with A.\ Alekseev, M.\ Ertl,  V.\ Frolov, G.\ Kunstatter, S.\
Solodukhin, W.\ Thirring and in particular also W.\ Kummer. 

\newpage \cleardoublepage

\end{document}